\documentclass[aps,rmp,reprint,amsmath,superscritaddress,amssymb,graphicx]{revtex4-1}

\usepackage{bm}
\usepackage[retainorgcmds]{IEEEtrantools}
\usepackage{graphicx}
\usepackage{mathrsfs}
\usepackage{amsmath}
\usepackage{amssymb}
\usepackage{color}
\usepackage{amsfonts}
\usepackage{times,txfonts}
\usepackage{nicefrac}
\usepackage[colorlinks=true,linkcolor=blue,urlcolor=blue,citecolor=blue,pdfusetitle]{hyperref}
\usepackage{hyperref}
\usepackage[normalem]{ulem}

\newtheorem{theorem}{Theorem}
\newtheorem{lemma}[theorem]{Lemma}

\newcommand{\ud}{\,\mathrm{d}}

\DeclareMathOperator{\tr}{tr}

\newcommand{\trans}{^{\text{T}}}

\newcommand{\avrg}[1]{\langle#1\rangle}
\newcommand{\Tr}{{\rm Tr}}

\newcommand{\sand}[3]{\langle{#1}\vert{#2}\vert{#3}\rangle}
\newcommand{\ket}[1]{\vert#1\rangle}
\newcommand{\bra}[1]{\langle#1\vert}

\newcommand{\bs}{\bm{\sigma}}
\definecolor{orange}{rgb}{0.98, 0.6, 0.01}

\newcommand{\kk}{\bm{\gamma}}

\begin{document}

\title{Irreversible entropy production, from quantum to classical}

\author{Gabriel T. Landi$^1$, Mauro Paternostro$^2$}
\email{entropyproductionreview@gmail.com}
\affiliation{$^1$Instituto de F\'isica da Universidade de S\~ao Paulo,  05314-970 S\~ao Paulo, Brazil\\
$^2$Centre for Theoretical Atomic, Molecular and Optical Physics, School of Mathematics and Physics, Queen's University Belfast, Belfast BT7 1NN, United Kingdom}

\date{\today{}}

\begin{abstract}
Entropy production is a key quantity in any finite-time thermodynamic process. It is intimately tied with the fundamental laws of thermodynamics, embodying a tool to extend thermodynamic considerations all the way to non-equilibrium processes. It is also often used in attempts to provide the quantitative characterization of logical and thermodynamic irreversibility, stemming from processes in physics, chemistry and biology. Notwithstanding its fundamental character, a unifying theory of entropy production valid for general processes, both classical and quantum, has not yet been formulated. Developments pivoting around the frameworks of stochastic thermodynamics, open quantum systems, and quantum information theory have led to substantial progress in such endeavour. This has culminated in the unlocking of a new generation of experiments able to address stochastic thermodynamic processes and the impact of entropy production on them. This paper aims to provide a compendium on the current framework for the description, assessment and manipulation of entropy production. We present both formal aspects of its formulation and the implications stemming from the potential quantum nature of a given process, including a detailed survey of recent experiments.

\end{abstract}

\maketitle{}

\tableofcontents

\newpage

\section{\label{sec:int}Introduction}

In every finite-time process, entropy may flow from one system to another. 
However, entropy does not satisfy a continuity equation so that it may also be irreversibly produced~\cite{Carnot1824,Clausius1854,Clausius1865}. 
Such \emph{entropy production}, $\Sigma$, 
is always non-negative and zero only in the limiting case where the process is reversible. 
It therefore serves as the key quantity behind the $2^{\rm nd}$ law of thermodynamics, which can be stated mathematically as
\begin{equation}\label{intro_2nd_law}
    \Sigma \geq 0.
\end{equation}
Albeit compact, this expression has far reaching consequences, as it places severe restrictions on the types of transformations allowed in a physical process. At a foundational level, the statement embodied by Eq.~\eqref{intro_2nd_law} manifests the lack of time-reversal in finite-time processes and stems from the existence of an {\it arrow of time}~\cite{Eddington1927,Schnakenberg1976,Pomeau1982,Luo1984,Mackey1989,Mackey1992,Qian2001,Jian2003,Maes2003,Gaspard2004,Costa2005,Seifert2005,Porporato2007,Blythe2008,Parrondo2009,Batalhao2015}. 
Hence, the characterization and assessment of irreversible entropy production is one of the most important tasks in non-equilibrium physics. 

The formulation of the entropy production problem, however, is not universal. 
It depends on the underlying physical system, as well as its governing dynamical laws.  
Despite this, during the last century  several widely applicable frameworks have been developed, from Onsager's reciprocity theory~\cite{Onsager1931,DeGroot1961}  to the celebrated fluctuation theorems~\cite{Esposito2009, Campisi2011,Vinjanampathy2016,Goold2016}. 
More recently, the demonstrated possibility to control elementary quantum systems has drawn attention to the potential for thermodynamic applications in the quantum domain.
This is the primary drive towards a formulation of a theory of entropy production capable of encompassing both classical and quantum features. 

The goal of this review is to provide an overview of the progress in this formulation. 
Our approach will be centered around a unified picture of the $2^{\rm nd}$ law, described in terms of global  system-environment quantum  unitary interactions [cf. Sec.~\ref{sec:Qgen}]. 
This allows us to establish a link with information theory, and construct entropy production solely in terms of information-theoretic quantities. 
The result is a generalized form of the $2^{\rm nd}$ law, valid beyond the standard paradigms of thermodynamics, but with classical results  recovered in the suitable limit.
This approach also has a clear operational interpretation, with irreversibility emerging from the restrictions on the allowed set of operations for a given thermodynamic process. 
{\color{black}For a broader perspective on the developments in Quantum Thermodynamics over the last two decades, see~\cite{Binder2018a}.}

The results of Sec.~\ref{sec:Qgen} are central to this review. Before arriving there, we briefly establish the notation and jargon in Sec.~\ref{ssec:irr_thermo}, and then discuss -- in Sec.~\ref{sec:Why} -- why the entropy production problem is relevant.
We then move to explore the consequences and ramifications of such unified formulation. 
Sec.~\ref{sec:information} focuses on information-theoretic corrections to Landauer principle and the role of classical and quantum correlations in heat flow. 
Sec.~\ref{sec:classical} embodies another essential part of the review. We use the concepts developed in Sec.~\ref{sec:Qgen} as building blocks to assess the entropy production in more general types of dynamics, constructed in terms of a collisional model. This allows us to address the classical limit  as a particular case of the quantum formulation. 

The link between information and thermodynamics has other far-reaching consequences, as it allows information to be cast as a \emph{resource}, on equal footing to traditional thermodynamic resources, such as heat and work. 
That is, information can be consumed, stored or interconverted into other resources. 
And it can be used to fuel thermodynamic tasks, such as Maxwell Demon Engines.
This is the topic of Sec.~\ref{sec:thermal_ops}. 
In the quantum domain this acquires additional significance due to the possibility of manipulating  quantum coherence, as well as quantum correlations such as discord and entanglement.
How these features are implemented within a quantum formulation of the entropy production problem, is a central theme of this review. 

Finally, Secs.~\ref{sec:apps} and \ref{sec:exp} discuss applications and experiments. 
There is an inevitable arbitrariness on the choice of papers to cover, and we have chosen to address those which we believe are (i) representative of the types of problems the community is currently interested in; and (ii) have the potential to open unexplored avenues of research. 
Concerning the experiments, we have also tried to focus on those contributions which specifically characterize the entropy production at the quantum level.

We finish this review in Sec.~\ref{sec:conc}, by taking a step back to look at the bigger picture. 
We compare the formulation put forth in  Sec.~\ref{sec:Qgen} with other approaches, both historical and modern. 
The main argument we make is that the 2nd law is always formulated by starting with a basic physical principle, such as those of Carnot, Clausius and Kelvin, or statements such as ``the entropy of the universe never decreases.'' 
One then asks what are the overarching  consequences of this principle, and which other principles can be derived from it. 
This provides a measure of how general it is. 
The information theoretic formulation of Sec.~\ref{sec:conc} falls under this category. 
However, its main advantage is that it starts by assuming full knowledge of all degrees of freedom involved, thus allowing for  precise mathematical statements. Irreversibility is then constructed  operationally, by specifying which sources of information can, or cannot, be known in a given process. 
This feature  greatly generalizes the breadth and scope of the $2^{\rm nd}$ law. It not only contains classical statements as particular cases, but can also go much further, removing the constraints in the standard thermal paradigms, such as the need for macroscopically large thermal baths.

\subsection{\label{ssec:irr_thermo}Irreversible thermodynamics}
%






In order to clarify the basic ideas, as well as fix the notation, we will start with a brief textbook review of entropy production in classical thermodynamics~\cite{Fermi1956,Callen1985}. 
We consider the simplest scenario of a system $S$ interacting with  multiple reservoirs $E_1,E_2,\ldots$, each with a temperature $T_i$. 
The flow of entropy from $S$ to $E_i$ during a given process is given by the famous Clausius expression~\cite{Clausius1854,Clausius1865}. 
\begin{equation}\label{intro_Phi}
    \Phi_i = \frac{Q_{E_i}}{T_i}
\end{equation}
where $Q_{E_i}$ is the heat that entered $E_i$ (positive when energy leaves the system).\footnote{We always define heat in this way, as the change in energy of the environment. The reason is that, as will become clear in Sec.~\ref{sec:Qgen}, this helps to avoid  ambiguities concerning the distinction between heat and work, something which is quite delicate in the quantum domain. 
}
According to the Clausius principle, the corresponding change in the system entropy $S_S$ will be bounded by 
\begin{equation}\label{intro_clausius}
\Delta S_S \geqslant -\sum\limits_i \frac{Q_{E_i}}{T_i},
\end{equation}
Motivated by this inequality, one then defines the entropy production as 
\begin{equation}\label{intro_sigma_clausius}
\Sigma = \Delta S_S + \sum\limits_i \frac{Q_{E_i}}{T_i} \geqslant 0.
\end{equation}
The entropy of a system may either increase or decrease during a process, so that $\Delta S_S$ does not have a well defined sign. 
This is due to the terms $Q_{E_i}/T_i$, since heat can flow both ways. 
The only quantity which has a well defined sign is the entropy production $\Sigma$. 

In the past, the terms \emph{``entropy''} and \emph{``entropy production''} were often used interchangeably, but nowadays they have evolved to have entirely different meanings. 
Entropy refers to a property of the system  whereas entropy production refers to  \emph{transformations} underwent by the system.
Thus, interestingly, entropy production is actually closer in meaning to the original use of the word entropy, as first coined by Clausius in the 1860s \cite{Clausius1854,Clausius1865}, in which \emph{``trop\'e"} refers to the word ``transformation" in Ancient Greek.

The first law of thermodynamics states that the total change in internal energy of the system will be given by 
\begin{equation}\label{first_law}
    \Delta H_S = W - \sum\limits_{i} Q_{E_i},
\end{equation} 
where $W$ is the work performed by an external agent, with $W>0$ meaning work was performed \emph{on} the system. 
Alternatively, one may also simply view $W$ as the mismatch between the local energy changes  $\Delta H_S$ and $Q_{E_i}$ in system and baths. 
Focusing on the case where there is a single reservoir present, if we substitute $Q_E = W-\Delta H_S$ in Eq.~(\ref{intro_sigma_clausius}) we may write the entropy production as 
\begin{equation}\label{sigma_free_energy}
    \Sigma = \beta (W - \Delta F_S), 
\end{equation}
where $\beta = 1/T$ ($k_B = 1$) and $\Delta F_S= \Delta H_S - T \Delta S_S$ is the change in free energy of the system. For multiple baths at different temperature, it is in general not possible to express $\Sigma$ in this way and one must use Eq.~\eqref{intro_sigma_clausius}. 

It is often useful to express the results in terms of the entropy production \emph{rate} $\dot{\Sigma} = d\Sigma/dt$. In this case the second law is usually written as 
\begin{equation}\label{intro_2nd_law_rates}
    \frac{dS_S}{dt} = \dot{\Sigma} - \dot{\Phi}, \qquad \dot{\Sigma} \geqslant 0,
\end{equation}
with $\dot{\Phi} = \sum_i \dot{Q}_{E_i}/T_i$ being the entropy flow rate. 
This formula is particularly suited for studying 
non-equilibrium steady-state (NESSs) which occur when a system is coupled to two or more reservoirs kept at different temperatures. 
The typical scenario to have in mind is a piece of metal coupled to a hot bath at one end a cold one at the other. 
In this case, after a long time has passed the system will eventually reach a steady-state where $dS_S/dt = 0$. 
This, however, does not mean the system is in equilibrium. 
It simply means $\dot{\Sigma} = \dot{\Phi}$; that is, entropy is continually being produced in the system, but all of it is being dumped  to the reservoirs. 
A NESS is therefore characterized by a finite and constant entropy production rate $\dot{\Sigma}$. 
Thermal equilibrium, on the other hand, occurs only when $\dot{\Sigma} = \dot{\Phi} = 0$.

Irrespective of the definitions of entropy production and entropy production rate, the second law of thermodynamics can ultimately be summarized by the statement that both $\Sigma \geqslant 0$ and $\dot{\Sigma} \geqslant 0$. 
Next we discuss some of the far reaching consequences of this seemingly simple statement.

\section{\label{sec:Why}Why entropy production matters}

The goal of this Section is to illustrate, by means of famous examples, why entropy production is relevant in characterizing non-equilibrium systems. 

\subsection{\label{ssec:why_heat_engines}Operation of heat engines}

Consider a system interacting continuously with two reservoirs at temperatures $T_h$ and $T_c < T_h$, plus an external agent on which the system can perform work on. 
The first and second laws, Eqs.~(\ref{first_law}) and (\ref{intro_2nd_law_rates}), then become
\begin{IEEEeqnarray}{rCl}
\label{1st_law_2baths}
\frac{d H_S}{\ud t} &=&  \dot{W}-\dot{Q}_h - \dot{Q}_c , \\[0.2cm]
\label{Pi_multi_bath_2baths}
\dot{\Sigma} &=& \frac{d S_S}{\ud t} + \frac{\dot{Q}_h}{T_h} + \frac{\dot{Q}_c}{T_c}.
\end{IEEEeqnarray}
Writing the results in terms of rates makes the analysis simpler. 
One may picture this, for instance, as a continuously operated machine; or it may also be a stroke-based machine, but where the strokes happen very fast that we may write all thermodynamic quantities as rates (like a car engine). 
Following~\cite{Marcella1992}, we now show how the usual statements of the 2nd law can all be viewed as a consequence of Eqs.~(\ref{1st_law_2baths}) and  ~(\ref{Pi_multi_bath_2baths}).

If the machine is operated for a sufficiently long time, it will eventually reach a steady-state (limit cycle) where  $d H_S/d t = d S_S/d t = 0$. This therefore means that all quantities in 
Eqs.~(\ref{1st_law_2baths}) and  ~(\ref{Pi_multi_bath_2baths}) balance out: 
\begin{IEEEeqnarray}{rCl}
\label{1st_law_2baths_2}
\dot{W}&=&\dot{Q}_h + \dot{Q}_c  , \\[0.2cm]
\label{Pi_multi_bath_2baths_2}
\dot{\Sigma} &=&  \frac{\dot{Q}_h}{T_h} + \frac{\dot{Q}_c}{T_c}.
\end{IEEEeqnarray}
The steady-state is therefore characterized by a steady conversion of heat into work, accompanied by a steady production of entropy. 

In the standard operation of a heat engine, heat flows from the hot bath to the system ($\dot{Q}_h<0$) and work is extracted ($\dot{W}<0$). 
Using Eqs.~(\ref{1st_law_2baths_2}) and  ~(\ref{Pi_multi_bath_2baths_2}) one may  write the efficiency of the engine as 
\begin{equation}\label{engine_operation_efficiency}
    \eta =  \frac{\dot{W}}{\dot{Q}_h} = 1 + \frac{\dot{Q}_c}{\dot{Q}_h} = 1 - \frac{T_c}{T_h} + \frac{T_c}{\dot{Q}_h} \dot{\Sigma}.
\end{equation}
The first two terms on the right hand side are nothing but Carnot's efficiency $\eta_C = 1-T_c/T_h$.
Since $\dot{Q}_h<0$, the 2nd law~(\ref{intro_2nd_law}) implies that the last term in Eq.~(\ref{engine_operation_efficiency}) will be strictly non-positive. 
Hence, the efficiency of an engine is always \emph{reduced} from Carnot's efficiency by an amount proportional to the entropy production
 $  \eta = \eta_C - {T_c} \dot{\Sigma}/{|\dot{Q}_h|}$. 
This is Carnot statement of the $2^{\rm nd}$ law~\cite{Carnot1824}: \emph{``The efficiency of any quasi-static or reversible cycle between two heat reservoirs depends only on the temperatures of the reservoirs themselves, and is the same, regardless of the working substance. An engine operated in this way is the most efficient possible heat engine using those two temperatures.''}

It is also useful to cast Eq.~(\ref{engine_operation_efficiency}) in terms of the output power, $P = - \dot{W}$, which leads to
\begin{equation}\label{engine_Pi_output_power}
    \dot{\Sigma} = \frac{P}{T_c} \frac{(\eta_C-\eta)}{\eta}.
\end{equation}
We therefore see that, \emph{for fixed power output}, the closer we are to Carnot efficiency, the smaller is the entropy production rate. 
This nicely illustrates  why entropy production is often used as a quantifier of the degree of irreversibility.

Next suppose we only have access to a single bath, so $\dot{Q}_c = 0$.
Eq.~(\ref{1st_law_2baths_2}) then reduces to $\dot{W} = \dot{Q}_h$, so that Eq.~(\ref{Pi_multi_bath_2baths_2}) becomes
\begin{equation}
    \dot{\Sigma} = \frac{\dot{Q}_h}{T_h} = \frac{\dot{W}}{T_h} \geqslant 0.
\end{equation}
Positive work means work is injected into the system instead of extracted. 
Whence, work cannot be extracted from a single bath. 
This is precisely the  Kelvin-Planck statement of the 2nd law~\cite{Thomson1851,PlanckTreatise}: \emph{``It is impossible to devise a cyclically operating device, the sole effect of which is to absorb energy in the form of heat from a single thermal reservoir and to deliver an equivalent amount of work.''}

Lastly, suppose there is no work involved, $\dot{W} = 0$, but only heat flow between the two reservoirs. Eq.~(\ref{1st_law_2baths_2}) then yields $\dot{Q}_h = - \dot{Q}_c$ which, plugging in Eq.~(\ref{Pi_multi_bath_2baths_2}), leads to 
\begin{equation}\label{engine_heat_flow}
    \dot{\Sigma} = \left(\frac{1}{T_c} - \frac{1}{T_h}\right) \dot{Q}_c \geqslant 0.
\end{equation}
If $T_c < T_h$, we must then necessarily have $\dot{Q}_c \geqslant 0$; i.e., \emph{heat flows from hot to cold}. 
This is Clausius' statement of the 2nd law~\cite{Clausius1854,Clausius1865}: \emph{``Heat can never pass from a colder to a warmer body without some other change, connected therewith, occurring at the same time.''}

\subsection{Heat and particle flow}

Continuing with the assumption that $\dot{W}=0$, let us now assume that the environments also allow for particle flow. 
The first law~(\ref{1st_law_2baths}) is  modified to 
\begin{equation}
     \frac{dH_S}{dt} = -\dot{Q}_h - \dot{Q}_c + \mu_h \dot{N}_h + \mu_c \dot{N}_c,
\end{equation}
where $\mu_i$ are the chemical potentials of each bath and $\dot{N}_i$ the corresponding particle fluxes from bath to system (i.e., $N_i>0$ when particles enter the system).  
The last two terms represent chemical work. 

Particle conservation implies that, in the steady-state, $\dot{N}_c = - \dot{N}_h$. 
But, crucially, this does not mean that $\dot{Q}_h = - \dot{Q}_c$. 
Indeed, their mismatch  is precisely, 
\[
\dot{Q}_h = -\dot{Q}_c + (\mu_h - \mu_c)\dot{N}_h,
\]
which is non-zero whenever there is a chemical potential difference. 
Using this to eliminate $\dot{Q}_h$ allows us to write Eq.~(\ref{Pi_multi_bath_2baths}) as 
\begin{equation}\label{Pi_particle_flow}
    \dot{\Sigma} = \left(\frac{1}{T_c} - \frac{1}{T_h}\right) \dot{Q}_c + \frac{\mu_h - \mu_c}{T_c} \dot{N}_h.
\end{equation}
If we assume $T_c = T_h$, then the second law implies that if $\mu_h > \mu_c$, one must have  $\dot{N}_h >0$; that is, particle flows from high chemical potential to low chemical potential. 

We see in Eq.~(\ref{Pi_particle_flow}) the appearance of both a gradient of temperature and a gradient of chemical potential. 
These are called thermodynamic affinities, or generalized forces, as they are the ones responsible for driving the system out of equilibrium. 
Each current has a corresponding conjugated affinity;  heat $\dot{Q}_c$ is conjugated to the affinity $(1/T_c)-(1/T_h)$ while particle current $\dot{N}_h$ is conjugated to $(\mu_h-\mu_c)/T_c$.
The entropy production in Eq.~(\ref{Pi_particle_flow}) is thus simply the product between currents and affinities. 

For concreteness, suppose $(T_c,\mu_c) = (T,\mu)$ and $(T_h, \mu_h) = (T+\delta T, \mu+\delta \mu)$, where $\delta T$ and $\delta \mu$ are small. 
Eq.~(\ref{Pi_particle_flow}) then becomes
\begin{equation}\label{Pi_Onsager_linear_response}
    \dot{\Sigma} = \frac{\delta T}{T^2} \dot{Q}_c + \frac{\delta \mu}{T} \dot{N}_h.
\end{equation}
Intuitively, we expect that the currents should be zero when the affinities are zero. 
Moreover, if the affinities are small, the currents should also be proportionally small. 
Hence, in macroscopic systems it is natural to expect a linear dependence of the form~\cite{DeGroot1961}
\begin{equation}
\label{onsager_linear_response}
\begin{pmatrix}
\dot{Q}_c\\
\dot{N}_h 
\end{pmatrix}
= \frac{1}{T}
\begin{pmatrix}
L_{qq}&L_{qn}\\
L_{nq}&L_{nn}
\end{pmatrix}
\begin{pmatrix}
\frac{\delta T}{T}\\
{\delta \mu}
\end{pmatrix},
\end{equation}
where $L_{ij}$ are called the Onsager transport coefficients \cite{Onsager1931a,Onsager1931}. 
This kind of relation is not a consequence of the 2nd law~(\ref{Pi_Onsager_linear_response}); 
it is an additional assumption which  relies on the underlying dynamics of the system. 

The coefficient $L_{qq}$ represents Fourier's law of heat conduction. 
Similarly, $L_{nn}$ represents either Fick's law of diffusion in the case of particle transport (e.g. chemical solutions) or Ohm's law in the case of electric transport. 
The cross coefficients $L_{qn}$ and $L_{nq}$ are the Peltier and Seebeck coefficients, which are the basis for thermoelectrics. They describe the flow of heat due to a chemical potential gradient and the flow of particles due to a temperature gradient.
Onsager showed that due to the underlying time-reversal invariance of the dynamics, the cross coefficients actually coincide, $L_{qn} = L_{nq}$. As a consequence, the matrix $L$ is symmetric.  

Inserting Eq.~(\ref{onsager_linear_response}) in Eq.~(\ref{Pi_Onsager_linear_response}) we  find that in the linear response regime the entropy production will be a quadratic form in the vector of affinities $\bm{x} = (\delta T/T^2, \delta \mu/T)$:
\begin{equation}\label{Onsager_quadratic_form}
\dot{\Sigma} = \bm{x}\trans L \bm{x} \geqslant 0. 
\end{equation}
Since this must be true for all $\bm{x}$, it then follows that $L$ must be positive semi-definite.
Thus, even though the 2nd law does not predict the linear response relations~(\ref{onsager_linear_response}), it places strict restrictions on the values that the transport coefficients may take. 

\subsection{\label{sec:matters_Landauer}Landauer's erasure}

Consider again the Clausius inequality~(\ref{intro_clausius}), but focusing on the case of a single bath at a temperature $T$:
\begin{equation}\label{landauer}
    Q_E \geqslant - T \Delta S_S. 
\end{equation}
It is important to realize how this bound relates quantities from two different systems:
It bounds the heat absorbed by the bath to a quantity related to the entropy change of the system. 
While initially constructed within the realm of macroscopic thermodynamics, it turns out that this same inequality also holds true when the system is microscopic, with the entropy now being the system's information-theoretic entropy (either Shannon's or von Neumann's; to be properly defined below).\footnote{
Landauer's principle is often stated in terms of the heat cost to erase one bit of information, which
is $Q_E \geqslant T \ln 2$. 
This is actually a particular case of Eq.~\eqref{landauer} for dichotomic (binary) variables.}
In this context, Eq.~(\ref{landauer}) places restriction on the \emph{heat cost of erasing information}, which is called  Landauer's principle \cite{Landauer1961}.

We say information is erased when $\Delta S_S < 0$~\cite{Shannon1949}.
This is a bit counter-intuitive at first because large entropy means little information, so that  $\Delta S_S < 0$ means the information after interacting with a bath is  larger than what we initially had (it looks like information is acquired, not erased). 
But what is acquired is information about the \emph{final} state of the system, not the initial one. 
Before interacting with the bath the system had some information stored in it, which the experimenter simply did not know about (hence the large entropy). 
The act of interacting with a bath irreversibly erases this information~\cite{Plenio2001}.

Landauer's erasure therefore fits very naturally within the entropy production framework since erasing information is an inherently irreversible operation.
In fact, it is suggestive to interpret Landauer's principle as a direct consequence of the 2nd law~\eqref{intro_2nd_law}, written as $\Sigma = \beta Q_E + \Delta S_S \geqslant 0$.
This connection is subtle, however: 
In the 2nd law, $S_S$ is the thermodynamic entropy (see Sec.~\ref{sec:conc} for a more precise definition), whereas in~\eqref{landauer} it is the information-theoretic entropy. 
Notwithstanding, it turns out that, indeed,  Landauer's principle {\color{black}can be rederived using the more modern formulation of the 2nd law, which will be the subject of this review.} 
This connection was firmly established in~\cite{Esposito2010a,Reeb2014}, and is one of the hallmarks of the modern formulation of quantum thermodynamics. 
It will be reviewed in detail in Secs.~\ref{sec:Qgen} and~\ref{sec:landauer}.

\subsection{Thermodynamic Uncertainty Relations}

In the examples above, all thermodynamic quantities  were treated as simple numbers, that could not fluctuate. In macroscopic systems this is usually a good approximation due to the large number of particles involved. 
But in meso- and microscopic systems, fluctuations play an important role. 
It has recently been discovered that some properties of the fluctuations are also largely bounded by the \emph{average} entropy production.
Consider the transport of heat from a hot to a cold system and let $\dot{Q}$ denote the \emph{average} heat rate. 
In addition, let us define $\Delta_Q^2$ as the time-averaged variance of the heat current. 
In Refs.~\cite{Barato2015,Pietzonka2015} it was shown that for certain classical Markovian systems, the signal-to-noise ratio $\Delta_Q^2/Q^2$ satisfies a Thermodynamic Uncertainty Relation (TUR)
\begin{equation}\label{TUR_barato}
\frac{\Delta_Q^2}{\dot{Q}^2} \geqslant \frac{2}{\dot{\Sigma}},
\end{equation}
where $\dot{\Sigma}$ is the average entropy production rate. TUR shows that fluctuations are bounded by the average entropy production. 
And albeit simple, this bound is actually quite counter-intuitive: Since $\dot{\Sigma}$ appears in the denominator, in order to curb fluctuations (reduce the left-hand side) one must actually increase the entropy production. 
More irreversible processes therefore fluctuate less.

TUR can also be adapted to autonomous engines \cite{Pietzonka2017}. In this case one studies instead the average output power $P = -\dot{W}$, as well as its corresponding variance $\Delta_P^2$. 
A TUR of the same shape as~(\ref{TUR_barato}) also holds for $P$. That is, $\Delta_P^2/P^2 \geqslant 2/\dot{\Sigma}$.
However, in this case one can go further and relate $P$ and $\dot{\Sigma}$ using Eq.~(\ref{engine_Pi_output_power}).
Writing also 
$P = -\eta \dot{Q}_h$ (which simply follows from the definition of efficiency as $\eta =  \dot{W}/\dot{Q}_h$), one then finds 
\begin{equation}\label{TUR_engine}
\Delta_P^2 \geqslant 2 T_C P \frac{\eta}{\eta_C - \eta},
\end{equation}
Hence, we see that for fixed average power $P$,  as one approaches Carnot's efficiency, the fluctuations in the power must diverge.\footnote{
Strictly speaking, the divergence never actually occurs since $P$ implicitly depends on $\eta$ and, in particular, is zero for a Carnot engine (since a Carnot engine must operate quasi-statically and hence will have zero output power).
Notwithstanding, there will in general be ranges of the engine's parameter space where one can vary $\eta$ for fixed $P$. 
}
This therefore reflects a fundamental trade-off between operation efficiency and fluctuations. In real devices, particularly at the nanoscale, fluctuations could have a deleterious effect in the engine's operation. 
Eq.~(\ref{TUR_engine}) therefore provides guidelines on how to curb them.
{\color{black}For a recent overview on the latest developments in TURs, c.f.~\cite{Horowitz2019}.}

{
\subsection{Fluctuation theorems}

TURs illustrate the benefits of 
looking at fluctuations of thermodynamic quantities. 
Such benefits are even more evident owing to celebrated fluctuation theorems (FTs)~\cite{Gallavotti1995,Evans1993,Crooks1998,Jarzynski1997,Esposito2009, Campisi2011}, which have been a central topic of research over the last two decades.
FTs address the probability distribution of  thermodynamic quantities such as  work~\cite{Jarzynski1997,Crooks1998} or heat~\cite{Jarzynski2004a} and can be framed in a unifying language in terms of entropy production, which is thus placed at the centre of the investigations on thermodynamics of microscopic systems.

The basic idea is to study the probability distribution $P_F(\sigma)$ of the entropy production in a certain process, such as work extraction, heat exchange and so on (the subscript F stands for ``forward''). 
This is to be compared with the corresponding time-reversed (``backward'') distribution $P_B(\sigma)$.
FTs reflect a symmetry of these two distributions, constraining the forward and backward distributions,  which usually have the form
\begin{equation}\label{matters_FT}
    \frac{P_F(\sigma)}{P_B(-\sigma)} = e^\sigma.
\end{equation}
This is known as a detailed FT. 
And it immediately implies that
\begin{equation}\label{matters_integral_FT}
    \langle e^{-\sigma} \rangle = \sum\limits_\sigma P_F(\sigma) e{-\sigma} = 1, 
\end{equation}
called an integral FT. 
In turn, Eq.~\eqref{matters_integral_FT}, combined with Jensen's inequality, implies that
\begin{equation}
    \langle \sigma \rangle \geqslant 0. 
\end{equation}
Thus, \emph{on average}, the entropy production is always non-negative. 
The idea, therefore, is that when the entropy production is described as a fluctuating quantity, the second law is only valid on average, and may eventually be violated at the stochastic level. 
In this sense, FTs contain the second law.

FTs have been addressed in detail in~\cite{Esposito2009,Campisi2011,Jarzynski2011a,Seifert2012}.
In Sec.~\ref{sec:qgen_FT} we focus on reviewing some more recent developments, particularly those concerned with quantum processes. 
We also discuss some subtleties raised in~\cite{Manzano2017a} on how to define the backward process.

An intuition into what Eq.~\eqref{matters_FT} entails is gathered by considering the scenario of~\cite{Jarzynski2004a}, which consists of two thermal systems, prepared at temperatures $T_A$ and $T_B$, which are then put in contact and allowed to exchange heat. 
As will be discussed in Sec.~\ref{sec:qgen_FT}, the entropy production in this case is given by
$\sigma = \beta_A Q_A + \beta_B Q_B$ (see also  Eq.~\eqref{Pi_multi_bath_2baths_2}).
If one assumes there is no work involved, $Q_A = - Q_B \equiv Q$ and we may write $\sigma = \Delta \beta Q_A$, where $\Delta \beta = \beta_A - \beta_B$.
Moreover, in this particular scenario, it turns out that the forward and backward processes are actually the same (this would not be case, for instance, if an external agent was explicitly performing work). 
Eq.~\eqref{matters_FT} therefore reduces to 
\begin{equation}\label{matters_JW}
    \frac{P(Q)}{P(-Q)} = e^{\Delta \beta Q}.
\end{equation}
The FT therefore directly compares the probability of exchanging heat $Q$ or $-Q$. 
Suppose $T_B > T_A$ so that $\Delta \beta = \beta_A - \beta_B >0$. 
In this case we expect heat should flow from $B$ to $A$, so we expect $Q= Q_A\geqslant 0$.
Due to fluctuations, however, it is possible to eventually observe $Q<0$. 
What Eq.~\eqref{matters_JW} says is that the probability of observing negative heats is \emph{exponentially smaller} than that of observing a flow in the ``right'' direction: $P(-Q) = e^{-\Delta \beta Q} P(Q)$. 
Note also that heat is an extensive quantity. 
Hence, for macroscopic systems, the exponent $e^{-\Delta \beta Q}$ tend to be incredibly small; 
only for meso- and nanoscopic systems, where fluctuations are significant, will $P(-Q)$ be non-negligible. 
}

{
\subsection{Stochastic thermodynamics}

Consider a system interacting with one or more reservoirs, and undergoing some generic thermodynamic process. 
At the microscopic level, the system is described by a stochastic trajectory, which would be different 
each time the dynamics of the system is considered. 
Hence, one may construct a probability distribution for each individual trajectory.
For classical systems, the sole knowledge of such trajectories is sufficient to formulate the entropy production resulting from the stochastic  dynamics~\cite{Seifert2005}. 
This is a significant feature in the description of classical microscopic processes.
The reason is that often one does not have a physical model for the global dynamics, but only an effective reduced description. 
Being able to express the entropy production solely through this effective description thus provides a major advantage. 

This approach, called stochastic thermodynamics has been reviewed in detail in a substantive body of literature, including ~\cite{Seifert2012,VANDENBROECK20156}
([cf. also Secs.~\ref{sec:stoch_thermo} and~\ref{ssec:classical_phase_space}]. 
In contrast, a major difficulty in the formulation of entropy production for quantum processes is that, in general, the reduced description does \emph{not} suffice to unambiguously determine the entropy production. 
In other words, the latter can only be defined by having knowledge of the global system-environment interaction, whose lack might lead to inconsistencies, including the apparent violation of the second law~\cite{Levy2014}. 
Note that a reduced description may very well provide a good approximation for the dynamics; but this does not imply it also approximates well the \emph{thermo}dynamics. 
A major theme of this review, particularly in Sec~\ref{sec:classical}, will be to address in detail under which conditions does a reduced description suffice, as far as the second law is concerned.
}

{
\subsection{Maxwell, Szilard and information thermodynamics}

In his famous treatise ``The theory of heat''~\cite{Maxwell1888}, Maxwell describes a thought experiment where a demon, capable of knowing the precise position and moment of all particles in a gas, uses that information to violate the second law. 
It does that by inserting a partition in a box and selectively opening a small hatch  when a hot particle comes through. After a sufficient time, all hot particles will be on one side and all cold ones on the other. 
Szilard used the same idea to make an engine cyclically extract work from a single reservoir (thus apparently violating Carnot's statement, Sec.~\ref{ssec:why_heat_engines})~\cite{Szilard1929}.
Recently, these ideas have seen a surge of interest, with several experiments providing 
physical implementations of Maxwell's demon~\cite{Toyabe2010,Camati2016,Elouard2017,Naghiloo2017,Masuyama2017,Peterson2016} and proof-of-principle demonstrations of Szilard's engine~\cite{Koski2014,Koski2015,Paneru2018}.

The problem can be phrased in terms of \emph{information gain and feedback control}.
That is, information is acquired about the system through measurements, which is in turn used to perform some action on it (the feedback). 
To ``exorcise'' the demon (i.e., reinstate the validity of the second law), this information has to be included in a description of the entropy production. 
This was first done by Bennett~\cite{Bennett1973}, who used Landauer's principle  (Sec.~\ref{sec:matters_Landauer}) 
to show that the heat cost associated with erasing information exactly counterbalances the work extracted by the demon. 

A stochastic description of these ideas, in terms of fluctuation theorems, was first put forth in a series of seminal papers by Sagawa and Ueda~\cite{Sagawa2009,Sagawa2009a,Sagawa2010}.
The basic idea is that the stochastic entropy production $\sigma$ must now be modified to $\sigma \to \sigma + I$, where $I$ is an information-theoretic term accounting for how much information was gained about the system, during the process. 
Eq.~\eqref{matters_integral_FT} is then changed to $\langle e^{-\sigma - I} \rangle = 1$, which in turn implies $\langle \sigma \rangle \geqslant - \langle I \rangle$. 
For $\langle I \rangle > 0$, the average entropy production may thus be negative.

When extending these ideas to the quantum domain,  the inevitable backaction caused by quantum measurements should be considered. 
Acquiring information about the system is no longer without consequences and may, in fact, severely degrade it. 
The recent developments in such interplay between information and thermodynamics will be reviewed in  Sec.~\ref{sec:information}.
}

%
%
\section{\label{sec:Qgen}Entropy production in quantum processes}
\label{sec:foundations}
%
%

\subsection{\label{sec:qgen_global}Global unitary dynamics for system + environment}

A unified formulation for entropy production in open quantum systems, which holds for arbitrary non-thermal environments and arbitrary dynamics, can be made by analyzing the global system-environment unitary evolution. 
We consider the interaction of a system $S$ with an environment $E$, prepared in arbitrary states $\rho_S$ and $\rho_E$,  by means of a global unitary $U$. 
The final state of the composite SE system after the interaction will be given by
\begin{equation}\label{Qgen_global_map}
    \rho_{SE}' =   U (\rho_S \otimes \rho_E) U^\dagger.
\end{equation}
This map is incredibly general.
All information about the types of interactions involved are encoded in $U$, which therefore may contemplate both weak and strong coupling, as well as time-dependent Hamiltonians and work protocols.
The map also makes no assumptions about the structure of $E$, which does not need to be macroscopic and may very well have dimensions comparable to those of $S$. 
One could therefore have $S$ and $E$ to be two qubits. 
Or to have $S$ be a hot pan and $E$ a large bucket of water. Both cases will be described by the same map~\eqref{Qgen_global_map} (admittedly, in the latter the unitary $U$ would be a bit more complicated). 

The reduced state of the system can  be obtained by tracing over the  environment, which leads to the quantum operation
\begin{equation}\label{Qgen_reduced_map}
    \rho_S' = \mathcal{E}(\rho_S) = \tr_E \rho_{SE}' =  \tr_E \bigg\{ U (\rho_S \otimes \rho_E) U^\dagger \bigg\}.
\end{equation}
On a conceptual level, tracing over the degrees of freedom of the environment can be pinpointed as the origin of irreversibility in this process. 
After all, the map~\eqref{Qgen_global_map} is unitary and hence reversible by construction. 
But tracing over (discarding) the environment  embodies the assumption that after the interaction one  no longer has access to its degrees of freedom or is able to perform on it any local operation. 
Irreversibility thus emerges from discarding any information contained locally in the state of $E$, as well as the non-local information shared between $S$ and $E$.

The entropy production separately quantifies these two contributions, being given by 
\begin{equation}\label{Qgen_Sigma}
    \Sigma = \mathcal{I}_{\rho_{SE}'}(S\! : \! E) + S(\rho_E'||\rho_E).
\end{equation}
To our knowledge, this formula was first put forth in Ref.~\cite{Esposito2010a}. 
Its justification and ramifications will be the central topic of this Section. 
This will culminate with a description in terms of fluctuation theorems, as first put forth in  Ref.~\cite{Manzano2017a} and which will be reviewed in Sec.~\ref{sec:qgen_FT}.

The first term in Eq.~(\ref{Qgen_Sigma}) is the mutual information (MI) developed between system and environment due to their interaction, where the mutual information of any bipartite system $AB$ is defined as 
\begin{equation}\label{Qgen_MI}
    \mathcal{I}_{\rho_{AB}}(A\! : \! B) = S(\rho_{AB} ||\rho_A\otimes \rho_B) = S(\rho_A) + S(\rho_B) - S(\rho_{AB}),
\end{equation}
with $S(\rho) = - \tr(\rho \ln \rho)$ being the von Neumann entropy.
$\mathcal{I}_{\rho_{SE}'}(S\! : \! E)$ thus quantifies the amount of shared information which is lost if one no longer has access to the state of $E$.
The second term in Eq.~(\ref{Qgen_Sigma}), on the other hand, is the quantum relative entropy, defined as 
\begin{equation}\label{relative_entropy}
    S(\rho|| \sigma) = \tr \bigg\{ \rho \ln \rho - \rho \ln \sigma\bigg\},
\end{equation}
which is a type of distance between two density matrices.\footnote{Strictly speaking it is not a distance since it does not satisfy the triangle inequality. Notwithstanding, it is such that $S(\rho||\sigma) \geqslant 0$ and $S(\rho||\sigma) = 0$ iff $\rho = \sigma$.} 
The term $S(\rho_E'||\rho_E)$ thus quantifies how  the environment was pushed away from equilibrium, a process which is irreversible since we are assuming one can no longer perform local operations on it.
In both formulas $\rho_E' = \tr_S \rho_{SE}'$ is the reduced density matrix of the environment after the map~(\ref{Qgen_global_map}). 
Combining the definitions in~\eqref{Qgen_MI} and~\eqref{relative_entropy}, it is also possible to rewrite~\eqref{Qgen_Sigma} as 
\begin{equation}
    \Sigma = S\Big( \rho_{SE}' \;|| \;\rho_S' \otimes \rho_E\Big). 
\end{equation}
Notice the asymmetry in this formula: the quantity on the right is a tensor product between the final state $\rho_S'$ of the system with the initial state $\rho_E$ of the bath. 
The interpretation for this will be discussed in Sec.~\ref{sec:qgen_FT}.

For a generic environment, the entropy production in Eq.~(\ref{Qgen_Sigma}) will no longer be given by the Clausius expression Eq.~(\ref{intro_sigma_clausius}). 
Notwithstanding, it is still reasonable to \emph{define} a similar splitting and write
\begin{equation}\label{Qgen_sigma_phi}
    \Sigma = \Delta S_S + \Phi, 
\end{equation}
where $\Phi$ is called the entropy flux, from the system to the environment. 
This equation can actually be viewed  as the definition of $\Phi$. Of course, as we will see, for thermal systems one recovers $\Phi = Q_E/T$. But in general the expression for $\Phi$ will be different. 

The reason why it makes sense to call $\Phi$ a flux can be seen as follows. 
Since  the system and environment are initially uncorrelated, one has that $S(\rho_{SE}') = S(\rho_S) + S(\rho_E)$. Thus, the mutual information may be expressed as 
\begin{equation}\label{Qgen_MI_DeltaS}
\mathcal{I}_{\rho_{SE}'}(S\! : \! E) = \Delta S_S + \Delta S_E,
\end{equation}
where $\Delta S_S = S(\rho_S') - S(\rho_S)$ and similarly for $\Delta S_E$. 
Eq.~(\ref{Qgen_Sigma}) can then be written  as 
\begin{equation}\label{Qgen_sigma_trace_expression}
\Sigma = \Delta S_S + \tr_E\bigg\{(\rho_E - \rho_E') \ln \rho_E\bigg\}.
\end{equation}
Comparing with Eq.~(\ref{Qgen_sigma_phi}), one finds that the entropy flux is 
\begin{equation}
\label{Qgen_flux}
\Phi = S(\rho_E') - S(\rho_E) + S(\rho_E' || \rho_E)
=\tr_E\left\{(\rho_E - \rho_E') \ln \rho_E \right\}.
\end{equation}
The entropy flux thus depends \emph{solely} on the local state of the environment.
The entropy production is thus split in two terms, $\Delta S_S$, which refers only to the system, and $\Phi$, which refers only to the bath.

Eq.~(\ref{Qgen_Sigma}) can be viewed as a general proposal for the entropy production in any system-environment interaction. 
It is clearly non-negative as both terms are individually non-negative. But, of course, that does not suffice for it to be considered as a physically consistent definition.
In order to do so this formula must acquire operational significance, which can be done by specializing it to specific contexts. 
This will be our focus in the following Sections.

\subsection{\label{sec:Qgen_thermalE}Thermal environments}

Let us assume that the environment is thermal, 
$\rho_E = \rho_E^{\text{th}} = e^{-\beta H_E}/Z_E$.
Again, we do not assume it is necessarily macroscopic. Only that initially it is in a thermal state. 
Inserting this in Eq.~(\ref{Qgen_sigma_trace_expression}), but only in the logarithm, leads to 
\begin{equation}\label{Qgen_Sigma_thermal}
\Sigma = \Delta S_S + \beta Q_E,
\end{equation}
where 
\begin{equation}\label{Qgen_QE}
     Q_E = \tr\bigg\{ H_E (\rho_E'-\rho_E^{\text{th}})\bigg\}
\end{equation}
is the total change in energy of the environment during the unitary $U$. 
Eq.~(\ref{Qgen_Sigma_thermal}) thus coincides with the standard form of the second law, Eq.~(\ref{intro_sigma_clausius}).
This is quite remarkable: $\rho^\text{th}_E$ is the only assumption required to convert the general, and fully information-theoretic expression [Eq.~(\ref{Qgen_Sigma})], into the traditional thermodynamic expression in Eq.~(\ref{Qgen_Sigma_thermal}). 

There is a subtlety, however.
Namely that the heat entering  Eq.~(\ref{Qgen_Sigma_thermal}) refers to the change in energy of the environment [Eq.~(\ref{Qgen_QE})]. 
It hides the fact that the process may also involve work, which is encoded in the unitary $U$.
The heat $Q_E$ will therefore in general not coincide with the change in system energy $\Delta H_S$. 
This allows us to \emph{define} work as their mismatch
\begin{equation}\label{Qgen_first_law}
W := \Delta H_S + Q_E.
\end{equation}
This formula is valid whether or not the Hamiltonian of the system changed during the process. For simplicity, we are assuming that it remains the same, but the results also hold if it does not. 
Substituting this for $Q_E$ in Eq.~(\ref{Qgen_Sigma_thermal})  then leads to the second law in the form of Eq.~(\ref{sigma_free_energy}); viz.,
\begin{equation}\label{Qgen_Sigma_DeltaF}
    \Sigma = \beta (W - \Delta F_S),
\end{equation}
where  $\Delta F = F(\rho_S') - F(\rho_S)$ is the change in  \emph{non-equilibrium free energy} 
\begin{equation}\label{Qgen_non_eq_F}
    F(\rho_S) = \tr( H_S \rho_S) - T S(\rho_S) = F_\text{eq} + T S(\rho_S || \rho_S^\text{th}),
\end{equation}
which is defined for any state $\rho_S$, with  $\rho_S^\text{th} = e^{-\beta H_S}/Z_S$ being  reference  a thermal state of the system at the same temperature $T$ as the bath (if the final Hamiltonian is $H_S'$ then $F(\rho_S')$ should be defined with respect to $H_S'$). 
We thus conclude that the general proposal~(\ref{Qgen_Sigma}) for the structure of the entropy production reduces \emph{exactly} to the expected thermal results whenever the bath is assumed to start in thermal equilibrium.
Even the form~\eqref{Qgen_non_eq_F} remains the same, provided one now works instead with the non-equilibrium free energy.

Eq.~(\ref{Qgen_sigma_trace_expression}) can also be specialized to the case where $E$ is  composed of multiple parts, $E_1, E_2, \ldots$, with $\rho_E = \rho_{E_1} \otimes \rho_{E_2} \otimes \ldots$ and 
each prepared in a thermal state 
$\rho_{E_i} = e^{-\beta_i H_{E_i}}/Z_{E_i}$ at different inverse temperatures $\beta_i$. 
In this case an identical calculation leads to
\begin{equation}\label{Qgen_Sigma_multiple_baths}
    \Sigma = \Delta S_S + \sum\limits_i \beta_i Q_{E_i}, 
\end{equation}
which is Eq.~(\ref{intro_sigma_clausius}).
Even though Eq.~\eqref{Qgen_Sigma_multiple_baths} involves only the local changes in energy of each bath, the map~\eqref{Qgen_global_map} will still generate correlations between the different $E_i$, since they all interact with a common system. 
In order to see how these correlations affect $\Sigma$, one may start from Eq.~(\ref{Qgen_Sigma}) and add and subtract a term $\sum_i S(\rho_{E_i})$. 
This then allows us to write it as 
\begin{equation}\label{Qgen_Sigma_multipartite_correlations}
    \Sigma = \mathcal{I}_{\rho_{SE}'}(S\! :\! E_1 \! :\! E_2 \! : \!\ldots) + \sum\limits_{i} S(\rho_{E_i}'|| \rho_{E_i}),
\end{equation}
where $\mathcal{I}_{\rho_{SE}'}(S\! :\! E_1 \! :\! E_2 \! : \!\ldots) = S(\rho_S') + \sum_{i} S(\rho_{E_i}) - S(\rho_{SE}')$ is the so-called \emph{total correlations}~\cite{Goold2015a} between system and the individual environmental components. 
This quantity captures not only the correlations between $S$ and $E$, but also correlations between $E_i$ and $E_j$. 
It therefore shows that entropy is also produced due to the accumulation of multipartite correlations between the different parts of the bath, as a consequence of their common interaction with the system.

\subsection{\label{sec:non_thermal_fixed_points}Maps with global fixed points}

Next let us specialize to a different scenario. 
We consider once again the map in Eq.~(\ref{Qgen_global_map}) and no longer assume that $\rho_E$ is thermal. 
Instead, we look into those cases where the map has a \emph{global fixed point}; that is, a special state $\rho_S^*$ satisfying
\begin{equation}\label{Qgen_fixed_point_rhoSstar}
    U(\rho_S^* \otimes \rho_E) U^\dagger = \rho_S^* \otimes \rho_E.
\end{equation}
Notice that this condition is much stronger than $\rho_S^* = \mathcal{E}(\rho_S^*)$, which would be a local fixed point (global implies local, but the converse is seldom true). 
An example of maps with global fixed points are the so-called thermal operations, which will be reviewed in Sec.~\ref{sec:Qgen_thermal_ops}.

We now focus on the entropy flux~\eqref{Qgen_flux}. 
Expanding the trace over $E$ to be over $S+E$ allows us to write it as 
$\Phi = \tr_{SE} \big\{(\rho_S \rho_E - \rho_{SE}') \ln\rho_E\big\}$ (we omit the tensor product symbol for simplicity).
Next we take the logarithm on both sides of  Eq.~\eqref{Qgen_fixed_point_rhoSstar}, which allows us to write 
\[
U^\dagger (\ln\rho_E) U - \ln \rho_E = - U^\dagger (\ln\rho_S^*) U + \ln \rho_S^*.
\]
Plugging this in the expression for $\Phi$ and then carrying out the trace over $E$, one then finally finds 
\begin{equation}
    \Phi = \tr_{S}\big\{ (\rho_S' - \rho_S) \ln \rho_S^*\big\}.    
\end{equation}
For systems with a global fixed point, the entropy flux can thus be written solely in terms of system-related quantities. 

Plugging this in Eq.~(\ref{Qgen_sigma_trace_expression}) then allows us to express the entropy production as
\begin{equation}\label{Qgen_Sigma_nonThermal_FixedPoint}
    \Sigma = S(\rho_S || \rho_S^*) - S(\rho_S' || \rho_S^*).
\end{equation}
Quite nicely, this is written solely in terms of local quantities of the system. 
This is only possible for systems with global fixed points; {\color{black}for local fixed points}, the entropy production will be an intrinsically non-local quantity.\footnote{
At first glance, Eq.~(\ref{Qgen_Sigma_DeltaF}) also seem to be written solely in terms of local quantities of the system.
But that is not true because the work $W$, as defined in Eq.~(\ref{Qgen_first_law}), still involves quantities pertaining to the environment.
}

The positivity of Eq.~(\ref{Qgen_Sigma_nonThermal_FixedPoint}) is guaranteed by its definition in Eq.~(\ref{Qgen_Sigma}). 
But within the optics of Eq.~(\ref{Qgen_Sigma_nonThermal_FixedPoint}), positivity can also be viewed as a consequence of the data processing inequality: 
\begin{equation}\label{data_processing}
    S(\mathcal{E}(\rho) || \mathcal{E}(\sigma)) \leqslant S(\rho|| \sigma),
\end{equation}
which holds for any quantum channel $\mathcal{E}$. 
But since $\rho_S^*$ is a fixed point of $\mathcal{E}$, it then follows that 
\begin{equation}\label{Qgen_data_proc_sigma}
    S(\rho_S' || \rho_S^*) = S(\mathcal{E}(\rho_S) || \mathcal{E}(\rho_S^*)) \leqslant S(\rho_S || \rho_S^*), 
\end{equation}
which therefore implies $\Sigma \geqslant 0$.
Entropy production can thus be viewed as quantifying the map's ability to process information and hence reduce the distinguishability between the initial state $\rho_S$ and the fixed point $\rho_S^*$.
This result neatly emphasizes the interpretation of the entropy production~(\ref{Qgen_Sigma}) as a purely informational quantity, defined without any reference to the energetics of the system, such as the separation between heat and work. 

\subsection{\label{sec:Qgen_thermal_ops} Strict energy conservation and thermal operations}

Thermal operations, first introduced in~\cite{Janzing2000} and later popularized in \cite{Horodecki2013,Brandao2013,Brandao2015}, are maps which involve a thermal environment \emph{and} have a global fixed point (thus combining the results of the two previous subsections). 
One way to ensure that the map has a global fixed point when interacting with a thermal bath is to impose that the unitary global $U$ in~(\ref{Qgen_global_map}) should satisfy the so-called \emph{strict energy conservation} condition
\begin{equation}\label{Qgen_strict_energy_conservation}
    [U, H_S + H_E] = 0,
\end{equation}
(note that, in general $U$ does not commute with $H_S$ and $H_E$ individually,  only with their sum). 
This implies that
\begin{equation}\label{Qgen_strict_energy_conservation_exponential}
    U e^{-\beta (H_S + H_E)} U^\dagger = e^{-\beta (H_S + H_E)},
\end{equation}
so that $\rho_S^\text{th}$ is a global fixed point of the dynamics, provided it is defined with the same $\beta$ as the environment. 
As a consequence, the entropy production reduces to Eq.~(\ref{Qgen_Sigma_nonThermal_FixedPoint}):
\begin{equation}\label{Qgen_sigma_thermal_op}
\Sigma = S(\rho_S|| \rho_S^\text{th}) - S(\rho_S' || \rho_S^\text{th}).
\end{equation}
Naively, one may think that any map involving a thermal environment would necessarily have the thermal state $\rho_S^\text{th} = e^{-\beta H_S}/Z_S$ as a fixed point. This, however, is in general not true.
But when strict energy conservation holds, it is. 
Thermal operations enjoy a wide range of nice properties and have been extensively studied in the literature, within the context of quantum resource theories. These will be reviewed in Sec.~\ref{sec:thermal_ops}.

It is important to clarify the meaning of Eq.~(\ref{Qgen_strict_energy_conservation}). 
Its key implication is that 
all energy that leaves the system enters the environment and vice-versa (nothing stays ``trapped'' in the interaction); viz., 
\begin{equation}\label{Qgen_strict_DeltaH}
    \Delta H_S = - \Delta H_E \equiv Q_E.
\end{equation}
This kind of condition is  seldom met in practice\footnote{Unitaries of the form~(\ref{Qgen_strict_energy_conservation}) can be generated by resonant-type interactions. 
For instance, if $S$ and $E$ are qubits with $H_{i} = \Omega_{i}\sigma_i^{z}/2$ (here $i = S,E$) and if the interaction is generated by a potential $V = g( \sigma_S^+ \sigma_E^- + \sigma_S^- \sigma_E^+$), then the unitary will be energy conserving only when $\Omega_E = \Omega_S$.
} and should thus be viewed as an idealized scenario where drawing thermodynamic conclusions is much easier. 
Despite this apparent artificiality, Eq.~\eqref{Qgen_strict_energy_conservation} is actually incredibly similar to the \emph{weak-coupling approximation} present in the vast majority of open quantum system studies
{\color{black}(a discussion on how violations of this condition affect thermodynamics of strongly coupled systems can be found in~\cite{Hilt2011}).
}
Weak coupling assumes the interaction energy is small. 
Eq.~\eqref{Qgen_strict_energy_conservation} assumes the interaction can be arbitrarily large, but nothing stays trapped in it. 
To a great extent, this is essentially the same thing. 
The big difference is that weak coupling is imposed as an approximation, whereas Eq.~(\ref{Qgen_strict_energy_conservation}) is postulated \emph{a priori}.

Comparing Eq.~(\ref{Qgen_strict_DeltaH}) with Eq.~(\ref{Qgen_first_law}) also shows that in a thermal operation there is no work involved,  $W = 0$.
Indeed, Eq.~(\ref{Qgen_sigma_thermal_op}) can also be rewritten in terms of the non-equilibrium free energy~\eqref{Qgen_non_eq_F}, as
\begin{equation}
    \Sigma = - \beta \Delta F.
\end{equation}
The expenditure of work does not have to be associated with a work protocol, but may simply be related to the cost of turning the system-environment interaction on and off. 
To elucidate this point, let us suppose that the unitary $U$ was generated by turning on an interaction  $V_{SE}$ for a certain length of time $\tau$.
Rigorously speaking, since we turn this interaction on and off, the total Hamiltonian must be time-dependent and will have the form $H_{SE}(t) = H_S + H_E + \lambda(t) V_{SE}$, where $\lambda(t)$ is the unit-box function between $t\in [0,\tau]$. 
Since the composite $S+E$ system evolves unitarily, any work that is performed can be unambiguously associated with the total change in energy of $S+E$:
\begin{equation}
    W = \int\limits_{-\infty}^\infty dt \; \left\langle \frac{\partial H_{SE}(t)}{\partial t}\right\rangle = \langle V_{SE}\rangle_0 - \langle V_{SE}\rangle_\tau. 
\end{equation}
We therefore see that, in general, there is a work cost associated with turning the interaction on and off. 
But when strict energy conservation holds, $\Delta H_S = -\Delta H_E$ and hence $W = 0$.  

This on-off work is usually negligible for macroscopic systems, so that classical studies never really worry about it. 
This is because the energies $H_S$ and $H_E$ are proportional to the number of atoms in the bulk, whereas the interaction $V_{SE}$ is usually proportional to the number of atoms on the surface, which is usually negligible compared to the bulk. 
In most of statistical mechanics, the system is therefore always assumed to be \emph{weakly} coupled to a bath. 
But in microscopic systems this may very easily break down since $V_{SE}$ may be of the same order as $H_S$ (even if it is still much smaller than $H_E$). 
As a consequence, the on-off work may be significant.
For instance, the SWAP engine, analyzed in \cite{Campisi2015},  operates with two qubits and is based precisely on the extraction of on-off work
(see Sec.~\ref{sec:swap_engine} for more details).

Properly accounting for all sources and sinks of energy is an important part of thermodynamics at the quantum level. It has also been the source of significant debate.
Additional methods for dealing with this will be reviewed in Sec.~\ref{sec:fluc_work}.

\subsection{\label{sec:qgen_FT}Fluctuation theorems}

The proposal of a general form of the entropy production in Eq.~(\ref{Qgen_Sigma}) gains solidity by analyzing it from multiple perspectives. 
In this sense, great insight can be gained by analyzing the corresponding fluctuation theorem at the quantum trajectory level.
This problem was solved in Ref.~\cite{Manzano2017a} where the authors also showed how the two terms in Eq.~(\ref{Qgen_Sigma}) are related to the definition of what is the backward stochastic process. 
Crucially, shattering previous beliefs, \emph{the backward process is not unique}. 
Different choices of backward process lead to different expressions for the entropy production, which quantifies the information that is assumed to be lost between forward and backward protocols~\cite{Manzano2017a}. 
This therefore attributes a clear operational significance to the entropy production.

We consider here the same map as in Eq.~(\ref{Qgen_global_map}). No assumptions are made about either the environment or the unitary. 
Let $\rho_S = \sum_n p_n |n\rangle\langle n|$ and $\rho_E = \sum_\nu q_\nu |\nu \rangle\langle \nu|$ denote the eigendecompositions of the initial states of $S$ and $E$.
Moreover, we introduce bases for the final reduced states $\rho_S' = \sum_m p_m' |\psi_m\rangle\langle \psi_m|$ and $\rho_E' = \sum_\mu q_\mu' |\phi_\mu\rangle\langle \phi_\mu|$, which will in general differ from the bases $|n\rangle$ and $|\nu\rangle$. 
At the stochastic level, we now consider the following protocol. We first measure both $S$ and $E$ in their respective eigenbasis $|n\rangle\otimes |\nu\rangle$. 
Next, we evolve them according to a global unitary $U$ and finally we measure them in the bases $|\psi_m\rangle\otimes |\phi_\mu\rangle$. 
The last measurement is performed  in the eigenbases of the reduced density matrices $\rho_S'$ and $\rho_E'$. 
This choice ensures that the ensemble entropy of $\rho_S'$ remains unaffected by the measurement backaction \cite{Santos2019,Elouard2017a}, even though it kills any quantum correlations present in $\rho_{SE}'$. 
For other choices of the final measurement scheme, see~\cite{Manzano2017a} and also \cite{Park2017}.

\begin{figure}
    \centering
    \includegraphics[width=0.35\textwidth]{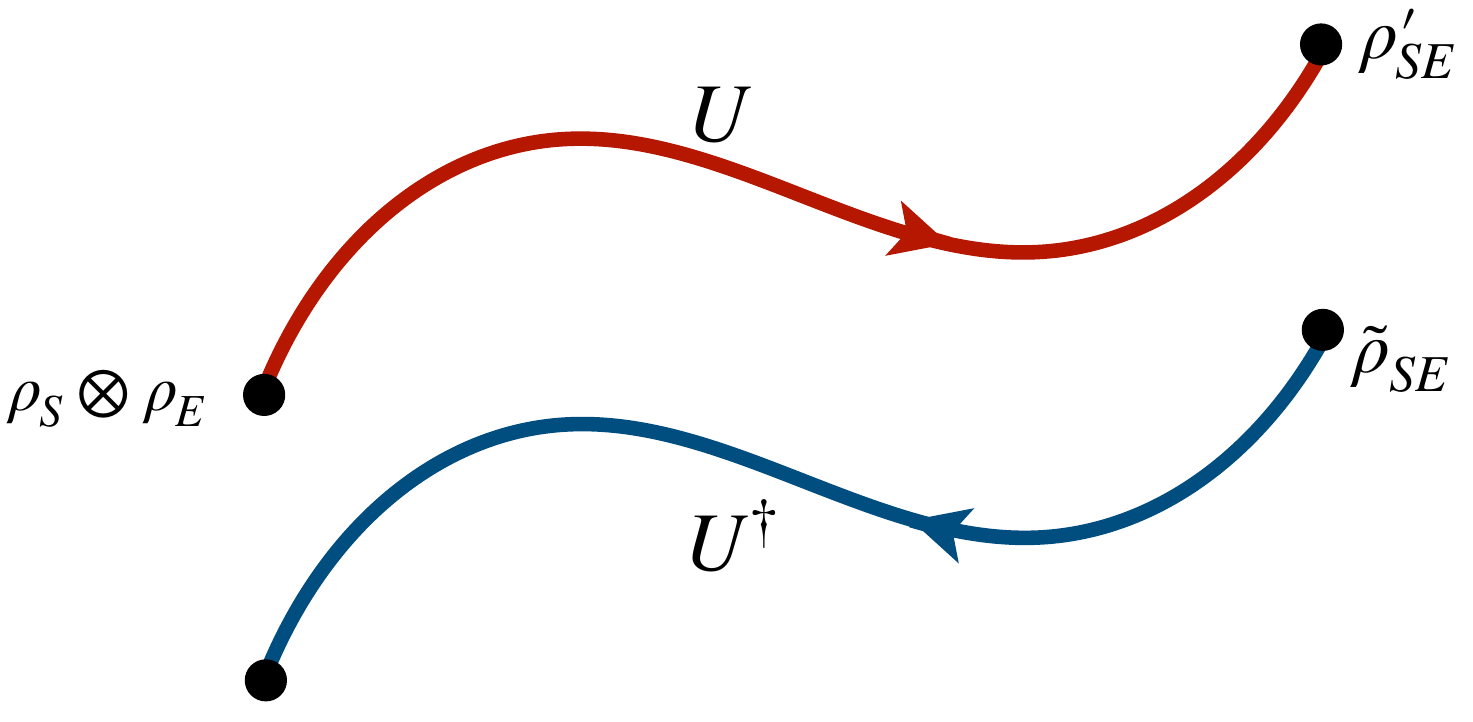}
    \caption{General schematics for the forward and backward trajectories for the fluctuation theorems.}
    \label{fig:FT_Manzano}
\end{figure}

The quantum trajectory is specified by the four measurement outcomes $\gamma = \{n,\nu,m,\mu\}$, which occurs with path probability 
\begin{equation}\label{Qgen_P_forward}
    P_F[\gamma] = |\langle \psi_m,\phi_\mu| U | n,\nu\rangle|^2 p_n q_\nu. 
\end{equation}
In order to build a fluctuation theorem one must now establish the backward process, corresponding to the time-reverse evolution with unitary $U^\dagger$. 
The key observations of Ref.~\cite{Manzano2017a}, however, is that this backwards process is not unique. 
The arbitrariness comes from the choice of  initial state $\tilde{\rho}_{SE}$ for the backwards evolution (see Fig.~\ref{fig:FT_Manzano}).
Different choices, as we now show, lead to different expressions for the entropy production. 
This is also intimately related to the notion of Petz recovery map, a systematic way of building reverse processes for general quantum channels, as considered in~\cite{Kwon2018c}.

For the moment, let us leave $\tilde{\rho}_{SE}$ unspecified. 
We consider a backward process where $\tilde{\rho}_{SE}$ is first measured in the basis $|\psi_m\rangle\otimes |\phi_\mu\rangle$, then  put to evolve with $U^\dagger$ and finally measured one more time, now in the basis  $|n\rangle\otimes |\nu\rangle$. The corresponding backward trajectory probability will thus be
\begin{equation}\label{Qgen_P_backward}
    P_B[\gamma] = |\langle n,\nu| U^\dagger | \psi_m,\phi_\mu\rangle|^2 \tilde{\rho}_{m \mu}, 
\end{equation}
where $\tilde{\rho}_{m \mu} = \langle \psi_m,\phi_\mu | \tilde{\rho}_{SE} | \psi_m,\phi_\mu\rangle$.

Armed with $P_F$ and $P_B$, the entropy production is then defined as usual, as~\cite{Gallavotti1995,Evans1993, Crooks1998}:
\begin{equation}\label{Qgen_sigma_stochastic_fundamental}
    \sigma[\gamma] = \ln \frac{P_F[\gamma]}{P_B[\gamma]}.
\end{equation}
By construction, this quantity satisfies an integral fluctuation theorem, $\langle e^{-\sigma[\gamma]}\rangle = 1$. 
Using Eqs.~(\ref{Qgen_P_forward}) and (\ref{Qgen_P_backward}) the dynamical term cancels out, leaving us only with the boundary term,
\begin{equation}\label{Qgen_sigma_stochastic}
    \sigma[\gamma] = \ln \frac{p_n q_\nu}{\tilde{\rho}_{m\mu}}.
\end{equation}
As we will now discuss, depending on the choice of $\tilde{\rho}_{SE}$, this expression will unravel in different ways.

First, suppose we choose $\tilde{\rho}_{SE} = \rho_S' \otimes \rho_E$. This means the system is taken at the final state~(\ref{Qgen_reduced_map}), whereas the bath is \emph{reset} to the initial state $\rho_E$. 
In this case $\tilde{\rho}_{m\mu} = p_m' q_\mu$ and Eq.~(\ref{Qgen_sigma_stochastic})  becomes 
\[
\sigma[\gamma] = \ln \frac{p_n q_\nu}{p_m' q_\mu}.
\]
The average entropy production is computed as $\langle \sigma \rangle = \sum_\gamma \sigma[\gamma] P[\gamma]$. Carrying out the sum, one finds 
\begin{equation}\label{Qgen_FT_aveSigma_basic}
    \langle \sigma \rangle = \mathcal{I}_{\rho_{SE}'}(S\! : \! E) + S(\rho_E'||\rho_E) = S\Big( \rho_{SE}' \;|| \;\rho_S' \otimes \rho_E\Big),
\end{equation}
which is precisely the definition of $\Sigma$ in Eq.~(\ref{Qgen_Sigma}). 
Notice how $\langle \sigma\rangle$ is just the relative entropy between the final state $\rho_{SE}'$ of the forward process and the initial state $\rho_S'\otimes \rho_E$ of the backwards process. 
This provides a solid physical basis for this expression, as being related to the  act of tracing over the environment: 
The two terms in Eq.~(\ref{Qgen_FT_aveSigma_basic}) appear because we reset $E$ in the backward process, meaning we lost all access to both the correlations developed between $S$ and $E$, as well as the changes that were made in the state of $E$. 

As a second choice, suppose $\tilde{\rho}_{SE} = \rho_S' \otimes \rho_E'$. 
That is, $S$ and $E$ are initialized in the backward process at the  final states of the forward process, but marginalized to destroy any correlations between them. 
Arguably, correlations are the most difficult part to access, since they require global operations on $S$+$E$.
In this case Eq.~(\ref{Qgen_sigma_stochastic}) becomes
$\sigma[\gamma] = \ln (p_n q_\nu)/(p_m' q_\mu')$
which, upon averaging, yields 
\begin{equation}
    \langle \sigma \rangle = \mathcal{I}_{\rho_{SE}'}(S\! : \! E) = \Delta S_S + \Delta S_E.
\end{equation}
Hence, irreversibility stems solely from the $SE$ correlations that are no longer accessible. 

As a third choice, one may take the post-measurement state 
\begin{IEEEeqnarray}{rCl}
\nonumber
    \tilde{\rho}_{SE} = \Delta (\rho_{SE}') 
    &:=& \sum_{m\mu}  |\psi_m,\phi_\mu\rangle \langle \psi_m,\phi_\mu | \rho_{SE}' | \psi_m,\phi_\mu\rangle\langle \psi_m,\phi_\mu | \\[0.2cm]
    &=& \sum_{m,\mu} \rho_{m\mu}' |\psi_m\rangle\langle \psi_m| \otimes | \phi_\mu\rangle\langle \phi_\mu|,
\end{IEEEeqnarray}
which is obtained from the final state $\rho_{SE}'$ after measuring  in the 
$|\psi_m\rangle\otimes |\phi_\mu\rangle$ basis.
Thus, it corresponds to the maximally dephased state in the basis $|\psi_m,\phi_\mu\rangle$ (note that albeit dephased, this state is  still classically correlated).
The entropy production~(\ref{Qgen_sigma_stochastic}), upon averaging, reduces in this case to 
\begin{equation}
\langle \sigma \rangle = S(\Delta(\rho_{SE}')) - S(\rho_{SE}') = \mathcal{C}(\rho_{SE}'), 
\end{equation}
which is the relative entropy of coherence~\cite{Streltsov2016a}.
We thus conclude that, for this choice of backward protocol, the irreversibility stems solely from the decoherence of the measurement backaction in the final basis $|\psi_m,\phi_\mu\rangle$.

In order to perform a final measurement with absolutely no backaction, one would have to measure $S+E$ in the global basis diagonalizing $\rho_{SE}'$. In this case the entropy production would, on average, be identically zero and the process is reversible. However, this requires assessing fully non-local degrees of freedom of $S$ and $E$,  which quickly becomes prohibitive even for small quantum systems. 

As a final choice of measurement, we can assume that both system and environment are completely reset, so $\tilde{\rho}_{SE} = \rho_S \otimes \rho_E$ is exactly the initial state. 
Eq.~(\ref{Qgen_sigma_stochastic}) then becomes $\sigma = \ln (p_n q_\nu)/(p_m q_\mu)$ which, upon averaging, becomes
\begin{IEEEeqnarray}{rCl}
\langle \sigma \rangle 
&=& \mathcal{I}_{\rho_{SE}'}(S\! : \! E)
+ S(\rho_S'||\rho_S) + S(\rho_E' || \rho_E).
\IEEEeqnarraynumspace
\label{Qgen_FT_JW_entropy_manipulation}
\end{IEEEeqnarray}
The first and last terms are exactly the original definition of $\Sigma$ in  Eq.~(\ref{Qgen_Sigma}).
However, we now get the additional term $S(\rho_S'||\rho_S)$, quantifying how much the system was pushed away from equilibrium. 
This is a consequence of the fact that in the backward process, we also reset the system to its original thermal state, thus introducing an additional degree of irreversibility. 

In the particular case where both system and environment start in thermal states, but at different temperatures, $\rho_S = e^{-\beta_S H_S}/Z_S$ and $\rho_E = e^{-\beta_E H_E}/{Z_E}$, Eq.~\eqref{Qgen_FT_JW_entropy_manipulation} reduces to 
\begin{equation}\label{Qgen_sigma_JW_noConservation}
    \langle \sigma\rangle = \beta_S \Delta H_S + \beta_E \Delta H_E,
\end{equation}
where $\Delta H_{S(E)}$ are the changes in energy in the system and environment respectively. 
This choice of $\tilde{\rho}_{SE}$ therefore corresponds to the famous  exchange fluctuation theorem~\cite{Jarzynski2004a}. 
If, on top of all this, the unitary satisfies strict energy conservation [Eq.~\eqref{Qgen_strict_energy_conservation}], then we may define $Q_E:= \Delta H_E = - \Delta H_S$, in which case the entropy production reduces to 
\begin{equation}\label{Qgen_sigma_JW}
    \langle \sigma \rangle = (\beta_E - \beta_S) Q_E, 
\end{equation}
which is the expression appearing in~\cite{Jarzynski2004a}. 

A summary of these results is presented in Table~\ref{tab:sigmas_FTs}.
The main message from this Section is that the definition of entropy production is actually not unique, but depends on the assumptions about which aspects of the system-environment dynamics become inaccessible or irretrievable. 
The definition~(\ref{Qgen_Sigma}), which we have focused on most of this Section, contemplates the most general scenario where everything pertaining to the environment is assumed to be lost after the interaction. 
If the environment is macroscopic, highly chaotic and etc. (e.g. a bucket of water), this will inevitably be the case, so that Eq.~(\ref{Qgen_Sigma}) becomes the only relevant definition of entropy production. 
But in the quantum domain, comparing the different definitions may be quite relevant. 

One may also attempt to compare the relative importance of each term in these expressions. 
Let us assume that the bath is much larger than the system so that the process only pushes it slightly away from equilibrium. 
That is, such that  $\rho_E' = \rho_E + \mathcal{O}(\epsilon)$, for some small parameter $\epsilon$. 
Using standard perturbation theory one then finds that $\Delta S_E \propto \epsilon$ while $S(\rho_E'|| \rho_E) \propto \epsilon^2$~\cite{Rodrigues2019}. 
Thus, it becomes irrelevant whether to include or not the relative entropy term, since the mutual information tends to dominate. 
This, however, is not always the case, as recently elucidated in ~\cite{Ptaszynski2019b}. 
As the authors discuss, the mutual information is actually bounded by the Araki-Lieb inequality, 
\[
\mathcal{I}_{\rho_{SE}'}(S\! : \! E) \leqslant 2 \text{min} \big\{ S(\rho_S'), S(\rho_E')\big\}. 
\]
For small $S$ and large $E$, $\mathcal{I}$ will be essentially capped by $S(\rho_S')$. 
On the other hand, the relative entropy $S(\rho_E'||\rho_E)$ is unbounded and can thus increase indefinitely over time. 
This will be the case, for instance, in non-equilibrium steady-states of systems connected to multiple baths.  

{\color{black}The above discussion can also be extended to multiple measurements. One way to accomplish this is through a collisional model approach, as will be discussed in Sec.~\ref{sec:CM}. 
This will  simply lead to a composition of the results presented in this section. 
Alternatively, one may also analyze it from the perspective of stochastic master equations, describing continuously measured systems. 
This was done in~\cite{Horowitz2013a,Horowitz2014} and yields the entropy production as a function of the entire trajectory of quantum jumps. 
The exploration of different choices for the reverse trajectory, however, is not discussed as the framework is based solely on the reduced description of the system, in terms of a master equation. 
However, at the ensemble level, the authors obtain an entropy production consistent with Eq.~\eqref{Qgen_sigma_thermal_op}, which should thus correspond to the bath reset choice (first line in Table~\ref{tab:sigmas_FTs}). 
}

\begin{table}
\begin{center}
\caption{\label{tab:sigmas_FTs}Different choices for the initial state $\tilde{\rho}_{SE}$ of the backward process and the corresponding formula for the average entropy production $\langle \sigma \rangle$.}
\begin{tabular}{c|c}
     $\tilde{\rho}_{SE}$    &   $\langle \sigma \rangle$ \\[0.2cm]
     \hline
     \hline\\
     $\rho_S' \otimes \rho_E$ & $\mathcal{I}_{\rho_{SE}'}(S\! : \! E) + S(\rho_E'||\rho_E) \equiv \Sigma$ \\[0.2cm]
     (bath reset) & (Eq.~\eqref{Qgen_Sigma}) \\[0.2cm]
     \hline\\
     $\rho_S' \otimes \rho_E'$ & $\mathcal{I}_{\rho_{SE}'}(S\! : \! E)$ \\[0.2cm]
     (correlations destroyed) &  $\big( = \Delta S_S + \Delta S_E\big)$\\[0.2cm]
     \hline\\
     $\Delta(\rho_{SE}')$ & $C(\rho_{SE}')$ \\[0.2cm]
     (post-measurement state) & (relative entropy of coherence) \\[0.2cm]
     \hline\\
     $\rho_S \otimes \rho_E$ & $\mathcal{I}_{\rho_{SE}'}(S\! : \! E) + S(\rho_S'||\rho_S) + S(\rho_E'||\rho_E)$ \\[0.2cm]
      & $=(\beta_S - \beta_E)Q_E$ \\[0.2cm]
     (both  reset) & \cite{Jarzynski2004a}\\[0.2cm]
     \hline
\end{tabular}
\end{center}
\end{table}

%
\subsection{\label{sec:lag}Non-equilibrium lag}

\begin{figure}[b]
    \centering
    \includegraphics[width=0.4\textwidth]{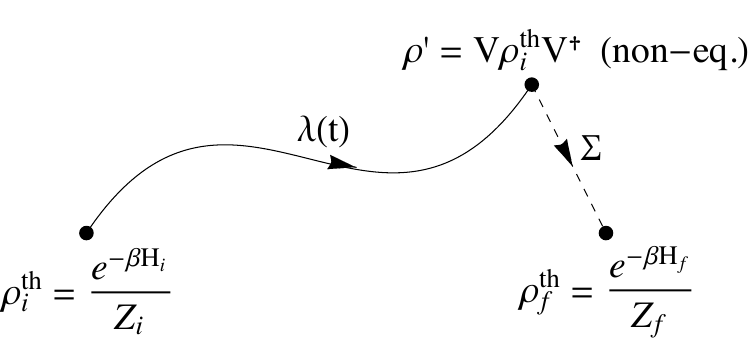}
    \caption{The typical non-equilibrium lag scenario. A system initially prepared in a thermal state $\rho_i^\text{th}$ is driven unitarily by a work protocol to a non-equilibrium state $\rho'$. After the protocol, the system is allowed to thermalize from $\rho'$ to $\rho_f$.}
    \label{fig:nonEq_lag}
\end{figure}

A scenario which is deeply related to the above, and which has been the subject of considerable research, is the non-equilibrium lag that occurs when an isolated quantum system undergoes a work protocol. 
This has been covered in detail in~\cite{Campisi2011}.
Here we focus only on the most recent developments. 

We consider a system $S$ initially prepared in the equilibrium state
$\rho_i^\text{th} = e^{-\beta H_i}/Z_i$, at temperature $\beta$ and  Hamiltonian $H_i$. 
The system is then driven by a work protocol $\lambda(t)$ which changes the Hamiltonian from $H_i = H(\lambda(0))$ to $H_f = H(\lambda(\tau))$, where $\tau$ is the duration of the protocol.
The drive causes the system to evolve unitarily to a non-equilibrium state $\rho' = V \rho_i^\text{th} V^\dagger$, where $V$ is the time-evolution operator generated by  $H(\lambda(t))$.
After the protocol is applied, the system is then placed in contact with a bath and allowed to fully thermalize towards a new equilibrium state $\rho_f^\text{th} = e^{-\beta H_f}/Z_f$ (see Fig.~\ref{fig:nonEq_lag}).

The unitary drive produces no entropy since the dynamics is closed. 
Irreversibility stems solely from the thermalization step.
The entropy production for this relaxation process will be given, in the simplest scenario, by Eq.~(\ref{Qgen_sigma_thermal_op}). 
Since the thermalization is total, the second term vanishes, leaving us with 
\begin{equation}\label{Qgen_nonEqLag_Sigma}
\Sigma = S(\rho' || \rho_f^\text{th}), 
\end{equation}
Despite being associated to the thermalization process, it turns out this quantity is also of significance to the unitary evolution in itself.
In fact, usually this is defined without even mentioning the thermalization. 
The reason is that  Eq.~\eqref{Qgen_nonEqLag_Sigma} is also directly associated with the irreversible work produced by the unitary $V$: 
\begin{equation}\label{Qgen_nonEqLag_irr_work}
    \Sigma \equiv \beta W_\text{irr} = \beta (\langle W \rangle - \Delta F),
\end{equation}
where $\langle W \rangle = \tr(H_f \rho') - \tr(H_i \rho_i^\text{th})$ is the average work and $\Delta F = - T \ln Z_f/Z_i$ is the change in equilibrium free energy.
For this reason, Eq.~\eqref{Qgen_nonEqLag_Sigma} is also called the \emph{non-equilibrium lag}. 
For all intents and purposes, ``non-equilibrium lag'' can be taken as a synonym of entropy production. 
The reason to introduce this terminology is simply to emphasize that it refers to the unitary protocol, for which no entropy is produced.
In the past years, significant attention has been given to the non-equilibrium lag, particularly in the  context of quantum phase transitions. These will be reviewed in Sec.~\ref{sec:inf_quenches}.
 
The non-equilibrium lag can also be studied from a stochastic perspective, using the two-point measurement scheme; the first measurement is done in the eigenbasis $|n_i\rangle$ of $H_i$ and the second in the eigenbasis $|m_f \rangle$ of $H_f$.
The stochastic entropy production associated to this process is then~\cite{Campisi2011} 
\begin{equation}
    \sigma[n_i,m_f] = \ln p_{n_i}^\text{th}/p_{m_f}^\text{th} = \beta (E_{m_f}^f - E_{n_i}^i - \Delta F), 
\end{equation}
where $E^{i(f)}$ are the energies of $H_{i(f)}$ and $\Delta F = F_f - F_i = - T \ln Z_f/Z_i$ is the change in non-equilibrium free energy.
Moreover, 
$p_{n_i}^\text{th} = e^{-\beta E_{n_i}^i}/Z_i$ is the initial thermal probability and $p_{m_f}^\text{th} = e^{-\beta E_{m_f}^f}/Z_f$ is a thermal probability associated with the final Hamiltonian $H_f$.
The probability distribution of $\sigma$ is thus
\begin{equation}\label{Qgen_nonEqLag_P_sigma}
    P(\sigma) = \sum\limits_{n_i,m_f} p(m_f|n_i) p_{n_i} \delta \bigg(\sigma - \sigma[n_i,m_f]\bigg), 
\end{equation}
where $p(m_f|n_i) = |\langle m_f | V | n_i \rangle|^2$ is the transition probability from $|n_i\rangle \to |m_f\rangle$. 
By construction, this is such that $\langle \sigma \rangle = \Sigma$ [Eq.~\eqref{Qgen_nonEqLag_Sigma}].

It is convenient to study the cumulant generating function $K(\lambda) = \ln \langle e^{-\lambda \sigma} \rangle$, which can be conveniently written as~\cite{Talkner2007,Esposito2009}
\begin{equation}\label{Qgen_nonEqLag_CGF}
    K(\lambda) = \ln \tr \bigg\{ V^\dagger e^{-\beta \lambda(H_f - F_f)} V e^{\beta \lambda (H_i - F_i)}  \rho_i^\text{th}\bigg\}. 
\end{equation}
The cumulants may be computed from $K(\lambda)$ through the relation
\begin{equation}\label{Qgen_nonEqLag_cumulants_deriv}
    \kappa_n(\sigma) = (-1)^n\frac{\partial^n K}{\partial \lambda^n} \bigg|_{\lambda = 0}.
\end{equation}
The first cumulant is the average and is  given by Eq.~\eqref{Qgen_nonEqLag_Sigma}. 
Similarly, the second cumulant is the variance and can be written as 
\begin{equation}
    \text{var}(\sigma) = \tr\big\{ \rho' (\ln \rho' - \ln \rho_f^\text{th})^2\big\} - S(\rho'|| \rho_f^\text{th})^2,
\end{equation}
which is sometimes called the relative entropy variance. 

The CGF~\eqref{Qgen_nonEqLag_CGF} can also be expressed in terms of the so-called  R\'enyi divergences, which will be discussed further in Sec.~\ref{sec:thermal_ops} and are defined as
\begin{equation}\label{Renyi_divergence}
    S_\lambda(\rho||\sigma) = \frac{1}{\lambda-1} \ln \tr \big\{ \rho^\lambda \sigma^{1-\lambda}\big\}
\end{equation}
They correspond to a generalization of the relative entropy~\eqref{relative_entropy}, which is recovered from $S_\lambda(\rho||\sigma)$ in the limit $\lambda \to 1$.  
Comparing~\eqref{Renyi_divergence} with Eq.~\eqref{Qgen_nonEqLag_CGF} one then sees that~\cite{Guarnieri2019}:\footnote{
This can also be equivalently written as $K(\lambda) = - \lambda S_{1-\lambda}(\rho' || \rho_f^\text{th})$.}
\begin{equation}\label{Qgen_nonEqLag_CGF_Renyi}
    K(\lambda) = (\lambda-1) S_\lambda(\rho_f^\text{th} || \rho'),
\end{equation}
This expression has been used in several recent studies. 
Following~\cite{Guarnieri2018}, we will review in Sec.~\ref{sec:resource_reconciliation} how~\eqref{Qgen_nonEqLag_CGF_Renyi} can be used as a connection to the resource-theoretic formulation of thermodynamics, which is the subject to  Sec.~\ref{sec:thermal_ops}. 
In Sec.~\ref{sec:inf_quenches} we review Refs.~\cite{Miller2019,Scandi2019}, which use~\eqref{Qgen_nonEqLag_CGF_Renyi} as a tool to extract the contribution from quantum coherence in slow processes. 

%
%
\section{\label{sec:information}Information-theoretic aspects}
%
%

\subsection{\label{sec:landauer}Corrections to Landauer's principle}

Landauer's principle was introduced in Sec.~\ref{sec:matters_Landauer} and is based on the idea that information erasure is an irreversible process, with a fundamental heat cost associated to it. 
This is synthesized by Eq.~\eqref{landauer}, representing a lower bound on the heat $Q_E$ dissipated to the environment, in terms of the change in entropy $\Delta S_S$  of the system. 
Being a lower bound,  one can then conclude that  changes in entropy  must be accompanied by a fundamental heat cost. 

In Sec.~\ref{sec:matters_Landauer} we hinted at the subtle nature of Landauer's principle: in classical thermodynamics, Eq.~\eqref{landauer} is a direct consequence of the 2nd law, but with $S_S$ being the thermodynamic entropy of the system. Landauer's original bound, on the other hand, concerns the information theoretic entropy.
The framework put forth in Sec.~\ref{sec:Qgen}, however, unifies both views, as it reformulates the 2nd law in terms of the system's von Neumann entropy. 
Indeed, Eqs.~(\ref{Qgen_Sigma}) and~(\ref{Qgen_Sigma_thermal}) imply that:
\begin{equation}\label{info_landauer_0}
\Sigma 
= \mathcal{I}_{\rho_{SE}'}(S\! :\! E) + S(\rho_{E}'|| \rho_{E})
= \Delta S_S + \beta Q_E.
\end{equation}
The 2nd law $\Sigma \geqslant 0$ then yields 
$Q_E \geqslant - T \Delta S_S$, which is precisely Landauer's bound~\eqref{landauer}. 
Equality is achieved when $\Sigma = 0$; i.e., for reversible processes.
These results are present already in~\cite{Esposito2010a}, but the link with Landauer's principle was strengthened in in~\cite{Reeb2014}, a publication which greatly popularized this subject. 

Eq.~(\ref{landauer}) is important because it is universal. 
The only hypothesis is that the bath is initially thermal (and uncorrelated from the system). 
Other than that, the bath may have arbitrary dimension and arbitrary Hamiltonian;
the system may be prepared in any initial state;
and the interaction $U$ can be any unitary  whatsoever.

This universality, however, has the downside that,  the bound is in general quite loose. 
Tighter bounds can be obtained by assuming additional information about the environment and/or the process.
We now discuss several such formulations, taking care to properly state which additional pieces of information are assumed in each case.
First, we consider the case where the only additional piece of information one has is that the environment is finite dimensional, with a Hilbert space dimension $d_E$. 
In this case, when $\Delta S_S < 0$, the following correction to~(\ref{landauer}) holds~\cite{Reeb2014}:
\begin{equation}\label{info_landauer_finite_d}
    Q_E \geqslant - T \Delta S_S + \frac{2 T (\Delta S_S)^2}{4 + \ln^2(d_E-1)}.
\end{equation}
This shows that finite dimensions impose more strict constraints on heat dissipation. 
The correction vanishes when $d_E \to \infty$; however, notice that the dependence is logarithmic and therefore extremely slow. 
Additional finite-size bounds are also presented in~\cite{Reeb2014}, although they depend on more complicated functions.

The original bound~(\ref{landauer}), or its finite size correction~(\ref{info_landauer_finite_d}) become trivial in the limit $T\to 0$. 
This is clearly unsatisfactory: can erasure really be performed with zero dissipation when $T\to 0$? 
The bound trivializes in this case due to the term  $S(\rho_E'||\rho_E)$ in Eq.~(\ref{info_landauer_0}), which diverges when $T\to 0$. 
To bypass this difficulty, in~\cite{Timpanaro2019a} it was shown how to derive a tighter bound starting only from the mutual information term $\mathcal{I}_{\rho_{SE}'}(S\! :\! E)$. 
The bound in this case acquires the form 
\begin{equation}\label{info_landauer_zeroT}
    Q_E \geqslant \mathcal{Q}(\mathcal{S}^{-1}(-\Delta S_S)), 
\end{equation}
where the functions $\mathcal{Q}(T')$ and $\mathcal{S}(T')$ are defined as
\begin{equation}\label{info_finiteT_QS}
    \mathcal{Q}(T') = \int\limits_T^{T'} C_E(\tau) d\tau, 
    \qquad
    \mathcal{S}(T') = \int\limits_{T}^{T'} \frac{C_E(\tau)}{\tau} d\tau,
\end{equation}
with $C_E(T)$ being the equilibrium heat capacity of the environment. 
In these expressions $T$ is the actual initial temperature of the environment, whereas $T'$ is merely the argument of the functions. 
This bound requires only one additional piece of information; namely the environment's heat capacity $C_E(T)$. 
This is to be compared with~\eqref{landauer}, which requires only a single number, $T$, or with Eq.~\eqref{info_landauer_finite_d}, which requires two numbers, $T$ and $d_E$. 
Admittedly, knowing an entire function $C_E(T)$ is definitely more difficult, although the heat capacity is in general an easy quantity to measure experimentally, even at extremely low temperatures. However, one can also show that the bound is always tighter than both~\eqref{landauer} and~\eqref{info_landauer_finite_d}.
To provide an example, if we happen to have $C_E = a T$, for some constant $a$, Eq.~\eqref{info_landauer_zeroT} becomes
\begin{equation}
     Q_E \geqslant - T \Delta S_S + \frac{\Delta S_S^2}{2a}.   
\end{equation}
As in~\eqref{info_landauer_finite_d}, the  correction also involves a term proportional to $\Delta S_S^2$, but with a coefficient that is temperature independent. Thus, in the limit $T\to 0$ the last term still survives, showing that a fundamental heat cost still exists even when $T = 0$.



Tighter bounds can also be derived when information about the $SE$ unitary $U$ and the system initial state $\rho_S$ are available~\cite{Goold2014b,Lorenzo2015,Guarnieri2017}.
Here we review the approach in~\cite{Goold2014b}, which derives a bound using the fluctuating properties of heat. 
The key idea is to interpret the global map~(\ref{Qgen_global_map}) as a quantum channel for the environment, instead of the system, as described by the Kraus map
 \begin{equation}
\begin{aligned}
\rho'_{E} = {\Tr}_{S}[U(\rho_{S} \otimes\rho_{E}) U^\dag]
= \sum_{l}  A_{l} \, \rho_E \,  A_{l}^\dag,
\label{operation}
\end{aligned}
\end{equation}
where $A_{l = jk} = \sqrt{\lambda_j} \sand{s_k}{U}{s_j}$, with  $\{\lambda_j\}$ and $\{\ket{s_j}\}$ being the eigenvalues and eigenstates of $\rho_S$. Trace-preservation implies  $\sum_l A^\dag_l  A_l = \openone_E$. 
Letting $E_n$ and $|r_n\rangle$ denote the eigenvalues and eigenvectors of $H_E$, the heat distribution of the environment (via a two-point measurement) can now be written as~\cite{Talkner2009}. 
\begin{equation}
\label{PofQ}
P(Q_E) = \sum_{l,m,n} \sand{r_n}{A_l}{r_m} (\rho_E)_{mm}\sand{r_m}{A^\dag_l}{r_n} \delta(Q_E-(E_{n}- E_m))
\end{equation}
with $(\rho_E)_{nm}=\bra{r_n}\rho_E\ket{r_m}$.
From this one may now show that 
$\avrg{e^{-\beta Q}} = \Tr[\mathbf{M} \, \rho_S]$,
where $\mathbf{M} = \Tr_E[U^\dag \, (\openone_S \otimes \rho_E) \, U]$. 
Using Jensen's inequality then leads to 
\begin{equation}\label{info_boundtotal}
\avrg{Q_E} \geqslant
-T \ln(\Tr[\mathbf{M} \,\rho_S ]).
\end{equation}
This result  establishes a bound on $\langle Q_E \rangle$, which depends on both the state of the system as well as the unitary $U$. 
It therefore naturally encompass also a dependence on the size of $E$, in line with Eq.~(\ref{info_landauer_finite_d}).

Using the formalism of full counting statistics~\cite{Esposito2009}, one can also extend these results to obtain an entire single-parameter family of bounds~\cite{Guarnieri2017}. 
We first introduce the cumulant generating function of $P(Q_E)$, \begin{equation}
\label{fullcount}
\Theta(\eta, \beta) \equiv \ln \left\langle e^{-\eta Q_E}\right\rangle=\ln \int P(Q_E) e^{-\eta Q_E} d Q_E.
\end{equation}
H\"older's inequality then implies that for $\eta >0$, 
\begin{equation}
\label{Giacomo}
\beta \langle Q_E\rangle \geqslant-\frac{\beta }{\eta} \Theta(\eta, \beta) 
\quad(\eta>0),
\end{equation}
which contains Eq.~\eqref{info_boundtotal} as a particular case. 
Conversely, for $\eta <0$ we obtain  the upper bounds
$\beta \langle Q_E \rangle\leqslant\beta\Theta(\eta,\beta)/|\eta|$.
In the limit $|\eta| \to 0$ both bounds coincide with $\beta \langle Q_E \rangle$.

\subsection{Conditional entropy production}

We consider once again the general map~\eqref{Qgen_global_map} of Sec.~\ref{sec:Qgen}. 
But now we suppose that after the map we measure the environment, or at least a part of it.
\cite{Funo2013} studied how the information acquired from this measurement affects the entropy production.
Since it is only the bath that is measured, there can be no backaction to the system, as this would violate no-signaling.
As a consequence, one expects that learning the outcomes of the measurements should always make the the process more reversible; that, is part of the ignorance captured by the entropy production should be resolved.  

To formalize this idea, we consider a generalized measurement on $E$ described by Kraus operators $\{M_k\}$ and labeled by a set of outcomes $k$. We denote the local states of $S$ and $E$ after the map, conditioned on an outcome $k$, by  
\begin{equation}
    \rho_{E|k}' = \frac{M_k \rho_E' M_k^\dagger}{p_k},
    \qquad 
    \rho_{S|k}' = \frac{1}{p_k} \tr_E \big( M_k \rho_{SE}' M_k^\dagger\big), 
\end{equation}
where $p_k = \tr(M_k \rho_E' M_k^\dagger)$ is the probability of  outcome $k$ (as before, primed quantities always refer to states after the map).
One may also  verify that  
$\sum_k p_k \rho_{S|k}' = \rho_S'$, thus confirming that the measurement in $E$ causes no backaction in $S$. 
But there may, of course, be a backaction in $E$ so $ \tilde{\rho}_E := \sum_k p_k \rho_{E|k}'  \neq \rho_E'$.

We now ask how to construct the entropy production conditioned on a given outcome. 
The goal is to define, in analogy with  Eq.~(\ref{Qgen_sigma_phi}), a conditional entropy production $\Sigma_k$ and a conditional flux $\Phi_k$, which are related by 
\begin{equation}\label{info_Sigma_k_def}
    \Sigma_k = S(\rho_{S|k}') - S(\rho_S) + \Phi_k.
\end{equation}
This is still merely a definition, and will only acquire meaning once $\Sigma_k$ and $\Phi_k$ are defined. 
Averaging  over all outcomes $k$ then yields a relation between the \emph{conditional} average entropy production and flux:
\begin{equation}\label{info_cond_sigma_def}
    \Sigma_c = \sum\limits_k p_k S(\rho_{S|k}') - S(\rho_S) + \Phi_c, 
\end{equation}
where $\Sigma_c = \sum_k p_k \Sigma_k$ and similarly for $\Phi_c$.
The entropy difference on the first two terms of the right-hand side is known as the Ozawa-Groenewold quantum-classical information~\cite{Groenewold1971,Ozawa1986,Funo2018}.
Notice also that $\Sigma_k$ and $\Phi_k$ are not necessarily linear functions of $\rho_{SE}'$, so that, in general, their averages $\Sigma_c$ and $\Phi_c$ do not have to coincide with the unconditional quantities $\Sigma$ and $\Phi$. 

Eq.~(\ref{info_Sigma_k_def}) is merely a definition of $\Sigma_k$ and $\Phi_k$. 
The relevant question is how to properly define these quantities in a way that is physically consistent. 
We first analyze the flux. 
A look at Eq.~(\ref{Qgen_flux}) shows that a natural generalization to the case of conditional states is
$\Phi_k = S(\rho_{E|k}') - S(\rho_E) + S(\rho_{E|k}'|| \rho_E)$,
which therefore simply amounts to replacing $\rho_{E}'$ with $\rho_{E|k}'$.
Averaging over all $p_k$ and using the second line in Eq.~(\ref{Qgen_flux}), one then finds 
\begin{equation}
    \Phi_c := \sum\limits_k p_k \Phi_k = \tr\big\{ (\rho_E - \tilde{\rho}_E) \ln \rho_E\big\}, 
\end{equation}
where $\tilde{\rho}_E = \sum_k p_k \rho_{E|k}'$. 
If the measurement is performed on the initial eigenbasis of $\rho_E$, it then follows that $\Phi_c = \Phi$ (even though $\tilde{\rho}_E \neq \rho_E'$). 
This result has a clear and beautiful physical interpretation: the entropy flux refers only to the flow of information to the environment. It should therefore be independent on whether or not we condition on any measurement outcomes. 
The flux should therefore only change if there is backaction from the measurement.
In other words, the difference $\Phi_c - \Phi$ has nothing to do with the system nor the $SE$ interaction, but only with the backaction caused by the measurement. 
For this reason, we henceforth assume that the measurement is such that $\Phi_c = \Phi$. 
Interestingly, this assumption has also been used implicitly in Ref~\cite{Breuer2003}, which defines entropy production from the perspective of quantum jump trajectories.

Using $\Phi_c = \Phi$ in Eq.~(\ref{info_cond_sigma_def}) and comparing with Eq.~(\ref{Qgen_sigma_phi}) allows one to conclude that 
\begin{equation}\label{info_Sigma_c_Sigma_Holevo}
    \Sigma_c = \Sigma - \chi_M(\rho_S'),
\end{equation}
where 
\begin{IEEEeqnarray}{rCl}\label{info_Holevo}
    \chi_M(\rho_S') &=& S(\rho_S') - \sum\limits_k p_k S(\rho_{S|k}')=\sum\limits_k p_k S(\rho_{S|k}' || \rho_S'),\IEEEeqnarraynumspace
\end{IEEEeqnarray}
is the Holevo quantity~\cite{Nielsen}, which is always non-negative. 
Eq.~(\ref{info_Sigma_c_Sigma_Holevo}) beautifully illustrates the idea of reducing irreversibility through measurement: Conditioning on the measurement outcomes \emph{reduces}, on average, the entropy production by an amount proportional to the Holevo quantity, an object with numerous applications in information theory.

The Holevo quantity $\chi_M$ is  a basis-dependent version of the classical information used in quantum discord theory~\cite{Modi2012}. It thus follows that, for any choice of measurement operators $\{M_k\}$, one should have
$\chi_M(\rho_S') \leqslant I_{\rho_{SE}'}(S\!:\! E)$.
Comparing with the definition of $\Sigma$ in Eq.~(\ref{Qgen_Sigma}), one then concludes from
Eq.~(\ref{info_Sigma_c_Sigma_Holevo}) that  
\[
\Sigma_c =I_{\rho_{SE}'}(S\!:\! E) +   S(\rho_E'|| \rho_E)- \chi(\rho_S') \geqslant  S(\rho_E'|| \rho_E).
\]
Hence, even though $\Sigma_c \leqslant \Sigma$, it is nonetheless still strictly non-negative. 
This occurs because the interaction irreversibly pushes the bath away from equilibrium, so that even if all possible information was to be acquired, the dynamics would still be irreversible.  

\subsection{\label{sec:heat_flow_corr}Heat flow in the presence of correlations}

Another key manifestation of information in thermodynamics is the influence of initial correlations in the heat flow between two bodies. 
According to the second law, if we put in contact two systems $A$ and $B$, initially prepared in equilibrium at different temperatures, heat will always flow hot to cold [Eq.~\eqref{engine_heat_flow}]. 
This assumes, however, that the two bodies are initially uncorrelated. 
If that is not true, heat may eventually flow from cold to hot. 
This problem was first considered in the quantum scenario in~\cite{Partovi2008}, who discussed only the case where the global state of $AB$ is pure. 
This was then generalized in Refs.~\cite{Jennings2010} and \cite{Bera2017b}, who also addressed some of the information-theoretical aspects of the problem.
An experimental demonstration of this effect was recently performed in a nuclear magnetic resonance setup~\cite{Micadei2017}. 
In a broader sense, these ideas are ultimately related to the use of mutual information to reduce entropy, as first discussed in the seminal paper by~\cite{Lloyd1989}.

We consider two systems  with Hamiltonians $H_A$ and $H_B$, prepared in a global (generally correlated) state $\rho_{AB}$.
We assume, however, that the reduced density matrices of $A$ and $B$ are still thermal, $\rho_A = \tr_B \rho_{AB} = e^{-\beta_A H_A}/Z_A$, and $\rho_B = e^{-\beta_B H_B}/Z_B$, at different temperatures $\beta_A$ and $\beta_B$.
The two systems are then put to interact with a unitary $U$ satisfying strict energy conservation $[U,H_A+H_B] = 0$ (cf. Eq.~(\ref{Qgen_strict_energy_conservation})).
The state after the interaction is thus $\rho_{AB}' = U \rho_{AB} U^\dagger$, from which one can compute the corresponding marginals $\rho_A'$ and $\rho_B'$. 

The correlations between $A$ and $B$ are characterized by the mutual information $\mathcal{I}_{\rho_{AB}}(A\! : \! B)$ defined in Eq.~(\ref{Qgen_MI}). Since the dynamics is unitary it follows that $S(\rho_{AB}') = S(\rho_{AB})$, which allows one to show that 
\begin{equation}\label{ThermoCorr_Delta_MI}
    \Delta I(A\! :\! B) = \Delta S_A + \Delta S_B,
\end{equation}
where $\Delta I(A\! :\! B) =  I_{\rho_{AB}'}(A\! :\! B)-I_{\rho_{AB}}(A\! :\! B)$ is the change in the mutual information between $A$ and $B$. 

Next, consider the quantity
\begin{equation}\label{ThermoCorr_sigma_Jennings}
    \mathcal{S} = S(\rho_A' || \rho_A) + S(\rho_B' || \rho_B) \geqslant 0,
\end{equation}
which is non-negative  because the relative entropies are non-negative. 
This quantity is a part of the entropy production, when cast in terms of the  Jarzynski-W\'ojcik scenario [cf. Eq.~(\ref{Qgen_FT_JW_entropy_manipulation})].  
What is important for the present purposes is that this quantity is purely local, depending only on the reduced density matrices of $A$ and $B$ before and after the interaction.
Substituting the initial thermal forms of $\rho_A$ and $\rho_B$, together with Eq.~(\ref{ThermoCorr_Delta_MI}), then leads to \cite{Jennings2010}:
\begin{equation}\label{ThermoCorr_sigma_general_expr}
    \mathcal{S} = \beta_A \Delta H_A + \beta_B \Delta H_B - \Delta  I(A\! :\! B) \geqslant 0.
\end{equation}
Let us assume $T_A > T_B$. 
Due to strict energy conservation, the average heat exchanged is simply defined as 
\begin{equation}\label{ThermoCorr_heat_exchanged}
    Q_B = \Delta H_B  =  \tr\big\{H_B(U\rho_{AB}U^\dagger - \rho_{AB})\big\} = - \Delta H_A,
\end{equation} 
so that Eq.~(\ref{ThermoCorr_sigma_general_expr}) becomes
\begin{equation}\label{ThermoCorr_bound_DeltaI}
    (\beta_B-\beta_A) Q_B \geqslant \Delta  I(A\! :\! B).
\end{equation}
This can be viewed as a generalization of the bound~\eqref{engine_heat_flow} to take into account initial correlations. 

If $A$ and $B$ are initially uncorrelated then $\Delta I(A\! :\! B) =  I_{\rho_{AB}'}(A\! :\! B) \geqslant 0$, which implies  $Q_B$ must have the same sign as $\beta_B - \beta_A$ (i.e., heat flows from hot to cold).
But if they are initially correlated and the process is such that this correlation is consumed ($\Delta I(A\! :\! B)<0$), then it is possible for heat to flow from cold to hot. 
This is thus an example of a situation where an information theoretic resource is being consumed to perform a thermodynamic task that would not naturally occur. 
This is akin to refrigerators, where heat also flows from cold to hot, but the resource being used is work from the electrical plug.
The result can also be formulated in the language of Maxwell's Demons. 
A demon, in this context, has access to additional information, in the form of global correlations shared between $A$ and $B$. These correlations can then be consumed as a thermodynamic resource. 


Correlations, of course, will not always make heat flow from cold to hot. 
They may very well have the opposite effect, accelerating the heat from hot to cold. 
An illustrative example is the problem studied experimentally in~\cite{Micadei2017}. 
Consider two qubits with $H_i = \Omega |e\rangle\langle e|_i$ ($i = A,B$) and initially prepared in a correlated state of the form 
\begin{equation}
\rho_{AB} = \rho_A^\text{th} \otimes \rho_B^\text{th} + \chi
\end{equation}
where $\rho_i^\text{th} = (1-f_i) |g\rangle\langle g| + f_i |e\rangle\langle e|$, with $f_i = (e^{\Omega/T_i}+1)^{-1}$, are the local thermal states of each qubit and $\chi = \alpha e^{i\theta} |g,e\rangle\langle e,g| + \alpha e^{- i \theta} |e,g\rangle\langle g,e|$
represents the correlations, with $\alpha$ and $\theta$ being real parameters.  
The two qubits are then put to interact with an energy-preserving unitary $U = \exp\big\{-i g t( e^{i \phi} |g,e\rangle\langle e,g| + e^{-i \phi} |e,g\rangle\langle g,e|)\big\}$, where $\phi$ is an arbitrary phase and $g$ is the interaction strength.  
The heat $Q_B= \Delta H_B$ that enters system $B$ at time $t$ will  be given by 
\begin{equation}
    Q_B(t) = \Omega \sin(gt) \bigg[ (f_A - f_B) \sin(gt) - 2 \alpha \sin(\theta-\phi) \cos(gt)\bigg].
\end{equation}
We again assume $T_A > T_B$ for concreteness. 
Since $f_i$ is monotonically increasing with $T_i$, when $\alpha= 0$ we always get $Q_B \propto (f_A - f_B) > 0$, so that heat will flow from hot to cold. 
But when $\alpha \neq 0$, the direction of the heat flow will actually depend on a fine interplay between the phases $\theta$ and $\phi$ appearing in $\chi$ and $U$, respectively. These phases may combine either constructively, reversing the heat flow, or destructively, accelerating the already natural flow direction.

\subsection{\label{ssec:XFT_corr}Fluctuation theorem under classical and quantum correlations}

The problem treated in Sec.~\ref{sec:heat_flow_corr} can also be analyzed from a quantum trajectories perspective, which will serve to highlight the non-trivial role of quantum~vs.~classical correlations. 
We begin by considering the case of two-point measurements (TPM), where both $A$ and $B$ are measured at the beginning and the end of the process. 
\cite{Jevtic2015a} discusses the implications of measuring  in the local energy bases $|n_A\rangle$ and $|n_B\rangle$ of the Hamiltonians $H_A$ and $H_B$. 
A quantum trajectory will be specified by four quantum numbers, 
$\gamma = (n_A,n_B,m_A,m_B)$ and occurs with probability 
\begin{equation}\label{ThermoCorr_pathProb_TPM}
    \mathcal{P}[\gamma] = |\langle m_A m_B | U | n_A n_B \rangle|^2 p_{n_A n_B},
\end{equation}
where $p_{n_A n_B} = \langle n_A n_B | \rho_{AB} | n_A n_B\rangle$. 
Crucially, since $\rho_{AB}$ is not a product state, in general  $p_{n_A n_B} \neq p_{n_A} p_{n_B}$. 

The probability that a heat $q_B[\gamma] = E_{m_B} - E_{n_B}$ enters system $B$ will then be given by 
$    P(q_B) = \sum_{\gamma} \delta (q_B - q_B[\gamma]) \;\mathcal{P}[\gamma]$.
Using this to compute the average heat $\langle q_B\rangle$, we find 
\begin{equation}\label{ThermoCorr_aveQ_TMP}
    \langle q_B \rangle = \tr\big\{ H_B \big[ U \Delta(\rho_{AB}) U^\dagger - \Delta(\rho_{AB})\big]\big\},
\end{equation}
where 
$\Delta(\rho_{AB}) = \sum_{n_A, n_B} |n_A n_B\rangle\langle n_A n_B| \rho_{AB} |n_A n_B\rangle\langle n_A n_B|$
is the operation of fully  dephasing $\rho_{AB}$ in the basis $|n_An_B\rangle$.

The important point to realize now is that Eq.~(\ref{ThermoCorr_aveQ_TMP}) is, in general, \emph{different} from the average heat in Eq.~\eqref{ThermoCorr_heat_exchanged}. 
The difference is due to the presence of the dephasing operator $\Delta$ and is thus a consequence of the measurement backaction, which dephases $\rho_{AB}$. The two quantities will only coincide when $\rho_{AB}$ is already diagonal in $|n_A n_B\rangle$. 
Put it differently, when $\rho_{AB}$ is not diagonal, the TPM scheme used here will fundamentally change the amount of heat exchanged between the two systems, producing an entirely  different dynamics when compared with the bare unitary evolution. 
The entropy production is thus \emph{extrinsic}; that is, dependent not only on the systems $A$ and $B$, but also on the details on how one performs the experiment. 

This highlights the fundamental difference between correlations present in the populations (i.e., which are diagonal in $|n_A n_B\rangle$) and correlations which are present in the coherences (off-diagonals). 
The latter can be viewed as a basis-dependent quantum discord; i.e., as the amount of discord present in the energy basis (the energy basis appears as a preferred basis due to the energy-conserving nature of the unitary $U$; as will be reviewed in Sec.~\ref{sec:emergence}). 

Returning to Eq.~(\ref{ThermoCorr_pathProb_TPM}), let us introduce the reverse process, where both $A$ and $B$ start at the same state, but one applies the unitary $U^\dagger$ instead (this is the Jarzynski-W\'ojcik scenario of Sec.~\ref{sec:qgen_FT}). 
The probability for the backward trajectory $\gamma^* = (m_A,m_B,n_A,n_B)$ will  be given by 
$\mathcal{P}[\gamma^*] = |\langle n_A n_B | U^\dagger | m_A m_B \rangle|^2 p_{m_A m_B}$.
The ratio of the two processes  reduce to $\mathcal{P}[\gamma]/\mathcal{P}[\gamma^*] = p_{n_A n_B}/p_{m_A m_B}$, since the dynamical term cancels out (as usual).  
To make the physics of this ratio more evident, we introduce the stochastic mutual information 
$ I_{n_A n_B} = \ln p_{n_A n_B}/p_{n_A} p_{n_B}$,
where $p_{n_A} = \sum_{n_B} p_{n_A n_B}$ (and similarly for $p_{n_B}$) are the marginal distributions of the initial state, which we chose to be thermal, $p_{n_A} = e^{-\beta_A E_{n_A}}/Z_A$. 
The average of $I_{n_A n_B}$ over $p_{n_A n_B}$ yields the mutual information of the dephased state
\begin{equation}\label{ThermoCorr_stochastic_MI_TPM}
    \langle I_{n_A n_B} \rangle = \sum\limits_{n_A, n_B} p_{n_A n_B} \ln \frac{p_{n_A n_B}}{p_{n_A} p_{n_B}} = \mathcal{I}_{\Delta(\rho_{AB})} (A\! : \! B). 
\end{equation}
where $\mathcal{I}_{\rho}(A\! : \! B)$ is defined in Eq.~(\ref{Qgen_MI}).

Writing $p_{n_A n_B} = p_{n_A} p_{n_B} e^{I_{n_A n_B}}$ allows us to express
$\mathcal{P}[\gamma]/\mathcal{P}[\gamma^*] =
\big(p_{n_A} p_{n_B}/p_{m_A} p_{m_B}\big) e^{ -\Delta I[\gamma]}
$,
where $\Delta I[\gamma] = I_{m_A m_B} - I_{n_A n_B}$. 
But since the reduced states are thermal, $p_{n_\alpha}/p_{m_\alpha} = e^{\beta_\alpha (E_{m_\alpha} - E_{n_\alpha})}$,  and we may finally write
\begin{equation}\label{ThermoCorr_FT_TPM}
    \frac{\mathcal{P}[\gamma]}{\mathcal{P}[\gamma^*]} = e^{(\beta_B - \beta_A) q_B[\gamma] - \Delta I[\gamma]},
\end{equation}
where we also used the fact that $E_{m_A} - E_{n_A}= - (E_{m_B} - E_{n_B})$.

Eq.~(\ref{ThermoCorr_FT_TPM}) represents a modified exchange fluctuation theorem, generalizing the results of~\cite{Jarzynski2004a} to the case where $A$ and $B$ have  initial correlations. 
Eq.~(\ref{ThermoCorr_FT_TPM}) implies a non-equilibrium equality $\langle e^{(\beta_B - \beta_A) q_B[\gamma] - \Delta I[\gamma]} \rangle = 1$, which yields the bound
\begin{equation}\label{ThermoCorr_bound_DeltaI_TMP}
    (\beta_B - \beta_A) \langle q_B[\gamma]\rangle \geqslant  \langle \Delta I[\gamma]\rangle.
\end{equation}
This is structurally similar to Eq.~(\ref{ThermoCorr_bound_DeltaI}). 
However, as discussed before, they cannot be directly compared since they pertain to different processes due to the dephasing action of the first measurement. 

The above results show clearly that, when constructing fluctuation theorems,  quantum correlations are fundamentally hampered by the backaction of the two-point measurement scheme. 
A way to circumvent this is to use the notion of augmented trajectories, first discussed by Dirac~\cite{Dirac1945} and used more recently in~\cite{Park2017,Micadei2019}.  
We decompose the initial (correlated) state of $AB$ as $\rho_{AB} = \sum_s p_s |s\rangle\langle s |$, where $|s\rangle$ are eigenvectors living on the composite Hilbert space of $AB$. 
Before the dynamics, we  perform instead a measurement in the basis $|s\rangle$. The second measurement can be in the energy basis, as in Sec.~\ref{ssec:XFT_corr}, since it does not matter if we destroy the correlations after the end of the protocol. 

The quantum trajectory will therefore be described in this case by the quantum numbers $\gamma = (s,m_A,m_B)$ and the corresponding path probability will be given, instead of Eq.~(\ref{ThermoCorr_pathProb_TPM}), by
$\mathcal{P}[\gamma] = |\langle m_A m_B | U | s\rangle|^2 p_s$.
Knowing the outcome $s$ of the first measurement, however, does not uniquely specify which energy eigenstates $|n_A n_B\rangle$ the two systems were initially in. 
In order to account for this, we augment the trajectories by considering the conditional probability $p_{n_A n_B|s} = |\langle n_A n_B | s \rangle|^2$ that $AB$ are found in $|n_A n_B\rangle$ given that globally they are in $|s\rangle$. 
The augmented trajectory $\tilde{\gamma} = (s,n_A,n_B,m_A,m_B)$ will then have a path probability 
\begin{equation}\label{ThermoCorr_pathProb_augmented_augmented}
\tilde{\mathcal{P}}[\tilde{\gamma}] = |\langle m_A m_B | U | s\rangle|^2 p_s p_{n_A n_B|s}.
\end{equation}
This formulation fixes the issues that arise from the backaction of the first measurement. 
For instance, as shown in Ref.~\cite{Micadei2019}, it leads to the full identity~(\ref{ThermoCorr_bound_DeltaI}) and not its dephased version~(\ref{ThermoCorr_bound_DeltaI_TMP}). 

Eq.~(\ref{ThermoCorr_pathProb_augmented_augmented}) also illustrates well a recurring problem in extending thermodynamics to the quantum regime.
Thermodynamics does not deal with states, but with processes; i.e., with transformations between states. 
Assessing these transformations therefore touches on the  inevitable measurement backaction. 
Eq.~(\ref{ThermoCorr_pathProb_augmented_augmented}) circumvents this by constructing a distribution free from any backaction. 
This distribution, however, has to be constructed using full state tomography. 
An alternative approach, put forth in~\cite{Levy2019}, formulates the problem using instead the notion of quasiprobabilities; that is, probabilities which can take on negative values.
As the authors show, these negativities are directly related  to the notion of contextuality.  

\section{\label{sec:classical}Quantum dynamics and the classical limit}
%
%

The global unitary map~(\ref{Qgen_reduced_map}) is extremely general and represents the basic structure behind most open system dynamics (the only assumption in it is that $S$ and $E$ are initially uncorrelated). 
To make it practical, however, this map has to be specialized to specific paradigms. 
The usual paradigm in open quantum systems \cite{Gardiner2004,Breuer2007,Rivas2012}
is to assume that the environment is macroscopically large and the unitary is left turned on for an arbitrary time. 
Eq.~(\ref{Qgen_global_map}) is then naturally reinterpreted as the continuous time map 
\begin{equation}\label{M_continuous_map}
    \rho_S(t) = \mathcal{E}_t (\rho_S(0)) = \tr_E \bigg\{ U(t) \big(\rho_S(0) \otimes \rho_E \big) U^\dagger(t) \bigg\}.
\end{equation}
Common questions in the theory of open quantum systems, such as whether or not the map will be divisible, are all contained in the properties of $\rho_E$ and $U(t)$. 

All results derived in Sec.~\ref{sec:Qgen} for the entropy production remain valid in this case, although it becomes more natural to study the entropy production rate $\dot{\Sigma} = d\Sigma/dt$. 
An important observation, however, is that even though $\Sigma \geqslant 0$ by construction, this is not in general guaranteed for $\dot{\Sigma}$. 
This, of course, is expected to happen for macroscopic environments, but  has to be analyzed in a case-by-case basis. 
In fact, temporary negativities in $\dot{\Sigma}(t)$ can be used as a measure of  non-Markovianity~\cite{BreuerRMP,deVega2017}, as they represent instances of time where information backflows to the system (which fits well with the interpretation of $\dot{\Sigma}$ as a measure of irreversibility).  This will be reviewed in Sec.~\ref{ssec:non_mark}.

More serious difficulties may arise, however, when one is interested in quantum master equations derived from the map~(\ref{M_continuous_map}).
The problem is that master equations use several approximations  to describe the dynamics solely from the optics of the reduced state of the system. They therefore have no information about the global $S+E$ state, which is paramount for quantifying  entropy production. 
Thus, while these approximations may be reasonable for describing the dynamics, they can be disastrous for the \emph{thermo}dynamics. 
Of course, in many situations no issues arise. And, in fact, master equations are routinely employed in the study of thermodynamics, e.g. in the context of transport. 
However, there are situations where one may arrive at inconsistencies.
For instance, in Ref.~\cite{Levy2014}  it was shown how local master equations seem to violate the second law (allowing, e.g., heat to flow from cold to hot). 
If one has access also to the global dynamics, this would never happen by construction. 
This was used in~\cite{DeChiara2018} to reconcile local master equations with thermodynamics.

The thermodynamics of quantum master equations has to be analyzed in a case-by-case basis. 
Instead, we have opted to focus in this review on an alternative paradigm of open system, called collisional models (also called ``repeated interactions'').
These models, to be detailed below, have been used for a long time in different contexts \cite{Rau1963,Englert2002,Scarani2002}. 
However, they recently gained a surge in popularity~\cite{Karevski2009,Giovannetti2012,McCloskey2014,Landi2014b,Barra2015,Lorenzo2015,Strasberg2016,Pezzutto2016,Cusumano2018,Pereira2018}, largely because they allow full control over the approximations being employed.
{\color{black}
We will review the thermodynamics of collisional models, which were laid out on firm grounds in~\cite{Strasberg2016}. Throughout the section, we will also connect them with master equations, following~\cite{Barra2015,DeChiara2018}.
In fact, we will discuss how some of the famous results for the thermodynamics of master equations can actually be derived as limiting cases of such models.
This includes the famous result by Spohn~\cite{Spohn1978}, as well as the formulation of continuous measurements in~\cite{Horowitz2013a}. 
We also review how collisional models can be used to see the emergence of a classical limit and the classical rules of stochastic thermodynamics~\cite{Cwikli2015}. 
}

\subsection{\label{sec:CM}Collisional models}

Collisional models draws inspiration from Boltzmann's original \emph{Stosszahlansatz} (molecular chaos hypothesis).
The open system dynamics is envisioned as a series of sequential interactions, where in each time interval the system only interacts with a tiny fraction of the environment (which we shall henceforth refer to as an \emph{ancilla}).
After this interaction the ancilla is discarded and a fresh new one is introduced, again prepared in a thermal state. 
This is what happens, for instance, in classical Brownian motion: at each moment the particle only interacts with a small number of molecules.
Moreover, after they interact, the molecules return to the bath and never interact with the system again. 

The collisions may be assumed to happen at random times or be sequential. We focus on the latter for concreteness and assume each event lasts for a time $\tau$.  
If we let $\rho_{A_n}$ denote the density matrix of the $n$-th ancilla, then the collisional model can be described by the map 
\begin{equation}
    \rho_S^{n+1} = \tr_{A_n} \bigg\{ U_{SA_n} \big(\rho_S^{n} \otimes \rho_{A_n} \big) U_{SA_n}^\dagger \bigg\} := \mathcal{E}_n (\rho_S^{n}),
    \label{M_CM}
\end{equation}
where $\rho_S^n = \rho_S (n \tau)$ is the state of the system \emph{before} interacting with the  $n$-th ancilla.
As can be seen, this map is  nothing but a composition of the original map~(\ref{Qgen_reduced_map}). 
Hence, \emph{all} thermodynamic properties derived in Sec.~\ref{sec:Qgen} also hold for each stroke of the collisional model. 
Moreover, since the ancillas are assumed to be independent, it is trivial to compose the properties of multiple strokes. 
From a thermodynamic perspective, this offers a \emph{monumental} advantage. 

\begin{figure}
    \centering
    \includegraphics[width=0.5\textwidth]{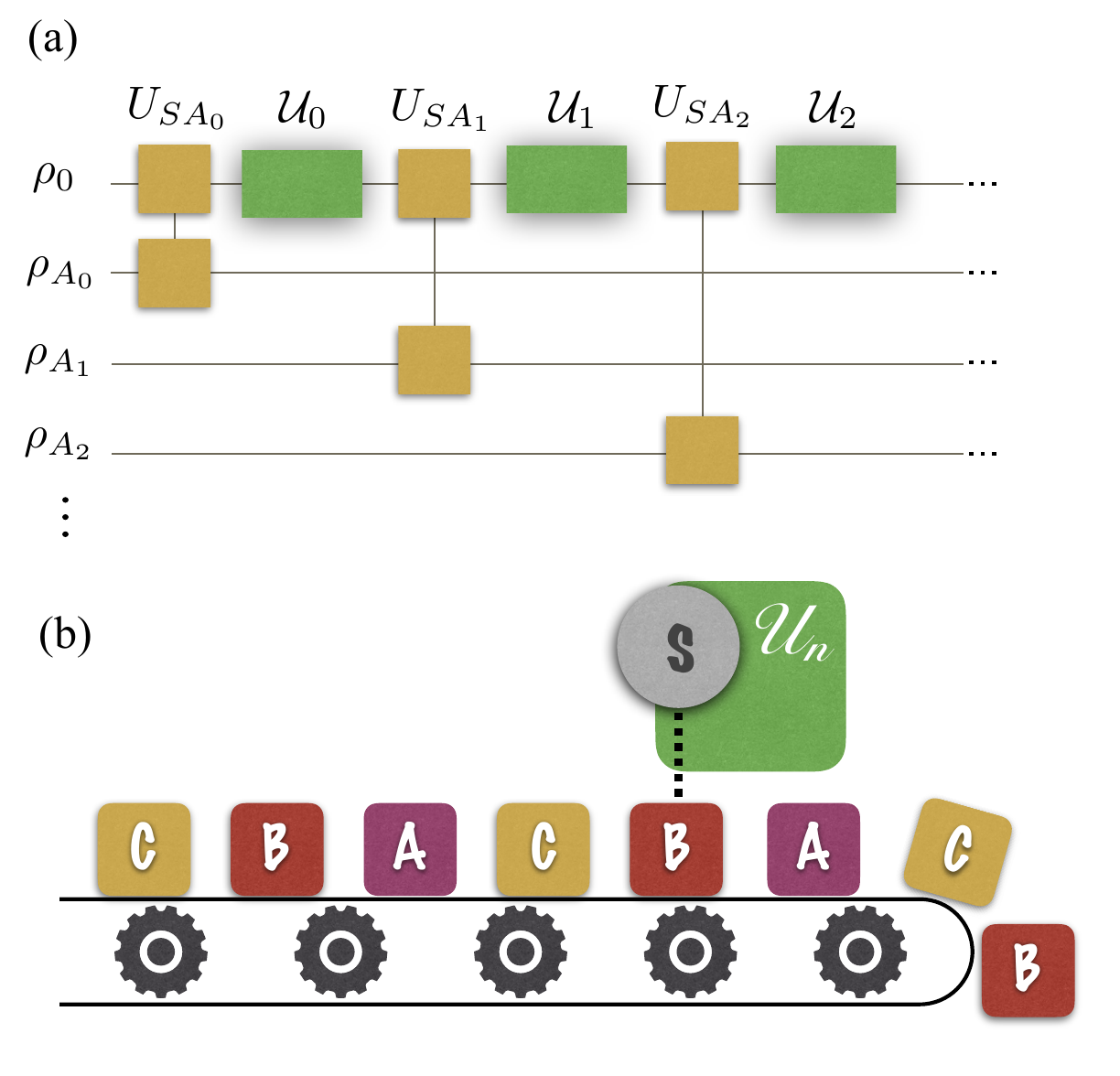}
    \caption{
    (a) Diagramatic illustration of the collisional model in Eq.~(\ref{M_CM_unitary}).
    (b) Scheme for studying non-equilibrium steady-states (NESSs) obtained when a system is coupled to multiple reservoirs. In this case the ancillas cycle through an alphabet of states, $\rho_A$, $\rho_B$, $\rho_C$, $\rho_A$, $\rho_B$, $\ldots$ so that the system will never reach  equilibrium, even when $\mathcal{U}_n = 1$ (no unitary strokes).}
    \label{fig:collisional_model}
\end{figure}

We can also increment the collisional model with the additional assumption that in between each $SA$ stroke, the system also undergoes a unitary evolution (see Fig.~\ref{fig:collisional_model}(a)). 
The map~(\ref{M_CM}) is then updated to 
\begin{equation}
    \rho_S^{n+1} = \mathcal{U}_n (\mathcal{E}_n (\rho_S^{n}))
    \label{M_CM_unitary}
\end{equation}
where $\mathcal{U}_n(\rho_S) = U_n \rho_S U_n^\dagger$ is a unitary stroke described by an arbitrary unitary $U_n$ acting only on $S$. 
The situation where the system is always close to equilibrium was recently analyzed in~\cite{Scandi2019}.
Since the unitary strokes $\mathcal{U}_n$ involve no heat by construction, this kind of map composition is a useful way of separating between heat and work, a quantum generalization of the type of splitting used, e.g., in~\cite{Crooks1998}. 
Of course, as discussed in Sec.~\ref{sec:Qgen_thermal_ops}, the ancilla strokes $\mathcal{E}_n$ may also contain a contribution due to work, depending on whether or not $U_{SA_n}$ satisfies strict energy conservation, Eq.~(\ref{Qgen_strict_energy_conservation}).

The states of the ancillas in the collisional model~(\ref{M_CM_unitary}) do not have to be identical. 
This can be  used  to implement non-trivial limit cycles. 
The basic idea is illustrated in Fig.~\ref{fig:collisional_model}(b). 
It consists in setting the ancillas to cycle through an alphabet of $m$ states, such as $\rho_A$, $\rho_B$, $\rho_C$, $\rho_A$, $\rho_B$, $\rho_C$, $\ldots$ in the example of the figure. 
If the composite map
\begin{equation}
    \Phi(\rho_S) = \mathcal{U}_m \circ \mathcal{E}_m \circ \ldots \circ \mathcal{U}_1 \circ \mathcal{E}_1 (\rho_S), 
    \label{M_CM_composite_alphabet}
\end{equation}
is applied for a sufficiently long time, any transients related to the system's initial conditions will vanish and the system will reach a limit cycle, characterized by $\rho_S^* = \Phi(\rho_S^*)$. 
Because the ancillas are always changing, however, this limit cycle will not be a fixed point of the individual maps~(\ref{M_CM_unitary}), only of the composite map~(\ref{M_CM_composite_alphabet}). 
As a consequence, the system will never reach a steady-state but will, instead,  keep bouncing back and forth within the limit cycle. This can be used to generate a diverse set of rich dynamics describing engine-like behaviour. 

For concreteness, we assume that within the system-ancilla strokes the system Hamiltonian remains fixed at $H_S^n$. 
During the subsequent unitary stroke, on the other hand, it changes from $H_S^n$ to $H_S^{n+1}$. 
The precise way through which this change takes place is encoded in the unitary $\mathcal{U}_n$. Heat is then defined, as in Sec.~\ref{sec:Qgen}, as the change in energy of the ancillas [cf. Eq.~(\ref{Qgen_QE})]; i.e., 
$Q_{A_n} := \tr\big\{ H_{A_n} \big( \rho_{A_n}' - \rho_{A_n})\big\}$
This is to be compared with the total change in energy of the system, 
\begin{equation}
    \label{CM_DeltaHSn}
    \Delta H_S^n = \tr\big( H_S^{n+1} \rho_S^{n+1} - H_S^n \rho_S^n \big).
\end{equation}
The mismatch between $Q_{A_n}$ and $\Delta H_S^n$ is then attributed entirely to work. 
This work, however, may have a contribution from the on/off work of the system-ancilla interaction and a contribution from the unitary $U_n$:
\begin{IEEEeqnarray}{rCl}
\label{CM_work_on_off}    
    W_n^\text{on/off} &=& \tr \bigg\{ H_S^n\big[ \mathcal{E}_n(\rho_S^n) - \rho_S^n\big]\bigg\} +Q_{A_n}, \\[0.2cm]
\label{CM_work_u}
    W_n^u &=& \tr\bigg\{ H_S^{n+1} \rho_S^{n+1} - H_S^n \mathcal{E}_n (\rho_S^n)\bigg\}.
\end{IEEEeqnarray}
The first law therefore decomposes as 
\begin{equation}
    \Delta U_S^n  = W_n^{\text{u}} + W_n^{\text{on/off}} + Q_{A_n}. 
\end{equation}
Notice how $\Delta H_S^n$ in Eq.~(\ref{CM_DeltaHSn}) is a function of state, whereas $Q_{A_n}$ and $W_n$ are not. 

As for the second law, based on the results of Sec.~\ref{sec:Qgen}, we have three tiers of possible expressions {\color{black} for the entropy production}: 
\begin{IEEEeqnarray}{rCl}
\label{CM_Sigma_line1}
\Sigma_n &=& \mathcal{I}_{\rho_{SA_n}'} (S\! : \! A_n) + S(\rho_{A_n}' || \rho_{A_n}) \\[0.2cm]
\label{CM_Sigma_line2}
&=& \Delta S_S^n + \beta_n Q_{A_n} \\[0.2cm]
\label{CM_Sigma_line4}
&=& S(\rho_S^n || \rho_S^\text{n,th}) - S(\rho_S^{n+1} || \rho_S^\text{n,th}),
\end{IEEEeqnarray}
where $\Delta S_S^n = S(\rho_S^{n+1}) - S(\rho_S^n)$ 
is the change in the entropy of the system in the map~(\ref{M_CM_unitary}). 
The first line is the general definition~(\ref{Qgen_Sigma}) and holds for any ancillary state. 
The second line is only true if  the ancillas are thermal, not necessarily at the same temperature [Eq.~(\ref{Qgen_Sigma_thermal})]. 
Finally, the third line is only true for thermal operations (i.e., if the ancillas are thermal \emph{and} the unitary satisfies the strict energy conservation condition~(\ref{Qgen_sigma_thermal_op})).
{\color{black}If the ancillas are identical, some of the indices $n$ may be dropped and the expressions simplify a bit.}

\subsection{\label{sec:emergence}The emergence of a preferred basis}

The classical limit is usually associated with the emergence of a preferred basis in which coherence among the basis elements tend to be suppressed. 
In the so-called ``einselection'' paradigm~\cite{Zurek1981}, this basis emerges due to the contact with a heat bath. 
Thermal operations (Sec.~\ref{sec:Qgen_thermal_ops}) provide a clear illustration of this principle and also highlight some of the subtle issues that arise in  the classical limit.  

We consider here the collisional model in Eq.~(\ref{M_CM_unitary}) and assume that the ancillary stroke~(\ref{M_CM}) is a thermal operation (Sec.~\ref{sec:Qgen_thermal_ops}). 
During the unitary stroke, the Hamiltonian is assumed to change from $H_S^n = H_S(\lambda_n)$ to $H_S^{n+1} = H_S(\lambda_{n+1})$, where $\lambda$ represents a generic work parameter. 
For simplicity, however, we  assume that this change is much faster than the system-ancilla stroke, so that we may set $\mathcal{U}_n \simeq 1$.

Let $H_S^n = \sum_i E_i^n |i_n\rangle\langle i_n|$ denote the spectral decomposition of $H_S^n$ in terms of the eigenvalues $E^n_i$ and corresponding eigenstates $\ket{i_n}$ at each given time $n$.
We assume that the eigenvalues $E_i^n$ are non-degenerate.
As shown in~\cite{Cwikli2015}, if the map~(\ref{M_CM}) is a thermal operation,  the populations $\langle i_n | \rho_S^n | i_n \rangle$ at the instantaneous eigenstates will evolve according to the classical Markov chain
\begin{equation}\label{CL_Markov_step}
    \langle i_n | \rho_S^{n+1} | i_n \rangle = \sum\limits_j M_n(i|j) \langle j_n | \rho_S^n | j_n \rangle,
\end{equation}
where $M_n(i|j) = \sum_{\mu,\nu} q_\nu^n |\langle i_n \mu_n | U_{SA_n} | j_n \nu_n \rangle |^2$, with $q_\nu^n$ and $|\nu_n\rangle$, $|\mu_n\rangle$ being the initial populations and eigenstates of ancilla $A_n$; i.e., $\rho_{A_n} = e^{-\beta H_E^n}/Z_E^n =   \sum_\nu q_\nu^n |\nu_n\rangle\langle \nu_n|$. 
We call attention to the fact that, in order to make sure each step is a thermal operation, the states of the ancillas and the unitaries $U_{SA_n}$ have to change to adjust to the strict energy conservation condition~(\ref{Qgen_strict_energy_conservation}); this makes the above construction difficult to realize exactly, although it is  realizable approximately, as we discuss below. 

Notice how the left hand side of~\eqref{CL_Markov_step} contains $\langle i_n | \rho_S^{n+1} | i_n \rangle$, which differs in general from $\langle i_{n+1} | \rho_S^{n+1} | i_{n+1} \rangle$.
This highlights a unique property of quantum dynamics; namely that the action of the work agent may change not only the populations $E_i^n$ of the system, but also rotate the eigenbasis $|i_n\rangle$. 
The notion of ``population and coherences'' in Eqs.~(\ref{CL_Markov_step}) and (\ref{CL_coherences_step}) should thus be interpreted with care, as they change with each step.
As a consequence, even quasi-static dynamics, which are usually somewhat dull for classical systems, may present interesting and highly non-trivial effects, which are of genuine quantum nature. 
This was recently explored in Ref.~\cite{Miller2019,Scandi2019} and will be reviewed in Sec.~\ref{sec:inf_quenches}.

For the remainder of this Section, we focus on the case where $[H_S^n,H_S^{m}] = 0$ for all $n,m$. 
This means that during the work strokes, the energy levels of the system may change, but the orientation of the eigenbasis $|i\rangle$ remains fixed.
One may then define the populations $p_i^n = \langle i | \rho_S^n |i \rangle$, so that Eq.~\eqref{CL_Markov_step} is converted into the classical Markov chain
\begin{equation}\label{continuous_MarkovChain}
    p_i^{n+1} = \sum\limits_j M_n(i|j) p_j^n. 
\end{equation}
The $M_n(i|j)$ are simply transition probabilities (their time-dependence comes from the fact that the Hamiltonian may be changing in time). 
Moreover, as the ancillas are thermal, they satisfy the detailed balance condition
\begin{equation}
    M_n(i|j) = M_n(j|i) e^{-\beta (E_i^n - E_j^n)}. 
\end{equation}
Thus, by all standards, the populations evolve according to an entirely classical evolution. 
Most importantly, the evolution of populations and coherences are completely decoupled. 
Indeed, the latter
are found to evolve according to 
\begin{equation}\label{CL_coherences_step}
 \langle i | \rho_S^{n+1} | j \rangle=
\mathcal{R}_{ij}^n~ 
\langle i | \rho_S^{n} | j \rangle
,
\end{equation}
where
$\mathcal{R}_{ij}^n = \sum_{\mu,\nu} q_\nu^n \langle i \nu_n | U_{SA_n} | i \mu_n \rangle \langle j \nu_n |U_{SA_n}^\dagger | j \mu_n\rangle$.
One may verify that $|\mathcal{R}_{ij}^n| < 1$~\cite{Cwikli2015}, 
so the off-diagonals are suppressed further and further with each collision, until eventually vanishing. 

This example clearly shows the emergence of a preferred basis. 
Due to the strict energy conservation property of thermal operations, the energy basis of the system is selected as a preferred basis by the environment, a process called ``environment-induced'' selection, or einselection.
Crucially, this effect is clearly manifested in the entropy production.  
The entropy produced at each stroke will be given by Eq.~(\ref{CM_Sigma_line4}), with $\rho_S^{n,th} = e^{-\beta H_S^n}/Z_S^n$. 
We may now split the relative entropy as 
\begin{equation}\label{CL_entropy_coherence_split}
    S(\rho_S^n || \rho_S^\text{n,th}) = S(\bm{p}^n || \bm{p}^\text{n,th}) + C(\rho_S^n), 
\end{equation}
where $S(\bm{p}^n || \bm{p}^\text{n,th})$ is the \emph{classical relative entropy} between the probability distributions $p_i^n$ and $p_i^{n,th} = e^{-\beta E_i^n}/Z_S^n$; the classical relative entropy is defined as
\begin{equation}\label{KL}
    S(\bm{p} || \bm{q} ) = \sum\limits_i p_i \ln p_i/q_i. 
\end{equation}
The second term in Eq.~(\ref{CL_entropy_coherence_split}), on the other hand, is the relative entropy of coherence in the energy eigenbasis $|i\rangle$,  $C(\rho_S^n) = S(\bm{p}^n) - S(\rho_S^n)$.
Plugging this in Eq.~(\ref{CM_Sigma_line4}) allows us to split the entropy production of each step in two parts~\cite{Santos2019,Mohammady2020} 
\begin{equation}
\Sigma_n = \Sigma_n^\text{cl} + \Sigma_n^\text{qu},
\end{equation}
where
\begin{IEEEeqnarray}{rCl}
    \Sigma_n^\text{cl} &=& S(\bm{p}^n || \bm{p}^\text{n,th}) - S(\bm{p}^{n+1} || \bm{p}^\text{n,th}),  \\[0.2cm]
    \Sigma_n^\text{qu} &=& C(\rho_S^n) - C(\rho_S^{n+1}).
\end{IEEEeqnarray}
The term $\Sigma_n^\text{cl}$ is a purely classical contribution and coincides with the formulation used in classical stochastic processes~\cite{Schnakenberg1976}. 
It describes the irreversibility associated with the system having to adapt its populations to those imposed by the environment. 
In addition to it, however, we also have an extra term $\Sigma_n^\text{qu}$ describing the irreversibility due to the way the environment process  quantum coherences. This thus represents a genuinely quantum contribution to the entropy production. 
Both terms are also individually non-negative~\cite{Santos2019}.

\subsection{\label{sec:CM_continuous}Continuous-time limit}

When the interaction time $\tau$ of each collision is small, the stroboscopic dynamics in Eq.~(\ref{M_CM_unitary}) can usually be converted into a continuous-time master equation for the system~\cite{Englert2002,Strasberg2016}.
In view of the importance of quantum master equations, we briefly review here the basic procedure. 
The  idea is to  construct a generator $\mathcal{L}$ according to 
\begin{equation}
    \frac{d \rho_S}{d t} := \lim\limits_{\tau\to 0} \frac{\rho_S^{n+1} - \rho_S^n}{\tau} = \mathcal{L}(\rho_S),
    \label{M_CM_limit_def}
\end{equation}
where clearly $\rho^{n+1}_S$ and $\rho^n_S$ are separated by the collision time $\tau$. The limit process in Eq.~\eqref{M_CM_limit_def}, however, has to be interpreted with care: Strictly speaking, one cannot take $\tau \to 0$, as this would imply no interaction at all.
Instead, this is to be interpreted as a leading order contribution to a series expansion. 
In a nutshell, the main idea  is to take $\tau$ sufficiently small to ensure that $d\rho_S/d t$ becomes  a \emph{smooth function}. 
Ultimately, this is a coarse-graining argument, which is actually ubiquitous in stochastic process. It also arises, for instance, in the classical Langevin equation describing Brownian motion [cf. Ref.~\cite{Cresser2017} for a critical assessment of the coarse-graining approach].

We shall focus on two distinct scenarios. 
First we will assume the Hamiltonian is time-independent but the collisions are not energy preserving. 
Then we consider the case where the Hamiltonian is time-dependent and the collisions are thermal operations (which is the same scenario discussed in Sec.~\ref{sec:emergence}).
The starting point for both cases is actually the same.
We thus remain general here and specialize the results in Sec.~\ref{sec:dynamics_on_off}.

We focus  on a single system-ancilla collision, where the Hamiltonian is given by  $H = H_S + H_A + V$ and the initial states are $\rho_S$ and $\rho_A$ for system and ancilla (all indices $n$ are omitted for now). 
The evolution of the system in this single collision will be given by 
\[
\rho_S' = \tr_A \bigg\{ e^{- i \tau H} (\rho_S \otimes \rho_A ) e^{i \tau H} \bigg\}. 
\]
Expanding the exponentials in a power series and  dividing by $\tau$ on both sides leads to 
\begin{equation}\label{M_CM_limit_naive}
    \frac{\rho_S'-\rho_S}{\tau} = - i \big[H_S+ \tr_A(V\rho_A),\rho_S) \big] - \frac{\tau}{2} \tr_A [V,[V,\rho_S\otimes \rho_A]].
\end{equation}
This formula illustrates well the physical meaning of the limit~(\ref{M_CM_limit_def}). 
If we naively take $\tau\to 0$, only the first term would survive. But this term contains only the original system Hamiltonian plus a unitary contribution (Lamb-shift) $\tr_A (V\rho_A)$. 
Moreover, this Lamb-shift is often zero for most choices of ancilla states and interactions (see~\cite{Rivas2012} for more details and ~\cite{Rodrigues2019} for a counterexample). 
Indeed, we shall henceforth assume that $\tr_A (V\rho_A)=0$.

The actual dissipative contribution, which is what we are interested in, corresponds to the second term in Eq.~(\ref{M_CM_limit_naive}). 
But this is still of order $\tau$ and would hence vanish if $\tau \to 0$. 
The limit~(\ref{M_CM_limit_def}) should therefore  correspond to a limit where $(\rho_S' - \rho_S)/\tau$ is sufficiently smooth to be interpreted as a derivative, but the last term is nonetheless not vanishingly small. 
A more systematic way of implementing this is to introduce a fictitious scaling  of the potential by changing $V \to V/\sqrt{\tau}$. 
This means that while we take the interaction time to be very short, we also take it to be very strong in the same proportion. 
{\color{black} This scaling is not physical but helps 
to identify 
the terms to neglect in the series expansion.} 
An identical situation also appears in classical Brownian motion:
{\color{black}
the 
white noise entering the Langevin equation can also be seen as resulting from a sequence of independent kicks, each occurring 
for an infinitesimal time $\Delta t$ and whose magnitude scales as $1/\sqrt{\Delta t}$.
}

With such rescaling 
Eq.~(\ref{M_CM_limit_naive}) becomes
\begin{equation}\label{M_CM_discrete_step}
    \rho_S^{n+1} = \rho_S^n - i \tau [H_S^n, \rho_S^n] + \tau \mathcal{D}_n(\rho_S^n), 
\end{equation}
where we already reintroduced all indices $n$. We also defined
\begin{equation}\label{M_CM_dissipator}
\mathcal{D}_n(\rho_S) = - \frac{1}{2} \tr_{A_n} [V_n,[V_n,\rho_S\otimes \rho_{A_n}]]. 
\end{equation}
Taking the limit $\tau \to 0$ then finally leads to 
\begin{equation}\label{CM_simplest}
    \frac{d \rho_S}{d t} = -i [H_S(t),\rho_S] +  \mathcal{D}_t(\rho_S),
\end{equation}
where $H_S(t=n\tau) = H_S^n$ and similarly for $\mathcal{D}_t$. 

{\color{black} Eq.~(\ref{M_CM_dissipator}) can always be put in Linbdlad form~\cite{Breuer2007} by decomposing the interaction as $V_n = \sum_k M_k F_k = \sum_k F_k^\dagger M_k^\dagger $, where $M_k$ and $F_k$ are Hermitian operators of system and ancilla respectively. 
This leads to 
\begin{equation}
\mathcal{D}(\rho_S) = \sum\limits_{k,q} \langle F_q^\dagger F_k\rangle_n \bigg[ M_k \rho_S M_q^\dagger - \frac{1}{2} \{ M_q^\dagger M_k, \rho_S\} \bigg],
\end{equation}
where $\langle F_q^\dagger F_k \rangle_n = \tr(F_q^\dagger F_k \rho_{A_n})$ is
, by construction, positive semi-definite. The evolution is thus Markovian and can always be put in canonical form.

To provide another example, we consider 
An 
interaction appearing often in the literature, i.e.
\begin{equation}\label{CM_V_simple}
    V = \sum\limits_k g_k (L_k^\dagger A_k + L_k A_k^\dagger),
\end{equation}
where $L_k$ and $A_k$ are operators for the system and ancilla respectively. 
We assume  
$\langle A_k A_q \rangle = 0, \langle A_k^\dagger A_q \rangle = \delta_{k,q} \langle A_k^\dagger A_k\rangle$.
Eq.~(\ref{M_CM_dissipator}) then acquires the familiar form 
\begin{equation}\label{CM_dissipator_standard}
    \mathcal{D}(\rho_S) = \sum\limits_k \bigg\{ \gamma_k^+ D[L_k] + \gamma_k^- D[L_k^\dagger]\bigg\},
\end{equation}
where $D[L] = L \rho_S L^\dagger - \frac{1}{2} \{L^\dagger L, \rho_S\}$ and $\gamma_k^+ = g_k^2 \langle A_k A_k^\dagger \rangle$, $\gamma_k^- = g_k^2 \langle A_k^\dagger A_k \rangle$.}

A further specialization is to the case where the $A_k$ are eigenoperators of the ancilla Hamiltonian. That is, they satisfy $[H_A,A_k] = - \omega_k A_k$ for some set of Bohr (transition) frequencies $\omega_k$. 
If the state of the ancillas is a thermal state $\rho_A^\text{th} = e^{-\beta H_A}/Z_A$, then this property will ensure that the coefficients $\gamma_k^\pm$  satisfy detailed balance
\begin{equation}\label{CM_detailed_balance}
    {\gamma_k^-}/{\gamma_k^+} = e^{-\beta \omega_k}.
\end{equation}
{\color{black}By considering the Fermi-Dirac distribution with $f_k:=\langle A_k^\dagger A_k\rangle = (e^{\beta \omega_k}+1)^{-1}=1-\langle A_k A_K^\dagger \rangle$, 
Eq.~\eqref{CM_dissipator_standard} becomes
$\mathcal{D}(\rho_S) = \sum_k g_k^2 \big\{ (1-f_k) D[L_k] + f_k D[L_k^\dagger]\big\}$. 
We might instead take the Bose-Einstein distribution with $n_k:=\langle A_k^\dagger A_k\rangle = (e^{\beta \omega_k}-1)^{-1}=\langle A_k A_k^\dagger\rangle-1$ 
to get 
$\mathcal{D}(\rho_S) = \sum_k g_k^2 \big\{ (n_k+1) D[L_k] + n_k D[L_k^\dagger]\big\}$. The entropy production in this case must be computed using Eq.~\eqref{CM_Sigma_line2}, since the reservoirs are assumed to be thermal, but $V$ is not necessarily a thermal operation.
Since we are interested in the continuous-time limit, we compute instead the entropy production rate
\begin{equation}\label{CM_sigma_dot_temporary}
    \dot{\Sigma} = \lim\limits_{\tau \to 0} \frac{\Delta S_S^n}{\tau} + \beta Q_{A_n}. 
\end{equation}
The first term clearly tends to $dS(\rho_S)/dt$, the rate of change of the system's von Neumann entropy. 
But the last term still involves a quantity related to the ancillas. 
We will now discuss under which conditions Eq.~\eqref{CM_sigma_dot_temporary} can be recast solely in terms of quantities related to the system.
}

\subsection{\label{sec:dynamics_on_off}On/off work and Spohn's separation}

{\color{black}
Eq.~\eqref{CM_sigma_dot_temporary} highlights the  need for addressing under which conditions can the entropy production be written solely in terms of system-related quantities. 
This was already broadly discussed in Sec.~\ref{sec:Qgen}. 
However, here it acquires additional significance  since master equations are often used as phenomenological models of open system dynamics, without knowledge of the baths and system-bath interactions. 
}
To gain insight into this non-trivial question, consider first the case where the system Hamiltonian is time-independent, $H_S^n = H_S$ and the ancillas are all identically prepared, $\rho_{A_n} = \rho_A$.
The change in energy of the system and ancilla in one collision can be found from Eq.~(\ref{M_CM_discrete_step}) and the corresponding analogous equation for the evolution of $\rho_{A_n}$:
\begin{IEEEeqnarray}{rCl}
\label{CM_CL_DeltaHS}
    \Delta H_S^n &=& - \frac{\tau}{2} \tr\bigg\{ [V,[V,H_S]] \rho_S^n \otimes \rho_{A}\bigg\},  \\[0.2cm]
\label{CM_CL_DeltaHA}    
    \Delta H_{A_n} &=& - \frac{\tau}{2} \tr\bigg\{ [V, [V, H_{A}]] \rho_S^n \otimes \rho_{A}\bigg\} \equiv Q_{A_n}, 
\end{IEEEeqnarray}
where $Q_{A_n}$ is precisely the quantity appearing in Eq.~\eqref{CM_sigma_dot_temporary}.
In general, the violation of strict energy conservation, $[V, H_S+H_A] \neq 0$ implies that 
$\Delta H_S^n \neq -Q_{A_n}$ and hence there will be a finite amount of on/off work (Sec.~\ref{sec:Qgen_thermal_ops}).

This is where the difficulties in dealing with the thermodynamics of master equations start. 
If one has  only access to Eq.~(\ref{CM_simplest}), it is not clear how to split $\Delta H_S^n$ into heat and work. 
For, according to~(\ref{CM_simplest}), one should  have 
\begin{equation}\label{CM_CL_energy_splitting}
    \frac{d \langle H_S\rangle}{d t} = \tr_S\big\{ H_S \mathcal{D} (\rho_S) \big\}, 
\end{equation}
and it is not at all obvious which part of this expression is heat and which part is work (something which is evident from the global dynamics). 
The problem is that, in general, $Q_{A_n}$ simply cannot be written in terms of quantities pertaining solely to the system. 

There is, however, an important case where this turns out to be possible. 
Namely, \emph{when the violation of strict energy conservation is caused by an operator of the system, not the ancilla}. 
That is, when it is possible to decompose the system Hamiltonian as $H_S = H_{S,0} + H_{S,1}$ such that 
\begin{equation}\label{CM_work_local_condition}
    [V,H_{S,0}+ H_A] = 0
    \quad \text{but} \quad 
    [V,H_{S,1}] \neq 0.
\end{equation}
If this is true, then we may substitute $[V,H_A] = - [V,H_{S,0}]$ in Eq.~(\ref{CM_CL_DeltaHA}), leading to 
\begin{equation}
    Q_{A_n} = \frac{\tau}{2} \tr\bigg\{ [V, [V, H_{S,0}]] \rho_S^n \otimes \rho_{A}\bigg\}.
\end{equation}
As a consequence, we can now split Eq.~(\ref{CM_CL_energy_splitting}) as 
\begin{IEEEeqnarray}{rCl}
\label{CM_CL_on_off_sep}
    \frac{d \langle H_S\rangle}{d t} 
    &=& \tr_S\big\{ H_{S,0} \mathcal{D} (\rho_S) \big\} +  \tr_S\big\{ H_{S,1} \mathcal{D} (\rho_S) \big\}, \\[0.2cm]
    &=& - \dot{Q}_A + \dot{W}, 
\end{IEEEeqnarray}
hence allowing us to unambiguously identify the first term as  heat and  the second as  on/off work.
{\color{black}In this case, Eq.~\eqref{CM_sigma_dot_temporary} may therefore be written as 
\begin{equation}\label{CM_sigma_correct_dot}
    \dot{\Sigma} = \frac{dS_S}{dt} - \dot{Q}_A
    = \frac{dS_S}{dt} + \tr_S \Big\{ H_{S,0} \mathcal{D}(\rho_S)\Big\},
\end{equation}
which is thus expressed solely in terms of quantities of the system. 
Note that these results hold also if the ancillas are not prepared in thermal states. 
This was used, for instance, in Ref.~\cite{Rodrigues2019} to study collisional models with weakly coherent ancillas. 
We also mention that the approach taken here starts with a discrete model and eventually reaches a coarse-grained, continuous-time limit for the entropy production rate. 
The opposite route can also be taken. That is, the entropy production rate of a continuous process can also be discretized in small time steps, which will then be depicted by a collisional model. 
This was used, for instance, by~\cite{Monsel2018}, to construct a method for measuring the entropy production of a driven autonomous system.
}

The situation described above happens often when the system is composed of multiple interacting parts, but with only one of the parts coupled to the ancillas~\cite{DeChiara2018,Barra2015,Pereira2018}. 
For instance, suppose the system is composed of two subsystems, $S_1$ and $S_2$ with a total Hamiltonian $H_S = H_{S_1} + H_{S_2} + V_{S_1,S_2}$, where $V_{S_1,S_2}$ is the interaction between them. 
Moreover, suppose there is only one bath and it is coupled only to $S_1$.
The interaction $V_{A,S_1}$ between $S_1$ and the ancillas $A_n$ is assumed to be \emph{locally} energy preserving, $[V_{A,S_1}, H_{S_1} + H_A] = 0$.
Notwithstanding, in general  $[V_{A,S_1}, V_{S_1,S_2}] \neq 0$. 
Thus, \emph{albeit locally energy preserving,  the collision may not be globally energy preserving due to the interaction between $S_1$ and $S_2$.}
The term $V_{S_1,S_2}$  will therefore play the role of $H_{S,1}$ in Eq.~(\ref{CM_CL_on_off_sep}) and will be responsible for the on/off work. 

{\color{black}To provide a concrete example,  consider a minimal model consisting of two qubits, with $H_S = \omega_1\sigma_z^1 + \omega_2 \sigma_z^2 + \lambda (\sigma_+^1 \sigma_-^2 + \sigma_-^1 \sigma_+^2)$~\cite{Barra2015}.
For simplicity, we assume only qubit 1 coupled to a bath.
The extension to two baths, one coupled to each qubit, is straightforward. 
We also take the  bath to be described by a collisional model, where the ancillas are made of thermal qubits with frequency $\omega_1$ (i.e., resonant with qubit 1). 
The system will then evolve according to 
\begin{equation}
    \frac{d\rho_S}{dt} = - i[H_S,\rho_S] + g^2 (1-f) D[\sigma_-^1] + g^2 f D[\sigma_+^1]. 
\end{equation}
(c.f. the discussion below~\eqref{CM_detailed_balance}). 
Eq.~\eqref{CM_work_local_condition} will be satisfied in this case, with $H_{S,0} \to \omega_1 \sigma_z^1 + \omega_2\sigma_z^2$ and $H_{S,1} = \lambda (\sigma_+^1 \sigma_-^2 + \sigma_-^1 \sigma_+^2)$.
As a consequence, there will be work involved.
The heat exchanged with the ancillas is going to be $-Q_{A} = \omega_1 \tr_S \big(\sigma_z^1 \mathcal{D}(\rho_S))$, while the work will be $W = \lambda \tr\big((\sigma_+^1 \sigma_-^2 + \sigma_-^1 \sigma_+^2) \mathcal{D}(\rho_S)\big)$.
It is the heat $Q_A$ which should enter Eq.~\eqref{CM_sigma_correct_dot}. 
If this is done, then one will guaranteed find $\dot{\Sigma} \geqslant 0$ for all times. 
Conversely, if one uses instead $\tr\big(H_S \mathcal{D}(\rho_S)\big)$ as a definition of heat, this will lead to violations of the second law, as discussed in~\cite{Levy2014}. 
}

Let us now change scenario and consider Eq.~(\ref{CM_simplest}) when $H_S(t)$ is explicitly time-dependent, but with the interactions engineered to be thermal operations. 
This means there is no on/off work involved, and $[V_n, H_{A_n}] = - [V_n, H_S^n]$.
As a consequence, the heat exchanged to the ancillas, Eq.~(\ref{CM_CL_DeltaHA}), becomes 
\begin{equation}\label{CM_CL_heat_thermal_ops}
    Q_{A_n} =\frac{\tau}{2} \tr\bigg\{ [V_n, [V_n, H_S^n]] \rho_S^n \otimes \rho_{A_n}\bigg\},
\end{equation}
which is written solely in terms of system-related quantities. 
From the master equation~(\ref{CM_simplest}) we now find the energy balance
\begin{equation}
    \frac{d \langle H_S \rangle}{d t} = \tr\bigg\{ \frac{\partial H_S}{\partial t} \rho_S\bigg\} + \tr\big\{ H_S(t) \mathcal{D}_t(\rho_S)\big\}. 
\end{equation}
Comparing this with Eq.~(\ref{CM_CL_heat_thermal_ops}) then leads to the celebrated Spohn separation of work and heat~\cite{Spohn1978}, 
\begin{IEEEeqnarray}{rCl}
    \dot{Q}_A = - \tr\left\{ H_S(t) \mathcal{D}_t(\rho_S)\right\},\quad \dot{W} = \tr\left\{\dot{H}_S(t) \rho_S\right\}.
\end{IEEEeqnarray}
Spohn's separation is usually employed phenomenologically: it is used when one has access to a master equation of the form~(\ref{CM_simplest}) and wishes to split the changes in energy into heat and work.
The above result shows that this separation is not at all universal. 
Quite the contrary, notice that for it to hold we had to assume that, even though the Hamiltonian is changing at each time step, the system-ancilla interaction and the state of the ancilla were adjusted to guarantee that the map was always a thermal operation. 
This would require considerable fine tuning and is very difficult to realize in practice.

\subsection{\label{sec:pauli}Pauli master equations and Schnackenberg's approach}

The Markov chain~(\ref{CL_Markov_step}) can be viewed as the classical dynamics emerging from  quantum collisional model~(\ref{M_CM_unitary}), in the case of thermal operations. 
Similarly, one may also consider the classical limit of the continuous-time master equation~(\ref{CM_simplest}).
All issues discussed in Sec.~\ref{sec:emergence} also remain in this case. In particular the non-trivial distinction between population and coherences in the case where the eigenbasis of $H_S(t)$ is time-dependent. 

In order to simplify the problem, we thus consider the scenario where only the eigenvalues of $H_S(t)$ are allowed to depend on time: $H_S(t) = \sum_i E_i(t) |i\rangle\langle i|$. 
The populations will then evolve according to Eq.~\eqref{continuous_MarkovChain}.
To obtain the short-time limit, we assume $U_{SA_n} = \exp\{-i \tau(H_S^n + H_{A_n} + V_n/\sqrt{\tau})\}$ and expand it in a power series in $\tau$. 
This leads to 
\begin{equation}
    M_n(i|j) = \delta_{ij} + \tau \bigg( W_{ij} - \delta_{ij} \sum\limits_{k} W_{kj}\bigg),
\end{equation}
where we have introduced the transition probabilities $W_{ij}(t) = \sum_{\mu,\nu} q_\nu^n |\langle i, \mu| V_n | j, \nu\rangle|^2$ and its time-dependence will be omitted for clarity when possible. 
Plugging this in Eq.~(\ref{continuous_MarkovChain}) and taking $\tau \to 0$ then leads to the classical Pauli master equation~\cite{Breuer2007}
\begin{equation}\label{Pauli_pauli}
    \frac{d p_i}{d t} = \sum\limits_j \bigg\{ W_{ij} p_j(t) - W_{ji} p_i(t) \bigg\}. 
\end{equation}
This procedure shows how, under specific conditions, one can recover the classical master equation evolution from the underlying quantum dynamics.

We now proceed to study the entropy production from the perspective solely of the Pauli master equation Eq.~(\ref{Pauli_pauli}). 
We review the framework put forth by Schnackenberg~\cite{Schnakenberg1976}. 
This approach is interesting because it also contemplates scenarios beyond the standard thermal-bath interaction. 
Master equations of the form~(\ref{Pauli_pauli}) also find a plethora of applications, from biomolecular processes to financial markets. 
And Schnackenberg's approach allows one to construct the entropy production rate and an entropy flux rate, irrespective of what physical system the master equation represents. 
Of course, the physical interpretation of $\dot{\Sigma}$ and $\dot{\Phi}$ is not necessarily evident, in general. 
Notwithstanding, it reproduces the thermal results as a particular case, as one should expect. 

This is an advantage of classical systems and unfortunately cannot be extended to the quantum case. 

The starting point is to consider the evolution of the Shannon entropy 
\begin{equation}\label{Pauli_Shannon}
    S(\bm{p}) = - \sum\limits_i p_i \ln p_i. 
\end{equation}
Differentiating with respect to time and inserting Eq.~(\ref{Pauli_pauli}) yields
\begin{equation}\label{Pauli_dSdt}
\frac{d S}{dt} = \frac{1}{2} \sum\limits_{i,j} \big( W_{ij} p_j - W_{ji} p_i\big) \ln p_j/p_i.
\end{equation}
Schnakenberg then proposed that the following quantity be associated with an entropy production:
\begin{equation}\label{Pauli_Pi}
    \dot{\Sigma}(t) = \frac{1}{2} \sum\limits_{i,j} \big( W_{ij} p_j - W_{ji} p_i\big) \ln \frac{W_{ij} p_j}{W_{ji} p_i}. 
\end{equation}
This expression is always non-negative as it has the form $(x-y)\ln(x/y) \geqslant 0$. 
That, of course, is in principle not enough to label a quantity as the entropy production. 
To scrutinize the correctness of this formula, one must analyze it from different perspectives. 

The difference between $\dot{\Sigma}$ and $dS/dt$ is associated with an entropy flux rate $\dot{\Phi}$ according to Eq.~(\ref{intro_2nd_law_rates}).
Using Eqs.~(\ref{Pauli_dSdt}) and (\ref{Pauli_Pi})  one then arrives at 
\begin{equation}\label{Pauli_Phi}
    \dot{\Phi}(t) = \frac{1}{2} \sum\limits_{i,j} \big( W_{ij} p_j - W_{ji} p_i\big) \ln \left(\frac{W_{ij}}{W_{ji}}\right).
\end{equation}
The entropy flux is thus seen to be \emph{linear} in the probabilities $p_i$. 

Additional justification for Eqs.~(\ref{Pauli_Pi}) and (\ref{Pauli_Phi}) can be given if we assume that the dynamics satisfies detailed balance \cite{VanKampen2007,Tome2014,Gardiner2010}; viz.,
\begin{equation}\label{detailed_balance}
    W_{ij} p_j^* = W_{ji} p_i^*,
\end{equation}
where $p_i^*$ is the steady-state distribution of Eq.~(\ref{Pauli_pauli}) (not necessarily a thermal state). 
In this case, Eq.~(\ref{Pauli_Pi}) may be rewritten in terms of the classical Kullback-Leibler divergence~(\ref{KL}) as 
\begin{equation}\label{Pauli_Pi_KL}
    \dot{\Sigma} = - \frac{\ud S(\bm{p} || \bm{p}^*) }{\ud t}, 
\end{equation}
which is the continuous-time and classical analog of Eq.~(\ref{Qgen_Sigma_nonThermal_FixedPoint}). 
The entropy flux~(\ref{Pauli_Phi}), on the other hand, can be rearranged as 
\begin{equation}
    \dot{\Phi} = \sum\limits_i \frac{dp_i}{dt} \ln p_i^*.
\end{equation}
In the particular case where the steady-state distribution is also the thermal equilibrium state, $p_i^* = e^{-\beta E_i}/Z$, this becomes 
\begin{equation}
    \dot{\Phi} = -\beta \sum\limits_i E_i \frac{\ud p_i}{\ud t} = - \beta \dot{Q},
\end{equation}
so that we recover the well-known thermodynamic result~(\ref{intro_2nd_law_rates}).

Returning to the general expression~(\ref{Pauli_Pi}), it is also interesting to define the \emph{probability current} 
\begin{equation}\label{Pauli_prob_current}
    J_{ij} = W_{ij} p_j - W_{ji} p_i, 
\end{equation}
which represents the current of probability flowing from $j$ to $i$. 
If we then define the so-called \emph{conjugated force}, 
\begin{equation}
    X_{ij} = \ln \frac{W_{ij} p_j}{W_{ji} p_i},
\end{equation}
then the entropy production can be cast as 
\begin{equation}
    \dot{\Sigma} = \frac{1}{2}\sum\limits_{ij} J_{ij} X_{ij}. 
\end{equation}
which is a stochastic version of Onsager's form~(\ref{Pi_Onsager_linear_response}); i.e., the entropy production is a product of fluxes times forces.
The difference is that here these are not macroscopic fluxes (like the flow of energy, for instance), but rather microscopic currents of probability. 

\subsection{\label{sec:stoch_thermo}Pauli master equation for multiple baths}

{\color{black}When extending the Pauli master Eq.~(\ref{Pauli_pauli}) to multiple baths, one usually assumes that transition rates $W_{ij}$ from different reservoirs contribute \emph{additively}~\cite{McConnell2019,Maguire2019}.
That is, they can be split as }
\begin{equation}\label{StochThermo_rate_split}
    W_{ij} = \sum\limits_\alpha W_{ij}^\alpha, 
\end{equation}
where $\alpha$ represents the different reservoirs present in the problem. 
Thus, for instance, if each reservoir is thermal, at temperature $T_\alpha$, each rate in Eq.~(\ref{StochThermo_rate_split}) would individually satisfy detailed balance
\begin{equation}
    \frac{W_{ij}^\alpha}{W_{ji}^\alpha} =e^{-\beta_\alpha (E_i-E_j)}. 
\end{equation}

This assumption is known to describe well a broad range of mesoscopic systems, from  biological engines to nanoscale junctions~\cite{VANDENBROECK20156}. 
However, when viewed as a limiting case of quantum processes, it is  extremely strong. 
First and foremost, the  Liouvillian of the master equation will, itself, not be separable in general. 
But even if it is (e.g. in the case of local master equations), this does not mean that the corresponding Pauli equation will have additive rates since the preferred basis of one bath may not coincide with the preferred basis of the other.
As a consequence, understanding under which conditions Eq.~(\ref{StochThermo_rate_split}) can be viewed as the limiting case of a quantum process is not trivial and, to the best of our knowledge, is still an open problem. 

Notwithstanding these difficulties, Eq.~(\ref{StochThermo_rate_split}) provides an interesting platform to characterize  entropy production. Starting from Eq.~(\ref{Pauli_dSdt}) and plugging Eq.~(\ref{StochThermo_rate_split}) leads to 
\begin{equation}
    \frac{dS}{dt} = \frac{1}{2} \sum\limits_{i,j,\alpha} \big( W_{ij}^\alpha p_j - W_{ji}^\alpha p_i\big) \ln p_j/p_i.
\end{equation}
Following~\cite{Esposito2010d} the correct way of identifying the entropy production is to add and subtract  $\ln W_{ij}^\alpha/W_{ji}^\alpha$ in each term of the sum. 
The entropy production rate is then identified as 
\begin{equation}\label{StochThermo_Pi}
    \dot{\Sigma} =  \frac{1}{2} \sum\limits_{i,j,\alpha} \big( W_{ij}^\alpha p_j - W_{ji}^\alpha p_i\big) \ln \frac{W_{ij}^\alpha p_j}{W_{ji}^\alpha p_i}. 
\end{equation}
Notice that this expression is not equivalent to Eq.~(\ref{Pauli_Pi}), which we would have obtained if we added and subtracted $\ln W_{ij}/W_{ji}$ instead. 
The expression~\eqref{StochThermo_Pi} is the correct one, as it yields proper thermodynamic expressions for the fluxes. 
Indeed, as shown in~\cite{Esposito2010d}, if this identification is not properly made, one will in general be underestimating the entropy produced.  
More details on the formulation of entropy production in this scenario can be found in~\cite{VANDENBROECK20156}. 
An extension to account for information flows was done in~\cite{Horowitz2014a}.

\subsection{\label{ssec:classical_phase_space}Classical phase space}

Stochastic thermodynamics can also be formulated for systems described by continuous degrees of freedom (e.g. position and momenta). In this case Eq.~\eqref{Pauli_pauli} is replaced by a Fokker-Planck equation. 
The formulation of the second law for such systems has recently been reviewed in detail in Ref.~\cite{Seifert2012}. 
Here, with Sec.~\ref{sec:QPhaseSpace} in mind, we shall focus on just two illustrative examples.

The first is the so-called colloidal particle~\cite{Seifert2012}, described by a single random variable $x$ evolving according to the Langevin equation 
\begin{equation}\label{PhaseSpace_Langevin}
\dot{x}=f(x)+B~\dot{\xi}(t),
\end{equation}
where $f(x) = - \partial_x V(x)$ is a conservative force, stemming from a potential $V(x)$, $B$ is a constant and $\xi(t)$ is a standard Wiener (i.e. Gaussian) process. 
One may equivalently describe the dynamics in terms of a Fokker-Planck equation for the probability density $P_t(x)$, which in this case reads
\begin{equation}
    \label{PhaseSpace_classical_Fokker_Planck}
    \frac{\partial P_t(x)}{\partial t} = - \frac{\partial J}{\partial x} = - \frac{\partial}{\partial x} \left[ f(x) P_t(x) - D_c \frac{\partial P_t(x)}{\partial x}\right],
\end{equation}
where $D_c = B^2/2$ is the diffusion constant.
The Fokker-Planck equation can be viewed as a continuity equation for $P(x)$, with 
$J(x) = f(x) P_t(x) - D_c\, \partial_x P_t(x)$ representing a probability current. 
The noise in Eq.~\eqref{PhaseSpace_Langevin} is ascribed to a thermal bath at a temperature $T$. 
As a consequence, one may verify that, in order for  the system to properly thermalize, one must choose $D_c \propto T$. 
In this case, the unique steady-state of~\eqref{PhaseSpace_classical_Fokker_Planck} will be the thermal state $P_\text{th} = e^{-\beta V(x)}/Z$, where $Z$ is the partition function. 

The definition of the entropy production associated to the Fokker-Planck equation Eq.~\eqref{PhaseSpace_classical_Fokker_Planck} was discussed extensively in Ref.~\cite{Seifert2012}, including its stochastic formulation and the associated fluctuation theorems. 
Extensions to more general Fokker-Planck equations were discussed in~\cite{Qian2002,Tome2010} and a more robust framework, based on path integrals, can be found in~\cite{Spinney2012}. Here we wish to point to a complementary approach, namely that with Eq.~\eqref{Pauli_Pi_KL} in mind, one may propose to define the entropy production as 
\begin{equation}\label{PhaseSpace_classical_sigma_KL}
    \dot{\Sigma} = - \frac{d}{dt} S(P_t || P_\text{th}),
\end{equation}
where $S(P_t || P_\text{th}) = \int dx \; P_t(x) \ln P_t(x)/P_\text{th}(x)$ is the continuous analog of Eq.~\eqref{KL}. 
Inserting Eq.~\eqref{PhaseSpace_classical_Fokker_Planck} into the above definition for $\dot{\Sigma}$, one finds
\begin{equation}
\dot{\Sigma} = \int dx \; \frac{\partial J}{\partial x} \ln P_t(x)/P_\text{th}. 
\end{equation}
Next we integrate by parts. Boundary terms are assumed to vanish as $P_t(x)\to 0$ for $x\to \pm \infty$. 
Moreover, using the definition of $J(x)$, together with the fact that $P_\text{th} \propto e^{-\beta V(x)}$,  one may verify that 
\begin{equation}
\frac{\partial}{\partial x} \ln P_t(x)/P_\text{th} = - \frac{J(x)}{D_cP_t(x)}.
\end{equation}
Therefore Eq.~\eqref{PhaseSpace_classical_sigma_KL} becomes
\begin{equation}\label{PhaseSpace_classical_J2}
    \dot{\Sigma} = \frac{1}{D_c}\int dx\;\frac{J(x)^2}{P_t(x)}, 
\end{equation}
which is the same result as in Ref.~\cite{Seifert2012}. 
This has a clear physical interpretation: the quantity $v(x) = J(x)/P(x)$ can be interpreted as a velocity in phase-space. The entropy production~\eqref{PhaseSpace_classical_J2} is thus seen to be associated with a mean-squared velocity. 
Thus, by construction, it is always non-negative and null if and only if the current itself vanishes. 
A method for estimating $\dot{\Sigma}$ using machine learning on the stochastic trajectory $x(t)$ was recently put forth in~\cite{Seif2021}. 

Finally, it is also worth mentioning that this approach, where Eq.~\eqref{PhaseSpace_classical_sigma_KL} is taken as the starting point for defining the entropy production, is not always possible, in particular when the system is connected to multiple baths. 
We chose to present it here, nonetheless, because it attributes a clear information-theoretic meaning to the entropy production, specially in light of the discussion in Sec.~\ref{sec:Qgen}. 

Next we consider a generalization of Eq.~\eqref{PhaseSpace_Langevin} to the case of multiple modes, so that $x = (x_1,\ldots, x_n)$ is now a vector of random variables. This could mean, for instance, a collection of position and momenta.
The vector $x$ continues to be described by a Langevin equation of the form~\eqref{PhaseSpace_Langevin}. 
However, now $f(x)$ is a $n$-dimensional vector and $\xi(t)$ is a $m$-dimensional vector of independent Wiener processes. 
As a consequence, $B$ is taken to be an $n\times m$ matrix.
We assume $B$ is independent of $x$, thus making this a problem with additive noise (multiplicative noise introduces significant mathematical complications~\cite{Spinney2012}). 

Here, we now focus on the special case of linear forces, $f(x) = - A x$, where $A$ is a $n\times n$ matrix. 
We no longer assume that $f(x)$ is a conservative force. But may very well contain damping terms. 
We do assume, though, that its eigenvalues have positive real parts, thus guaranteeing the stability of the problem.
Linear systems of Langevin equations, of this form, appear often in quantum optical experiments, as a semi-classical description of fluctuations in optical fields. 

For such systems, it is more convenient to recast the dynamical equation in terms of the first moments $\bar{x}$ and the covariance matrix (CM), defined as $\Theta = \langle x x\trans \rangle - \langle x \rangle\langle x\trans \rangle$.
One may verify that $\Theta$ evolves according to a \emph{Lyapunov equation} 
\begin{equation}
\label{PhaseSpace_dynLya}
\dot{\Theta}=-\left(A \Theta+\Theta A^{\mathrm{T}}\right)+2 D,
\end{equation}
where we have introduced the 
diffusion matrix $D=B B^{\mathrm{T}}/2\ge0$. The equilibrium solution of Eq.~\eqref{PhaseSpace_dynLya} satisfies the condition 
$A \Theta+\Theta A^{\mathrm{T}}=2 D$.
Continuous-time Lyapunov equations of this form have found significant applications in the fields of linear systems, control theory, and quantum optics~\cite{Brogan}. 
The formulation of the entropy production for this kind of problem can be constructed by introducing the distinction between even or odd functions under time-reversal. 
Intuitive instances of even variables include position of mechanical systems and voltages in circuits, their odd counterparts being velocities and currents. 

it is then possible to identify the {\it reversible} parts of Eq.~\eqref{PhaseSpace_dynLya}, that is the part that is even under time reversal, from the {\it irreversible} one that changes sign upon inversion of the sign of time. 
We call $A^\text{irr}$ the irreversible part of $A$ such that $A=A^\text{rev}+A^\text{irr}$.
Convenient expressions for the entropy production and flux rates [cf. Eq.~\eqref{intro_2nd_law_rates}] were derived for this scenario in Refs.~\cite{Landi2013b,Brunelli2018}, under the assumption of Gaussian states and dynamics. 
As already mentioned, this is often the case in many quantum optical experiments. 
Indeed, such expressions have been instrumental to the interpretation of the experiments reported in Ref.~\cite{Brunelli2018}, which will be reviewed in Sec.~\ref{QCS}.

%
%
\subsection{\label{sec:QPhaseSpace}Quantum phase space}
%
%

Many aspects of the transition from quantum to classical can be neatly visualized by moving to quantum phase space. 
The role of quantum effects in the entropy production is one of them. 
In this Section, we consider semiclassical formulations of the entropy production problem based on quantum phase space. 
The  idea is to replace the von Neumann entropy with a generalized entropy function, associated to the distribution in phase space. 
This yields a semiclassical formulation, which coincides with standard thermodynamics at high temperatures, but leads to valuable new insights otherwise. 
The approach, as we will show, can also be naturally extended to non-equilibrium reservoirs, such as dephasing and squeezed baths (which is also reviewed in Sec.~\ref{sec:SqueezedBaths}).

We consider a system of $n$ (in general interacting) harmonic oscillators (bosonic modes) whose positions and momenta (quadratures) we label as $q_i$ and $p_i$, respectively ($i=1,..,n$). We arrange them in the $2n$-dimensional vector $X^T=(q_1,p_1,q_2,p_2,..q_n,p_n)$.
We also define the corresponding annihilation operators as $a_i = (q_i + i p_i)/\sqrt{2}$. Moreover, within this Section we will assume -- for simplicity -- units such that $\hbar=k_B=1$.

We will discuss here two of the most widely used approaches for quantum phase space: the Wigner and the Husimi function~\cite{Lee1995}. 
Given a density matrix $\rho$, the former is defined as 
\begin{equation}\label{PhaseSpace_Wigner_def}
    \mathcal{W}({\bf x}) = \frac{1}{\pi^{2n}} \int d^{2n} \lambda \; e^{-\sum_i (\lambda_i \alpha_i^* - \lambda_i^* \alpha_i)} \tr \Big\{ \rho e^{ \sum_i (\lambda_i a_i^\dagger - \lambda_i^* a_i)}\Big\},
\end{equation}
where the integral is over the entire complex plane of each  $\lambda_i$ i.e. $d^{2n} \lambda = \prod_i d\text{Re}(\lambda_i) d \text{Im}(\lambda_i)$. 
Moreover, the argument ${\bf x}$ of the Wigner function stands for a $2n$-dimensional vector with entries $x_{2i-1} = (\alpha_i + \alpha_i^*)/\sqrt{2}$ and $x_{2i} = i (\alpha_i^* - \alpha_i)/\sqrt{2}$. One could equivalently interpret $\mathcal{W}$ as a function of the $2n$ complex variables $(\alpha_i, \alpha_i^*)$.
We will actually use both representations interchangeably in what follows. 

An alternative, equivalent formulation, is in terms of the Husimi-Q function, defined as 
\begin{equation}\label{PhaseSpace_Husimi_def}
    {\cal Q}(\alpha)=\frac{1}{\pi^n}\langle\alpha|\rho|\alpha\rangle,
\end{equation}
where $|\alpha\rangle =\otimes^n_{i=1} |\alpha_i\rangle$ and each $|\alpha_i\rangle$ is a coherent state of mode $i$, i.e. $a_i |\alpha_i \rangle = \alpha_i |\alpha_i \rangle$.
{\color{black}The Husimi function is interpreted as the probability distribution for the outcomes of a homodyne measurements; that is, simultaneous (but noisy) measurements of both position and momentum~\cite{Arthurs1965,Braunstein1991}. 
}

While $\mathcal{W}$ can be negative for certain states, $\mathcal{Q}$ is always strictly non-negative. The relation between the Wigner and Husimi functions is via a Gaussian convolution
\begin{equation}
    \label{PhaseSpace_Wigner_Husimi_relation}
    \mathcal{Q}(\alpha) = \frac{2^{n}}{\pi^n} \int d^{2n} \lambda\; \mathcal{W}(\lambda) e^{- 2 \sum_i |\alpha_i - \lambda_i|^2}. 
\end{equation}
This therefore shows how $\mathcal{Q}$ can be viewed as a type of coarse-grained version of the Wigner function,
which has often been used to explore the classical-quantum boundary~\cite{Takahashi1985}. 
This coarse-graining is just enough to make $\mathcal{Q} \geqslant 0$, for all $\rho$. 
The Wigner function  is in one-to-one correspondence with the state $\rho$. Surprisingly, despite this coarse-graining, the same is also true of $\mathcal{Q}$. 
This is a consequence of the overcompleteness of the coherent states basis.

A particularly important class of states, in the context of quantum phase space, are those which are Gaussian~\cite{Ferraro2005}. 
Gaussian states are completely characterized by their first moments $\bar{x_i} = \langle X_i \rangle$ and CM, whose elements we rewrite for convenience as 
\begin{equation}
\Theta_{ij}= \frac{1}{2}\left\langle \{X_i, X_j\}\right\rangle-\langle X_i\rangle\langle X_j\rangle.
\end{equation}
Thus, for Gaussian states, the correspondence between $\rho$ and Wigner/Husimi function is extended to the first moments and the covariance matrix, which now fully characterize the properties of the system.
The Wigner function for Gaussian states has the form of a multivariate normal
\begin{equation}
    \label{PhaseSpace_Husimi_gaussian}
    \mathcal{W}(x)=\frac{e^{-\frac{1}{2}(x-\bar{x})^{\mathrm{T}} \Theta^{-1}(x-\bar{x})}}{\sqrt{(2 \pi)^{n}\text{det}(\Theta)}},\quad   \mathcal{Q}(x)=\frac{e^{-\frac{1}{2}(x-\bar{x})^{\mathrm{T}} \Theta_Q^{-1}(x-\bar{x})}}{\sqrt{(2 \pi)^{n}\text{det}(\Theta_Q)}},
\end{equation}
where $\Theta_Q = \Theta + \mathbb{I}/2$ is the original CM, incremented by vacuum fluctuations. This is directly associated to the coarse-grained nature of $\mathcal{Q}$, 
which causes the CM associated with $\mathcal{Q}$ to be larger by a factor of 1/2.

Gaussian states are useful for systems undergoing Gaussian processes.
That is, processes which preserve the Gaussian character of a given input state. This, in turn, implies that the operation is linear in the phase-space variables and thus generated by a Hamiltonian that is a bilinear form of position and momentum. Such a class of states and operations is particularly useful to illustrate the general context that we aim at addressing. They play a crucial role in quantum optics and quantum information processing as  important resources for quantum communication protocols~\cite{Braunstein2005,Cerf2007,Serafini} and representations for the ground or thermal equilibrium states of linear systems. 
Gaussian states are also routinely prepared in many experimental settings, from linear optics to platforms exploiting (general) light-matter interactions~\cite{Serafini,Cerf2007}. 
The formulation of entropy production in terms of quantum phase space is greatly simplified for Gaussian states and operations; the formalism, however, is not restricted to this case, and below we will discuss both Gaussian and non-Gaussian processes in parallel.

Given the interpretation of $\mathcal{W}$ and $\mathcal{Q}$ as quasi-probability distributions in phase space, one may now naturally contemplate the possibility of using their associated Shannon entropies as quantifiers of information. 
The Shannon entropy of ${\cal W}({\bf x})$ is called the \emph{Wigner entropy}
\begin{equation}
\label{PhaseSpace_Wigner_entropy_def}
    {\cal S}_{W}=-\int \,d^{2n}{\bf x}\; {\cal W}({\bf x})\ln{\cal W}({\bf x}).
\end{equation}
An operational interpretation for $\mathcal{S}_W$ was given in Refs.~\cite{Buek1995}, where it was shown that it can be viewed as a sampling entropy via homodyne measurements.  
For general non-Gaussian states, $\mathcal{W}$ may be negative, so that the integral in Eq.~\eqref{PhaseSpace_Wigner_entropy_def} delivers a complex-valued entropy, which is  clearly unsuited as a measure of information. 
For Gaussian states, however, $\mathcal{S}_W$ acquires a very nice interpretation. 
First, an explicit calculation using $\mathcal{W}(x)$ in Eq.~\eqref{PhaseSpace_Husimi_gaussian} leads to~\cite{Buek1995,Landi2013b,Adesso2012}
\begin{equation}
\label{PhaseSpace_Wigner_entropy_Gaussian}
    \mathcal{S}_{W}=\frac{1}{2} \ln \text{det}(\Theta) + \frac{n}{2} \log (2 \pi e).
\end{equation}
This result shows that the entropy is determined solely by the determinant of the CM, therefore providing an extremely efficient way of evaluating the entropy of the system.
What is even more interesting is that, for Gaussian states, $\mathcal{S}_W$ is directly connected to the R\'enyi-2 entropy. 
Recalling the definition $S_\alpha=(1-\alpha)^{-1}\ln\tr\rho^\alpha$ of the  R\'enyi-$\alpha$ entropy, it was shown in~\cite{Adesso2012} that
$S_2=\frac{1}{2} \ln \text{det}(\Theta)$.
Whence,
\begin{equation}
\label{PhaseSpace_Wigner_entropy_Renyi2}
    \mathcal{S}_{W} = S_2 + \text{const}.
\end{equation}
This result links the Wigner entropy to $S_2$,  an important information-theoretic quantity~\cite{Renyi1960} of strong thermodynamic relevance~\cite{Baez2011}. 

For states whose Wigner functions are not necessarily positive, one may alternatively study the Shannon entropy of the Husimi function,
\begin{equation}
\label{PhaseSpace_Wehrl_Husimi_entropy_def}
    \mathcal{S}_Q = -\int d^{2n}\alpha \;{\cal Q}\ln{\cal Q},
\end{equation}
known as \emph{Wehrl's entropy}~\cite{Wehrl1978,Wehrl1979}.
Since $\mathcal{Q} \geqslant 0$,  Wehrl's entropy is always well defined and real. 
It can also be given an operational interpretation as a coarse-graining of the von Neumann entropy, stemming from convoluting the system's state with Gaussian noise induced by a heterodyne measurement~\cite{Wodkiewicz,Buek1995}. As a consequence, $\mathcal{S}_Q$ upper bounds the von Neumann entropy, $\mathcal{S}_Q \geqslant S(\rho)$~\cite{Lieb1978}.
Another advantage of the Husimi function and the Wehrl entropy is that they can be extended to spin systems in terms of spin coherent states. 
This will be discussed further below. 

We are now in the position to introduce the formulation of  entropy production within the context of the Wigner and Wehrl entropies. 
The main advantage of moving to quantum phase space is that any master equation can be mapped into a Quantum Fokker-Planck equation for $\mathcal{W}$ or $\mathcal{Q}$. 
Tools of classical stochastic processes can then be employed in order to obtain simple expressions for the entropy production rate and flux. Quite remarkably, this can be done for a wide variety of environments interacting with the system of interest~\cite{Santos2017}, including non-equilibrium baths. In what follows, we shall present a brief account of possible approaches towards the derivation of explicit expressions for such quantities. 

For the purpose of illustration, we begin by considering a single bosonic mode described by a standard Lindblad master equation of the form 
\begin{equation}
    \label{PhaseSpace_M_single_mode}
\partial_{t} \rho=-i[H, \rho]+\mathcal{D}(\rho),
\end{equation}
where $H=\omega(a^\dag a+1/2)$ and 
\begin{equation}\label{PhaseSpace_thermal_dissipator}
\mathcal{D}(\rho) = \gamma(\bar{n}+1) D[a] + \gamma \bar{n} D[a^\dagger],
\end{equation}
with $D[L] = L \rho L^\dagger - \frac{1}{2} \{L^\dagger L, \rho\}$, $\gamma$ is the damping rate and $\bar{n} = (e^{\beta \omega}-1)^{-1}$ is the Bose-Einstein distribution.
Using standard correspondence tables~\cite{Gardiner2004} one can convert~\eqref{PhaseSpace_M_single_mode} into a quantum Fokker-Planck equation for either $\mathcal{W}$ or $\mathcal{Q}$.
In the case of the Wigner function, this becomes
\begin{equation}
    \label{PhaseSpace_FPWigner}
    \partial_t\cal W= \mathcal{U}(\mathcal{W}) + \partial_\alpha J({\cal W})+ \partial_{\alpha^*} J^*(\mathcal{W}),
\end{equation}
where $\mathcal{U}(\mathcal{W}) = i \omega \Big[ \partial_\alpha(\alpha\mathcal{W}) - \partial_{\alpha^*} (\alpha^*\mathcal{W})\Big]$ is a differential operator associated with the unitary part of~\eqref{PhaseSpace_M_single_mode} and 
\begin{equation}\label{PhaseSpace_J_def}
    J({\cal W})=\frac{\gamma}{2}\left[\alpha{\cal W}+(\bar{n}+1/2)\partial_{\alpha^*}{\cal W}\right],
\end{equation}
is a complex-valued phase-space current associated with the irreversible part of the dynamics. 
Eq.~\eqref{PhaseSpace_FPWigner} can be viewed as a continuity equation in quantum phase space, where the changes in $\mathcal{W}$ stem from gradients of unitary and irreversible currents. 
In particular, the current $J$ vanishes if and only if $\mathcal{W}$ is a thermal state with occupation $\bar{n}$, that is for  
$\mathcal{W}_\text{eq} = e^{-\frac{|\alpha|^2}{\bar{n}+1/2}}/[{\pi (\bar{n}+1/2)}]$.
While Eq.~\eqref{PhaseSpace_FPWigner} vanishes for such a thermal state, the fact that the individual currents vanish is a stronger statement, which in classical systems is usually attributed to detailed balance. 
It also provides an alternative interpretation for the thermal equilibrium state, as being the unique state for which no quasiprobability currents flow. 

The problem can be equivalently expressed as a Fokker-Planck equation for the Husimi function.
The equation will have the exact same form as~\eqref{PhaseSpace_FPWigner}, with small modifications.
For the choice of Hamiltonian in~\eqref{PhaseSpace_M_single_mode}, the unitary part turns out to be same with $\mathcal{W}$ replaced by $\mathcal{Q}$. But this is a coincidence of this simple Hamiltonian, as the unitary parts in general may differ significantly. 
The shape of the irreversible currents $J(Q)$ will look exactly like Eq.~\eqref{PhaseSpace_J_def}, except that $\bar{n}+1/2$ is replaced by $\bar{n}+1$. This reflects the additional vacuum fluctuations that naturally appear in the Husimi function, similar to what was found in Eq.~\eqref{PhaseSpace_Husimi_gaussian}. 

The formalism for the calculation of the entropy production rate set forth in Sec.~\ref{sec:foundations}, in particular  Eq.~\eqref{Qgen_Sigma_nonThermal_FixedPoint}, suggest that a meaningful definition for the Wigner entropy production could be [cf. Eq.~\eqref{Pauli_Pi_KL}]
\begin{equation}
    \label{PhaseSpace_dot_sigma_W}
    \dot{\Sigma}_{W}(t)=-\frac{d}{dt} \mathcal{S}_W({\cal \mathcal{W}}(t)|| \mathcal{W}_\text{eq}),
\end{equation}
$\mathcal{S}_W(\mathcal{W}_1 || \mathcal{W}_2) = \int d^2 \alpha\; \mathcal{W}_1 \ln \mathcal{W}_1/\mathcal{W}_2$ is the Wigner analog of the Kullback-Leibler divergence.
As shown in~\cite{Adesso2012}, for Gaussian states this coincides with the R\'enyi-2 mutual information.
By using the RHS of Eq.~\eqref{PhaseSpace_FPWigner} in the definition of $\dot{\Sigma}_W$ and integrating by parts over the phase space, we get~\cite{Santos2017}
\begin{equation}
\label{PhaseSpace_PiWigner}
    \dot{\Sigma}_W(t)=\frac{4}{\gamma(\bar{n}+1 / 2)} \int d^{2} \alpha \frac{|J({\cal W})|^{2}}{\cal W}.
\end{equation}
This expression has several nice properties and a clear physical interpretation. 
First, clearly $\dot{\Sigma}_W \geqslant 0$, as expected for any second law. 
Second, $\dot{\Sigma}_W = 0$ iff the currents vanish, which happens iff $\mathcal{W} = \mathcal{W}_\text{eq}$. Thus, the entropy production is zero only when the system is in thermal equilibrium with the bath. 
Third, Eq.~\eqref{PhaseSpace_PiWigner} directly links entropy production with the existence of irreversible currents in phase space. 
In particular, one can derive a phase-space velocity $J(\mathcal{W})/\mathcal{W}$~\cite{Seifert2012}, so that 
$\dot{\Sigma}_W$ is interpreted as the mean-squared phase-spaced velocity.

Next we turn to the entropy flux, which can be computed from $\dot{\Phi}_W = \dot{\Sigma}_W- d\mathcal{S}_W/dt$ [cf.~Eq.~\eqref{intro_2nd_law_rates}]. 
Using the explict form of $J(\mathcal{W})$ in Eq.~\eqref{PhaseSpace_J_def}, together with 
Eq.~\eqref{PhaseSpace_PiWigner}, one finds that 
\begin{equation}\label{PhaseSpace_Flux_single_mode}
    \dot{\Phi}_W = \frac{\gamma}{\bar{n}+1/2} \Big( \langle a^\dagger a \rangle - \bar{n} \Big). 
\end{equation}
The interpretation of this equation is straightforward as well. 
Starting from Eq.~\eqref{PhaseSpace_M_single_mode}, one may compute the energy flow to the bath, which reads 
    $\dot{\langle H \rangle} = \omega \gamma ( \bar{n} - \langle a^\dagger a \rangle)$. 
For simplicity, we assume that this can be attributed to heat entering the bath,  $\dot{\langle H \rangle} \equiv - \dot{Q}_E$ (cf. Sec.~\ref{sec:dynamics_on_off}).
As a consequence, comparing with Eq.~\eqref{PhaseSpace_Flux_single_mode}, one finds that 
\begin{equation}\label{PhaseSpace_Phi_Heat_relation}
    \dot{\Phi}_W = \frac{\dot{Q}_E}{\omega (\bar{n}+1/2)}.
\end{equation}
This can be compared with the standard thermodynamic result, $\dot{\Phi} = \dot{Q}_E/T$ [Eq.~\eqref{intro_Phi}]. 
We see that formulating the problem in terms of the Wigner function leads to a modification of the standard thermodynamic result, where the heat flux is now weighted by a new prefactor $\omega(\bar{n}+1/2)$, instead of the temperature $T$.
When $T \gg \omega$, however, a series expansion leads to $\omega(\bar{n}+1/2) \simeq T$. Thus, one recovers the standard thermodynamic results at high temperatures. 

A particularly important special case of the above formalism is to describe photon losses in optical cavities. 
The standard dissipator used to describe this, $\mathcal{D}(\rho) = \gamma \left[a \rho a^\dagger - \frac{1}{2} \{a^\dagger a, \rho\}\right]$, corresponds to a zero-temperature ($\bar{n}\to0$) limit of~\eqref{PhaseSpace_thermal_dissipator}. 
The problem with this is that the standard formulation of the second law breaks down in this limit, since the relative entropy in Eq.~\eqref{Qgen_Sigma} diverges when the environment is in a pure state. 
The  phase space approach, on the other hand, remains perfectly well defined in this limit, thanks to the factors of 1/2 in Eqs.~\eqref{PhaseSpace_PiWigner} and~\eqref{PhaseSpace_Phi_Heat_relation}. 
The reason, therefore, is because the phase space approach also takes into account vacuum fluctuations, which persist even at zero temperature. 

Eqs.~\eqref{PhaseSpace_PiWigner} and~\eqref{PhaseSpace_Phi_Heat_relation} provide solid physical grounds to the choice of~\eqref{PhaseSpace_dot_sigma_W} as a basic definition of entropy production in the context of quantum phase space. 
In Ref.~\cite{Santos2017}, two additional approaches to the derivation these results were also put forth, one of them based on the complex-plane averaging of stochastic trajectories. 
The fact that all approaches agree, corroborate the correctness of the framework. One should mention, however, that Eq.~\eqref{PhaseSpace_dot_sigma_W} is not expected to hold for all types of phase-space open dynamics.
It fails, for instance, in the case of a linear lattice connected to multiple baths~\cite{Malouf2018a}. 
Hence, the above construction should ultimately be performed on a case-by-case basis. 

We also mention, in passing, that the results above remain valid if one uses instead the Husimi function. The only difference is that all factors of $\bar{n}+1/2$ should be replaced by $\bar{n}+1$. This apparent similarity between the two approaches, however, is deceiving, as it only happens for the the simple models considered here. In more complicated scenarios, the two approaches may differ significantly. A nice example is the case of two-photon losses, described by a Lindblad dissipator $a^2 \rho a^{\dagger 2} - \frac{1}{2} \{a^{\dagger 2} a^2, \rho\}$ (a highly non-Gaussian process). The Fokker-Planck equation associated with this dissipator is \emph{completely} different if one employs either the Wigner or the Husimi functions, as one may verify. 
The same is also true for more complicated unitary contributions. In fact, due to the coarse-grained nature of the Husimi function, unitary terms may contribute for the evolution of $\mathcal{S}_Q$. 
These terms may be particularly important in systems undergoing dissipative phase transitions~\cite{Goes2019}, which will be reviewed in Sec.~\ref{sec:dissipative}.
But they also persist even in completely isolated systems undergoing unitary dynamics, as studied in~\cite{Goes2020a}. 

The approach presented above can also be flexibly extended to master equations describing non-equilibrium reservoirs. 
We consider two examples. 
The first  is a squeezed thermal bath which, in addition to the thermal occupation $\bar{n}$, is also described by a squeezing parameter $z = r e^{i\theta}$. The full form of the dissipator in this case is presented below, in Eq.~\eqref{squeezed_M}. 
The calculations in this case are analogous and amount solely to the substitution 
\begin{equation}\label{PhaseSpace_J_squeezing}
    J({\cal W})\to J({\cal W}) \cosh r+\left[\gamma \alpha^{*} {\cal W}-J^{*}({\cal W})\right] e^{i\left(\theta-2 \omega_{s} t\right)} \sinh r,
\end{equation}
where $\omega_s$ is the central frequency of the broadband bath and accounts for non-resonant energy 
exchanges with the system. 
Squeezed baths will be reviewed further in Sec.~\ref{sec:SqueezedBaths}. 

The second example we discuss is that of a dephasing bath, which describe the loss of quantum coherence without the exchange of excitations. 
The effects of a dephasing bath can be accounted for in Eq.~\eqref{PhaseSpace_M_single_mode} by using the super-operator 
\begin{equation}
    {\cal D}^\text{deph}(\rho)=-\frac{\lambda}{2}\left[a^\dag a,\left[a^\dag a,\rho\right]\right],
\end{equation}
with $\lambda$ the dephasing rate. 
A similar procedure in this case reveals that the flux is identically zero, $\dot{\Phi}_W \equiv 0$. 
This is in agreement with the idea that this sort of environmental effect is not associated with a flux of excitations to or from the system.
As a consequence, one may identify the rate of change of the Wigner entropy of the system with the entropy production rate, which takes the form~\cite{Santos2017}
\begin{equation}\label{PhaseSpace_sigma_dephasing_bosonic}
    \dot{\Sigma}^\text{deph}_W(t)=\frac{2}{\lambda}\int \frac{|J^\text{deph}({\cal W})|^2}{|\alpha|^2{\cal W}}\,d^2\alpha,
\end{equation}
where we have introduced the dephasing current $J^\text{deph}({\cal W})=\lambda\alpha[\alpha^*\partial_{\alpha^*}{\cal W}-\alpha\partial_\alpha{\cal W}]/2$. 
We therefore see that a similar structure emerges, but now associated with the irreversible currents generated by the dephasing bath. There is also an additional factor of $|\alpha|^2$ in the numerator, which tend to favor currents near the origin of the complex plane. 

Ref.~\cite{Santos2018a} has extended the formalism of phase-space approaches to entropy production to the case of spin-like systems making use of the useful spin-coherent state representation~\cite{Radcliffe1971,Takahashi1985}. 

\section{\label{sec:thermal_ops}Resource theoretic approach}
%
%



Quantum features can be exploited to provide advantages for a series of applications. 
Different applications, however, exploit different features. 
For instance, quantum communications may exploit entanglement, while  metrological applications may exploit radiation squeezing. 
Each of these features therefore represent a resource, which can be consumed to yield a quantum advantage for certain tasks. 
Resource theories provide a mathematical formulation of this idea. 
Initially focused on entanglement~\cite{Horodecki2009}, they were subsequently extended to several other resources, such as purity~\cite{Horodecki2003}, asymmetry~\cite{Horodecki2003} and coherence~\cite{Streltsov2016a}. 
A recent review can be found in~\cite{Chitambar2019}. 

Thermodynamics can also be cast in this framework, known as the \emph{resource theory of athermality}, first pioneered by~\cite{Brandao2013}.
In this case, the resources are all quantum states  which are not in thermal equilibrium. 
The reason is that such states can be used to extract work, which is the most fundamental task of thermodynamics. 
Whence \emph{athermality} (i.e., how ``far'' a system is from equilibrium) is the resource which is consumed to extract work. 
{\color{black}
The earlier works on the resource theory of athermality are reviewed in~\cite{Goold2016}. 
In this section we focus on some of the more recent developments, as well as aspects which pertain specifically to entropy production.
}

The starting point for any resource theory is the definition of what are the allowed  \emph{free operations}. That is, operations which only consume a resource and never create it. 
In the case of thermodynamics, this means  no associated work. 
Moreover, the idea is to focus on operations that are physically meaningful and  endowed with interesting properties. 
While there is no unique proposal [cf.~\cite{Bera2017b}],  the most widely used so far are the thermal operations, discussed in Sec.~\ref{sec:Qgen_thermal_ops}.

Recall that a thermal operation (TO) is any map $\mathcal{T}(\rho)$ of the form~\cite{Brandao2013} 
\begin{equation}\label{resource_TO}
    \mathcal{T}(\rho_S) = \tr_E \bigg\{ U \bigg( \rho_S \otimes \rho_E^\text{th}\bigg) U^\dagger \bigg\}, 
    \qquad
    [U,H_S + H_E] = 0,
\end{equation}
with $\rho_E^{\text{th}} = e^{-\beta H_E}/Z_E$.
That is, a TO is a map where the system interacts with a thermal environment by means of a unitary that preserves the total energy [cf. Eq.~(\ref{Qgen_strict_energy_conservation})]. 
As discussed in Sec.~\ref{sec:Qgen_thermal_ops}, this kind of operation has a series of nice properties. 
First, the fixed point of the map is the thermal state $\rho_S^\text{th} = e^{-\beta H_S}/Z_S$. Hence, it  describes the partial (or full) thermalization of the system towards $\rho_S^\text{th}$. 
Second, $[U,H_S+H_E] = 0$ implies there is no work involved in coupling $S$ and $E$, so that the change in  energy of $S$ coincides with the heat that flows to $E$. 
And third, the entropy production of the process can be written solely in terms of system related quantities, as in Eq.~\eqref{Qgen_sigma_thermal_op}: 
\begin{equation}\label{resource_sigma}
    \Sigma = S(\rho_S || \rho_S^\text{th}) - S(\rho_S' || \rho_S^\text{th}) = - \beta \Delta F,
\end{equation}
where $\rho_S' = \mathcal{T}(\rho_S)$ is the state of the system after the map and $\Delta F = F(\rho_S') - F(\rho_S)$, where $F(\rho_S) = \tr(H_S \rho_S) - T S(\rho_S)$ is the \emph{non-equilibrium} free energy of $\rho_S$. 

In the resource theory of athermality, the state $\rho_S^\text{th}$ is called the free state. Any state which is not  $\rho_S^\text{th}$ is viewed as a resource (this includes thermal states at a different temperature $\beta'$).
The TOs~(\ref{resource_TO}) represent the free operations; they cannot create resources, but only consume it. Moreover, they do nothing to free states. 
Another key feature of resource theories is the idea of a \emph{monotone}; i.e., a c-number function $f(\rho_S)$ satisfying 
\begin{equation}\label{resource_monotone}
f\big( \mathcal{T}(\rho_S) \big) \leqslant f(\rho_S). 
\end{equation}
A natural monotone, in this case, is the relative entropy $S(\rho_S|| \rho_S^\text{th})$. 
This quantity is a monotone because  $\rho_S^\text{th}$ is a fixed point of $\mathcal{T}$, so that the data processing inequality implies $S(\rho_S'|| \rho_S^\text{th}) \leqslant S(\rho_S|| \rho_S^\text{th})$ [cf. Eq.~\eqref{Qgen_data_proc_sigma}].  
The entropy production~(\ref{resource_sigma}) and the second law ($\Sigma \geqslant 0$), therefore naturally appear as the monotones of the resource theory.

In the particular case of a system with zero Hamiltonian, $H_S = 0$, the free energy becomes simply the von Neumann entropy of the system, $F(\rho_S) = - T S(\rho_S)$ and one recovers the resource theory of purity~\cite{Horodecki2003}.

One of the basic questions of resource theories is: \emph{Given two states $\rho_S$ and $\rho_S'$, is there an operation $\mathcal{T}$ such that $\rho_S' = \mathcal{T}(\rho_S)$?}
Put it differently, is it possible to convert $\rho_S$ to $\rho_S'$ via thermal operations?
This means one has to search over all possible maps $\mathcal{T}$ (i.e. over all possible environments and all possible energy-preserving unitaries). 
The question is therefore highly non-trivial.
Notwithstanding, it is also extremely important, as it allows to establish a hierarchy of resources and thus determine how more resourceful a state is with respect to another. 
As we will show, it turns out that the entropy production plays a fundamental role in determining state interconversion. 

In the context of thermodynamics, state interconversion is directly associated to work~\cite{Aberg2013}; or, more specifically, the notions of \emph{work extraction} and \emph{work of formation}~\cite{Dahlsten2011,Horodecki2013}. These tasks can be accomplished, for instance, by coupling the system to an additional work qubit~\cite{Horodecki2013} or a continuous variable system (mimicking a classical weight)~\cite{Skrzypczyk2014,Chubb2018}. 
The maximum amount of work that can be extracted  occurs when the system is taken from a state $\rho_S$ to the thermal state $\rho_S^\text{th}$ (full thermalization). 
Work of formation, on the other hand, refers to the reverse problem: if the system starts in a thermal state $\rho_S^\text{th}$, what is the minimum amount of energy that must be invested to take it towards a certain state $\rho_S$?
Extraction and formation are therefore two particular examples of state interconversion.

\subsection{\label{sec:resource_majorization}Thermo-majorization}

We now turn to the question of state interconversion in the single-shot scenario. 
That is, given two states $\rho_1$ and $\rho_2$, we ask whether it is possible to convert $\rho_1 \to \rho_2$ using only thermal operations of the form~(\ref{resource_TO}). 
This problem was first addressed in~\cite{Horodecki2013} and is based on a criteria called thermo-majorization (which is a variation of the idea of majorization used in probability theory).
Let $H_S = \sum_i E_i |i\rangle\langle i|$.
For simplicity of presentation, we focus  on states which are diagonal in the basis $|i\rangle$; i.e., which are of the form $\rho_S = \sum_i p_i |i\rangle\langle i|$. 
The results of Ref.~\cite{Horodecki2013} also hold for states which are block-diagonal 
{\color{black}
(c.f. Eq.~\eqref{resource_block_diagonal_state} below)
}; but not for states which have arbitrary off-diagonal elements.
A treatment of the latter was put forth in~\cite{Lostaglio2015} and will be reviewed below. 

\begin{figure}[t]
    \includegraphics[width=\columnwidth]{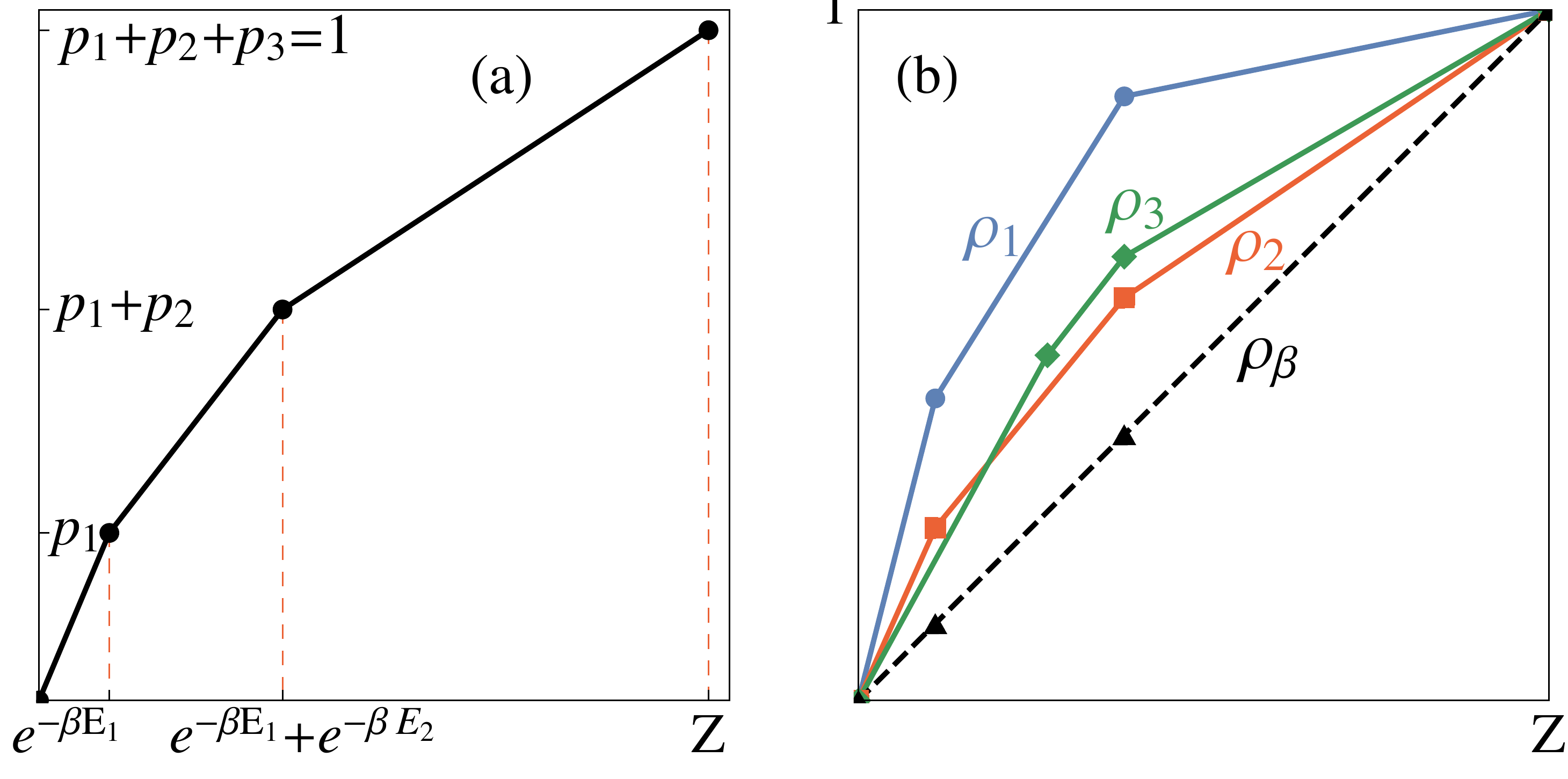}
    \caption{The thermo-majorization condition determining when two states are interconvertible using thermal operations. (a) The procedure for constructing the thermo-majorization curve is given in the main text and summarized in Eq.~(\ref{resource_thermo_points}). (b) This curve is then used to compare different states. The state $\rho_1$ thermo-majorizes $\rho_2$, $\rho_3$ and $\rho_\beta$. Conversely, $\rho_2$ and $\rho_3$ do not thermo-majorize each other. 
    Based on Ref.~\cite{Horodecki2013}.
    }
    \label{fig:thermo_majorization}
\end{figure}

The criteria of thermo-majorization can be formulated as follows.
For each given state $\rho_S = \sum_i p_i |i\rangle\langle i|$, we construct the so-called thermo-majorization curve of $\rho_S$.
First we relabel the probabilities so that 
\begin{equation}\label{resource_thermo_ordering}
p_1 e^{\beta E_1} \geqslant p_2 e^{\beta E_2} \geqslant \ldots \geqslant p_d e^{\beta E_d} , 
\end{equation}
where $d$ is the Hilbert space dimension, which we assume to be finite. 
This is called $\beta$-ordering. 
We then construct a special curve with points  \begin{equation}\label{resource_thermo_points}
\bigg\{ \sum\limits_{i=1}^k e^{-\beta E_i}, \;\; \sum\limits_{i=1}^k p_i \bigg\}, \qquad k = 1, \ldots, d.
\end{equation}
as illustrated in Fig.~\ref{fig:thermo_majorization}(a). 

This curve is used to compare different states, as we exemplify in Fig.~\ref{fig:thermo_majorization}(b).
If the curve for a certain state $\rho_1$ is always above another, say $\rho_2$, we say $\rho_1$ \emph{thermo-majorizes} $\rho_2$, which is written as 
\begin{equation}\label{resource_majorization_def}
    \rho_1 \succ_{\beta} \rho_2.
\end{equation}
In the example of Fig.~\ref{fig:thermo_majorization}(b) $\rho_1 \succ_{\beta} \rho_2$ but $\rho_2 \nsucc_{\beta} \rho_3$. 
By construction the thermal state $\rho_\beta$ at temperature $\beta$ is a straight line and is majorized by all other states. 
The majorization symbol therefore introduces a natural ordering between states. 
It is essential to note, however, that this ordering is made with reference to the  temperature $\beta$ of the bath. 
In particular, since any state thermo-majorizes $\rho_\beta$, it follows that this must also be true for other thermal states with different temperatures $\beta'$; i.e, $\rho_{\beta'} \succ_{\beta} \rho_\beta$ for any $\beta'$. 

The main result of  Ref.~\cite{Horodecki2013} can now be summarized by the following theorem: 
\begin{theorem}
\label{theorem2}
Given two block-diagonal states $\rho_1$ and $\rho_2$, if $\rho_1 \succ_{\beta} \rho_2$ then it is possible to convert $\rho_1$ to $\rho_2$ using thermal operations. 
\end{theorem}
Thermo-majorization thus offer an unambiguous way of ordering states within the context of thermal operations. 
By analyzing which curves are above the other, we can say which states can be converted to others by means of thermal operations. 
The proof, as well as the intuition, behind Theorem~\ref{theorem2} is based on the connection with majorization theory.
In addition to the original reference, we also refer the reader to~\cite{Lostaglio2015}, where the basic ideas are neatly summarized, and~\cite{Weilenmann2016}, which provides a  thorough discussion on the connection with majorization and the resource theory of purity. 

The basic rationale goes as follows. 
Given two $D$-dimensional probability vectors $\bm{\gamma}_1$ and $\bm{\gamma}_2$ we say that $\bm{\gamma}_1$ majorizes $\bm{\gamma}_2$, written 
$\bm{\gamma}_1 \succ \bm{\gamma}_2$, when 
\begin{equation}\label{resource_majorization}
    \sum\limits_{i=1}^k \gamma_{1i}^{\downarrow} \geqslant \sum\limits_{i=1}^k \gamma_{2i}^{\downarrow},
\end{equation}
for all $k = 1, \ldots, D$. 
Here $\bm{\gamma}^\downarrow$ means the probability $\bm{\gamma}$ sorted in descending order. 
In order to link majorization [Eq.~(\ref{resource_majorization})] to thermo-majorization [Eq.~(\ref{resource_majorization_def})], consider a system with $d$ levels and thermal distribution  $p_i^\text{th} = e^{-\beta E_i}/Z$, where $i = 1, \ldots, d$. 
For simplicity, we assume that the $p_i^\text{th}$ are rationals; i.e., they can be written as $p_i^\text{th} = k_i/D$, where $k_i$ and $D$ are integers such that $\sum_i k_i = D$ (to ensure normalization). In practice, one can always approximate the $p_i^\text{th}$ in this way, with arbitrary accuracy, by using sufficiently large integers. 

Given an arbitrary probability vector $\bm{p} = (p_1, \ldots, p_d)$, one may then define a mapping $\bm{\gamma}(\bm{p})$ that converts the $d$-dimensional vector $\bm{p}$ into the $D$-dimensional vector
\begin{equation}\label{resource_Gamma_map}
    \bm{\gamma}(\bm{p}) = \bigg(\frac{p_1}{k_1}, \frac{p_1}{k_1}, \ldots, \frac{p_2}{k_2}, \frac{p_2}{k_2}, \ldots, \frac{p_d}{k_d}, \frac{p_d}{k_d}, \ldots\bigg),
\end{equation}
where each term $p_i/k_i$ occurs $k_i$ times. 
Notice that the $k_i$'s implicitly depend on $\beta$, since they are defined from  $p_i^\text{th}$. 

As a particular case, we see that the map in Eq.~(\ref{resource_Gamma_map}) takes the thermal state $\bm{p}^\text{th}$ into a uniform distribution, 
\begin{equation}\label{resource_map_gamma_uniform}
    \bm{\gamma}(\bm{p}^\text{th}) = \bm{\eta}, 
\end{equation}
where $\eta_i = 1/D$ is a $D$-dimensional uniform distribution. 
This is similar in spirit to the mapping  between the canonical and microcanonical ensembles in statistical mechanics, {\color{black} in the sense that it maps a thermal distribution in a smaller space, into a uniform distribution (all states equally likely) in a higher-dimensional space.}

Thermal operations have $\bm{p}^\text{th}$ as a fixed point. 
In the larger space of dimension $D$, this is then converted into a map $\mathcal{R}(\bm{\gamma})$ having the uniform distribution $\bm{\eta}$ as the fixed point. 
Maps of this form are called \emph{noisy operations} and play a central role in the resource theory of purity~\cite{Horodecki2003}. 
The question posed in Theorem~\ref{theorem2} can now be converted into, under which conditions can  $\bm{\gamma}(\bm{p}_1)$ be converted into $\bm{\gamma}(\bm{p}_2)$ by means of noisy operations?
As shown in~\cite{Ruch1978}, this is possible precisely when $\bm{\gamma}(\bm{p}_1) \succ \bm{\gamma}(\bm{p}_2)$.
But because of the structure in~(\ref{resource_Gamma_map}), saying that $\bm{\gamma}(\bm{p}_1) \succ \bm{\gamma}(\bm{p}_2)$ is  equivalent to $\bm{p}_1 \succ_\beta \bm{p}_2$.
Hence Theorem~\ref{theorem2} follows. 

The above analysis also serves to emphasize the deep connection between athermality and purity. 
All results for majorization are recovered from thermo-majorization by setting $H_S = 0$;
$\beta$-ordering in Eq.~(\ref{resource_thermo_ordering}), for instance, simply becomes descending ordering and so on.
Thermo-majorization is thus the generalization of majorization theory for ``non-zero Hamiltonians''. 
This will acquire a deeper significance starting from the next section, when we discus monotones for athermality. 
In the resource theory of purity, all that matters are probabilities, so the von Neumann entropy appears as the natural monotone.
For athermality, however, energy also plays a role. 
And, as a consequence, the natural monotones will instead be related to the free energy $F = U - T S$, which is precisely a combination of  energy and entropy.

\subsection{The second laws of thermodynamics}
\label{RTsecondlaw}

The second law~(\ref{Qgen_sigma_thermal_op}) says that a transition from $\rho_S$ to $\rho_S'$ is only possible if the corresponding entropy production is non-negative. 
This, however, is only a necessary condition. 
There may, in principle, exist states which cannot be interconverted into one another, despite leading to a positive entropy production. 
For general maps, establishing sufficient and necessary conditions is unfeasible. 
But for the restricted class of thermal operations, this turns out to be possible, as first done in Ref.~\cite{Brandao2015} using  the idea of \emph{catalytic thermal operations}.

\begin{figure}[t]
    \centering
    \includegraphics[width=\columnwidth]{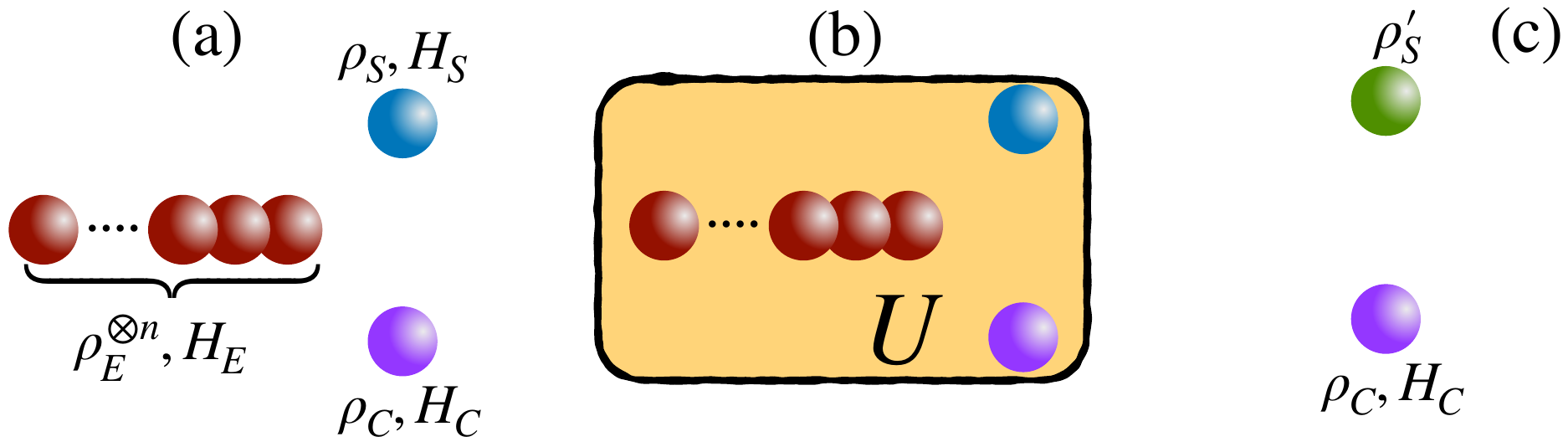}
    \caption{Illustration of the cyclic protocol based on catalytic thermal operations, used in Ref.~\cite{Brandao2015} to generalize the $2^{\rm nd}$ law of thermodynamic to the microscopic quantum domain. {(a)} A system $S$, catalyst $C$ and reservoir $E$, each endowed with their respective Hamiltonians $H_j~(j=S,C,E)$, are prepared in an uncorrelated state. The initial state of the reservoir  consist of the tensor product of $n$ copies of the same thermal state $\rho_{{\rm th},E}$. {(b)} The parties involved in the protocol evolve jointly via the catalytic thermal operation $U$ such that $[U,\sum_j H_j]=0$. {(c)} The process is such that, at the end of the evolution, the catalyst is brought back to its initial state, while  the system $S$ ends up in a state $\rho'_S$.}
    \label{fig:brandao}
\end{figure}

The scenario is the same as in the previous Subsections. However, in addition to the system $S$, one introduces an ancillary system, called the catalyst, with Hamiltonian $H_C$ and prepared in an arbitrary state $\rho_C$. 
The joint $SC$ system then undergoes a thermal operation (conserving the total energy $H_S + H_C + H_E$).
Crucially, though, the thermal operation is chosen such that the catalyst is brought back to its original state $\rho_C$ after the process (cf. Fig.~\ref{fig:brandao}). 
Given this setting, one then asks whether it is possible to convert a state $\rho_S$ into another $\rho_S'$. 

The usual $2^{\rm nd}$ law, written in the form~(\ref{Qgen_sigma_thermal_op}), states that this is possible when $S(\rho_S||\rho_S^\text{th}) \geqslant S(\rho_S'||\rho_S^\text{th})$, a condition which is necessary but not sufficient.  
Instead, as shown in Ref.~\cite{Brandao2015} a necessary and sufficient condition is provided by the following theorem.
\begin{theorem}
\label{theo:Brandao}
A state $\rho_S$, block diagonal in the energy basis, can be converted into $\rho_S'$ by means of catalytic thermal operations if 
\begin{equation}\label{resource_2ndLaws}
    \Sigma_\alpha := S_\alpha(\rho_S || \rho_S^\text{th}) - S_\alpha(\rho_S' || \rho_S^\text{th}) \geqslant 0, \qquad \forall \; \alpha \geqslant 0,
\end{equation}
where 
\begin{equation}\label{resource_Renyi_divergence}
    S_\alpha(\rho_S || \rho_S^\text{th}) = \frac{1}{\alpha-1} \ln \sum\limits_i p_i^\alpha (p_i^\text{th})^{1-\alpha}, 
\end{equation}
is the R\'enyi-$\alpha$ divergence. 
\end{theorem}
This result thus establishes a family of second laws, $\Sigma_\alpha \geqslant 0$, parametrized by the continuous parameter $\alpha \in [0,\infty)$. 
The usual second law in Eq.~(\ref{Qgen_sigma_thermal_op}) is a particular case, corresponding to $\alpha \to 1$.

Alternatively, one may also cast the second laws in terms of R\'enyi-$\alpha$ free energies, defined as
\begin{equation}\label{resource_free_energies}
    F_\alpha(\rho_S) = F_\text{th} + T S_\alpha(\rho_S|| \rho_S^\text{th}), 
\end{equation}
where $F_\text{th} = - T \ln Z_S$. This represents the R\'enyi generalization of the non-equilibrium free energy~(\ref{Qgen_non_eq_F}). 
Eq.~(\ref{resource_2ndLaws}) then becomes
\begin{equation}\label{resource_2ndLaws_F}
    F_\alpha(\rho_S) \geqslant F_\alpha (\rho_S'),\qquad \forall \; \alpha \geqslant 0.
\end{equation}
Recall from  Eq.~(\ref{Qgen_Sigma_DeltaF}) that in the absence of work, $\Sigma = - \beta \Delta F$. Thus, the statement $\Sigma \geqslant 0$ is tantamount to saying ``in order for a process to be possible, the free energy must go down.'' 
But, again, this  is only a necessary condition. 
Conversely, for quantum systems and thermal operations, ``all free energies must go down.''

As in the previous section, a macroscopic limit can be defined in which (i) the system's dimensions $d$ is large and (ii) the state of the system has an energy distribution sharply peaked around $\langle H_S \rangle$. 
In this case it can be shown that all $2^{\rm nd}$ laws stated in Eq.~(\ref{resource_2ndLaws}) converge to the usual one~(\ref{Qgen_sigma_thermal_op}).

\subsection{Coherence and the resource theory of asymmetry}

All results in the previous section hold only for states $\rho_S$ which are block diagonal in the energy basis. 
That is, states of the form 
\begin{equation}\label{resource_block_diagonal_state}
    \rho_S = \sum\limits_{i,j} \delta(E_i = E_j)\rho_{ij} |i\rangle\langle j| ,
\end{equation}
where $\delta(a=b)$ is the Kronecker delta and $\{|i\rangle\}$, $\{E_i\}$ are the eigenstates and eigenvalues of $H_S$.  
Coherences of this form are called non-energetic and play a much smaller role than coherences between different energy states (\emph{energetic coherences})  
due to the special role that energy plays in the dynamics an because -- with the exception of accidental degeneracies -- different energy states are usually associated with different macroscopic configurations.


In the context of thermodynamics, it was shown in Ref.~\cite{Lostaglio2015,Lostaglio2015a} that energetic coherences place additional constraints on the allowed transformations, on top of the second laws~(\ref{resource_2ndLaws}). 
This connection was made by showing that a resource theory of thermodynamics is actually composed of two parts: athermality and \emph{asymmetry}. 

The resource theory of asymmetry (also called quantum reference frames) concerns arbitrary transformations under a certain group~\cite{Gour2008}. 
Let $G$ denote a Lie group and $V_g$ a unitary corresponding to a representation $g\in G$ of the group. 
A state $\rho$ is called a free state if $V_g \rho V_g^\dagger = \rho$. That is, free states are invariant under $G$. 
Similarly, an arbitrary quantum channel $\mathcal{E}(\rho)$ is called a free operation if  
\begin{equation}
    \mathcal{E}\big(V_g \rho V_g^\dagger\big) = V_g \mathcal{E}(\rho) V_g^\dagger, \qquad \forall \rho, \forall g \in G.
\end{equation}
Such channels are also called \emph{covariant}.

Thermal operations [cf. Eq.~(\ref{resource_TO})] are covariant under the group generated by time-translations, i.e., where $V_t =  e^{-i H_S t}$, with  $H_S$ being the generator of the group.
This follows from the fact that $[U,H_S+H_E] = 0$ and $e^{-i H_E t} \rho_E^\text{th} e^{i H_E t} = \rho_E^\text{th}$. {\color{black} After straightforward manipulations, one has
\begin{IEEEeqnarray*}{rCl}
    e^{-i H_S t} \mathcal{T}(\rho_S) e^{i H_S t} 
    &=& \mathcal{T}\big( e^{-i H_S t} \rho_S e^{i H_S t}\big). 
\end{IEEEeqnarray*}
Thermal operations are thus also free operations with respect to asymmetry.} 
The standpoint of this approach is therefore that, by inducing the emergence of a directional arrow of time, thermodynamic irreversibility prevents time-translational invariance in general thermodynamic processes.
On the other hand, the free states will be those which are block diagonal in the basis of $H_S$, since these are the ones which satisfy $e^{-i H_S t} \rho_S e^{i H_S t} = \rho_S$. 
The free states are therefore those with no energetic coherences.

A monotone for coherence can be given by the relative entropy of coherence~\cite{Baumgratz2014}, 
\begin{equation}\label{resource_rel_ent_coh}
    C(\rho_S) = S(\rho_S || \Delta_{H_S}(\rho_S)) = S(\Delta_{H_S}(\rho_S)) - S(\rho_S), 
\end{equation}
where $\Delta_{H_S}(\rho_S)$ is the operation that fully dephases all entries of $\rho_S$ which are not block diagonal in the energy basis of $H_S$ [cf. the discussion in Sec.~\ref{sec:emergence}].
Notice, therefore, that  $\Delta_{H_S}(\rho_S)$ will be a free state from the perspective of asymmetry, for any $\rho_S$. 
Eq.~(\ref{resource_rel_ent_coh}) therefore measures the entropic distance between the state $\rho_S$ and its incoherent version, which is time-translation invariant. Whence, it provides a measure of the break down of time-translation invariance~\cite{RodriguezRosario2013}.  

Moreover, since thermal operations are free operations, they can only reduce the amount of coherence in a state, so that 
\begin{equation}\label{resource_entropy_coh_basic}
    C(\rho_S) \geqslant C(\rho_S'),
\end{equation}
where $\rho_S' = \mathcal{T}(\rho_S)$.
This statement therefore implies that thermal operations cannot generate additional time-translation asymmetry in a system. Further, it characterizes the depletion of coherence and the tendency of the system to equilibrate onto time-translation invariant states, thus elevating coherence to the role of a second important resource in thermodynamics, complementing athermality.

One may now draw here a parallel  with the distinction made in the previous section,  between the second law and the second \emph{laws}: Eq.~(\ref{resource_entropy_coh_basic}) is only a necessary criteria for  $\rho_S$ to be interconvertible into $\rho_S'$. 
Instead, in Ref.~\cite{Lostaglio2015} the authors have proven the following stronger result: 
\begin{theorem}
\label{theo:Lostaglio}
The set of thermal operations on a quantum system is a subset of the set of time-translation invariant operations. Moreover, for all $\alpha>0$ any thermal operation results in
\begin{equation}\label{resource_coherence_2ndLaws}
    S_\alpha(\rho_S|| \Delta_{H_S}(\rho_S)) \geqslant S_\alpha (\rho_S' || \Delta_{H_S}(\rho_S')) \geqslant 0, \qquad \forall \alpha \geqslant 0. 
\end{equation}
\end{theorem}
These conditions are independent of the second laws~(\ref{resource_2ndLaws}) and therefore represent additional constraints that must be satisfied in systems having coherence. 

It is also important to note that Theorems~\ref{theo:Brandao} and~\ref{theo:Lostaglio} cannot be combined into a single family of inequalities. 
For $\alpha = 1$, as already discussed in Sec.~\ref{sec:emergence}, one may split
\begin{equation}
    S(\rho_S || \rho_S^\text{th}) = S(\Delta_{H_S}(\rho_S)|| \rho_S^\text{th}) + C(\rho_S). 
\end{equation}
The first term represents  the quantity entering in Eq.~(\ref{resource_2ndLaws}), that is the block-diagonal part of the state. The second term, on the other hand, is the quantity appearing  in Eq.~(\ref{resource_coherence_2ndLaws}).
Hence, the case $\alpha=1$ can be combined into a single statement 
\begin{equation}
    S(\rho_S|| \rho_S^\text{th}) \geqslant S(\rho_S' || \rho_S^\text{th}),
\end{equation}
which is nothing but the data processing inequality. 
Theorems~\ref{theo:Brandao} and~\ref{theo:Lostaglio}, however, require that the inequalities be satisfied for all $\alpha \geqslant 0$. For $\alpha \neq 1$, there is no simple way of combining Eqs.~(\ref{resource_2ndLaws}) and (\ref{resource_coherence_2ndLaws}).

\subsection{\label{sec:fluc_work}Fluctuating work in the resource theory context}

The concept of work is not easily defined within a resource theory context. 
The reason is that one of the main paradigms in resource theories is to make all processes completely accounted for. 
For instance, the notion of an external agent, which changes the  system Hamiltonian  through a work protocol, must be \emph{internalized} within the description of the process, as this is the only way to guarantee that all changes in energy are accounted for. 
The same difficulties arise for the \emph{storage} of work.
The goal is therefore two-fold:
First, to allow for the Hamiltonian of the system to change during the process, starting at $H_S$ and ending at $H_S'$. 
Second, to provide a physical mechanism to store the work extracted from the process (the \emph{battery}). 
These two problems were addressed  in Ref.~\cite{Alhambra2016}. 
The former is solved  using the notion of a \emph{switch} $X$ and the latter using a continuous variable work storage ancilla, called a \emph{weight} $W$.

The thermodynamic processes in question therefore involves four parts: the system ($S$), weight ($W$), switch ($X$) and environment ($E$). 
The allowed operations are unitaries on $SWXE$ satisfying, as before, the strong energy conservation 
\begin{equation}\label{resource_energy_conservation_SWXE}
    [\mathcal{U}_{SWXE}, H_{SWXE}] = 0,
\end{equation}
where $H_{SWXE}$ is the total Hamiltonian (which will be specified below). 
We now discuss how  the switch and weight have to be constructed in order to yield consistent thermodynamic results. 

We begin with the switch. 
It is chosen as a qubit with computational basis $\{|0\rangle, |1\rangle\}$ and initially prepared in $|0\rangle\langle 0|$. 
We assume that the total Hamiltonian of $SWXE$ has the special form 
\begin{equation}\label{resource_H_SWXE}
    H_{SWXE} = H_S \otimes |0\rangle\langle 0|_X + H_S' \otimes |1\rangle\langle 1|_X + H_W + H_E, 
\end{equation}
where $H_S$ and $H_S'$ are the initial and final Hamiltonians of the system  and $H_W$ and $H_E$ are the Hamiltonians of the weight and environment respectively. 
In addition, one also assumes that all unitaries $\mathcal{U}_{SWXE}$  have the form of controlled operations on the switch: 
\begin{equation}\label{resource_V_SWXE}
    \mathcal{U}_{SWXE} = U_{SWE} \otimes |1\rangle\langle 0|_X + U_{SWE}^\dagger \otimes |0\rangle \langle 1 |_X, 
\end{equation}
where $U_{SWE}$ is a unitary acting only on $SWE$. 

Given an arbitrary initial state $\rho_{SWE}$ of $SWE$, this will therefore  produce the map
\begin{equation}\label{resource_SWXE_map}
    \mathcal{U}_{SWXE} \bigg(\rho_{SWE} \otimes |0\rangle\langle 0|_X\bigg) \mathcal{U}_{SWXE}^\dagger = \rho_{SWE}' \otimes |1\rangle\langle 1|_X,
\end{equation}
where
\begin{equation}\label{resource_SWE_map}
    \rho_{SWE}' = U_{SWE} \rho_{SWE} U_{SWE}^\dagger. 
\end{equation}
The switch therefore neatly \emph{internalizes} the idea of a changing Hamiltonian. 
In particular, it solves the issue of how to express strong conservation in the case when $H_S$ changes during the process: namely, at the level of $SWXE$, the condition  remains in the usual form~(\ref{resource_energy_conservation_SWXE}).
Conversely, at the level of $SWE$, plugging Eqs.~(\ref{resource_H_SWXE}) and~(\ref{resource_V_SWXE}) into Eq.~(\ref{resource_energy_conservation_SWXE}) leads to
\begin{equation}\label{resource_USWE_energy_conservation}
    U_{SWE} (H_S + H_W + H_E) = (H_S' + H_W + H_E) U_{SWE}, 
\end{equation}
which can be viewed as a statement of strong energy conservation for the case where the system Hamiltonian changes. 
In the particular case where $H_S' = H_S$, we recover the usual condition $[U_{SWE}, H_S+H_W+H_E] = 0$. 

As the changes in the switch are trivial [cf. Eq.~(\ref{resource_SWXE_map})], one may henceforth focus only on $SWE$ and its corresponding map. 
That is, the switch is in practice no longer necessary. We therefore now turn to the battery $W$. 
Instead of using a discrete battery, the authors of Ref.~\cite{Alhambra2016} discuss the use of a continuous degree of freedom. 
That is, the battery is assumed to be described by an operator $\hat{x}_W$ having continuous spectra (exactly like the position operator), $\hat{x}_W = \int \ud x \; x |x\rangle\langle x|_W$. 
This is intended to mimic a classical weight, which can be pulled up and down continuously. 
A similar approach is also used in the resource theory of coherence~\cite{Aberg2014}.
The Hamiltonian of the system is then taken to be $H_W = \epsilon \hat{x}$, where $\epsilon$ is just a scaling factor. 
For simplicity, we henceforth set $\epsilon =1$, thus making $\hat{x}_W$ have units of energy instead of position. 

An immediate critique for such a Hamiltonian is that its spectrum is not lower bounded. 
This, however, is usually not an issue: while most of the times the ground-state energy is not involved, when it is, one can always consider a regularized version of $H_W$.
For instance, one can  picture $H_W$ as being instead a displaced harmonic oscillator, but with a large mass and small frequency. 
The large mass makes inertial effects irrelevant and the small frequency represents a very loose trap, which has virtually no influence in the system. 
The spectrum of a displaced oscillator, however, is always lower bounded. 

Since the weight Hamiltonian is proportional to $\hat{x}_W$, displacements of the weight are generated by the corresponding conjugated momentum $\hat{p}_W$ (defined such that $[\hat{x}_W, \hat{p}_W] = i$). 
Based on this, the authors in~\cite{Alhambra2016} postulate that, in addition to Eq.~(\ref{resource_USWE_energy_conservation}), the unitary $U_{SWE}$ should also be constrained to satisfy 
\begin{equation}\label{resource_USWE_p}
    [U_{SWE},\hat{p}_W] = 0. 
\end{equation}
Physically, this implies \emph{translation invariance} for the weight: Pulling the weight before the process does not affect the dynamics. 
Under the constraints in Eqs.~(\ref{resource_USWE_energy_conservation}) and (\ref{resource_USWE_p}), the family of unitaries $U_{SWE}$ is drastically simplified, as shown by the following lemma~\cite{Aberg2014,Alhambra2016}: 
\begin{lemma}
A unitary $U_{SWE}$ satisfying Eqs.~(\ref{resource_USWE_energy_conservation}) and (\ref{resource_USWE_p}) can always be parametrized as 
\begin{equation}\label{resource_U_SWE_parametrization}
    U_{SWE} = e^{i (H_S' + H_E) \hat{p}_W} V_{SE} e^{-i (H_S+H_E)\hat{p}_W},
\end{equation}
where $V_{SE}$ is an arbitrary unitary acting only on $SE$. 
\end{lemma}
It is very important to note that the remaining unitary $V_{SE}$ is now \emph{completely} arbitrary; that is, it does not have to comply with any energy conservation requirements. 
In other words, $V_{SE}$ may perform an arbitrary amount of work on $SE$, because now this is appropriately stored in the weight $W$. 
This therefore represents a significant improvement in flexibility.

Using the representation $\hat{p}_W = \int dp \; p |p\rangle\langle p|_W$, we can also write 
\begin{equation}\label{resource_U_SWE_parametrization2}
    U_{SWE} = \int dp A_{SE}(p)\; |p\rangle \langle p|_W, 
\end{equation}
where
\begin{equation}\label{resource_A_SE}
    A_{SE}(p) = e^{i (H_S'+H_E)p} V_{SE} e^{-i (H_S + H_E)p}, 
\end{equation}
are a family of unitaries parametrized by $p$ (note how both $H_S$ and $H_S'$ appear in this expression). 
Let us now assume that the initial state of $SWE$ is of the form $\rho_{SE} \otimes \rho_W$.
We also assume, for concreteness, that  $\rho_W = |\psi\rangle\langle \psi|_W$ is pure. 
Plugging Eq.~(\ref{resource_U_SWE_parametrization2}) into Eq.~(\ref{resource_SWE_map}) and tracing over $W$ leads to the map
\begin{equation}\label{resource_rho_SE_prime_unital}
    \rho_{SE}' =  \int dp \; A_{SE}(p) \rho_{SE} A_{SE}^\dagger(p)\;  |\langle p |\psi \rangle|^2.
\end{equation}
At the level of $SE$, the dynamics is therefore given by a mixture of unitaries, weighted by probabilities $|\langle p |\psi \rangle|^2$~\cite{Masanes2014}. 
Channels of this type are called \emph{unital}. 
A special property of unital maps is that they always increase the entropy of $SE$. 
The presence of the weight $W$ therefore causes the dynamics of $SE$ to be unital, instead of unitary, introducing additional noise on $SE$.  

To proceed, we consider a slightly simpler  scenario. 
First, note that since $V_{SE}$ is arbitrary, 
the distinction between what is $S$ and what is $E$ becomes somewhat arbitrary. 
One may therefore label $SE$ as a new \emph{system}. 
Or, put it differently, Eq.~(\ref{resource_rho_SE_prime_unital}) also holds in the case when there is no environment present, in which case it can be written more explicitly as
\begin{equation}\label{resource_rho_S_prime_unital}
    \rho_S' = \int dp \; e^{i H_S' p} V_S e^{-i H_S p} \rho_S e^{i H_S p} V_S^\dagger e^{-i H_S' p} \;|\langle p | \psi\rangle|^2,
\end{equation}
where we used Eq.~(\ref{resource_A_SE}). 
This is now \emph{exactly} the usual Jarzynski-Crooks scenario: a system $S$, prepared in $\rho_S$, undergoes a work protocol characterized by a unitary $V_S$ and a change in the system Hamiltonian from $H_S$ to $H_S'$. 
To make this connection even stronger, we shall also assume that $\rho_S = e^{-\beta H_S}/Z_S$.
Eq.~(\ref{resource_rho_S_prime_unital}) then simplifies further to
\begin{equation}\label{resource_rho_S_prime_unital2}
    \rho_S' = \int dp \; e^{i H_S' p} V_S  \rho_S  V_S^\dagger e^{-i H_S' p}\; |\langle p | \psi\rangle|^2.
\end{equation}
Let us now introduce the eigendecompositions $H_S = \sum_n E_n |n\rangle\langle n|$ and $H_S' = \sum_m E_m' |m\rangle \langle m|$ where, in general, the bases $\{|n\rangle\}$ and $\{|m\rangle\}$ need not be the same. 
The evolution of the diagonal entries $p_m' = \langle m |\rho_S' |m\rangle$ 
is then found to be 
\begin{equation}\label{resource_rho_S_diagonal_diagonals}
    p_m' = \sum\limits_n |\langle m | V_S | n \rangle|^2 p_n, 
\end{equation}
where $p_n = \langle n |\rho_S | n\rangle = e^{-\beta E_n}/Z$.
This is thus independent of the weight and also exactly as one would intuitively hope.
For the off-diagonals, however, one finds
\begin{equation}\label{resource_rho_S_diagonal_off_diagonals}
    \langle m_1 | \rho_S' |m_2 \rangle = \langle m_1 |V_S \rho_S V_S^\dagger | m_2 \rangle \int dp \; e^{i (E_{m_1}'-E_{m_2}')p} |\langle p | \psi\rangle|^2.     
\end{equation}
The ``pure'' evolution $V_S \rho_S V_S^\dagger$ is thus \emph{dephased} by an amount which depends on the initial state $|\psi\rangle$ of the weight and the energy differences $E_{m_1}' - E_{m_2}'$. 

For concreteness, let us take as an example a Gaussian wavefunction,  $|\psi\rangle = \int dx \; \psi(x) |x\rangle$, with 
\begin{equation}\label{resource_psi_x}
    \psi(x) = \frac{e^{-x^2/4\delta^2}}{(2\pi\delta^2)^{1/4}},
\end{equation} 
where $\delta$  measures how localized  $\psi(x)$ is in  position space. 
{\color{black} The integral in Eq.~(\ref{resource_rho_S_diagonal_off_diagonals}) can be carried out exactly}, leading to 
\begin{equation}\label{resource_rho_S_diagonal_off_diagonals2}
    \langle m_1 | \rho_S' |m_2 \rangle = \langle m_1 |V_S \rho_S V_S^\dagger | m_2 \rangle e^{-(E_{m_1}' - E_{m_2}')^2/8\delta^2}.
\end{equation}
If $\delta \to 0$, the exponential makes all terms in the right-hand side vanish, except those where $E_{m_1}'=E_{m_2}'$. 
As a consequence, the dynamics takes $\rho_S$ to $\Delta_{H_S'}(V_S\rho_S V_S^\dagger)$, where $\Delta_{H_S'}$ is the full dephasing operator in the eigenbasis of $H_S'$; i.e., which makes $V_S \rho_S V_S^\dagger$ block-diagonal.
It is also important to bear in mind that $\delta \to 0$ corresponds to an ideal weight, since this is the scenario where the ``pointer'' of the weight is perfectly localized at $x = 0$. 

Conversely, when $\delta \to \infty$ the exponential in Eq.~(\ref{resource_rho_S_diagonal_off_diagonals2}) vanishes, leading to $\rho_S' = V_S \rho_S V_S^\dagger$. 
In this limit the evolution of the system is therefore \emph{completely unaffected} by the weight. 
However, the weight itself is now useless since it is initially spread around all positions $x$, so that there is no way of knowing how much work was extracted. 
{\color{black}Curiously, this very type of scenario  appears in voltage-biased Josephson junctions~\cite{Lorch2018}.}

It is therefore quite interesting to note that, as far as the diagonal entries are concerned, the initial state of the weight has no effect on the dynamics. 
Conversely, for the coherences, there is a trade-off between dephasing and the precision with which one can use the weight to extract work. 
This, of course,  is ultimately a consequence of the fact that the weight is performing a von Neumann measurement on the system and  therefore decoheres it in a preferred basis~\cite{Zurek1981}. 

Finally, if the initial state $\rho_S$ of the system is not diagonal, similar conclusions also hold. 
In this case Eq.~(\ref{resource_rho_S_prime_unital}) becomes, component-wise
\begin{equation}
\label{resource_rho_S_prime_components}
\begin{aligned}
    \langle m_1 | \rho_S' |m_2 \rangle 
    &= \sum\limits_{n_1,n_2} \langle m_1 |V_S | n_1 \rangle\langle n_1 |\rho_S | n_2 \rangle\langle n_2 | V_S^\dagger |m_2\rangle \\
   &\times e^{-(E_{m_1}'-E_{m_2}' - E_{n_1}+E_{n_2})^2/8\delta^2}.
   \end{aligned}
\end{equation}
The effect of the weight will only be invisible to those states for which $E_{m_1}'-E_{m_2}' = E_{n_1} - E_{n_2}$.

\subsection{\label{sec:resource_reconciliation}Reconciliation with the stochastic approach}

We are now  in the position to use the framework of Sec.~\ref{sec:fluc_work} to define work at the stochastic level. 
This will serve to reconcile the resource theory approach with the usual work statistics in the Jarzynski-Crooks scenario.
We will discuss this reconciliation using two complementary approaches, one based on the distribution of work~\cite{Alhambra2016} and the other on the cumulant generating function~\cite{Guarnieri2018}. 

The scenario is still the same as in the previous Section. 
We take $\rho_S$ to be initially thermal and consider a two-point measurement scheme. 
First, the system is measured in the basis $|n\rangle$ and the weight prepared in $|\psi\rangle$. 
One then applies the unitary $U_{SW} = e^{i H_S' \hat{p}_W} V_S e^{-i H_S \hat{p}_W}$ [Eq.~(\ref{resource_U_SWE_parametrization})] and, finally,  measure the system in the new energy basis $|m\rangle$ and the weight in the position basis $|x\rangle$. 
The reason for measuring $W$ in $|x\rangle$ is because the weight Hamiltonian is $H_W =\hat{x}_W$.
The position $x$ therefore directly determines the work stored in the weight. 
The conditional probability of obtaining $(m,x)$ given that initially the system was in $n$ is then 
\begin{equation}
    P(m,x|n) = |\langle m, x| U_{SW} | n,\psi\rangle|^2. 
\end{equation}
This is also conditional on $|\psi\rangle$, but we don't write this explicitly since $|\psi\rangle$ is fixed. 
This expression can be simplified further using $U_{SW} = e^{i H_S' \hat{p}_W} V_S e^{-i H_S \hat{p}_W}$. 
In terms of $q(x) = |\psi(x)|^2$, it becomes
\begin{equation}\label{resource_transition_prob_weight}
    P(m,x|n) = |\langle m | V_S | n\rangle|^2 q(x + E_m' - E_n), 
\end{equation}
We therefore see that the transition probability factors as a product of a standard transition pertaining only to the system and a term associated with the initial spread of the weight.

The work distribution can now be computed by  multiplying Eq.~(\ref{resource_transition_prob_weight}) by the initial probability $p_n = e^{-\beta E_n}/Z_S$ and summing over $n,m$, giving
\[
P_F(x) = \sum_{n,m} P(m,x|n) p_n,
\]
where the suffix $F$ stands for forward protocol (an identical construction can also be made for the backward case). 
To match with the standard notation, we will henceforth write $w$ instead of $x$, even though in our construction of the weight the two are the same thing. Substituting Eq.~(\ref{resource_transition_prob_weight}) we then arrive at
\begin{equation}\label{resource_PF_x}
    P_F(w) = \sum\limits_{n,m} |\langle m | V_S | n\rangle|^2 p_n \; q(w+ E_m' - E_n).
\end{equation}
This result can now be directly compared with the standard expression for the work distribution in a unitary protocol~\cite{Talkner2007}, 
\begin{equation}\label{resource_PF_W_id}
    P_F^\text{id}(w) = \sum\limits_{n,m} |\langle m | V_S | n\rangle|^2 p_n \; \delta(w + E_m' - E_n). 
\end{equation}
We see that the only difference is that the delta function is replaced by the probability distribution $q$ of the initial state of the weight. 
In fact, the two distributions are related by the convolution
\begin{equation}
    P_F(w) = \int dw' P_F^\text{id}(w') q(w-w').
\end{equation}
These results illustrate some of the fundamental limitations of thermodynamics in the quantum regime. 
By internalizing the work storage device, one pays the price of obtaining a \emph{noisy work distribution}, where the outcomes $P_F^\text{id}(w)$ are convoluted with the noise $q(w)$ stemming from the initial state of the battery. 
Thus, for instance, while $P_F^\text{id}$ satisfies a Crooks fluctuation theorem~\cite{Crooks1998}, the same is not true for $P_F$. 

To take an example, consider once again the Gaussian wavefunction  in Eq.~(\ref{resource_psi_x}). 
In this case $q(x) =  e^{-x/2\delta^2}/\sqrt{2\pi\delta^2}$, which approximates a delta function when $\delta$ is small.  
But what enters Eq.~(\ref{resource_PF_x}) is $q(w+ E_m' - E_n)$. 
Thus, we reach the important conclusion that in order for the weight to faithfully capture the work statistics, the value of $\delta$ must be much smaller than the typical energy spacings $E_m' -E_n$ entering the process. 
This makes intuitive sense: the precision of the weight must be compatible with the typical energetic transitions entering the process. 

These results help to gain intuition behind the resource theoretical formulation of the weight as an explicit part of the composite system. 
They also show how to reconcile the resource theory and stochastic approaches.
It is important to note, however, that the results summarized by Eq.~(\ref{resource_rho_SE_prime_unital}) cover a much broader set of scenarios, since they encompass (i) the presence of a bath, (ii) arbitrary initial system+bath states and (iii) arbitrary unitaries $V_{SE}$. 
This framework thus also covers joint fluctuation theorems for heat and work, as well as  quantum coherent and correlated scenarios, where the two-point measurement scheme becomes invasive.

Another way of reconciling the resource theoretic and stochastic approaches is by means of the cumulant generating function~\cite{Guarnieri2018};  
We consider again a closed system (no bath) undergoing a work protocol. 
The cumulant generating function associated with the ideal work distribution~(\ref{resource_PF_W_id}) is defined as 
\begin{equation}
\Phi_\eta\equiv\ln \left\langle e^{-\eta W}\right\rangle=\ln \int  P_F^\text{id}(W) e^{-\eta W}\,dW
\end{equation}
The $m$-th cumulant of $P_F^\text{id}(W)$ is then found from $(-1)^m(\partial^m/\partial\eta^m)\Phi_\eta\vert_{\eta=0}$.
Following lines akin to those presented in Sec.~\ref{sec:landauer} [cf. Eq.~\eqref{Giacomo}], one can use H\"older's inequality to obtain a family of lower bounds for the average work (first cumulant), which read
\begin{equation}
\label{lowerboundGiacomo}
\beta\langle W\rangle \geqslant-\frac{\beta}{\eta} \Phi_\eta, \quad \eta \geqslant 0,
\end{equation}
and 
\begin{equation}
\label{upperboundGiacomo}
    \beta \langle W \rangle \leqslant \frac{\beta}{|\eta|} \Phi_\eta, \qquad \eta \leqslant 0. 
\end{equation}
We will now connect this family of bounds to the notions of work extraction and work of formation, discussed in Sec.~\ref{sec:wit}.
To this end, we assume that $\rho_S = \rho_S^\text{th} =  e^{-\beta H_S}/Z_S$. The scenario will thus be akin to that of the work of formation, since we wish to \emph{form} the final state $\rho_S' = V_S \rho_S V_S^\dagger$ from an initially thermal state. 
There is one difference, though, which is that here, during the process, we are also changing the Hamiltonian from $H_S$ to $H_S'$.

Using Eq.~(\ref{resource_PF_W_id}) we may write $\Phi_\eta$ as~\cite{Esposito2009} 
\begin{equation}
    \Phi_\eta =\ln \tr_{S}\left[e^{-\frac{\eta}{2} H_S'} V_S e^{\frac{\eta}{2} H_S} \rho_S e^{\frac{\eta}{2}H_S} V_S^{\dagger} e^{-\frac{\eta}{2} H_S'}\right].
\end{equation}
Defining also ${\rho_S'}^\text{th} = e^{-\beta H_S'}/Z_S'$, as the thermal state at the final Hamiltonian $H_S'$, one may show that $\Phi_\eta$ can be written as 
\begin{equation}\label{Giacomo1}
    \Phi_\eta = - \frac{\eta}{\beta} S_{1-{\eta}/{\beta}}(\rho_S' || {\rho_S'}^\text{th}) - \eta \Delta F,
\end{equation}
where $\Delta F = - T \ln Z_S'/Z_S$ is the difference in equilibrium free energies. 
The cumulant generating function is thus directly associated with the R\'enyi divergences [Eq.~(\ref{resource_Renyi_divergence})], which are the central objects in the resource theory of thermodynamics (recall the discussion in Sec.~\ref{sec:lag}). 

\;
\section{\label{sec:apps}Applications}
%
%

\subsection{\label{sec:swap_engine}The SWAP engine}

\begin{figure*}
    \centering
    \includegraphics[width=\textwidth]{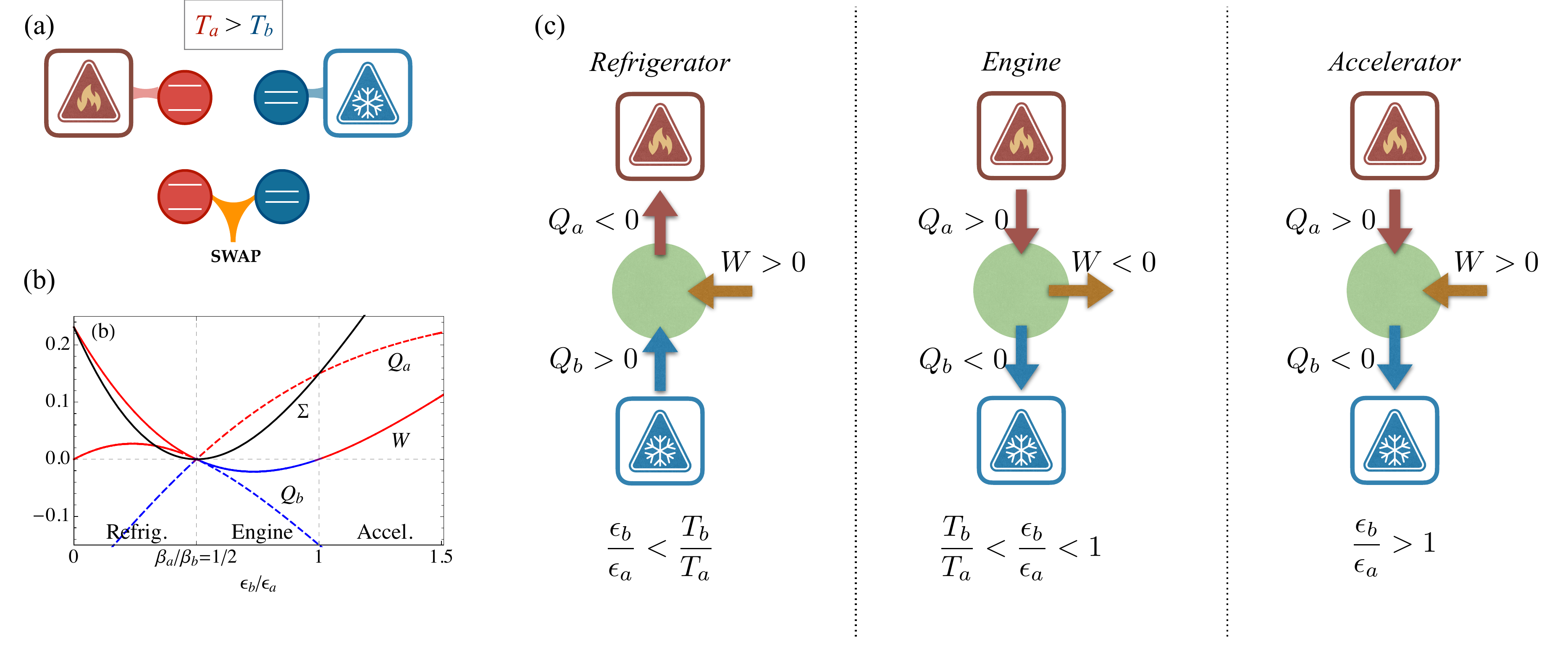}
    \caption{ 
    (a) The two strokes of a SWAP engine.
    (b) Average thermodynamic quantities and entropy production as a function of $\epsilon_b/\epsilon_a$ for $T_b/T_a = 1/2$. All energetic quantities are plotted in units of $\epsilon_a = 1$. 
    (c) Operation regimes of the SWAP engine.}
    \label{fig:swap_engine}
\end{figure*}

One of the prime applications of entropy production is in the description of quantum heat engines. 
Four-stroke engines will be reviewed  in~\ref{sec:app_stroke_based}. 
Here, we begin by describing a particularly simple model, called the SWAP engine~\cite{Allahverdyan2010,Uzdin2014,Campisi2014,Campisi2015}. 
The basic idea is summarized in Fig.~\ref{fig:swap_engine}. 
The working fluid is comprised of two non-resonant qubits, with energy gaps $\epsilon_a$ and $\epsilon_b$. 
The machine operates in two strokes, as depicted in  Fig.~\ref{fig:swap_engine}(a). 
In the first stroke each qubit interacts with its own environment, kept at temperatures $T_a$ and $T_b$ respectively. 
During this stroke the qubits do not interact. 
Moreover, it is assumed that this step is a thermal operation, so that the change in energy of each qubit is entirely associated with the heat that flows to each bath (Sec.~\ref{sec:Qgen_thermal_ops}).
In the second stroke, the baths are uncoupled and the qubits are put to interact by means of a partial SWAP. 
No heat is involved. 
However, since the qubits are not resonant, the partial SWAP will have an associated work cost. 


In the simplest case, one can assume that the thermalization in the first stroke is complete and the SWAP in the second stroke is full. 
Since thermalization is complete, after the first stroke the state of the system will  be $\rho_A^\text{th} \otimes \rho_B^\text{th}$. 
The partial SWAP then changes this to $\rho_B^\text{th} \otimes \rho_A^\text{th}$. 
The work associated with this process is the total change in energy of both qubits, $W = \Delta H_a + \Delta H_b$ which can be written as 
\begin{equation}\label{swap_work_full}
W = - (\epsilon_a - \epsilon_b) (f_a - f_b),
\end{equation}
where $f_i = (e^{\beta_i \epsilon_i}+1)^{-1}$ is the probability of finding each qubit in the excited state (the Fermi-Dirac function). 
The swapped state  $\rho_B^\text{th} \otimes \rho_A^\text{th}$ is then put to interact with the baths at temperatures $T_a$ and $T_b$, causing the system to go back to the original state $\rho_A^\text{th} \otimes \rho_B^\text{th}$. 
The heat exchanged with each bath in this case will then be
\begin{IEEEeqnarray}{rCl}
\label{swap_Qa_full}
    Q_a &=& \epsilon_a (f_a - f_b),   \\[0.2cm]
\label{swap_Qb_full}
    Q_b &=& \epsilon_b (f_b - f_a). 
\end{IEEEeqnarray}
Since the process is cyclic, one can verify that $W + Q_a + Q_b = 0$. 

The values of $W$, $Q_a$ and $Q_b$ are plotted in Fig.~\ref{fig:swap_engine}(b). We define heat and work to be positive when energy enters the system. 
Depending on the relation between $\epsilon_b/\epsilon_a$ and $T_b/T_a$, the engine can offer three regimes of operation:  refrigerator, engine and accelerator. The meaning of the different regimes is diagrammatically explained in Fig.~\ref{fig:swap_engine}(c). 

Since the thermalization strokes are thermal operations, the entropy produced in each cycle will be simply given by Eq.~(\ref{intro_sigma_clausius}), with $\Delta S_S = 0$
\begin{equation}\label{swap_sigma_full}
    \Sigma = - \beta_a Q_a - \beta_b Q_b = - (\beta_a \epsilon_a - \beta_b \epsilon_b) (f_a - f_b).
\end{equation}
This quantity is always non-negative since it has the form $-(x-y) (f(x)-f(y))$, where $f(x) = (e^x+1)^{-1}$ is monotonically decreasing in $x$. As a consequence $f(x)-f(y)$ will always have the opposite sign as $x-y$, for any $x,y$. Hence $\Sigma \geqslant 0$.
Eq.~(\ref{swap_sigma_full}) is plotted in black, in Fig.~\ref{fig:swap_engine}(b). 

Taking $T_a > T_b$, for concreteness, we can characterize the efficiency of the engine in each operating regime by~\cite{Callen1985} 
\begin{IEEEeqnarray}{rCCCCCl}
    \text{COP} &=& \frac{|Q_b|}{W} = \frac{\epsilon_b}{\epsilon_a - \epsilon_b},        &\qquad& \frac{\epsilon_b}{\epsilon_a} <& \frac{T_b}{T_a}&, \\[0.2cm]
    \eta &=& \frac{|W|}{Q_a} = 1 - \frac{\epsilon_b}{\epsilon_a},  
       &\qquad&  \frac{T_b}{T_a} < &\frac{\epsilon_b}{\epsilon_a}& < 1,\\[0.2cm]
    \text{COP}_h &=& \frac{Q_a}{W} = \frac{\epsilon_a}{\epsilon_b - \epsilon_a}, 
      & \qquad&  &\frac{\epsilon_b}{\epsilon_a}& > 1,
      \label{swap_coph_full}
\end{IEEEeqnarray}
where COP stands for coefficient of performance. The machine thus always operates at Otto efficiency. 
As shown recently in~\cite{Molitor2020}, there is an entire class of two-stroke engines for which this turns out to be the case. 

The Carnot point corresponds to  $\epsilon_b/\epsilon_a = T_b/T_a$. This point is special because, even though we get $\Sigma = 0$, we also get $Q_a = Q_b = W = 0$. Thus, at the Carnot point nothing happens (cf. Fig.~\ref{fig:swap_engine}(b)).
Another special point is at $\epsilon_b = \epsilon_a$, where $W = 0$, but  $Q_a = - Q_b \neq 0$.
At this point all heat that flows from the hot bath is converted into heat to the cold bath, so that no net output work occurs.

For $\epsilon_b/\epsilon_a > 1$ heat continues to flow from hot to cold \emph{and}, in addition, one also has to provide a finite work input ($W>0$). This regime is called an accelerator.
In the refrigerator regime, work is consumed to make heat flow from cold to hot. 
In an accelerator, work is consumed to make heat flow from hot to cold, but \emph{``faster''}. 
From a thermodynamical point of view {\color{black} accelerators are interesting because their performance is directly related to the existence of an \emph{excess} entropy production, which turns out to have a clear interpretation.}
The following argument is general and not restricted to the SWAP engine. 
We begin by substituting $Q_b = -W - Q_a$ in Eq.~(\ref{intro_sigma_clausius}) for the entropy production, which yields 
$\Sigma = (\beta_b - \beta_a) Q_a + \beta_b W$. 
In an accelerator $Q_a>0$ and $W>0$. Hence, there is a minimum entropy production associated with it, which is when $W = 0$, which reads 
$\Sigma_\text{min} = (\beta_b - \beta_a) Q_a$.
This is thus the entropy production associated with the natural flow of heat from hot to cold. 
The coefficient of performance of the accelerator is defined as  the amount of heat that can be extracted from the hot bath divided by the associated work cost, $\text{COP}_h = Q_a/W$, as in Eq.~(\ref{swap_coph_full}). 
With some rearrangements, we can also write this as 
\begin{equation}
    \text{COP}_h = \frac{\beta_b Q_a}{\Sigma - \Sigma_\text{min}}. 
\end{equation}
Thus, we see that the efficiency of an accelerator actually depends on the \emph{excess} entropy production {\color{black}$\Sigma - \Sigma_\text{min}$, which represents the extra irreversibility introduced by the additional work used to pump the heat.}

\subsection{\label{sec:app_stroke_based}Stroke-based engines}

\begin{figure}
    \centering
    \includegraphics[width=0.45\textwidth]{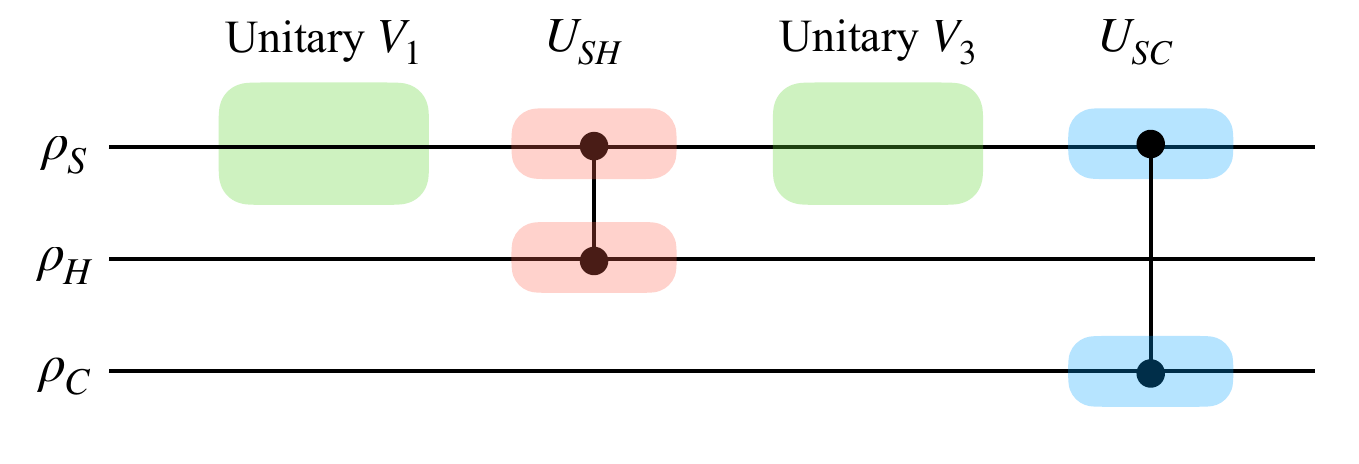}
    \caption{Circuit diagram of a 4-stroke heat engine, composed of two unitaries, $V_1$ and $V_2$ interlaced with two dissipative interactions with a hot and cold bath respectively.}
    \label{fig:stroke_engine_circuit}
\end{figure}

We now turn to a description of more general, four-stroke engines. 
We focus on how to apply the framework of Sec.~\ref{sec:Qgen}, and in particular the basic $SE$ map in Eq.~\eqref{Qgen_global_map}, to this specific problem. 
We consider a four-stroke engine, where unitary (work only) maps in the system are riffled between dissipative interactions with a hot and a cold bath (which may involve both heat and work). The corresponding circuit diagram is depicted in Fig.~\ref{fig:stroke_engine_circuit}.

We consider the engine's operation in a collisional model sense (Sec.~\ref{sec:CM}).
The system is initially prepared in an arbitrary state $\rho_S$. In each stroke, it interacts with two baths, $H$ and $C$,  prepared in states $\rho_H$ and $\rho_C$. For ease of mind, we imagine that these represent a hot and cold bath respectively. 
The results below, however, are actually  true for any bath state, not necessarily thermal. 
Each cycle of the engine is divided into four strokes, as follows.
The first and third strokes involve unitary interactions $V_1$ and $V_3$ acting only on the system.
The second stroke refers to the interaction with the hot bath, by means of a unitary $U_{SH}$.
And similarly, the fourth stroke is between $S$ and $C$, with a unitary $U_{SC}$. 
The global state of $SHC$ after each stroke will then be given by 
\begin{equation}
\begin{aligned}
    \rho_{SHC}^{(1)} &= \big[V_1 \rho_S V_1^\dagger \big]\rho_H \rho_C  
    =\rho_S^{(1)} \rho_H \rho_C,\\ 
    \rho_{SHC}^{(2)} &= U_{SH} \big[ \rho_S^{(1)} \rho_H \big] U_{SH}^\dagger \rho_C 
=\rho_{SH}^{(2)} \rho_C,\\
    \rho_{SHC}^{(3)} &= \big[ V_2 \rho_{SH}^{(2)} V_2^\dagger\big] \rho_C 
    =\rho_{SH}^{(3)} \rho_C,\\
    \rho_{SHC}^{(4)} &= U_{SC} \rho_{SH}^{(3)} \rho_C U_{SC}^\dagger,
    \end{aligned}
\end{equation}
where care was taken in highlighting, at each step, in which Hilbert spaces the unitaries act and what is the structure of the resulting state.
Combining all  strokes, the state at the end of the cycle will thus be 
\begin{equation}
    \rho_{SHC}^{(4)} = U_{SC} V_2 U_{SH} V_1 \big( \rho_S \rho_H \rho_C\big) V_1^\dagger U_{SH}^\dagger V_2^\dagger U_{SC}^\dagger. 
\end{equation}
Notice that all unitaries have a common support on $S$ and therefore in general do not commute. 
Tracing over $H$ and $C$ leads to a stroboscopic map for the system, 
\begin{equation}
    \rho_S' := \Phi(\rho_S) = \tr_{HC} \bigg\{ U_{SC} V_2 U_{SH} V_1 \big( \rho_S \rho_H \rho_C\big) V_1^\dagger U_{SH}^\dagger V_2^\dagger U_{SC}^\dagger\bigg\}. 
\end{equation}
This state is then to be used as  input state for the next cycle, which is constructed with fresh new baths $\rho_H$ and $\rho_C$. 

Proceeding in this way, one can construct finite-time engines operating 
under arbitrary conditions. 
The state of the system after each complete cycle is obtained from the previous one by applying the map $\Phi$. 
After many cycles are performed, the system will usually reach a \emph{limit-cycle}  $\rho_S^*$ satisfying $\rho_S^* = \Phi(\rho_S^*)$.
Once the limit cycle is reached, the engine's operation becomes periodic 
and any function of state, such as the system energy or entropy, no longer change. 
The limit cycle shares many similarities with  non-equilibrium steady-states (NESSs). 
In fact, the limit cycle can be viewed as a \emph{stroboscopic NESS}, in the sense that if viewed only at integer steps, the system no longer changes. 
\emph{Internally} however (i.e., inside each cycle), its state is constantly changing.

The entropy production in each stroke is given by the general expression~\eqref{Qgen_Sigma}. 
The first and third strokes are unitary and no entropy is produced. 
Thus
  $\Sigma = \Sigma_H + \Sigma_C$, 
which can be further split as
\begin{equation}\label{stroke_Sigma_MIs}
    \Sigma = \mathcal{I}_{\rho_{SH}^{(2)}} (S\! : \! H) + S(\rho_H^{(2)} || \rho_H) + \mathcal{I}_{\rho_{SC}^{(4)}} (S\! : \! C) + S(\rho_C^{(4)} || \rho_C).
\end{equation}
This expression is useful if one is interested in analyzing the individual contributions of the mutual informations and relative entropies to the total entropy production. 
Instead, if one is interested only in $\Sigma$ itself, it is simpler to use Eq.~\eqref{Qgen_sigma_phi} to express it in terms of the entropy flux $\Phi$, which in this case becomes 
\begin{equation}\label{stroke_Sigma_fluxes}
    \Sigma = \Delta S_S + \Phi_H + \Phi_C, 
\end{equation}
where $\Delta S_S = S(\rho_S^{(4)}) - S(\rho_S)$ is the net change in entropy of the system in a full cycle and
\begin{equation}
    \Phi_i = \tr_i \bigg\{ (\rho_i - \rho_i')\ln \rho_i\bigg\}, \qquad i = H, C,
\end{equation}
is the entropy flux to baths $H$ and $C$, with $\rho_i'$ denoting the state of the bath after having interacted with the system.

Eq.~\eqref{stroke_Sigma_fluxes} shows that the familiar structure for the entropy production, in terms of changes in entropy of the system and fluxes to the bath, also holds quite generally for any stroke-based engine with the structure of Fig.~\ref{fig:stroke_engine_circuit}. 
What is important to realize, is that this also includes \emph{arbitrary} initial states for the environments, not necessarily thermal. 
In fact, note that no mention has to be made of heat and work, and the associated conundrums.
Eqs.~\eqref{stroke_Sigma_MIs} or~\eqref{stroke_Sigma_fluxes} thus provide a fully information-theoretic definition of irreversibility for a cyclic engine.
Of course, if the bath happens to be thermal, then Eq.~\eqref{stroke_Sigma_fluxes} reduces to the familiar result $\Sigma = \Delta S_S + \beta_H Q_H +\beta_C Q_C$.

In the limit cycle the first term in~\eqref{stroke_Sigma_fluxes} vanishes and we are left only with $\Sigma = \Phi_H + \Phi_C$. 
It is crucial, however, to notice that this does not imply $\Sigma_H = \Phi_H$ and $\Sigma_C = \Phi_C$. 
This would, in fact, be inconsistent, as one of the two fluxes is in general negative. 
The \emph{net} entropy production rate $\Sigma$ coincides in the limit cycle with the \emph{net} flux $\Phi_H + \Phi_C$. But individually they do not. 
The individual contributions $\Sigma_H$ and $\Sigma_C$ are interesting, as they quantify the contribution of each dissipation channel to the system's irreversibility. 
But the only way to assess them is through Eq.~\eqref{stroke_Sigma_MIs}.


\subsection{Squeezed baths}
\label{sec:SqueezedBaths}

In this section we discuss the thermodynamics of  squeezed reservoirs. 
These types of baths can be used, for instance, as a resource to operate  heat engines above Carnot efficiency, as discussed theoretically in Ref.~\cite{Ronagel2014,Abah2014} and implemented experimentally in Ref.~\cite{Klaers2017a}. 
Here we focus on how to formulate the entropy production for this problem, as first put forth in~\cite{Manzano2016}. 

We begin by briefly reviewing the basics of squeezing. 
Consider a single bosonic mode $b$ with Hamiltonian $H = \Omega (b^\dagger b+{1}/{2})$. 
We say $b$ is prepared in a \emph{squeezed thermal state} when its density matrix has the form 
\begin{equation}\label{squeezed_state_basic}
    \rho = S(z) \rho_\text{th} S^\dagger(z), 
\end{equation}
where $\rho_\text{th} = e^{-\beta H}/Z$ is the thermal state and 
\begin{equation}\label{squeezed_squeeze_operator}
    S(z) = e^{\frac{1}{2}(z^* b^2 - z b^{\dagger2})}, \qquad z = r e^{i \theta},
\end{equation}
is the squeezing operator, with complex parameter $z$.
The action of $S(z)$ on annihilation operators is given by 
\begin{equation}\label{squeezed_action_on_b}
    S(z) b S^\dagger(z) = b \cosh(r) + e^{i \theta} b^\dagger \sinh(r). 
\end{equation}
From this, one may readily compute the expectation values of the second moments in the state of Eq.~(\ref{squeezed_state_basic})
\begin{IEEEeqnarray}{rCl}
\label{squeezed_ave_bdb}
    \langle b^\dagger b \rangle +{1}/{2}&=& (\bar{n}+{1}/{2})\cosh(2r), \\[0.2cm]
\label{squeezed_ave_bb}    
    \langle b b \rangle &=&  (\bar{n}+{1}/{2})e^{i\theta} \sinh(2r),
\end{IEEEeqnarray}
where $\bar{n} = (e^{\beta \Omega}-1)$ is the Bose-Einstein distribution, related to the thermal part of~(\ref{squeezed_state_basic}).
In terms of quadratures $q = (b+b^\dagger)/\sqrt{2}$  and $p = i(b^\dagger - b)/\sqrt{2}$,
if $\theta = 0$ we get 
$\langle q^2\rangle = e^{2r}(\bar{n}+{1}/{2})$ and $\langle p^2\rangle = e^{-2r}(\bar{n}+{1}/{2})$, so the variance of $q$ is stretched by $e^{2 r}$, while that of $p$ is squeezed by $e^{-2r}$.
When $\theta \neq 0$ something analogous happens, but in a different direction of the $(q,p)$ plane.


From a thermodynamic perspective, the squeezed state in Eq.~(\ref{squeezed_state_basic}) can be viewed as a Generalized Gibbs Ensemble (GGE), akin to the grand-canonical state $e^{-\beta (H - \mu \hat{N})}$ (where $\mu$ is the chemical potential and $\hat{N}$ is the particle number operator). 
This can be made more transparent by noting that, from Eq.~(\ref{squeezed_action_on_b}), one has 
    $S(z) H S^\dagger(z) = \cosh(2r) H + \sinh(2r) A$, 
where 
  $A = {\Omega}(e^{i \theta} b^{\dagger2} + e^{-i \theta} b^2)/2$, 
is what we shall henceforth refer to as asymmetry~\cite{Manzano2016} (in the sense that $\langle A \rangle$ measures how asymmetric, or compressed, the compressed Gaussian in phase space is).
Eq.~(\ref{squeezed_state_basic}) can then be written in the GGE form
\begin{equation}\label{squeezed_GGE}
    \rho = \frac{1}{Z}e^{-\beta (\cosh(2r) H + \sinh(2r) A)}.
\end{equation}
There is, though, one fundamental difference with respect to the usual Grand canonical state: namely that, unlike $H$ and $\hat{N}$,  the operators $H$ and $A$ do not commute. 
GGEs of this form are called non-Abelian~\cite{Manzano2020}.

We now use the results developed in Sec.~\ref{sec:Qgen} to formulate the entropy production of a system interacting with a squeezed thermal bath.
We shall do so using the standard von Neumann entropy. 
This therefore represents an alternative to the phase-space approach discussed in~\eqref{sec:QPhaseSpace}. 
The system is assumed to be arbitrary (it does not have to be bosonic) and the bath is taken to be a collection of bosonic modes $b_k$, with Hamiltonian $H_E = \sum_k \Omega_k (b_k^\dagger b_k +{1}/{2})$ and  prepared in a squeezed thermal state of the form~(\ref{squeezed_state_basic}); viz.,
\begin{equation}\label{squeezed_rhoE}
    \rho_E = \prod\limits_k \rho_k = \prod\limits_k S_k(z_k) \rho_k^\text{th} S_k^\dagger(z_k), 
\end{equation}
where $\rho_k^\text{th} =  (1-e^{-\beta \Omega_k})e^{-\beta \Omega_k b_k^\dagger b_k}$ is the thermal state and $S_k(z_k)$ is the squeezing operator~(\ref{squeezed_squeeze_operator}) for mode $b_k$, with parameter $z_k = r_k e^{i \theta_k}$. 
For now we allow each $r_k$ to be different.

The system and bath are then put to interact via an arbitrary unitary $U$, according to the map~(\ref{Qgen_global_map}). 
The entropy produced in the process is given by  Eq.~(\ref{Qgen_sigma_trace_expression}). 
This  can be simplified by inserting Eq.~(\ref{squeezed_GGE}) for $\ln \rho_E$, leading to 
\begin{equation}
\label{squeezed_sigma}
\begin{aligned}
    \Sigma &= \Delta S_S +  \beta \sum\limits_k \bigg\{\Omega_k \cosh(2r_k) \Delta \langle b_k^\dagger b_k \rangle \\
    & +\frac{\Omega_k}{2} \sinh(2r_k) \bigg( \Delta \langle b_k^\dagger b_k^\dagger \rangle e^{i \theta_k} + \Delta \langle b_k b_k \rangle e^{-i \theta_k} \bigg)\bigg\}, 
    \end{aligned}
\end{equation}
where $\Delta \langle \mathcal{O}_E\rangle = \tr\big\{ \mathcal{O}_E (\rho_E' - \rho_E)\big\}$ is the change in the expectation value of a bath observable during the process. 
It is essential to note that, in line with what was  discussed in Sec.~\ref{sec:Qgen},  all terms except the first actually refer to changes in quantities of the bath, not the system.
For this reason, the entropy production \emph{cannot}, in general, be computed solely from knowledge of the changes that take place in $S$ (more about this below). 

Since the bath is not thermal, 
Eq.~(\ref{squeezed_sigma})  cannot be written in the  Clausius form $\Sigma = \Delta S_S + \beta \Delta Q_E$ Let us assume, for concreteness, that all modes are squeezed by the same amount, $r_k = r$, $\theta_k = \theta$.
The second term in Eq.~(\ref{squeezed_sigma}) then becomes proportional to the heat flux,
    $\Delta Q_E = \sum_k \Omega_k \Delta \langle b_k^\dagger b_k \rangle$. 
Moreover, the last term becomes proportional to the change in  asymmetry, 
\begin{equation}\label{squeezed_asymmetry}
    \Delta A_E = \sum\limits_k \frac{\Omega_k}{2} \bigg(e^{i \theta} \Delta \langle b_k^\dagger b_k^\dagger \rangle  + e^{-i \theta} \Delta \langle b_k b_k \rangle  \bigg).
\end{equation}
Eq.~(\ref{squeezed_sigma}) thus becomes
\begin{equation}\label{squeezed_sigma2}
    \Sigma = \Delta S_S + \beta \bigg(\cosh(2r) \Delta Q_E +  \sinh(2r) \Delta A_E \bigg).
\end{equation}
This expression resembles the entropy produced when interacting with a grand canonical bath. 
The last two terms represent the changes in the corresponding thermodynamic charges, $\Delta Q_E$ and $\Delta A_E$, each multiplied by the corresponding thermodynamic affinities $\beta \cosh(2r)$ and $\beta \sinh(2r)$. 
This matches the previously discussed intuition of the squeezed state as a GGE.
For instance, one could have a situation where no heat flows to the bath, $\Delta Q_E = 0$, but entropy is still produced due to a flow of asymmetry.

For generic system Hamiltonians and system-environment interactions, it is not possible to write Eq.~(\ref{squeezed_sigma2}) solely in terms of system quantities. 
The situation is entirely analogous to that of strict energy conservation, Eq.~(\ref{Qgen_strict_energy_conservation}).
To provide a concrete example,  suppose the system is a single bosonic mode, described by annihilation operator $a$ and $H_S = \omega a^\dagger a$, while the bath is also comprised of a single mode, with operator $b$ and $H_E = \omega b^\dagger b$ (i.e., resonant with $S$). 
As shown in~\cite{Manzano2020}, the \emph{only} Gaussian unitary which preserves both energy and asymmetry for 2 modes is of the form 
\begin{equation}\label{squeezed_V}
    U_{SE} = \exp\{g t(a^\dagger b - b^\dagger a)\}.
\end{equation}
The choice of phase here is crucial. 
A generic interaction of the form $g a^\dagger b + g^* b^\dagger a$ preserves the number of quanta (and hence the energy, since $S$ and $E$ are assumed to be resonant). 
But in general it does not preserve the asymmetry. 
Only for the specific choice of phase in~\eqref{squeezed_V} will we have both $[U_{SE},a^\dagger a + b^\dagger b] = 0$ \emph{and} $[U_{SE}, aa+bb] \neq 0$. 
In this case $\Delta Q_E = - \Delta Q_S$ and $\Delta A_E = - \Delta A_S$, so Eq.~\eqref{squeezed_sigma2} can be expressed solely in terms of system-related quantities. 
Alternatively, we can also write $\Sigma$ as in 
Eq.~\eqref{Qgen_Sigma_nonThermal_FixedPoint}, with $\rho_S^*$ now being the GGE~\eqref{squeezed_GGE},
which will be a global fixed point (Sec.~\ref{sec:qgen_global}) of the map.

Lastly, we discuss the continuous time version of the above process, where the system evolves instead according to the Lindblad master equation 
\begin{equation}
\label{squeezed_M}
\begin{aligned}
    \frac{d\rho_S}{dt} &= \gamma(N+1) \bigg[ a \rho_S a^\dagger - \frac{1}{2} \{a^\dagger a, \rho_S\}\bigg] + \gamma N \bigg[a^\dagger \rho_S a - \frac{1}{2} \{a a^\dagger, \rho_S\} \bigg]\\
    & - \gamma M \bigg[ a^\dagger \rho_S a^\dagger - \frac{1}{2}\{a^{\dagger 2}, \rho_S\} \bigg] - \gamma M^* \bigg[ a \rho_S a - \frac{1}{2}\{a^2,\rho_S\}\bigg]. 
\end{aligned}
\end{equation}
Here $\gamma \geq 0$ is the damping rate and $N +{1}/{2}=  (\bar{n}+{1}/{2})\cosh(2r)$ and $M = (\bar{n}+{1}/{2})e^{i\theta} \sinh(2r)$ are the parameters imposed by the squeezed thermal bath [cf. Eqs.~(\ref{squeezed_ave_bdb}) and (\ref{squeezed_ave_bb})]. 
This equation can be derived using the usual Born/Markov/Secular approximations~\cite{Breuer2007} or using a collisional model, exactly as described in Sec.~\ref{sec:CM_continuous}. 
From knowledge only of the master equation only~(\ref{squeezed_M}), it is not possible to define the entropy production. 
But if one assumes that the master equation was derived via interactions which are both energy \emph{and} asymmetry preserving (at least approximately), then we can use a continuous-time version of Eq.~\eqref{Qgen_Sigma_nonThermal_FixedPoint}; i.e.,
\begin{equation}\label{squeezed_Pi}
    \dot{\Sigma} = - \frac{d}{dt} S(\rho_S(t) || \rho_S^*). 
\end{equation}
It was also shown in~\cite{Manzano2018b} that, if the strong fixed-point hypothesis does not hold, Eq.~\eqref{squeezed_Pi} will nonetheless still  describe a part of the entropy production; namely the 
so-called non-adiabatic component, associated with the entropy production needed to reach the stationary state.

%
%
\subsection{Quantum heat}
%
%

We now turn to another application of thermodynamics beyond standard thermal systems.
In Ref.~\cite{Elouard2017a} the authors  considered a generalization of the 1st and 2nd laws of thermodynamic for a situation where the interaction with a heat bath is replaced by a set of quantum measurements.
In its simplest formulation, the process can be described as follows. 
The system starts in a pure state $|\psi_0\rangle$. 
At evenly spaced times $n \Delta t$, $n = 0,1,\ldots$, one applies  a projective measurement described by an orthonormal basis $\{|k_n\rangle\}$. 
These sets may be different at different times, which is left implicit in the additional index $n$ in $|k_n\rangle$. 

On the other hand, in between jumps, from $n\Delta t^+$ to $(n+1)\Delta t^-$, the system evolves unitarily from $|k_n\rangle$ to 
$|\psi_{n+1}^- \rangle = U_{n+1,n} |k_n\rangle$,
where $U_{n+1,n}$ is the unitary generating this evolution.
At time $(n+1)\Delta t$ it then undergoes another quantum jump to one of the states $|k_{n+1}\rangle$. 
The probability associated to this jump is 
\begin{equation}\label{measurement_projective_conditional}
    p(k_{n+1} | k_n) = |\langle k_{n+1} | \psi_{n+1} \rangle|^2 = |\langle k_{n+1} | U_{n+1,n} |k_n\rangle|^2, 
\end{equation}
which thus only depends on the previous state $|k_n\rangle$.
A quantum trajectory for this process, up to time  $n\Delta t$, is then specified by the set of quantum numbers $\kk_n = (k_0,\ldots,k_n)$. 
Using Eq.~(\ref{measurement_projective_conditional}), the corresponding path probability reads
\begin{equation}\label{measurement_projective_path_prob_forward}
    \mathcal{P}_F[\kk_n] = p(k_n| k_{n-1}) p(k_{n-1} | k_{n-2}) \ldots p(k_1| k_0) p(k_0), 
\end{equation}
where $p(k_0) = |\langle k_0 | \psi_0 \rangle|^2$. If the initial state is an element of $\{ |k_0\rangle\}$ then $p(k_0)$ becomes deterministic. 

From Eq.~(\ref{measurement_projective_path_prob_forward}) one can readily compute the probability of the final state, which reads
\begin{equation}\label{measurement_projective_marginal}
    p(k_n) = \sum\limits_{k_1, \ldots, k_{n-1}} \mathcal{P}_F[\kk_n]. 
\end{equation}
This can then used to define the reverse process, where the system starts in $|k_n\rangle$ with probability $p(k_n)$, and then evolves backwards by applying the time-reversed unitaries $U_{n,n+1}^\dagger$.
Since $|\langle k_n| U_{n,n+1}^\dagger | k_{n+1} \rangle|^2 = |\langle k_{n+1} | U_{n+1,n} | k_n\rangle|^2 = p(k_{n+1} | k_n)$,  the time-reversed path probability becomes
\begin{equation}
    \mathcal{P}_B[\gamma] = p(k_1| k_0) \ldots p(k_{n-1} | k_{n-2})p(k_n|k_{n-1}) p(k_n).
\end{equation}
The entropy production is defined as in Eq.~\eqref{Qgen_sigma_stochastic_fundamental}, which simplifies in this case to
\begin{equation}\label{measurement_projective_sigma}
    \sigma[\kk_n] = \ln \frac{\mathcal{P}_F[\kk_n]}{\mathcal{P}_B [\kk_n]} =\ln \frac{p(k_0)}{p(k_n)},
\end{equation}
since all conditional terms in $\mathcal{P}_F$ and $\mathcal{P}_B$ cancel out.

The two terms in Eq.~(\ref{measurement_projective_sigma}) are interpreted as follows. 
The contribution $\ln p(k_n)$ is the entropy production associated with the randomness that is built up by the stochastic jumps caused by the projective measurements. 
The term $\ln p(k_0)$, on the other hand, is related to the fact that even the first measurement is non-deterministic;  this randomness is of purely quantum origin, being associated with the fact that $|\psi_0\rangle$ has some finite coherence in the basis $\{|k_0\rangle\}$.
The exact same result, however, could also be obtained if we were to assume that the initial state of the system was an incoherent mixture. 
Thus, the term $\ln p(k_0)$ refers to the general randomness stemming from the first measurement, irrespective of whether this randomness is classical or quantum. 

The stochastic entropy production~(\ref{measurement_projective_sigma}) satisfies a fluctuation theorem by construction. 
Moreover, averaging it over the forward distribution~(\ref{measurement_projective_path_prob_forward}) one finds 
\begin{equation}\label{measurement_projective_ave_sigma}
    \langle \sigma[\kk_n] \rangle = S(p(k_n)) - S(p(k_0)) \geqslant 0,
\end{equation}
where $S(p)=-\sum_n p_n \ln p_n$ is the classical Shannon entropy.
The positivity of Eq.~(\ref{measurement_projective_ave_sigma}) is actually a subtle feature of projective measurements, related to the fact that $p(k_n)$ and $p(k_0)$ are linked through a doubly stochastic matrix~\cite{Nielsen}.
More specifically,   from~(\ref{measurement_projective_path_prob_forward}) and (\ref{measurement_projective_marginal}), we can write
$p(k_n) = \sum_{k_0} \mathcal{M}(k_n,k_0) p(k_0)$,
where
\begin{equation}
    \mathcal{M}(k_n,k_0) = \sum\limits_{k_1,\ldots, k_{n-1}} p(k_n| k_{n-1}) p(k_{n-1} | k_{n-2}) \ldots p(k_1| k_0),
\end{equation}
is doubly stochastic, $\sum_{k_n} \mathcal{M}(k_n,k_0) = \sum_{k_0} \mathcal{M}(k_n,k_0) = 1$. 
Due to the data processing inequality, it then follows that the entropy of $p(k_n)$ is always larger or equal than that of $p(k_0)$, which thus implies the positivity of the average entropy production in Eq.~(\ref{measurement_projective_ave_sigma}). 

Although Eq.~(\ref{measurement_projective_sigma}) provides a consistent definition of entropy production, it is not possible to expect any relation between  $\sigma$ and thermodynamic quantities such as heat and work, as appears in the original  Clausius inequality Eq.~(\ref{intro_sigma_clausius})~\cite{Mohammady2020}. 
While notions of heat and work can still be  defined~\cite{Elouard2017a}, as the states in question are never thermal in shape, entropy production and heat have no straightforward relation with each other. 
We also mention, in passing, that such notions of heat and work do not take into account the energy cost itself of performing a projective measurement, something which has recently been put under scrutiny~\cite{Guryanova2018}.

%
%
\subsection{\label{sec:inf_quenches}Infinitesimal quenches}
%
%

We continue here our review of the non-equilibrium lag, first discussed in Sec~\ref{sec:lag}.
But now we focus on the specific scenario of infinitesimal quenches. 
All ideas and notations are the same as in Sec~\ref{sec:lag}.
One of the difficulties with characterizing the non-equilibrium lag is its dependence on the form of the work protocol $H(\lambda(t))$. 
Or, what is equivalent, the form of the unitary $V$ in Fig~\ref{fig:nonEq_lag}. 
This can be simplified by considering  \emph{quantum quenches}~\cite{Fusco2014a}. 
That is, one assumes that the protocol taking $H_i \to H_f$ is much faster than the typical time-scales of the system, so that the evolution can be taken to be instantaneous. 
This therefore amounts to setting $V \simeq 1$, so that the final state coincides with the initial one, $\rho' = \rho_i^\text{th}$.  
The basic idea is therefore that the changes in the Hamiltonian are so fast that the system has no time to respond, so even though $H_i \to H_f$, the system stays frozen at $\rho_i^\text{th}$. 
Of course, after the quench, many things can happen. 
If the system is isolated, it will evolve according to the new Hamiltonian $H_f$~\cite{Calabrese2005a}. And if it is coupled to a bath, it will eventually thermalize, changing from $\rho_i^\text{th} \to \rho_f^\text{th}$.

All equations in Sec.~\ref{sec:lag} are simplified in this case.
In particular, Eq.~\eqref{Qgen_nonEqLag_Sigma} becomes 
\begin{equation}\label{InfQuenches_Sigma}
\Sigma = S(\rho_i^\text{th} || \rho_f^\text{th}), 
\end{equation}
while the CGF~\eqref{Qgen_nonEqLag_CGF_Renyi} transforms to 
\begin{equation}\label{InfQuenches_CGF}
    K(\lambda) = (\lambda-1) S_\lambda(\rho_f^\text{th} || \rho_i^\text{th}) = -\lambda S_{1-\lambda} (\rho_i^\text{th} || \rho_f^\text{th}). 
\end{equation}
The two expressions for $K(\lambda)$ coincide due to the properties of the R\'enyi divergences.

In the quantum quench scenario, the non-equilibrium lag  depends only on the initial and final work parameters $\lambda_i$ and $\lambda_f$; it becomes independent of the specific protocol $\lambda(t)$ taking one to the other. 
An additional simplification can be obtained for \emph{infinitesimal quenches}. 
That is, when $\lambda_i = \lambda$ and $\lambda_f = \lambda + \delta \lambda$, with $\delta \lambda$ taken to be very small. In this case Eqs.~\eqref{InfQuenches_Sigma} and \eqref{InfQuenches_CGF} can be expanded in a power series in $\delta \lambda$, greatly simplifying the problem.

We start with~\eqref{InfQuenches_Sigma}. 
It is convenient to write it in terms of the average work and equilibrium free energy, Eq.~\eqref{Qgen_nonEqLag_irr_work}. 
In the quench scenario this becomes
\begin{equation}\label{InfQuenches_Sigma_irr_work}
    \Sigma =  \beta \tr\big\{ (H_f - H_i) \rho_i^\text{th} \big\} - \beta \Delta F. 
\end{equation}
We can now series expand each term in powers of $\delta \lambda$. 
We write $H_i = H(\lambda) \equiv H$ and $H_f = H(\lambda + \delta \lambda)$, leading to
$H_f - H_i =  \frac{\partial H}{\partial \lambda} \delta \lambda + \frac{1}{2} \frac{\partial^2 H}{\partial \lambda^2} \delta\lambda^2 + \ldots$.
We also expand $\Delta F$ in a similar way.
From  equilibrium statistical mechanics, however, it follows that for thermal states
\begin{equation}\label{infQuench_stat_mech_relation}
    \left\langle \frac{\partial H}{\partial \lambda}\right\rangle = \frac{\partial F}{\partial \lambda}.
\end{equation}
Hence, the terms of order $\delta \lambda$ in Eq.~\eqref{InfQuenches_Sigma_irr_work} cancel out, meaning the first non-zero contribution will be of order $\delta \lambda^2$ (as it must, since $\Sigma \geq 0$): 
\begin{equation}\label{InfQuenches_Sigma_expansion_intermediate}
    \Sigma = \frac{\beta \delta \lambda^2}{2} \left\{
    \left\langle \frac{\partial^2 H}{\partial \lambda^2}\right\rangle - \frac{\partial^2 F}{\partial \lambda^2} \right\}.
\end{equation}
It is important to note how $\langle W \rangle \sim \delta \lambda$, while $\Sigma \sim \delta \lambda^2$. 
That is, the first order contribution to the average work is exactly canceled by the contribution from $\Delta F$. 

One can always choose the work protocol such that it appears linearly in the Hamiltonian. 
That is, such that $H(\lambda) = H_0 + \lambda H_1$. 
In this case the first term in Eq.~\eqref{InfQuenches_Sigma_expansion_intermediate} vanishes and one is left with the simpler expression
\begin{equation}\label{InfQuenches_Sigma_expansion}
     \Sigma = -\frac{\beta \delta \lambda^2}{2} \frac{\partial^2 F}{\partial \lambda^2},
\end{equation}
which shows that the non-equilibrium lag is nothing but the \emph{thermal susceptibility} to $\lambda$, a concept widely studied in equilibrium statistical mechanics. 

The relation to the susceptibility makes it particularly inviting to study infinitesimal quenches in systems presenting a quantum phase transition as a function of $\lambda$. 
This problem was first studied by~\cite{Dorner2012}, who analyzed the transverse-field Ising model. 
A quantum phase transition strictly occurs only at $T\to 0$, while the non-equilibrium lag scenario involves a thermal state at finite temperature. 
Notwithstanding,  reflections of the $T=0$ critical point can still be felt at low  temperatures. 
This is precisely what was observed in~\cite{Dorner2012}, which found that the entropy production diverges logarithmically at the critical point, in the limit $T\to 0$ (while showing a sharp peak for finite $T$). 


Since then, there has been several papers dedicated to an understanding of the critical properties of the non-equilibrium lag. 
An extension to the general XY model was given in~\cite{Bayocboc2015} and the more exotic XZY-YZX model was studied in~\cite{Zhong2015}.
Very recently a general group-theoretic framework suitable for arbitrary quadratic Hamiltonians was introduced in~\cite{Fei2019}, generalizing the above results.  
An analysis of the related Lipkin-Meshkov-Glick model (which can be viewed as the long-range analog of the transverse field Ising model) was studied in~\cite{Campbell2016c}.
All of these refer to continuous transitions. 
The extension to discontinuous transitions was discussed in~\cite{Mascarenhas2014}. 
Finally, the extension to consider the full statistics (instead of just the first moment~\eqref{InfQuenches_Sigma}) was recently put forth in~\cite{Fei2020}.

In order to shed further light on the physics behind Eq.~\eqref{InfQuenches_Sigma_expansion}, it is necessary to distinguish whether $H_i$ and $H_f$ commute or not.  
Or, what is equivalent, whether $H$ and $\partial H/\partial \lambda$ commute~\cite{Fusco2014a}. 
The reason why this matters is because  differentiating $F = - T \ln \tr( e^{-\beta H(\lambda)})$ with respect to $\lambda$ is not trivial if $H$ and $\partial H/\partial \lambda$ do not commute. 
In fact, this can be readily seen from the following Baker-Campbell-Hausdorff expansion, applicable to an arbitrary operator $M(\lambda)$ 
\begin{equation}\label{infQuenches_BCH}
    \partial_\lambda e^{M(\lambda)} = \bigg\{ M' + \frac{1}{2} [M, M'] + \frac{1}{3!} [M, [M, M']] + \ldots \bigg\} e^{M(\lambda)}, 
\end{equation}
where $M' = \partial_\lambda M$. 
Thus, if $H$ and $\partial_\lambda H$ commute, one can readily write $\partial_\lambda e^{-\beta H} = -\beta (\partial_\lambda H) e^{-\beta H}$. But if they do not, one must use~\eqref{infQuenches_BCH} instead, where new terms appear, associated with the commutator $[M,M']$.
Due to the cyclic property of the trace, this effect turn out to be irrelevant when computing the first derivative $\partial F/\partial \lambda$, which is why Eq.~\eqref{infQuench_stat_mech_relation} is actually always true. 
But for the second derivative in Eq.~\eqref{InfQuenches_Sigma_expansion}, this is crucial. 

Another way to deal with this is to introduce the following Feynman integral representation: 
\begin{equation}
    \frac{\partial}{\partial \lambda}e^{-\beta H} = \beta \int\limits_0^1 dy e^{-\beta y H} (\partial H/\partial \lambda) e^{-\beta (1-y) H}.
\end{equation}
Using this to compute $ - \partial^2 F/\partial \lambda^2$, one eventually finds the following result for  Eq.~\eqref{InfQuenches_Sigma_expansion}~\cite{Scandi2019}:
\begin{equation}\label{InfQuenches_Sigma_y_cov}
    \Sigma = \frac{\beta^2}{2}  \int\limits_0^1 dy \; \text{cov}_i^y(\delta H, \delta H),
\end{equation}
where $\delta H = H_f - H_i = \delta \lambda (\partial H/\partial \lambda)$ and 
\begin{equation}\label{adiabatic_y_covariance}
    \text{cov}_i^y(A, B) =  \tr\big[A (\rho_i^\text{th})^y B (\rho_i^\text{th})^{1-y}\big] - \tr(A\rho_i^\text{th}) \tr(B\rho_i^\text{th}), 
\end{equation}
is the so-called $y$-covariance.  
It represents a generalization of the notion of covariance to the case of non-commuting operators.
When $[H, \delta H] = 0$, the $y$-covariance simplifies to the usual covariance. In this case the integral in $y$ can be performed explicitly,  leading to 
\begin{equation}\label{infQuenches_Sigma_commuting}
    \Sigma = \frac{\beta^2}{2} \text{var}(\delta H).
\end{equation}
Conversely, when $[H, \delta H] \neq 0$, this is no longer true. 

This commutativity issue can also be analyzed  from the perspective of the probability distribution $P(\sigma)$ defined in Eq.~\eqref{Qgen_nonEqLag_P_sigma}. 
The transition probabilities in the case of quenches simplify to  $p(m_f | n_i) = |\langle m_f | n_i \rangle|^2$.
If $[H, \delta H] = 0$, they therefore  trivialize.  
But if $[H,\delta H] \neq 0$, one may still find non-trivial transitions.

The relevance of these results lies in their connection with quantum coherence~\cite{Miller2019}. 
The case $[H, \delta H] = 0$ represents a quench which changes the energy levels of the system, but keeps the same eigenbasis. 
Conversely, $[H, \delta H] \neq 0$ means that, in addition to the change in energy, the eigenbasis is also rotated, so that $\rho_i^\text{th}$ will be coherent in the basis of $H_f$. 
As a consequence, there will be an additional entropy production associated with the loss of coherence in the thermalization process~\cite{Santos2019}.  

This can be made more  patent by introducing the Wigner-Yanase-Dyson skew information~\cite{Hansen2008}
$    I_y(\rho, A) = - \frac{1}{2} \tr\big\{ [\rho^y, A] [\rho^{1-y}, A]\big\}$,
which quantifies the coherence between $A$ and $\rho$, in the sens that it gauges the degree with which $\rho$ and $A$ fail to commute. 
$I_y(\rho, A)$ is always non-negative and zero iff $[\rho, A] = 0$.
In terms of this, one can rewrite Eq.~\eqref{InfQuenches_Sigma_y_cov} as 
\begin{equation}\label{infQuenches_Sigma_Q}
    \Sigma = \frac{\beta^2}{2} \text{var}(\delta H) - \mathcal{Q}, 
\end{equation}
where 
    $\mathcal{Q} = \frac{\beta^2}{2}  \int_0^1 dy\;  I_y(\rho_i^\text{th}, H_f)$ 
is a new contribution measuring the incompatibility of the final Hamiltonian with the initial state of the system. 
Compared  with Eq.~\eqref{infQuenches_Sigma_commuting}, the result in Eq.~\eqref{infQuenches_Sigma_Q} shows how lack of commutativity modifies the average entropy production. 

The same analysis can also be made for the full CGF~\eqref{InfQuenches_CGF}, as done in ~\cite{Scandi2019}.
The result is compactly expressed as
\begin{equation}\label{InfQuenches_CGF_expansion}
    K(\lambda) = - \frac{\beta^2}{2}  \int\limits_0^\lambda dx \int\limits_x^{1-x} dy \; \text{cov}_i^y(\delta H, \delta H).
\end{equation}
When $[H, \delta H] = 0$, this reduces to 
\begin{equation}\label{InfQuenches_CGF_expansion_commuting}
    K_\text{comm}(\lambda) = - \frac{\beta^2 \lambda(1-\lambda)}{2}   \text{var}(\delta H).
\end{equation}
From this expression, one appreciates that $K$ satisfies the Jarzynski equation $K(\lambda = 1) = \ln\langle e^{-\sigma}\rangle = 0$. 
In addition, it also satisfies the stronger Gallavotti-Cohen symmetry $K(\lambda) = K(1-\lambda)$, which implies that $P(\sigma)$  obeys an exchange fluctuation theorem $P(\sigma)/P(-\sigma) = e^\sigma$; or, put it differently, that the probability distribution of the time-reversed process is the same as for the forward one. 
This is a consequence of the infinitesimal/quasi-static nature of this process and does not happen for non-infinitesimal quenches.

Since Eq.~\eqref{InfQuenches_CGF_expansion_commuting} is quadratic in $\lambda$, $P(\sigma)$ must be a Gaussian distribution whose 
mean is Eq.~\eqref{infQuenches_Sigma_commuting}, while the variance reads 
\begin{equation}\label{infQuenches_variance_sigma}
    \text{var}(\sigma) = \beta^2  \text{var}(\delta H). 
\end{equation}
Comparing with Eq.~\eqref{infQuenches_Sigma_commuting}, we arrive at the fluctuation-dissipation theorem~\cite{Callen1951} relating the mean and variance of the entropy production $\langle \sigma \rangle = \frac{1}{2} \text{var}(\sigma)$. 

When $[H,\delta H] \neq 0$, however, the FDR no longer holds. 
Eq.~\eqref{infQuenches_variance_sigma} for the variance turns out to remain unchanged, but the mean is modified to Eq.~\eqref{infQuenches_Sigma_Q}.
Whence, the two quantities are now related by 
\begin{equation}
    \langle \sigma \rangle = \frac{1}{2} \text{var}(\sigma) - \mathcal{Q}. 
\end{equation}
The FDR is therefore broken due to the presence of the coherent term~\cite{Miller2019}. 
We mention in passing that the FDR for general quantum processes was also recently discussed in~\cite{Mohammad2018}, which showed the non-trivial role of the so-called Symmetric Logarithmic Derivative, a concept widely used in quantum metrology. 

Returning now to the non-commuting case, Eq.~\eqref{InfQuenches_CGF_expansion}, it is also possible to rewrite the CGF as 
\[
    K(\lambda) = K_\text{comm}(\lambda) + \frac{\beta^2}{2}  \int\limits_0^\lambda dx \int\limits_x^{1-x} dy\; I_y(\rho_i^\text{th}, H_f). 
\]
This shows how the presence of coherence makes $P(\sigma)$ non-Gaussian, as the last term makes $K$ non-polynomial in $\lambda$. 
Another consequence of this result concerns cumulants of order 3 or higher.
Since $K_\text{comm}$ is quadratic, it will only contribute to the first two cumulants. 
All higher order cumulants will therefore come from the second term. 
In fact, using Leibniz' integral rule together with Eq.~\eqref{Qgen_nonEqLag_cumulants_deriv}, one arrives at 
\begin{equation}
    \kappa_n = -(-1)^n \beta^2 \frac{\partial^{n-2}}{\partial \lambda^{n-2}} I_\lambda(\rho_i^\text{th}, H_f), \qquad n \geq 3.
\end{equation}
Using this result, it was shown in Ref.~\cite{Scandi2019} that all higher order cumulants are actually positive, $\kappa_n >0$ (for $n\geq 3$). 

The above discussion refers to a single quench, from $H_i$ to $H_f$. 
But this can now be used as a building block to study coherence in more general quasi-static processes. We imagine a quasi-static process where $H(t)$ is changed very slowly, with the system permanently in contact with a heat bath at fixed temperature. 
Following~\cite{Nulton1985,Crooks1998}, this process can be divided into a series of discrete, infinitesimal steps. 
At each step $H$ changes slightly, from $H_i$ to $H_{i+1}$ (the quench). After this quench, the system is allowed to relax back to thermal equilibrium, but now at the new Hamiltonian $H_{i+1}$. 
Using this construction, one may build a quasi-static process, where the system is in thermal equilibrium throughout but, notwithstanding, the entropy production can still be quantified. 
In fact, the net entropy production will be simply the sum of the entropy produced in each quench: $\sigma = \sigma_1 + \ldots + \sigma_N$. 
And since the system fully thermalizes at each step, the $\sigma_i$'s are statistically independent.
The full CGF is hence $K_\sigma(\lambda) = \sum_{i=1}^N K_{\sigma_i}(\lambda)$, and
the intuition from a single quench directly carries over to quasi-static (non-infinitesimal) process.

\subsection{\label{sec:dissipative}Dissipative phase transitions: basic models}

We recall the notion of non-equilibrium steady-states (NESSs) discussed in Sec.~\ref{sec:int},
which occur when a system is coupled simultaneously to multiple reservoirs. The hallmark of such states is a finite entropy production rate $\dot{\Sigma}$. 
In certain situations, NESSs can also present phase transitions.
In the classical literature these usually go by the name of  ``non-equilibrium transitions'' and in the quantum literature by the name of ``dissipative phase transitions'' (for concreteness, we shall henceforth use the latter). 
Since NESSs are characterized by a finite $\dot{\Sigma}$, it is therefore only natural to ask how $\dot{\Sigma}$  behaves across a dissipative transition. 
This is the issue we shall explore in this section. 
For classical systems the situation is  somewhat well understood. 
Conversely, in the quantum case there are dramatically few studies on the topic. 
Here we will try to discuss both scenarios together. Before discussing the thermodynamics, though, we begin by reviewing some of the prototypical models of dissipative phase transitions, as these may not be so widely known by the community working in stochastic and quantum thermodynamics.


Classically, dissipative phase transitions are usually studied in lattice models described by stochastic thermodynamics. 
This is well illustrated by the model studied  in~\cite{Tome2012}, corresponding to a 2D classical Ising model coupled to two baths at different chemical potentials. One bath couples only to the even sites of the lattice and the other to the odd sites (thus forming a checkerboard pattern). 
The lattice has a total of $N$ sites,  each described by a classical spin variable $\sigma_i = \pm 1$. 
The configurations of the system are described by the vector $\bs = (\sigma_1, \ldots, \sigma_N)$, where $\sigma_i = \pm 1$ and the spins interact with the typical nearest-neighbor Ising energy 
$E = -J \sum_{\langle i,j\rangle} \sigma_i \sigma_{j}$, where $\langle i,j\rangle$ means a sum over nearest neighbors. 
The probability distribution $p(\bs)$ is assumed to evolve according to the Pauli equation
\begin{equation}\label{critNESS_M}
    \frac{d p(\bs)}{dt} = \sum\limits_{i=1}^N \bigg\{w_i(\bs^i) p(\bs^i) - w_i(\bs) p(\bs)\bigg\},
\end{equation}
where $\bs^i = (\sigma_1, \ldots, -\sigma_i, \ldots, \sigma_N)$ and $w_i(\bs)$ is the single spin-flip transition rate $\sigma_i \to -\sigma_i$ at site $i$, each characterized by a temperature $T_i$ and a chemical potential $\mu_i$.
The authors assumed all $T_i = T$, and used an alternating chemical potential pattern of $\mu_i = \mu$ for odd sites and $\mu_i = -\mu$ for even sites. 

In the quantum domain, lattice models can be constructed with unusual types of dissipation. 
This acquires particular relevance in the context of ultra-cold atoms in optical lattices.
For instance,  Ref.~\cite{Diehl2008} considered a 2D bosonic lattice, with each site characterized by a annihilation operator $a_i$, and evolving according to the Lindblad master equation
\begin{equation}\label{critNESS_BEC_M}
    \frac{d\rho}{dt} = -i[H, \rho] + \sum\limits_\ell \kappa_\ell \bigg[L_\ell \rho L_\ell^\dagger - \frac{1}{2} \{L_\ell^\dagger L_\ell, \rho\}\bigg],
\end{equation}
where $H = - J \sum_{\langle i, j \rangle} a_i^\dagger a_j + \frac{U}{2} \sum_i a_i^\dagger a_i^\dagger a_i a_i$ is the Bose-Hubbard Hamiltonian. 
The authors  discuss the non-trivial effects that come about from using jump operators acting on nearest-neighbor sites, of the form $L_{\ell} = L_{ij} = (a_i^\dagger + a_j^\dagger)(a_i - a_j)$. 
These operators do not change the number of particles. 
Instead, they cause only a \emph{phase-sensitive decoherence}: the term $(a_i - a_j)$ annihilates anti-symmetric superpositions of the pair $(i,j)$, whereas $(a_i^\dagger + a_j^\dagger)$ recycles it towards a symmetric state. 
This dissipator therefore induces \emph{phase locking}, which is characteristic of Bose-Einstein condensates. 
It thus represents a novel type of dissipation, with a clear quantum signature. 

\begin{figure}
    \centering
    \includegraphics[width=0.45\textwidth]{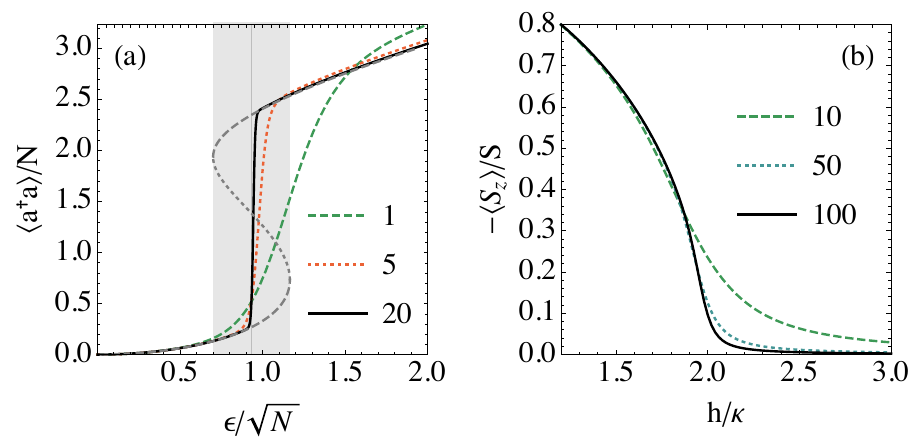}
    \caption{Examples of dissipative transition. (a) Average photon number for the Kerr bistability model~\eqref{critNESS_Kerr_M}, with $\kappa = 1/2$, $\Delta = -2$ and $U = 1/N$, where $N$ is a parameter used to tune the thermodynamic limit. The different curves correspond to $N = 1, 5$ and 20 and  were computed by numerically finding the steady-state of \eqref{critNESS_Kerr_M}.
    We also show in light gray the semiclassical result expected for optical bistability, showing that there is a region where there are two possible solutions. 
    (b) Steady-state magnetization of the macrospin model~\eqref{critNESS_spin_model}, computed for different values of $S$. 
    }
    \label{fig:critNESS_transitions}
\end{figure}

Notwithstanding this bout of interest in lattice systems, it turns out quantum models of dissipative phase transitions have actually been around for many decades, particularly  in the quantum optical community. 
The reason is that they often occur in non-linear optical systems coupled to optical cavities, such as the Dicke model~\cite{Dicke1954} or the Optical Parametric Oscillator~\cite{Drummond1981a}. 
These models are dissipative due to the characteristic photon losses of optical cavities. The transition, in this case, is driven by an external pump laser, which increases the number of photons in the cavity and thus the rate at which the  non-linear processes take place.
Criticality is marked by a threshold pump intensity, at which the quantum state of the cavity changes abruptly. 
This class of models are called \emph{driven-dissipative}. 
The simplest such model is that of Kerr bistability~\cite{Drummond1980,Casteels2017}, defined by a single bosonic model $a$ evolving according to the Lindblad equation in a rotating frame at the frequency of the pump
\begin{equation}\label{critNESS_Kerr_M}
    \frac{d \rho}{d t} = -i [H, \rho] + \kappa \bigg[a\rho a^\dagger - \frac{1}{2} \{a^\dagger a, \rho\}\bigg],
\end{equation}
where
$H = \Delta a^\dagger a + \frac{U}{2} a^\dagger a^\dagger a a + i \epsilon(a^\dagger - a)$.
Here $\Delta$ is the cavity detuning, $U$ is the non-linear interaction and $\epsilon$ is the external pump. 
For certain parameters, this model may exhibit a discontinuous transition as a function of the pump $\epsilon$, reminiscent of the phenomenon of optical bistability. 
This is illustrated in Fig.~\ref{fig:critNESS_transitions}(a).
For the transition to take place, one must define an appropriate thermodynamic limit, which corresponds to $U \to 0$, $\epsilon \to \infty$ but keeping $U\epsilon^2$ finite; or, what is equivalent, we introduce a fictitious integer $N$ such that $U\to U/N$ and $\epsilon \to \epsilon \sqrt{N}$. 
The curves in Fig.~\ref{fig:critNESS_transitions}(a) were computed numerically for different values of $N$. 
From a statistical mechanical point of view, driven-dissipative models are mean-field models, since the strong confinement of the optical cavity makes the interactions between the atoms in the non-linear medium to be long-ranged (``everyone interacts with everyone''). 

Dissipative phase transitions share many similarities with  quantum phase transitions, as well as important differences (see Tab.~\ref{tab:transitions}). 
As with any transition, they stem from a competition between different drives.
Thus, just like quantum phase transitions are driven by the competition between two non-commuting terms in the Hamiltonian, the drives in dissipative phase transitions can be any two (or more) terms generating the open system dynamics. 
Now, however, there are more possibilities. 
Not only can there be a competition between two dissipative mechanisms, such as two reservoirs at different temperatures, but also a competition between a dissipative and a unitary (and hence coherent) term, as is the case in Eq.~(\ref{critNESS_BEC_M}) and  Eq.~(\ref{critNESS_Kerr_M}).\footnote{This is not a quantum effect and may very well occur in classical stochastic systems, e.g. governed by a Fokker-Planck equation. } 
A simple but elegant example is a macrospin of size $S$, described by spin operators $S_x, S_y, S_z$ and evolving according to the Lindblad master equation
\begin{equation}\label{critNESS_spin_model}
    \frac{d \rho}{dt} = - i h [S_x, \rho] + \frac{2\kappa}{S}\bigg[ S_- \rho S_+ - \frac{1}{2} \{ S_+S_-, \rho\}\bigg]. 
\end{equation}
This describes a competition between a dissipative term favouring the south-pole (lowest eigenstate of $S_z$) and a unitary contribution corresponding to a transverse field.
This model is reminiscent of the Dicke model for collective atom interactions and has been studied since the 1970s [see, for instance~\cite{Schneider2002} and references therein].\footnote{ 
The steady-state $\rho_\text{ss}$ can actually be found analytically, as shown in Ref.~\cite{Puri1979}.}
In the thermodynamic limit (which in this case means $S\to \infty$) the model presents a phase transition at a critical field $h_c = 2\kappa$. 
This is illustrated in Fig.~\ref{fig:critNESS_transitions}(b), where we plot the order parameter  $\langle S_z \rangle_\text{ss}$ as a function of $h$.
For $h<h_c$ the dissipative part wins and the system tends to align towards the south pole, making $\langle S_z \rangle_\text{ss}$ non-zero and negative (when $h = 0$ the steady-state is precisely the south-pole). Conversely, for $h>h_c$ the two terms mix together to produce a disordered state with $\langle S_z \rangle_\text{ss} = 0$.

\begin{table}[!t]
\caption{Dissipative phase transitions, in comparison with quantum phase transitions. Based on~\cite{Kessler2012}.}
\label{tab:transitions}
\renewcommand{\arraystretch}{1.5} 
\setlength{\tabcolsep}{10pt} 
\begin{tabular}{c c c}
\multicolumn{1}{l|}{\textbf{}}                                                                                                                   & \textbf{Quantum}                                                                                              & \textbf{Dissipative}                                                                                                          \\ \hline
\multicolumn{1}{c|}{\textbf{Operator}}                                                     &  \begin{tabular}[c]{@{}c@{}}Hamiltonian\linebreak\\ $H(g)$\end{tabular}                        & \begin{tabular}[c]{@{}c@{}}Liouvillian \linebreak\\ $\mathcal{L}(g)$\end{tabular}                                      \\ \hline
\multicolumn{1}{c|}{\textbf{Spectra}} &  \begin{tabular}[c]{@{}c@{}}Energy eigenvalues \linebreak\\ $H(g)\ket{\psi_i} = E_i(g)\ket{\psi_i}$\end{tabular} & \begin{tabular}[c]{@{}c@{}}Eigenvalues \linebreak\\ $\mathcal{L}(g)\rho = \lambda_i(g)\rho$\end{tabular}              \\ \hline
\multicolumn{1}{c|}{\textbf{State}}                                                        &  \begin{tabular}[c]{@{}c@{}}Ground state \linebreak \\ $H(g)\ket{\psi_0} = E_0(g)\ket{\psi_0} $\end{tabular}         & \begin{tabular}[c]{@{}c@{}}NESS\\ $\mathcal{L}(g)\rho_\text{ss} = 0$\end{tabular}                    \\ \hline
\multicolumn{1}{c|}{\textbf{Gap}}  &  \begin{tabular}[c]{@{}c@{}}Energy gap \linebreak\\ $\Delta (g) = E_1(g) - E_0(g)$\end{tabular}            & \begin{tabular}[c]{@{}c@{}}Liouvillian gap\\ $\Re{[\lambda_1]}$\end{tabular} \\ \hline
\end{tabular}
\end{table}

\subsection{\label{sec:dissipativeentropy}Dissipative phase transitions: entropy production}


Having introduced some of the basic models and features of dissipative phase transitions, we now turn to the question of how the entropy production behaves as one crosses the critical point. 
We begin with classical systems. 
In this case much more is known since  the entropy production can be more readily computed. 
For systems described by a Pauli  master equation Eq.~(\ref{Pauli_pauli}), for instance, the entropy production can be computed from the general formula in Eq.~(\ref{StochThermo_Pi}), which even contemplates the presence of multiple heat baths (see Sec.~\ref{sec:stoch_thermo} for more details). 

\begin{figure}
    \centering
    \includegraphics[width=0.45\textwidth]{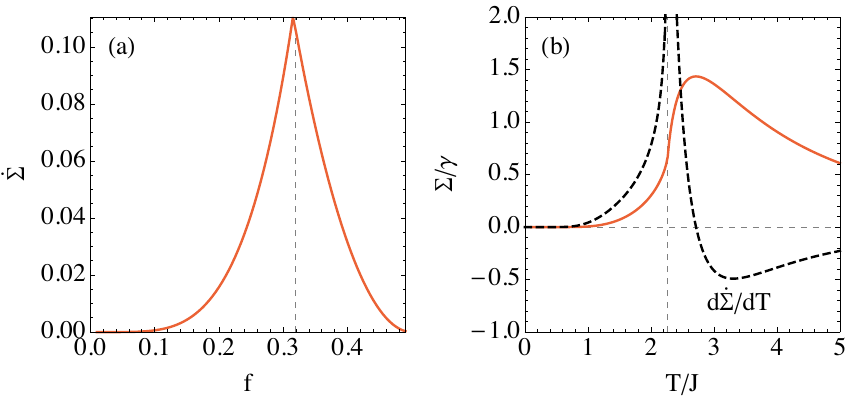}
    \caption{Example behavior $\dot{\Sigma}$ across a continuous dissipative transition.
    (a) For the mean-field Majority Vote model in Ref.~\cite{Noa2018}, where $f$ is the so-called misalignment parameter. 
    (b) For the two-bath Ising model of Ref.~\cite{Tome2012}, where the transition is driven by the temperature $T$.  
    In both cases $\dot{\Sigma}$ is continuous across the transition, but has a kink at the critical point, implying the derivative of $\dot{\Sigma}$ is discontinuous. 
    It is also possible, as shown in (b), that the derivative presents a logarithmic discontinuity. 
    }
    \label{fig:critNESS_Pi_classical}
\end{figure}

The entropy production rate in classical transitions is found to be \emph{always finite}, but becomes non-analytic at the critical point.
For continuous transitions, it always presents a kink, meaning its derivative with respect  to the driving parameter is discontinuous. This is illustrated in Fig.~\ref{fig:critNESS_Pi_classical}(a). 
It can also happen that the derivative  diverges logarithmically, as shown in Fig.~\ref{fig:critNESS_Pi_classical}(b) (the critical exponent of this divergence is associated with the equilibrium specific heat of the system). 
Notwithstanding, $\dot{\Sigma}$ itself is always finite. 
This behavior was found from both analytical, as well as numerical Monte Carlo simulations, in a variety of models~\cite{Tome2012,Shim2016,Crochik2005,Zhang2016,Herpich2019,Noa2018}.

\begin{figure}
    \centering
    \includegraphics[width=0.45\textwidth]{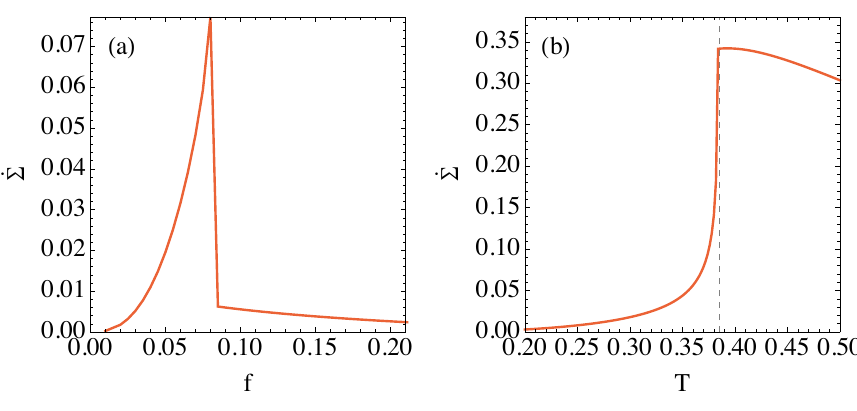}
    \caption{Similar to Fig.~\ref{fig:critNESS_Pi_classical}, but exemplifying $\dot{\Sigma}$ across discontinuous transitions. (a) Majority vote model, from Ref.~\cite{Noa2018}. This is the same model as in Fig.~\ref{fig:critNESS_Pi_classical}(a), which can be tuned from a continuous to a discontinuous transition depending on the parameter range. 
    (b) Ising model subject to an oscillating field, based on Ref.~\cite{Zhang2016}.
    }
    \label{fig:critNESS_Pi_classical_discont}
\end{figure}

For discontinuous transitions, on the other hand, $\dot{\Sigma}$ is finite but has a discontinuity at the phase coexistence region. 
This was also encountered in numerous models~\cite{Zhang2016,Herpich2018,Noa2018} and is exemplified in Fig.~\ref{fig:critNESS_Pi_classical_discont}. 
Image (a) corresponds to the same model as in Fig.~\ref{fig:critNESS_Pi_classical}(a), which can actually be tuned to present both continuous and discontinuous transitions across the critical point. 
Fig.~\ref{fig:critNESS_Pi_classical_discont}(b), on the other hand, was based in~\cite{Zhang2016} and corresponds to a classical Ising model subject to an oscillating magnetic field. 
This is therefore somewhat different from the NESS scenario that we have been discussing so far, as there is only one bath. But the  explicit time-dependent drive yields similar physics. 

The underlying mechanisms that lead to this kind of behavior have been established recently, in~\cite{Noa2018}.
They involve the stochastic fluctuations of the entropy production close to criticality which, due to the central limit theorem, can be approximated by a sum of Gaussians. 
These results show that the behaviors above, for both continuous and discontinuous transitions, are in fact universal for systems described by Pauli equations breaking a discrete $Z_2$ symmetry. 
Whether or not they extend to other types of discrete symmetries remain to be proved. The results in Ref.~\cite{Herpich2019}, however, which studied a $q$-state Potts model, seem to indicate that they do.

Next we turn to the quantum case. 
Very little is known about the behavior of the entropy production in quantum dissipative phase transitions. 
Not only are the models difficult to simulate/experiment with, but computing $\dot{\Sigma}$  presents an additional challenge. 
As discussed in Sec.~\ref{sec:Qgen}, the definition of entropy production requires knowledge of the specific system-bath interactions involved. 
With the exception of standard thermal baths, it is not possible to estimate $\dot{\Sigma}$ solely from the reduced dynamics. 
This acquires additional relevance in light of the fact that most dissipators studied in the context of dissipative phase transitions are actually not thermal. 
This is the case, for instance, of Eq.~\eqref{critNESS_BEC_M}. It is also true for driven-dissipative systems, such as for~\eqref{critNESS_Kerr_M} and \eqref{critNESS_spin_model}, which are effectively equivalent to zero temperature baths. 

To the best of our knowledge, the only studies on this issue have been  in driven-dissipative systems (driven optical cavity is loaded with a non-linear medium)~\cite{Brunelli2018,Goes2019}.
For such systems, even though the standard formulation of $\dot{\Sigma}$ is not available (since the baths are at zero temperature), one can approach the problem using the phase-space formulation discussed in  Sec.~\ref{sec:QPhaseSpace}. 

What is found is that the entropy production rate can be decompose in two terms, as
\begin{equation}\label{critNESS_Pi_Husimi_decomp}
    \dot{\Sigma} = \dot{\Sigma}_u + \dot{\Sigma}_d. 
\end{equation}
The first term is related to the unitary dynamics and behaves \emph{exactly} like the entropy production in classical systems, e.g. Figs~\ref{fig:critNESS_Pi_classical} and~\ref{fig:critNESS_Pi_classical_discont}.
The reason why this is so is not yet fully understood. 
The second term, $\dot{\Sigma}_d$, on the other hand, is related to the dissipative part and behaves like a susceptibility. 
As a consequence, it can \emph{diverge} at the critical point. 
These results therefore indicate that the entropy production in the quantum domain may have contributions that behave fundamentally different from their classical counterparts. 

\begin{figure}
    \centering
    \includegraphics[width=0.45\textwidth]{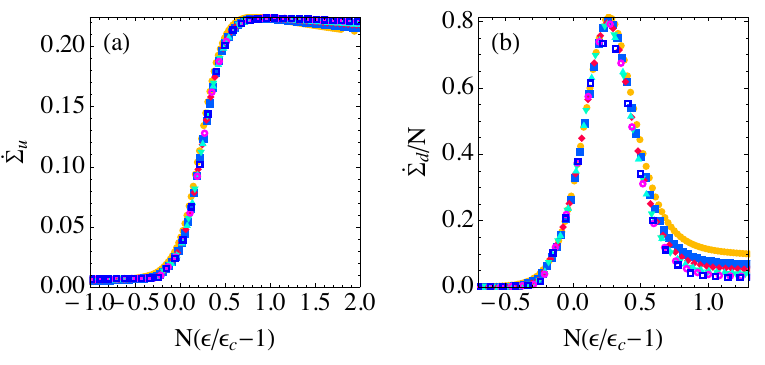}
    \caption{The two contributions in Eq.~\eqref{critNESS_Pi_Husimi_decomp} to the entropy production for the Kerr bistability model~\eqref{critNESS_Kerr_M}. The points correspond to different values of $N$ and have been plotted so as to yield a data collapse (see text for more details). }
    \label{fig:critNESS_Pi_kerr}
\end{figure}

We review two specific models of entropy production in dissipative phase transitions, studied in Ref.~\cite{Goes2019}. 
First, we look at the discontinuous transition of the Kerr model in Eq.~\eqref{critNESS_Kerr_M}.
Fig.~\ref{fig:critNESS_Pi_kerr} shows both contributions in Eq.~\eqref{critNESS_Pi_Husimi_decomp} as a function of the pump $\epsilon$ for several values of $N$ (the parameter controlling the thermodynamic limit; cf. Fig.~\ref{fig:critNESS_transitions}). 
The curves have been plotted so as to yield a data collapse, whose properties can help infer the nature of each contribution.  
The horizontal axes are rescaled to $N(\epsilon/\epsilon_c-1)$ whereas the vertical axis is not rescaled for $\dot{\Sigma}_u$, but rescaled by $1/N$ for $\dot{\Sigma}_d$. 
This means that $\dot{\Sigma}_u$ is intensive, while $\dot{\Sigma}_d$ is extensive. 
As a consequence, for large $N$ the dominant contribution will  be from $\dot{\Sigma}_d$. 

The behavior of $\dot{\Sigma}_u$ matches exactly what is found in classical sytems [cf. Fig.~\ref{fig:critNESS_Pi_classical_discont}(b)] and can be understood using the phenomenological 2-Gaussian model of Ref.~\cite{Noa2018}. 
Conversely, the behavior of $\dot{\Sigma}_d$ follows the variance of the order parameter, $\langle \delta a^\dagger \delta a\rangle$, where $\delta a = a - \langle a \rangle$. 
This contribution therefore behaves like a susceptibility. As it is a direct consequence of quantum fluctuations, it corresponds to an additional contribution to $\dot{\Sigma}$, of pure quantum origin.

\begin{figure}
    \centering
    \includegraphics[width=0.45\textwidth]{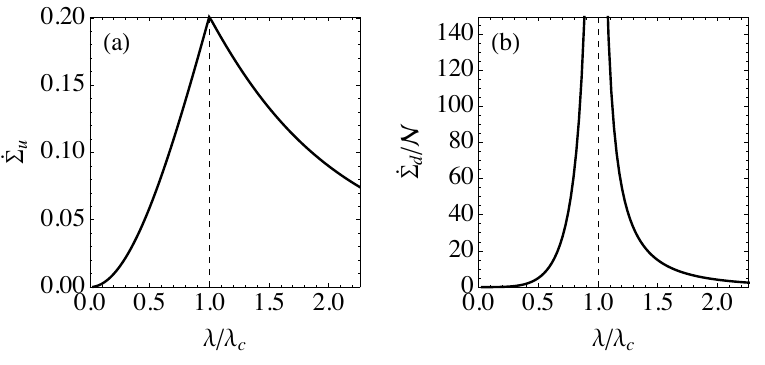}
    \caption{The two contributions in Eq.~\eqref{critNESS_Pi_Husimi_decomp} to the entropy production for the driven-dissipative Dicke model in Eq.~\eqref{critNESS_H_dicke}. 
    The critical point occurs at $\lambda_c = \sqrt{\omega_0(\kappa^2 + \omega^2)/\omega}$.
     }
    \label{fig:critNESS_Pi_dicke}
\end{figure}

The other model studied in Ref.~\cite{Goes2019} was the driven-dissipative Dicke model, described by a master equation identical to~\eqref{critNESS_Kerr_M}, but with Hamiltonian 
\begin{equation}\label{critNESS_H_dicke}
    H = \omega_s S_z + \omega a^\dagger a + \frac{2\lambda}{N} (a+a^\dagger) S_x, 
\end{equation}
where $S_i$ are macrospin operators of size $S = N/2$. 
The Dicke model describes an optical cavity with mode $a$ and loss $\kappa$ coupled to a non-linear medium, modeled as a macrospin $S$. 
{\color{black}
In this model the driving stems
from the Dicke interaction $(a+a^\dagger) S_x$ which is related to the field generating the optical lattice, and whose effect is to populate the cavity with a finite number of photons [cf. Ref.~\cite{Baumann2010} for more details]. 
}
The entropy production of this model was also studied experimentally in~\cite{Brunelli2018}, which will be reviewed in Sec.~\ref{sec:exp_brunelli}. 
The theoretical predictions for this model are shown in Fig.~\ref{fig:critNESS_Pi_dicke}. As can be seen, once again $\dot{\Sigma}_u$ behaves exactly like in the classical case [cf. Fig.~\ref{fig:critNESS_Pi_classical}(a)] whereas $\dot{\Sigma}_d$ behaves like a susceptibility and therefore diverges at the critical point. 

%
%
\subsection{\label{ssec:non_mark}Effects of non-Markovian dynamics on entropy production}
%
%

In this Section, we aim to explore potential connections between entropy production and the possible non-Markovian character of the system-environment dynamics. A flavour of such potential connections is already provided by Eq.~\eqref{Qgen_Sigma}. First, the assumptions that underlie it involve a certain degree of control over the environment $E$ which, as remarked in Sec.~\ref{sec:qgen_global}, might well have the same dimensions as $S$. This is well entailed by the finite-size corrections to these expressions, discussed in Sec.~\ref{sec:landauer}. Second,  Eq.~\eqref{Qgen_Sigma} implies the possibility that, due to the (globally unitary) system-environment interaction, \emph{both} $S$ and $E$ are affected. These features are strongly suggestive of influences of a potential backflow of information, from the environment back to the system, that has been pinpointed as one of the fundamental mechanisms for the emergence of non-Markovianity in the reduced dynamics of $S$~\cite{BreuerRMP,deVega2017}.

Specifically, Ref.~\cite{Breuer2009} defines a process as non-Markovian if there is a pair of initial states $\rho^{1,2}_S(0)$ of the system, and a time $t$ of its dynamics, such that
\begin{equation}
\label{BLP}
\frac{d}{d t} D\left(\rho^{1}_S(t), \rho_S^{2}(t)\right)>0.
\end{equation}
Here $D(\rho_1,\rho_2)=||\rho_1- \rho_2||/2$ is the trace distance between two states $\rho_{1,2}$ ($||\cdot||$ being the trace-1 norm of a matrix). The framework set in Ref.~\cite{Breuer2009} is based on the contractivity of the trace distance under positive trace-preserving maps: a break-down of contractivity makes the distance between two states grow(and thus Eq.~\eqref{BLP} hold), signaling non-Markovianity in the ensuing evolution. 

The identification of the reasons for the non-monotonic behaviour of the trace distance under non-Markovian dynamics is evidently key for the characterization of open-system dynamics. In this regard, one can demonstrate the following theorem~\cite{Mazzola2012} 
\begin{theorem}{For any quantum process described by a completely positive map, with associated system-environment interaction ruled by the propagator ${U}=e^{-i {H} t}$, we have 
\begin{equation}
\label{maz}
\frac{d}{d t} D\left(\rho^{1}_S(t), \rho_S^{2}(t)\right) \leqslant \frac{{\cal E}(t)+{\cal C}(t)}{2}
\end{equation}
with ${\cal E}(t)=\min _{k=1,2}\left\|{\Tr}_{E}\left[{H}, \rho^{k}_{S}(t) \otimes\big(\rho_{E}^1(t)-\rho_E^2(t)\big)\right]\right\|$ and 
${\cal C}(t)=\left\|{\Tr}_{E}\left[{H},\big(\chi_{S E}^1(t)-\chi_{S E}^2(t)\big)\right]\right\|$.
Here $\rho^k_{S(E)}(t)={\Tr}_{E(S)}[U\rho_{SE}^k U^\dag]$ are the reduced states of the system (environment) at time $t$, and
$\chi^k_{SE}(t)= \rho^k_{SE}(t)-\rho^k_S(t) \otimes \rho_E^k(t)$  are matrices that encode the correlations between $S$ and $E$.}
\label{maztheo}
\end{theorem}

Eq.~\eqref{maz} identifies the two mechanisms that underpin the occurrence of the backflow responsible for non-Markovian dynamics: namely, the possibility that, in light of the dynamical nature of the environment (as remarked above), the state of $E$ changes in time (as encompassed by ${\cal E}$); and the potential setting of system-environment correlations (here quantified by the boundary term ${\cal C}$). 
This result is to be compared with the entropy production rate, obtained by differentiating~\eqref{Qgen_Sigma} wrt to time: 
\begin{equation}
    \dot{\Sigma} = \frac{d}{dt} {\cal I}_{\rho'_{SE}}(S':E') + \frac{d}{dt} S(\rho'_E||\rho_E). 
\end{equation}
While $\Sigma \geqslant 0$, the same is not necessarily true for the rate $\dot{\Sigma}$.  
Theorem 6 resonates directly with this result for $\dot{\Sigma}$. The quantity $\mathcal{E}$ is in close correspondence to  $dS(\rho'_E||\rho_E)/dt$ and $\mathcal{C}$ with $d{\cal I}_{\rho'_{SE}}(S':E')/dt$.
This therefore shows that, even though the trace distance measure~\eqref{BLP} and the entropy production~\eqref{Qgen_Sigma} are defined in terms of different information-theoretic quantities, the mechanisms that underlie both are similar in spirit; put it differently, negativities in the entropy production rate can be viewed as a witness of non-Markovianity~\cite{Strasberg2019}. 
This is to be contrasted with the contractivity property of Markov processes, which enjoys a nice physical interpretation in the context of quantum (and indeed stochastic) thermodynamics, as it entails the positivity of the entropy production rate.



Next, consider a time-dependent system-bath interaction Hamiltonian reading 
$H_{\rm tot}(\lambda_t)=H(\lambda_t)+V+H_E$
with $H(\lambda_t)$ a driving term for the system, $H_{E}$ the Hamiltonian of the bath, $V$ their mutual coupling term, and $\lambda_t$ a work parameter. Due to the coupling between system and bath, which can well be strong, the equilibrium state of the system is not necessarily of the  Gibbs form with respect to $H(\lambda)$. Moreover, the initial state of the system-bath compound might not be factorized, thus entailing the potential emergence of non-Markovianity. 

Yet, we need to characterize such equilibrium state in order to be able to define the entropy production and its rate. To do so, it is convenient to introduce the so-called Hamiltonian of mean force~\cite{Kirkwood1935}
\begin{equation}
\label{meanforce}
    H^{\rm mf}(\lambda_t)=-\frac{1}{\beta}\ln\frac{\Tr_B\left[e^{-\beta H_{\rm tot}(\lambda_t)}\right]}{Z_B}
\end{equation}
with $Z_B$ the partition function of the equilibrium state of the bath. Eq.~\eqref{meanforce} describes the energy of the reduced state of the system if the global system-bath state is in equilibrium. Following Ref.~\cite{Strasberg2019}, we can introduce the non-equilibrium free energy 
    $F(t)=\left\langle H^{\rm mf}(\lambda_t)+\frac{1}{\beta}\ln\rho_S(t)\right\rangle$
with $\rho_S(t)$ an arbitrary state of the system at time $t$. This leads to the formal definition of work
\begin{equation}
    W(t)=\int^t_0 dt' \Tr_S\left[\frac{d H(\lambda_{t'})}{dt'}\rho_S(t')\right],
\end{equation}
which is structurally identical to the definition in the classical case. Using the definitions above, we have
\begin{equation}
    W(t)=\Tr_{SE}[\rho_{SE}(t)H_{\rm tot}(\lambda_t)]-\Tr_{SE}[\rho_{SE}(0)H_{\rm tot}(\lambda_0)]
\end{equation}
where $\rho_{SB}(t)$ is the instantaneous state of the total system-bath compound. 

If we now take the coarse-grained, mean-force version of the equilibrium state of the system 
\begin{equation}
    \pi_S(\lambda_t)=\frac{e^{-\beta H^{\rm mf}}(\lambda_t)}{Z^{\rm mf}(\lambda_t)},
\end{equation}
the entropy production rate may be defined as $   \dot{\Sigma}(t)=-\partial_t S(\rho_S(t)||\pi_S(\lambda_t))$, which is equivalent to
\begin{equation}
\label{resStrasberg}
\Sigma(t)=\beta(W(t)-\Delta F(t))=\delta S(t)-\delta S(0)
\end{equation}
with $\delta S(t)=S(\rho_{SE}(t)||\pi_{SE}(\lambda_t))-S(\rho_S(t)||\pi_S(\lambda_t))$. the variation in quantum relative entropy at a generic time $t$, with and without the inclusion of the bath (here $\pi_{SE}(\lambda_t)$ is the total system-bath Gibbs state and $\pi_S=\Tr_B(\pi_{SE})$)~\cite{Strasberg2019}. The monotonicity of the quantum relative entropy entails that $\delta S(t)\ge0\forall t$. Therefore, $\Sigma(t)\ge0$ provided that $\delta S(0)=0$, which happens in two noticeable cases: {\bf (a)} if the system-bath compound is initially prepared in the global Gibbs state $\pi_{SE}(\lambda_0)$ and {\bf (b)} for the class of zero-discord system-bath states (mathematically implying the condition $[H(\lambda_0),V]=0$). 

The entropy production rate thus cannot  be expressed as the relative entropy associated with the irreversible relaxation of the state of the system towards equilibrium. As a result, a relation between the sign of the entropy production rate and the occurrence of non-Markovian effects is not immediately apparent, not even for undriven systems. Notice that this is in contrast with the case of (undriven) classical open system dynamics, for which it is possible to establish that the negativity of the entropy production rate implies directly the non-Markovian nature of the dynamics under scrutiny. The fact that Eq.~\eqref{resStrasberg} requires the consideration of the bath. with which the system interacts, is a testament of the view, according to which a self-consistent formulation of the second law of thermodynamics for general (i.e. in principle non-Markovian) open quantum systems should not be based on the sole reduced-state dynamics of the system, as illustrated in Ref.~\cite{Marcantoni2017}. The relation between the conditions for the observation of non-Markovianity and the achievement of negative entropy production is currently an open question~\cite{Bhattacharya2017,Popovic2018}


A different approach to the inclusion of the global system-bath compound in the description of the thermodynamics of the system can be taken when considering the case of weak, yet non-negligible $S$-$E$ couplings~\cite{Rivas2012,Rivas2019}. In these conditions, the dynamics of the system might exhibit non-Markovian features, in light of the break-down of divisibility conditions. Let us assume the initial system-bath state to be the tensor product of the equilibrium states $\rho^{\rm th}_E=e^{-\beta H_E}/Z_E$ and $\rho^{\rm th}_S=e^{-\beta H_S}/Z_S$, i.e. $\rho_{SE}(0)=\rho^{\rm th}_S\otimes\rho^{\rm th}_E$, and a time-independent system Hamiltonian. Under the assumption of negligible system-bath coupling, the total Gibbs state of the compound is well approximated by such initial state, that is $\rho^{\rm th}_{SE}\simeq \rho_{SE}(0)$. As the global Gibbs state is a stationary state of the dynamics, the Gibbs state of the system $\rho^{\rm th}_S$ would be a steady state of the reduced dynamics in the refined weak coupling limit. However, this is not true at a finite time $t$. Let us call $\Lambda^R_t$ the map propagating the initial state of the system in the refined weak coupling limit. Following an argument similar to the one pursued when introducing the mean-force Hamiltonian, one can set
\begin{equation}
    \Lambda^R_{t}(\rho^{\rm th}_S)=\frac{e^{-\beta H^R_S(t)}}{Z^R_S(t)}
\end{equation}
with $Z^R_S(t)=Z_S$, in light of the trace preserving nature of the dynamical map. Moreover,  $H^R_S$ is the {\it refined} Hamiltonian of the system, which  reads
\begin{equation}
H^R_S(t)=-\frac{1}{\beta}\ln\Lambda^R_t\left (e^{-\beta H_S}\right ).
\end{equation}
In analogy with the standard formulation in the weak coupling limit, we define the refined average instantaneous energy $E^R(t)=\Tr_S[\rho_S(t)H^R_S(t)]$, which reduces to the standard weak-coupling value at $t=0$ and approaches $E(t)=\Tr_S[\rho_S(t)H_S]$ as $t\to\infty$, when $\Lambda^R_t$ approaches a Davies semi-group and $H^R_S(t)\to H_S$. As we are considering a time-independent process, the change in energy of the system equals the amount of refined heat $Q^R(t)$ flowing to/from the system, so that 
\begin{equation}
    Q^R(t)=\int^t_0dt' \dot{E}^R(t')=\int^t_0 dt'\Tr_S[\dot{\rho}_S(t')H^R_S(t')+{\rho}_S(t')\dot{H}^R_S(t')].
\end{equation}
As the map at hand is completely positive, the quantum relative entropy will satisfy contractivity upon application of $\Lambda^R_t$, so that 
\begin{equation}
    S\left(\Lambda^R_t(\rho_S(0))||\Lambda^R_t(\rho^{\rm th}_S)\right)\le S\left(\rho_S(0)||\rho^{\rm th}_S\right),
\end{equation}
which gives
\begin{equation}
  \Sigma^R(t) =\Delta S(t)-\beta Q^R(t)\ge0,
\end{equation}
with $\Delta S(t)=S(t)-S(0)$. The integral form of this relation is not accidental: as the map under scrutiny is in general non-divisible, the differential form of the $2^{\rm nd}$ law above is, in general, not valid, thus preventing a definite sign of the entropy production rate of the process~\cite{Rivas2019}. Albeit resulting from a much more intricate derivation, the same conclusion can be drawn for a general time-dependent process entailing the performance of work. 

More recent work has further elaborated on the approach above, providing a way to address the general case of a system coupled to a thermal bath through arbitrarily strong coupling rates~\cite{Seifert2016,Jarzynski2017,Miller2017,Strasberg2017}. In particular, in Ref.~\cite{RivasPRL}, a framework encompassing general initial states (including correlated ones of system and bath) has been proposed through which it is possible to show that a completely positive divisible map induced by a time-independent Hamiltonian, the entropy production rate is always positive. This plants the seeds for the clear-cut link between thermodynamic and Markovianity in the quantum regime, suggested in  Theorem~\ref{maztheo}.







\;
\section{\label{sec:exp}Experimental assessment of quantum entropy production}

In recent years there has been many experiments on non-equilibrium thermodynamics at the microscopic domain. 
In this section we have opted to review some representative results, with a focus on those contributions that specifically characterised the entropy production. 

\;
\subsection{Assessment at the level of quantum trajectories}

\begin{figure}
\includegraphics[width=\columnwidth]{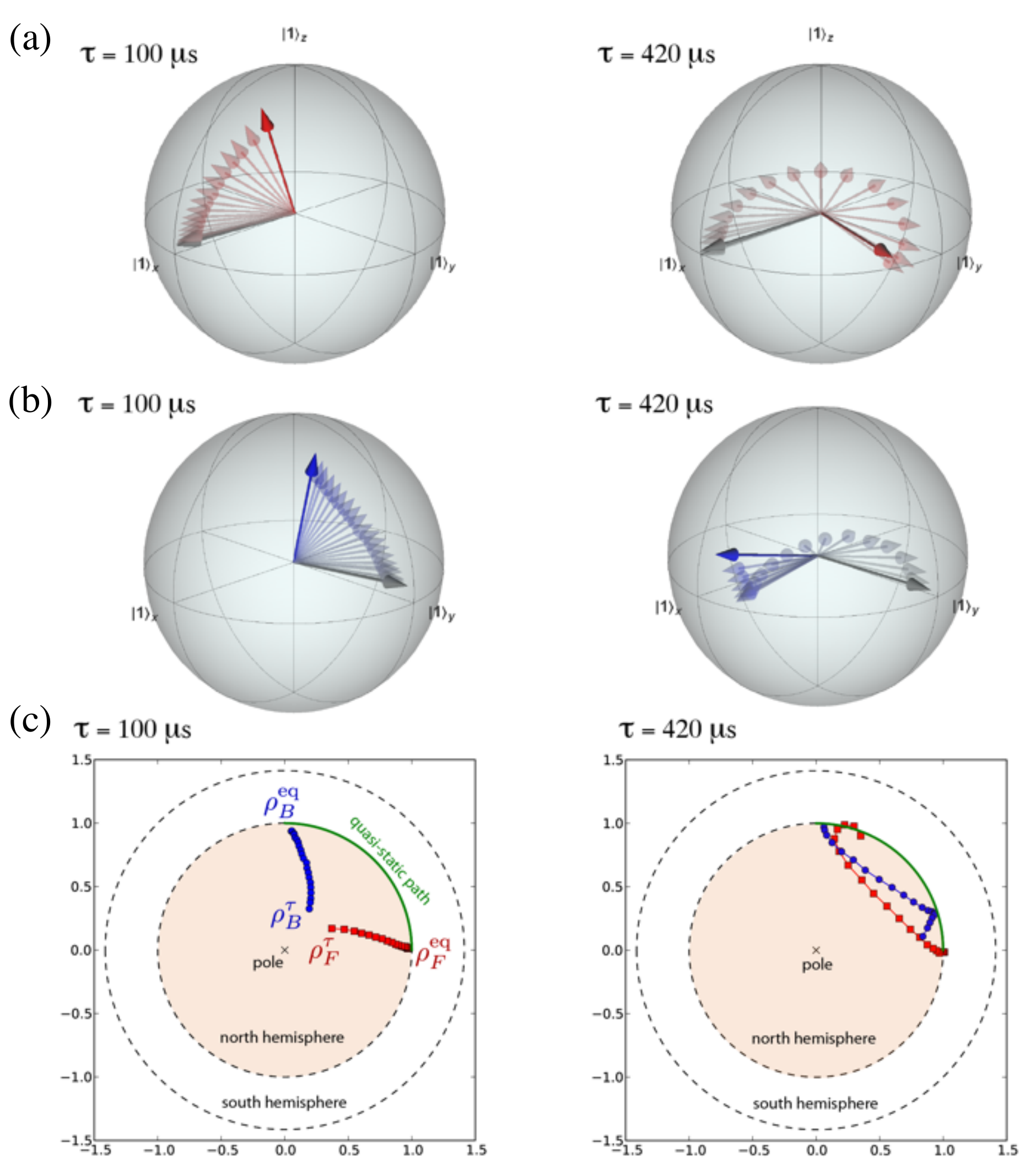}
\caption{
Evolution of the Bloch vector of the forward [backward] spin-1/2 state ${\rho}^\text{F}_t$ [${\rho}^{\text{B}}_{\tau-t}$] during a quench of the transverse magnetic field in the experiment reported in Ref.~\cite{Batalhao2015}, obtained via quantum state tomography. A sampling of 21 intermediate steps has been used. The initial magnetization (gray arrow)  is parallel to the external driven rf-field, aligned along positive $x$ [$y$] axis for the forward [backward] process. The final state is represented as a red [blue] arrow. Panel (c): Polar projection (indicating only the magnetization direction) of the Bloch sphere with the trajectories of the spin. Green lines represent the path followed in a quasistatic ($\tau\rightarrow \infty$) process.}
\label{fig:mag}
\end{figure}

In order to investigate the physical origin of irreversibility, Ref.~\cite{Batalhao2015} addressed the dynamics of a nuclear spin 1/2 system ($^{13}$C-labeled chloroform molecule in a liquid sample), initially prepared in a thermal state and driven out of equilibrium by a fast quench generated by a time-modulated radio-frequency (rf) field
producing  a time-dependent Hamiltonian $H^\text{F}_{t}$. 
A {\it backward} process was also realized by driving the system with the time-reversed Hamiltonian, $H^\text{B}_t= H^\text{F}_{\tau-t}$ with the system prepared in an equilibrium state of $H^\text{B}_0$. The work probability distributions of the forward and backward processes $P_{F,B}(W)$ are related via the Tasaki-Crooks fluctuation relation~\cite{Tasaki,Crooks1999}
\begin{equation}
\label{eq:Crooks}
 P_F\left( W \right)/P_B\left( -W \right)= e^{\beta \left(W - \Delta F\right)}.
\end{equation}
Eq.~\eqref{eq:Crooks} characterizes the positive and negative fluctuations of the quantum work $W$ along single realizations. It holds for arbitrary driving protocols, especially beyond the linear response regime, and is a generalization of the second law, to which it reduces on average as $\langle\Sigma\rangle=\beta(\langle W\rangle-\Delta F)\geq 0$.

\begin{figure}
\centering
\includegraphics[width=0.8\columnwidth]{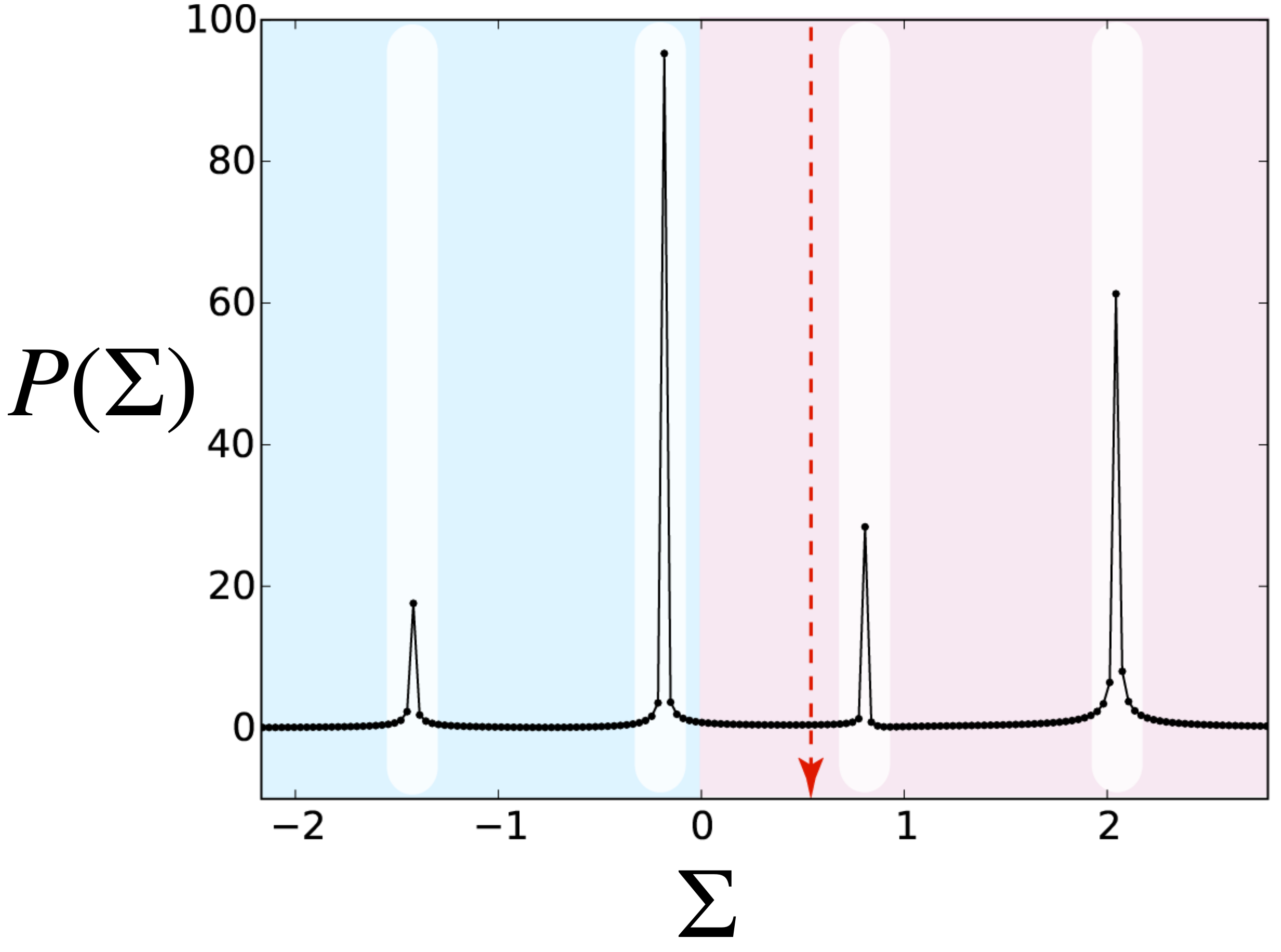}
\caption{Distribution of irreversible entropy production. Black dots represent the measured negative and positive values of the entropy production $\Sigma$ of  the spin-1/2 system after a quench of the transverse magnetic field of duration $\tau = 100$ $\mu$s. The mean entropy production (red line) is positive  in agreement with the second law.}
\label{fig:entropydist}
\end{figure} 

The Hamiltonian driving the forward process was taken to be 
${H}^\text{F}_t = 2\pi\hbar \nu\left( t \right)
 \left( \sigma_x\cos \phi (t)  + \sigma_y\sin \phi (t)   \right)$,
with $\phi (t)={\pi}t/({2\tau})$, ${\sigma}_{x,y,z}$ the  Pauli spin operators, and $\nu(t)=\nu_0 \left(1-t/{\tau}\right) + \nu_{\tau} t/{\tau}$ the (linear) modulation of the rf-field frequency over time $\tau$, from value $\nu_0 = 1.0\text{ kHz}$ to $\nu_{\tau} = 1.8\text{ kHz}$. Fig.~\ref{fig:mag} reports some of the trajectories followed by the system in both the forward and backward process.

The degree of irreversibility arising from such dynamics was quantified by measuring the probability distribution $P(\Sigma)$ of the irreversible entropy production  using the Tasaki-Crooks relation in Eq.~\eqref{eq:Crooks}. This was assessed using  NMR spectroscopy~\cite{Ivan} and the method described in Refs.~\cite{Batalhao2014,Dorner2013,Mazzola2013}. 
From this the forward and backward work distributions $P_\text{F,B}(W)$ can be determined and, from them, $\beta$, $W$ and $\Delta F$, and hence the  entropy produced during each process can be extracted. The measured nonequilibrium entropy distribution is shown in Fig.~\ref{fig:entropydist}. 
Both positive and negative values occur owing to the stochastic nature of the problem. However, the mean entropy production is positive (red arrow) in full agreement with the 2nd law, $\langle \Sigma \rangle \geq 0$.

\subsection{Assessment of the effects of quantum measurements}

\begin{figure}[t]
    \centering
    \includegraphics[width=\columnwidth]{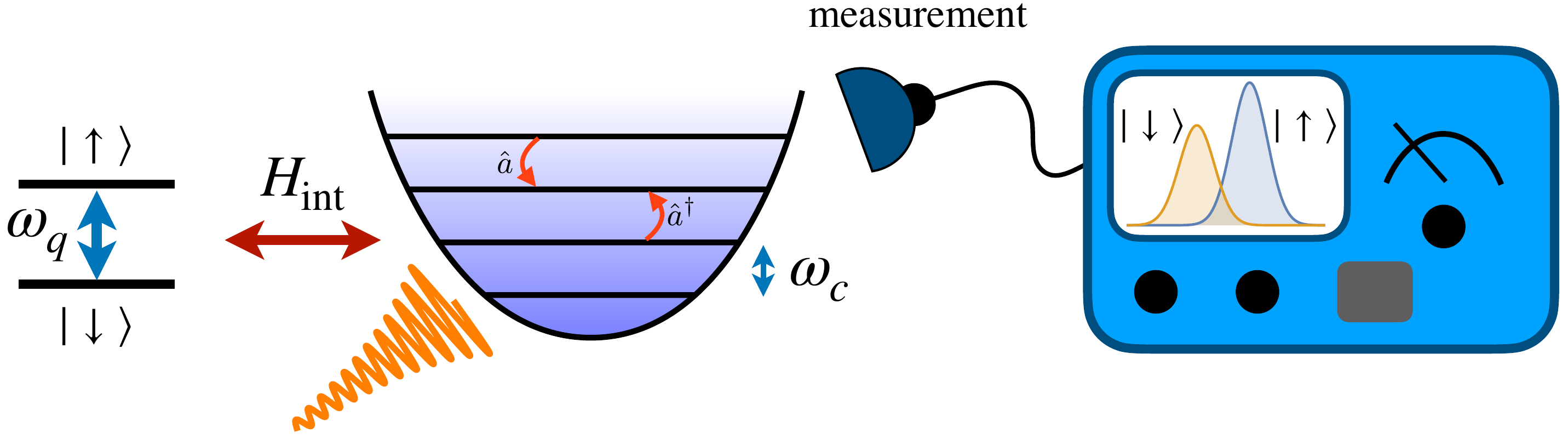}
    \caption{Scheme of principle of the experiment reported in Ref.~\cite{Harrington2019}. A two-level system, whose states $\{\ket{\uparrow}, \ket{\downarrow}\}$ differ in frequency by $\omega_q$, is coupled through the dispersive Hamiltonian $H_{\rm int}$ to a harmonic oscillator of frequency $\omega_c\neq\omega_q$. The interaction term correlates the states of the two-level system to a quadrature of the oscillator. The state of the two-level system is measured by probing the oscillator resonance with a (microwave) pulse, which acquires a phase shift that depends on whether the two-level system is prepared in $\ket{\uparrow}$ or $\ket{\downarrow}$.}
    \label{murch}
    \end{figure}

The experimental tracking of individual trajectories, probed by continuous measurements, allowed Ref.~\cite{Harrington2019} to assess irreversibility in a system affected by continuous weak measurements. A scheme of principle of the employed setting  is shown in Fig.~\ref{murch}. The experimental platform involved a superconducting transmon two-level system coupled to a microwave cavity through the dispersive-coupling interaction term $H_{\rm int}=-\chi a^\dag a\sigma_z$. Here $a$ and $a^\dag$ are the annihilation and creation operators of the cavity field, while $\sigma_z$ is the $z$-Pauli pseudo-spin operator for the two-level system. The rate $\chi$ determines a pseudo spin-dependent phase-shift $2|\chi|$, acquired by a microwave tone used to probe the field's resonance, which in turn is used to acquire information on the two-level system. The signal collected from the cavity field is probed in $n$ time steps $t_k~(k=0,1\ldots, n-1)$, providing the set of records $\{r_k\}$ that allow for the piece-wise reconstruction of individual trajectories of the two-level system. Formally, the statistics of the measurement records, and the corresponding measurement dynamics, can be described through the action of a generalized measurement, defined by operators $M_{r_k}$ that update the density matrix of the two-level system at step $k$, from $\rho_k$ to 
$\rho_{k+1} = M_{r_k} \rho_k M_{r_k}^\dagger/\tr[M_{r_k}\rho_k M^\dag_{r_k}]$.
A time-reversed measurement process is realized by reversing the dynamics for a single measurement update step. Formally, this is described by the time-reversed measurement operators $\tilde{M}_{r_k}$, defined such that 
$\tilde{M}_{r_k}{M}_{r_k}\rho_k{M}^\dag_{r_k}\tilde{M}^\dag_{r_k}=d(r_k)\rho_k$,
where $d(r_k)$ is a depletion coefficient, dependent on the value of the measurement record $r_k$ and entailing the effect of the irreversibility, which unbalances the forward and time-reversed trajectories. The corresponding statistical arrow of time has a length given by 
\begin{equation}
    {\cal A}=\sum_k\ln\frac{P(r_k|\rho_k)}{P(-r_k|\rho_{k+1})}
\end{equation}
with $P(r_k|\rho_k)dr_k=\tr[M_{r_k}\rho_k M^\dag_{r_k}]dr_k$ the probability density of measurement $r_k$, achieved when the system is prepared in $\rho_k$. The stochastic variable ${\cal A}$ is distributed according to a probability distribution ${\cal P}({\cal A})$ that satisfies the fluctuation theorem~\cite{Harrington2019} ${\cal P}({\cal A})=e^{\cal A}{\cal P}(-{\cal A})$.
A related result, aiming at quantifying the information-theoretic contribution to entropy production, resulting from the continuous measurement process~\cite{Belenchia2019} and implemented over a mesoscopic optomechanical system, has been reported in Ref.~\cite{Rossi2020}.

\subsection{Assessment of the non-equilibrium Landauer principle}
\label{experimentalLandauer}

\begin{figure}[t]
    \centering
    \includegraphics[width=\columnwidth]{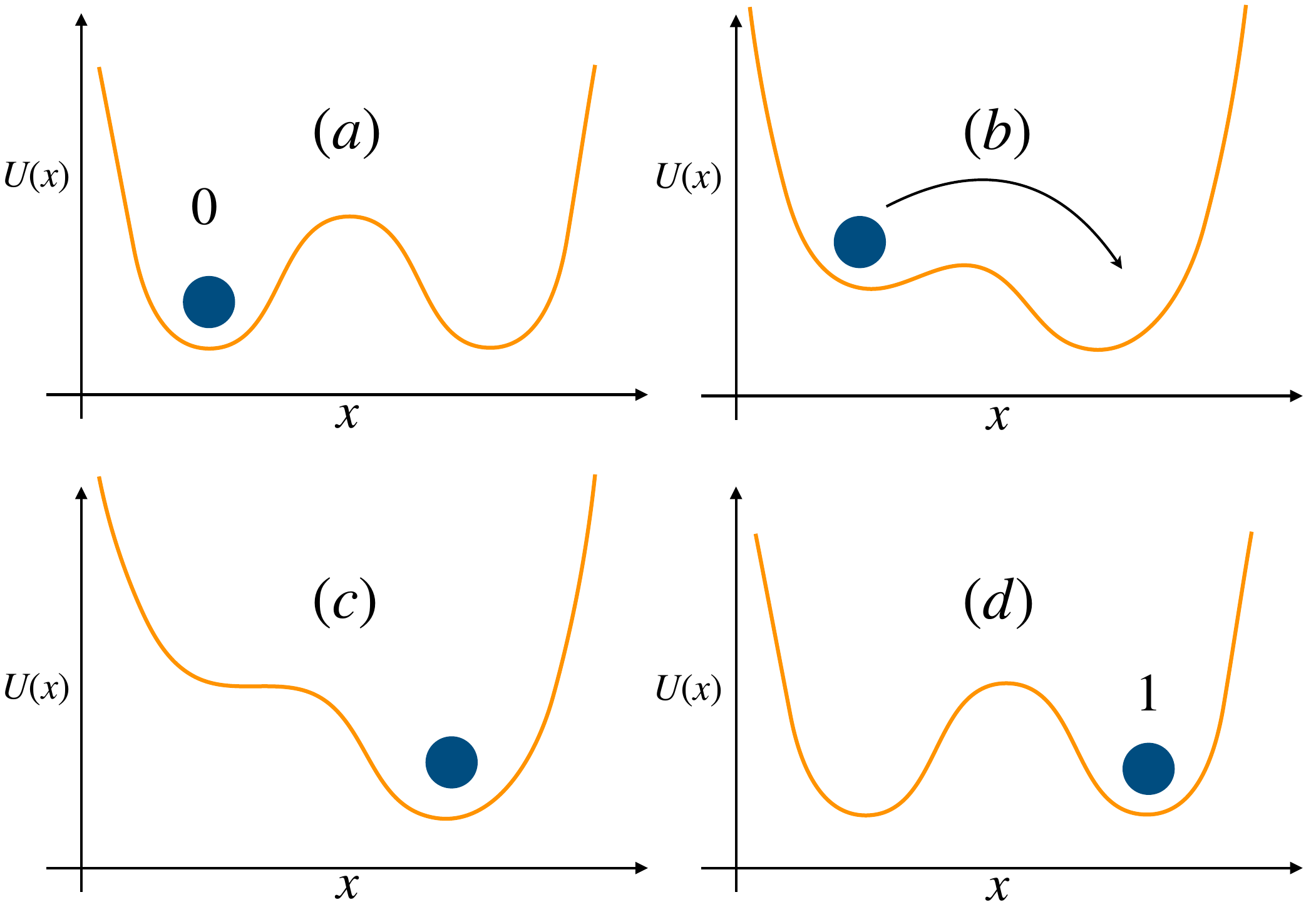}
    \caption{Scheme of the one-bit erasure process reported in Ref.~\cite{Berut2012}.  An overamped colloidal particle (a silica bead of 2$\mu$m in diameter) is trapped at the focus of a laser beam by an optical tweezer. The laser is focused at two distinct, but closely spaced places, alternately and at a very high switching rate. This provides the effective double-well potential into which the bead moves. 
    Initially, due to thermal fluctuations, the bead is equally likely to be in either of the two wells. The erasure process always takes the particle to the rightmost well, which corresponds to the logical state 1 of a classical bit. The initial entropy of the system is thus $S_i=\ln2$. The figure shows the erasure process where the particle is moved from the left to the right well. The barrier is initially high [panel {\bf (a)}] and then lowered and tilted to push the particle to the right well, thus switching the bit to the logical 1 state, which erases the memory [panels {\bf (b)} and {\bf (c)}]. {\bf (d)} By raising the barrier again, the erasure process is completed. The particle is now in the right well with certainty, so the initial side it was in, originally, has been irreversibly erased.
    As the process occurs in a finite time, it is stochastic in nature and the heat dissipated along a given trajectory $x(t)$ (with $x(t)$ the instantaneous position of the particle in the potential) is given by $Q=-\int_{0}^{\tau_{\rm cycle}} {d} t \dot{x}(t) \partial U(x, t) / \partial x$, where $U(x,t)$ is the analytical form of the trapping potential and $\tau_{\rm cycle}$ is the time taken to close an erasure cycle. The average dissipated heat is obtained by averaging $Q$ over 600 cycles, each started by randomly choosing the initial configuration. 
    }
    \label{fig:berut}
\end{figure}

Recent experiments have addressed erasure-like processes involving individual classical or quantum systems~\cite{Berut2012,Jun2014,Orlov2012,Peterson2018,Yan2018}. These experiments have contributed substantially to the resurgence of interest on the implications of Landauer's principle and its extension to general quantum contexts [cf. 
Sec.~\ref{sec:landauer}, particularly Eq.~\eqref{info_landauer_0}].

\begin{figure}[t]
    \centering
    \includegraphics[width=0.8\columnwidth]{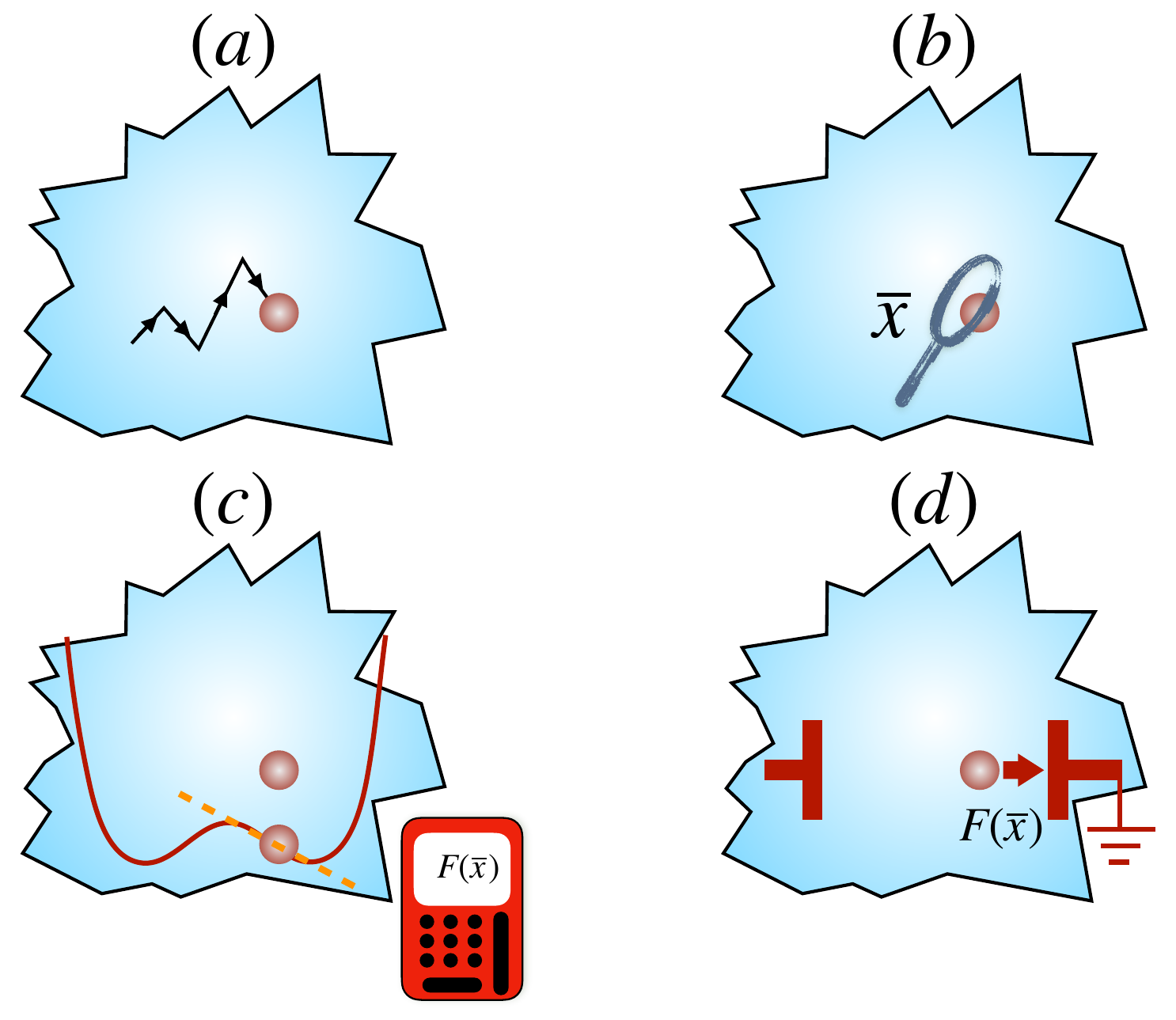}
    \caption{Scheme of the experimental assessment of Landauer principle reported in Ref.~\cite{Jun2014}. A 200nm fluorescent particle moves in an aqueous solution (blue-shaded area) {\bf (a)}, while being monitored by a camera {\bf (b)}. A computer reconstructs the position $\overline{x}$ from the images {\bf (c)} and generates a feedback (electric) force $F(\overline{x})$, applied via two electrodes {\bf (d)}. The force is chosen so as to create a {\it virtual} potential $V(\overline x)$, rather than an actual one, as in Fig.~\ref{fig:berut}, imposed by a computer algorithm and calculated at the estimated position $\overline x$ rather than $x$ itself. This is not a limiting feature of this implementation since, for feedback updates that are fast enough, the dynamics in such virtual potentials is known to converge asymptotically to the corresponding actual one~\cite{Jun2012}. In the experiment reported in Ref.~\cite{Jun2014}, the virtual potential and the erasure process were both along the lines of Fig.~\ref{fig:berut} [cf. Ref.~\cite{Dillenschneider2009}]. 
    }
    \label{fig:jun}
\end{figure}

At the classical level, the space of physical configurations physically accessible to a colloidal particle has been restricted to only two, thus implementing a {\it de facto} one-bit system, via the use of a modulated double-well potential~\cite{Berut2012} or a clever feedback-based trapping mechanism~\cite{Jun2014}. This was used to show that the mean dissipated heat resulting from a (stochastic) erasure process saturates at the (standard) Landauer bound. Details of these experiments are provided in the captions of Figs.~\ref{fig:berut} and~\ref{fig:jun}.

\begin{figure}[t]
    \centering
    \includegraphics[width=\columnwidth]{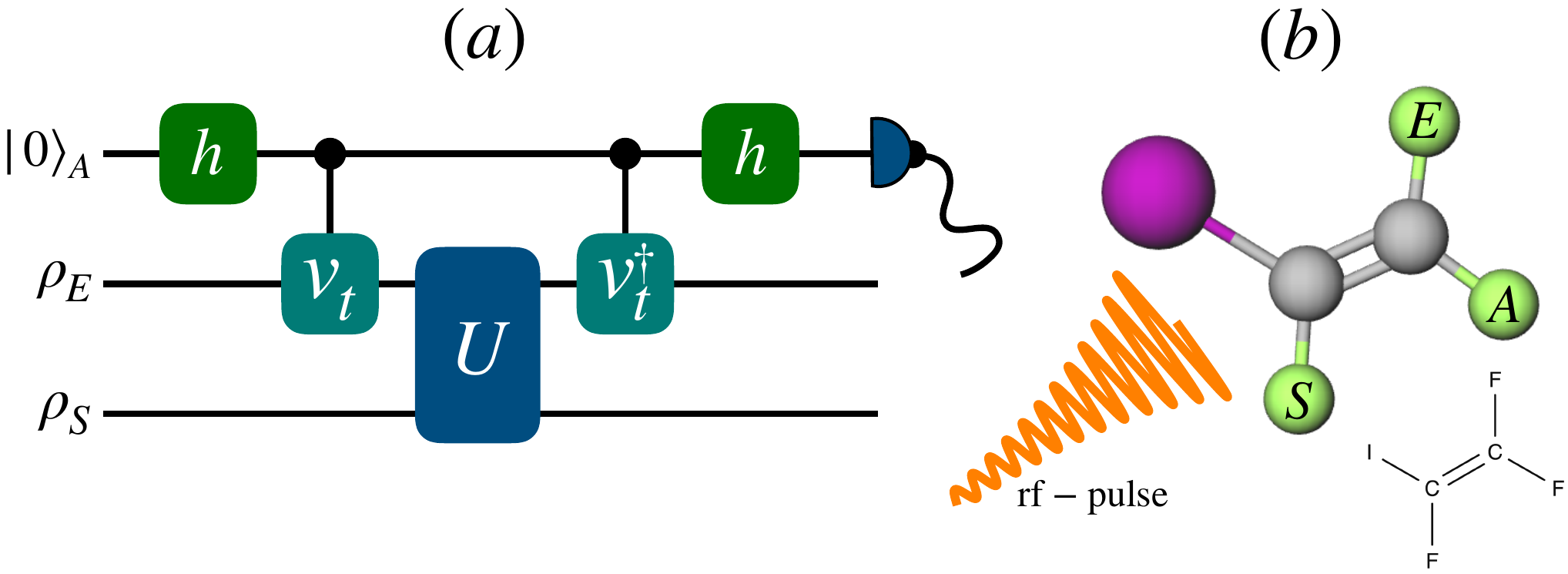}
    \caption{{\bf (a)} Quantum circuit for the reconstruction of the heat probability distribution in Eq.~\eqref{Petterson_PQ} [see also Eq.~\eqref{PofQ}]. Here, $h=(\sigma_z+\sigma_x)/\sqrt2$ is the Hadamard gate on the ancilla $A$, while $v_t=\exp[iH_E t]$ is the free evolution of the environment $E$. Finally, $U$ embodies the $S+E$ unitary governing the heat-dissipation process. {\bf (b)} 3D chemical structure of the Trifluoroiodoethylene molecule ($C_2F_3I$), accommodating the nuclear spins encoding the ancilla, system and environment.
    Suitably arranged rf pulses are employed to prepare and manipulate the state of such tripartite system. 
    }
    \label{fig:peterson}
\end{figure}

The non-equilibrium quantum scenario was addressed in~\cite{Peterson2016} and \cite{Yan2018}.
 Peterson {\it et al.}~\cite{Peterson2018} studied a Nuclear Magnetic Resonance (NMR) system comprising Trifluoroiodoethylene molecules in acetone, whose $^{19}F$ nuclear spins are used to encode three two-level systems. Two of them represent the system and environment of a non-equilibrium erasure process, while the third is used as an ancilla, that was instrumental in the reconstruction of the statistics of the dissipated ( Fig.~\ref{fig:peterson}). 
 
 The system is prepared in the maximally mixed state $\rho_S=\openone_S/2$ through a suitable set of radio-frequency (rf) pulses, thus carrying one bit of information and embodying a proper memory that we wish to reset. The environment is instead initialized in a thermal state $\rho_E=\exp[-\beta H_E]/Z_E$ (with $Z_E=\Tr[e^{-\beta H_E}]$), at an the inverse temperature $\beta$ that could be experimentally controlled. Here $H_E$ is the environment Hamiltonian. Finally, the ancilla is prepared in the logical state  $\ket{0}_A$. Following the approach put forward in Ref.~\cite{Goold2014}, which adapts to the statistics of heat $P(Q)$,  a methods first devised for the reconstruction of the work probability distribution~\cite{Dorner2013,Mazzola2013}, it is possible to show that $P(Q)=\int\Theta(t) e^{-i Q t}dt$ with 
 \begin{equation}\label{Petterson_PQ}
     \Theta(t)=
{\Tr}\left[{U} {\rho}_{{E}} {v}_{t}^{\dagger} \otimes {\rho}_{{S}} {U}^{\dagger} {v}_{t}\right]=
\left\langle{\sigma}_{x}(t)\right\rangle_{{A}}-i\langle{\sigma}_{y}(t)\rangle_{{A}}.
 \end{equation}
 This offers an operational method to infer $P(Q)$ via measurements performed on the ancilla. The latter are operated by amplifying, digitalising, and filtering the free induction decay signal collected from the NMR sample through a pickup coil~\cite{Peterson2018}. Needless to say, the features of $P(Q)$ depend on the joint dynamics encompassed by $U$, while the validity of Landauer principle clearly does not. Peterson {\it et al.} have chosen both a controlled-NOT and a SWAP gate as significant instances of $U$, the latter providing a realization of the paradigmatic erasure process where the state of the system is changed into the initial state of $E$ at every application of the protocol. By tomographically reconstructing the change of entropy in the state of the system following the erasure, Ref.~\cite{Peterson2018} thus demonstrated the validity of the Landauer bound in a genuinely quantum mechanical non-equilibrium scenario.  
 
Despite addressing quantum dynamics, such experiment was unable to quantitatively address the information theoretical contributions to the dissipated heat, arising from the non-equilibrium quantum evolution and highlighted in Eq.~\eqref{info_landauer_0}. The reason is simply because such contributions are negligible in the NMR sample used in~\cite{Peterson2018}. Such assessment was instead made possible by the exquisite control of the trapped-ion experiment reported in~\cite{Yan2018}.
 
\begin{figure}[b]
    \centering
    \includegraphics[width=\columnwidth]{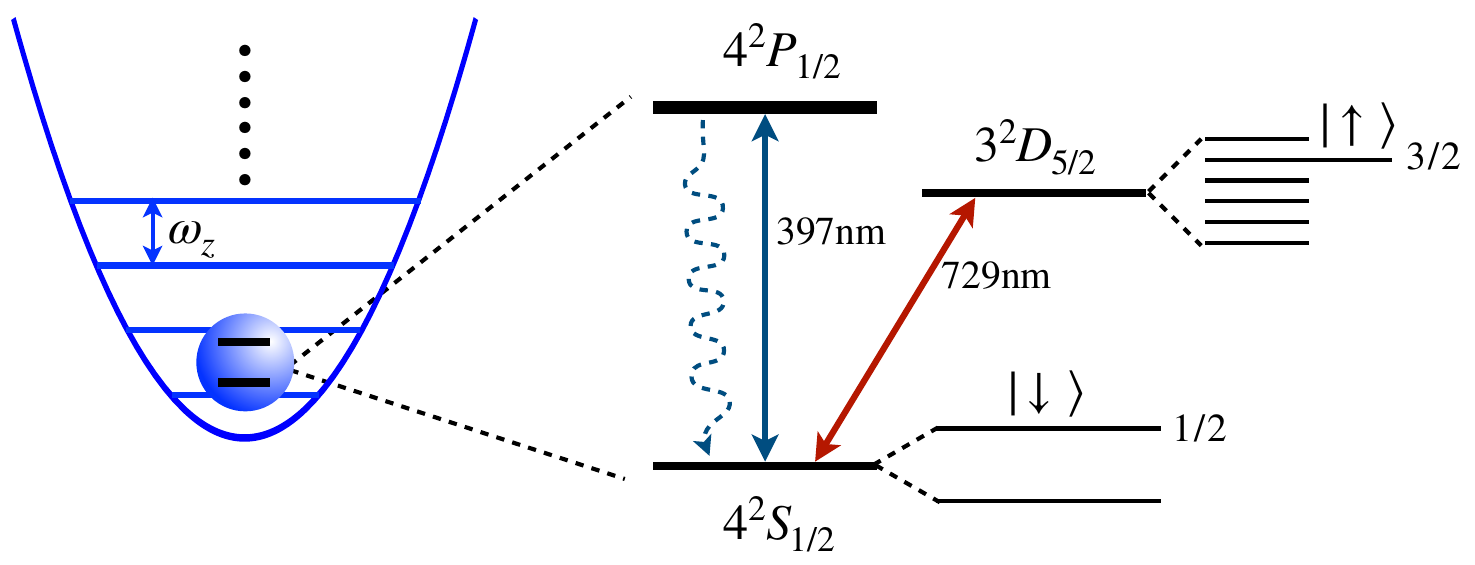}
    \caption{Schematic diagram of the system used for the verification of the information-theoretic contributions to quantum erasure, reported in Ref.~\cite{Yan2018}. A $^{40}{\rm Ca}^+$ ion is confined in a linear Paul trap that provides an axial $z$ pseudopotential of frequency $\omega_z$. The ion is subjected to a magentic field that Zeeman-splits the $4{}^2S_{1/2}$ and $3{}^2D_{5/2}$ atomic states into manifolds of hyperfine levels. Among them, the $\ket{4{}^2S_{1/2},1/2}$ and  $\ket{3{}^2D_{5/2},3/2}$ ones are chosen to encode the logical $\ket{\downarrow}_S$ and $\ket{\uparrow}_S$ pseudospin states of a two-level system embodying $S$ in the erasure protocol. The vibrational $z$ mode of the ion is used to encode the environment $E$, which is thus an infinite-dimensional system.
    }
    \label{fig:yan}
\end{figure}

In such implementation, the system is encoded in two hyperfine internal energy levels of a ${}^{40}{\rm Ca}^+$ ion confined in a linear Paul trap, while the environment is provided by one of the vibrational modes of the particles in the trapping parabolic pseudopotential (say that along the $z$ direction of the reference frame associated with the axes of the trap). Fig.~\ref{fig:yan} shows a diagram of the physical configuration and the relevant part of the energy spectrum of the ion. 
As in~\cite{Peterson2018}, the experiment starts with the system being prepared in a classical mixture of its logical state $\rho_S=\alpha\ket{\downarrow}\bra{\downarrow}_S+(1-\alpha)\ket{\uparrow}\bra{\uparrow}_S$ (with $\alpha\in[0,1]$ being experimentally adjustable), achieved by combining a rotation in the space of states of $S$ and spin dephasing (with no population loss). The vibrational $z$ mode is instead left to relax to a thermal state with an average phonon number $n_0$, by switching off the cooling lasers for an adjustable time. The joint $S$-$E$ evolution that provides the core part of the erasure protocol is given by the arrangement of a red-sideband coupling, induced by a laser field driving the $729$nm ${4{}^2S_{1/2},1/2}\leftrightarrow{3{}^2D_{5/2},3/2}$ transition and ruled by the Hamiltonian~\cite{Leibfried2003} 
\begin{equation}
\label{redside}
H_{SE}=\eta \hbar \Omega\left(a \sigma_{+} e^{i \phi}+a^{\dagger} \sigma_{-} e^{-i \phi}\right) / 2.
\end{equation}
Here $\Omega$ is the Rabi frequency of the coupling, $\phi$ is the phase of the driving field, $\eta\simeq0.09$ is the Lamb-Dicke parameter~\cite{Leibfried2003}, $a$ ($a^\dag$) is the annihilation (creation) operator of the $z$ vibrational mode and $\sigma_\pm$ are the two-level ladder operators. Eq.~\eqref{redside} associates the creation of a phonon to the $\ket{\uparrow}_S\to\ket{\downarrow}_S$ transition. The erasure protocol $U=e^{-i H_{SE}t}$ thus consists of the transformation $\rho_S\to\ket{\downarrow}_S$, accompanied by an increase in the energy of the environment $E$, which is interpreted as a process of heat dissipation from $S$. The setup allows for the experimental inference of the phonon number change, which gives direct access to the amount of dissipated heat and the $S(\rho'_E||\rho_E)$ term in Eq.~\eqref{info_landauer_0}. Similarly, the change of entropy in the state of the system can be directly assessed by straightforward measurements of the population of the pseudospin states. The mutual information $I_{\rho'_{SE}}(S':E')$, on the other hand, is not directly accessible, but can nonetheless be estimated, as discussed in Ref.~\cite{Yan2018}. While such estimation affects the uncertainty associated with the evaluation of the right-hand side of Eq.~\eqref{info_landauer_0}, resulting in relatively large error bars, the experiment was successful in demonstrating the compatibility between the amount of entropy produced in the erasure process and the joint contribution coming from the information theoretic terms.

\;
\subsection{\label{sec:exp_brunelli}Assessment of entropy production in non-equilibrium steady-states}
\label{QCS}

\begin{figure*}
\centering
\includegraphics[width=\textwidth]{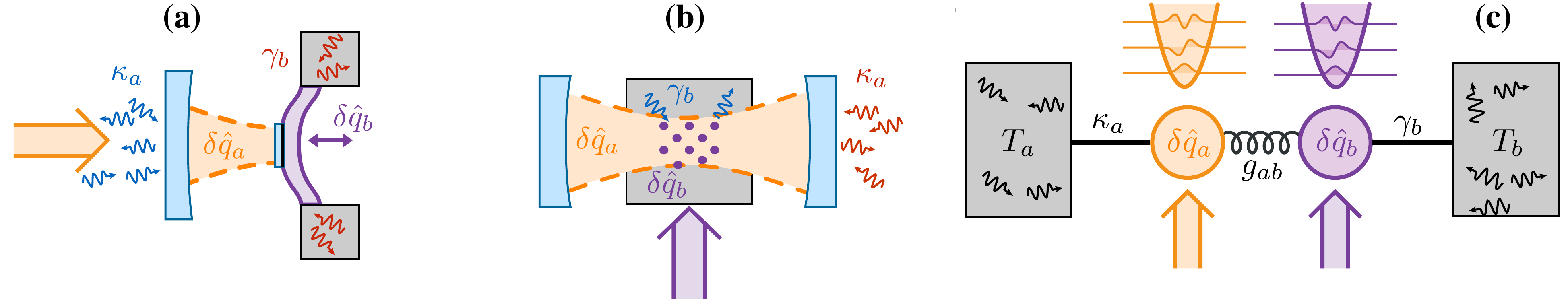} 
\caption{(a) Optomechanical setup: a micro-mechanical oscillator ($\delta\hat{q}_b$) is coupled to the field mode of an optical Fabry-Perot cavity ($\delta\hat{q}_a$). For this setup only the cavity is pumped. (b) 
Cavity-BEC setup: the external degree of freedom of a BEC ($\delta\hat{q}_b$) is coupled to the field mode of a cavity ($\delta\hat{q}_a$). For this setup only 
the atoms are pumped. Red and blue wiggly lines indicate heating or cooling of the subsystems via coupling to the baths. In both setups {the number of excitations in the optical bath is zero, i.e. $n_{T_a}=0$ }. {(c)} Both systems can be modelled as two quantum harmonic oscillators at frequencies $\omega_a$ and $\omega_b$, linearly coupled with 
a strength $g_{ab}$. Each oscillator is coupled to independent local baths at temperature $T_a$ and $T_b$, respectively. The corresponding coupling rates are 
$\kappa_a$ and $\gamma_b$. The oscillators can be pumped by an external field (purple and orange arrows in the figure).}
\label{setup}
\end{figure*}

Recent efforts have been deployed to the assessment of entropy production in non-equilibrium steady-states of mesoscopic quantum systems~\cite{Brunelli2018}. In particular, settings based on cavity optomechanics and ultra-cold atom systems have been used as paradigm of situations leading to non-trivial non-equilibrium steady states.

In cavity optomechanics, the position of a mechanical oscillator accommodated in an externally driven cavity is displaced by an amount directly proportional to the number of photons in the field of the cavity itself [cf. Fig.~\ref{setup}(a)]. This brings the state of the mechanical system  to an out-of-equilibrium steady state, resulting from the competition between the coupling of the cavity field to the zero-temperature electromagnetic environment, and the equilibrium phononic reservoir that affects the mechanical system~\cite{Aspelmeyer}.

The second experimental platform that has been studied in this context comprises  a Bose-Einstein condensate (BEC) loaded into a high-finesse optical cavity and illuminated by a transverse laser field [cf. Fig~\ref{setup}(b)]. 
The off-resonant photon 
scattering from the laser field into an initially empty cavity field mode couples the zero-momentum mode of the BEC to an excited momentum 
mode. The process mediates effective and tunable-in-strength (via the transverse laser beam) long-range atom-atom interactions~\cite{Mottl}. 
Such interaction can be brought to competition with the kinetic energy of the atoms, resulting 
in a structural phase transition~\cite{ZurichNew} akin to a 
Dicke phase transition~\cite{Baumann2010}. The cavity light field leaking through the mirrors with a heterodyne detection setup. 
The spectral analysis of this signal is used to infer the diverging amount of atomic density fluctuations accompanying the structural phase transition~\cite{ZurichNew}.

In both cases, the effective interaction between the fluctuations of the filed operators of the matter-like subsystems and their optical counterpart 
can be shown to be that of two harmonic oscillators coupled via the Hamiltonian [cf. Fig.~\ref{setup} {\bf (c)}] 
\begin{equation}\label{HamiltonianPaper}
\hat{H}=\frac{\hbar \omega_a}{2} (\delta\hat{q}_a^{2}+\delta\hat{p}_a^{2})+\frac{\hbar \omega_b}{2} (\delta\hat{q}_b^{2}+\delta\hat{p}_b^{2})+\hbar g_{ab}\delta\hat{q}_a
\delta\hat{q}_b.
\end{equation}
Here, $\delta \hat{q}_{a,b}$ and $\delta \hat{p}_{a,b}$ are the position and momentum fluctuation operators around the mean-field values of the two oscillators ($a$ and $b$ refer to the optical and mechanical/atomic oscillators, respectively), $\omega_p$ is the frequency of the driving pump fields, the oscillators have frequencies $\omega_a=\omega_c-\omega_p$ and $\omega_b$, $\omega_c$ is the frequency of 
the cavity field, and $g_{ab}$ is the coupling strength between the modes~\cite{Brunelli2018}. 
The  cavity mode is coupled to the surrounding electromagnetic vacuum with a decay rate $\kappa_a$. On the other hand, the nature of the mechanical/atomic bath is specific to the setup being considered. The optomechanical system considered in Ref.~\cite{Brunelli2018} consisted of a Fabry-Perot cavity with one of its mirrors being a doubly clamped, highly reflective, mechanical cantilever. The mechanical support of the cantilever thus provided a local heat bath at room temperature responsible for the quantum Brownian motion of the mechanical system. In the cavity-BEC system, dissipation is due to the collection of excited 
Bogolioubov modes, which provides a bath for the condensate. In both cases, we assume  oscillator $b$ to be in contact with a  bath at temperature $T_b$ and 
rate $\gamma_b$. The average number of excitations in the equilibrium state of oscillator $b$ is thus $n_{T_b}=(e^{\hbar \omega_b/ k_B T_b}-1)^{-1}$.

The linear dynamics undergone by the coupled oscillators allows for the use of the framework for the quantification of entropy production in phase space illustrated in Sec.~\ref{sec:QPhaseSpace}. The entropy production rate in the non-equilibrium steady-state of such respective systems 
thus takes the form
\begin{equation}\label{Pi_ss}
\Pi \equiv \dot{\Sigma} =2\gamma_b\left( \frac{ n_b+1/2}{n_{T_b}+1/2} - 1 \right) + 4\kappa_a n_a=\mu_b+\mu_a,
\end{equation}
where $n_a=\langle(\delta\hat{q}_a^{2}+\delta\hat{p}_a^{2}-1)\rangle_s/2$ and $n_b=\langle(\delta\hat{q}_b^{2}+\delta\hat{p}_b^{2}-1)\rangle_s/2$ are the average number of 
excitations in the non-equilibrium steady-state of the two oscillators in excess of the zero-point motion of the respective harmonic oscillator. 
In the cavity-OM expression for $\mu_b$, instead of the full phonon number $n_b$, only the momentum variance $\langle\delta\hat p_b^2\rangle_s$ enters as we assume Brownian motion damping. 

Eq.~\eqref{Pi_ss} quantifies the entropic contribution of quantum fluctuations that the system has to pay to remain in its 
non-equilibrium steady-state. 
It is directly determined by the individual entropy flows $\mu_j~(j=A,B)$ from the mechanical/atomic and optical oscillator to their respective environment. 
\par
\begin{figure}[t!]
\centering
\includegraphics[width=\columnwidth]{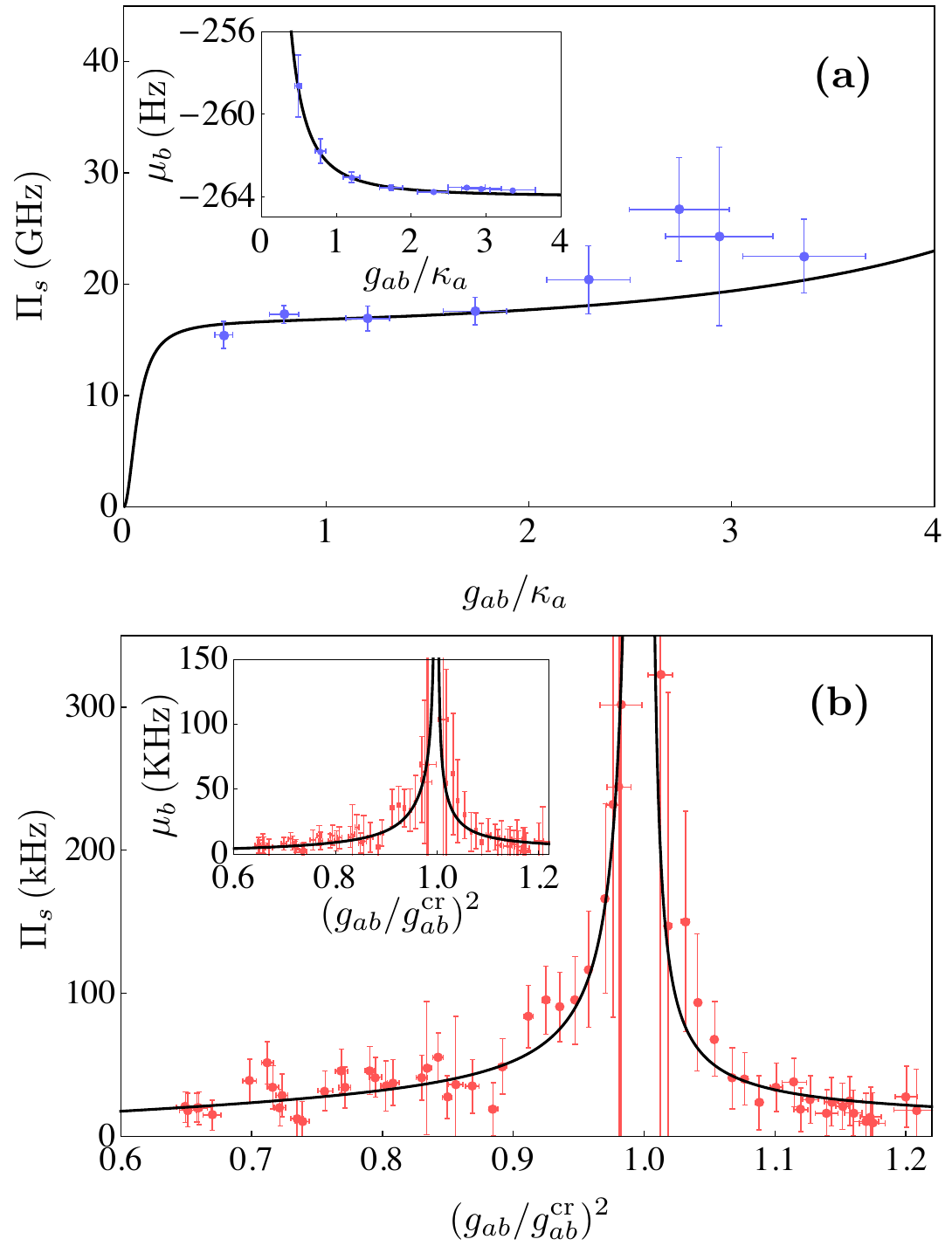}
\caption{Experimental assessment of the irreversible entropy production rate $\Pi_s$ at the non-equilibrium steady-state for {(a)} the cavity optomechanical system and {(b)} the cavity-BEC system reported in Ref.~\cite{Brunelli2018}. For the cavity-BEC 
setup, $g_{ab}^{\text{cr}}=\sqrt{(\kappa_a^2+\omega_a^2)\omega_b/4\omega_a}$ is the critical parameter of the model. The insets 
show the behaviour of $\mu_b$. In both panels, the solid black lines show the theoretical predictions {\color{black}based on parameters extracted from experiment~\cite{Brunelli2018}}. The blue and red dots show the 
experimental data for the optomechanical and cavity-BEC experiment, respectively. In panel {(a)}, the vertical error bars report statistical errors extracted from the fit, while the
horizontal ones show experimental error on the values of the parameter. In panel {(b)}, the vertical and horizontal error bars report the statistical errors from the fit and the 
determination of the critical point, respectively~\cite{ZurichNew}. }
\label{EntroSS}
\end{figure}

In Ref.~\cite{Brunelli2018} the two terms $\mu_a$ and $\mu_b$ have been separately reconstructed [cf. Fig.~\ref{EntroSS}, which displays the experimental data together with the theoretical model]. The  behaviour of $\mu_{b}$ observed for the optomechanical system is a signature of cooling: the entropy flow from the mechanical resonator 
to the cavity field grows with $g_{ab}$ as the effective temperature of the resonator decreases. As for the cavity-BEC system, the divergent behaviour of the entropy production rate 
at $g_{ab}=g_{ab}^{\text{cr}}\equiv\sqrt{(\kappa_a^2+\omega_a^2)\omega_b/4\omega_a}$ reflects the occurrence of the Dicke phase transition: at $g_{ab}^{\text{cr}}$, the populations of the two oscillators at the steady-state diverges, resulting in the singularity of $\mu_a$ and $\mu_b$. 



\section{\label{sec:conc}Conclusions}

The 2nd law has always been intimately linked with information theory. 
The underlying laws of physics are time-reversal invariant. Thus, how can the ensuing macroscopic dynamics be irreversible? 
This is perhaps one of the deepest questions in  physics, and a major source of confusion. 
The answer is that irreversibility is an \emph{emergent property}: 
It emerges from the fact that information easily becomes irretrievable, when the number of degrees of freedom involved is large.
A classical thermodynamic argument goes as follows: suppose one has a gas of $10^{23}$ particles, but can only monitor the position and momenta of $10^{23} - 1$ of them. 
Since the motion of the gas is highly chaotic, even this minuscule loss of information can lead to dramatic effects on the description of the remaining $10^{23}-1$ particles, causing their dynamics to be irreversible. 

This  argument, however,  conceals a much more dramatic effect, which only becomes clear in the full quantum treatment. 
In order to properly account for multiple degrees of freedom, it is not enough to monitor them individually; one must monitor them \emph{globally}. 
Consider a system with $N$ particles and density matrix $\rho_{1\ldots N}$ (mixed or pure). 
Local measurements on each subsystem only explore
the local corners of $\rho_{1\ldots N}$, and are not enough enough to reconstruct the full state. 
To do that, one would also have to perform global measurements (e.g. Bell-like). 
Such measurements are difficult, even for two qubits. 
For already a handful of degrees of freedom, it easily becomes surreal. 
In this quantum picture, therefore, information spreads not only from one degree of freedom to another, but also from the local to the global corners of a many-body density matrix. 
The basic definition~\eqref{Qgen_Sigma} naturally encompass both aspects:
The mutual information accounts for the spreading of information to the different corners of $\rho_{SE}$, while the relative entropy accounts for the local transfer of information, from the degrees of freedom of the system to those of the bath. 

Compared to the approach of Eq.~\eqref{Qgen_Sigma}, the historical formulations of Clausius, Carnot and Kelvin (Sec.~\ref{sec:Why}) were much more pragmatic, stating the 2nd law solely in terms of heat and work, which are palpable quantities.
But although pragmatic, their scope is much less evident at first sight. For instance, demonstrating that the different principles are equivalent requires complicated constructs, involving thermal machines operating in different ways~\cite{Fermi1956}.

A more general statement of the $2^{\rm nd}$ law thus comes at the expense of introducing the notion of entropy. 
At the thermodynamic level, entropy is defined as an abstract function of state, with the property that changes in entropy for reversible processes, close to equilibrium, satisfy $\Delta S =  Q/T$, where $Q$ is the heat exchanged. 
The $2^{\rm nd}$ law can then be formulated as ``the entropy of the universe never decreases''. 
For instance, if the universe is comprised of a system and bath only, which interact and exchange an amount of heat $Q$, then $\Delta S_S +\Delta S_E \geqslant 0$.
If, in addition, the bath is kept close to equilibrium, then $\Delta S_E = Q_E/T$ and the 2nd law becomes $\Delta S_S + Q_E/T \geqslant 0$, which is Eq.~\eqref{intro_sigma_clausius}. 
For this reason, historically the entropy production $\Sigma$ was often stated as representing the change in entropy of the universe. 
{\color{blue}
The same reasoning also appears in other contexts, such as Boltzmann's famous $H$-theorem.
For instance~\cite{Tolman1938} analyzes the scenario of an isolated gas (the universe) where the molecules may undergo collisions with each other, described phenomenologically using Boltzmann's equation. 
The entropy, in this case, is given in terms of the gas' phase space density. 
And, as a consequence of the choice of rules used for describing the collisions, it always increases.
}


A natural question, then, would be to ask whether one can carry over this interpretation of $\Sigma$ to the microscopic realm. 
This could be called a \emph{top-down} approach, where one starts with a macroscopic principle and then adapt it to the microworld. 
And it is opposite of the \emph{bottom-up} approach we have followed in this review, where we started with a fully microscopic definition of $\Sigma$, in terms of information-theoretic quantities, from which the classical principles emerged as particular cases. 

In addition to the progress on the bottom up approach, reported in this review, recent years have also seen significant advances in the top-down formulation of the 2nd law. 
The main challenge is in the definition of a thermodynamic entropy, something which the bottom-up formulation avoids, since it does not interpret entropy production as the change in entropy of the universe. 
Clearly, the von Neumann entropy is not a good candidate for thermodynamic entropy, since it is constant under unitary evolution. In statistical mechanics one often uses the Boltzmann entropy $S= \ln \Omega$, where $\Omega$ is the ``number of microstates associated to a given macrostate''. 
But this quantity is only reasonable  close to equilibrium and only defined for macroscopic systems. 
For micro- and mesoscopic systems, it fluctuates violently~\cite{Pathria2011,Gupta1947} and is also awkward to define explicitly.
Most advances in the top-down approach have therefore focused on alternative definitions of thermodynamic entropy, such as the diagonal entropy~\cite{Polkovnikov2011b} or, more generally, coarse-grained entropies~\cite{Safranek2019}. 

One of the basic features of the thermodynamic entropy is that it is extensive. 
This is what allows one to write the entropy of the universe as the sum of the entropies of its parts. 
Interestingly, in this regard, taking the \emph{local} von Neumann entropy works quite well. 
Consider a system of $N$ particles with generic density matrix $\rho_{1,\ldots,N}$. The sum of the von Neumann entropies of the reduced states $\rho_i$ can be written as 
\begin{equation}\label{conc_vN_universe}
\sum\limits_i S(\rho_i) = S(\rho_{1,\ldots,N}) + S(\rho_{1,\ldots,N} || \rho_1\otimes \ldots \otimes \rho_N),
\end{equation}
where the last term is the total correlations [cf. Eq.~\eqref{Qgen_Sigma_multipartite_correlations}], measuring the distance between the global and the maximally marginalized states. 
Now suppose initially the $N$ particles are in a product state, but are then put to interact according to a global unitary $U$, leading to a final, correlated state. 
The first term on the right-hand side of~\eqref{conc_vN_universe} does not change, since the dynamics is unitary. 
The second term was initially zero, but then evolves into something non-negative. 
Hence, one concludes that for any initially uncorrelated system under closed evolution 
$\sum\limits_i \Delta S(\rho_i) \geqslant 0$.
Thus, if one takes as thermodynamic entropy the local von Neumann entropy of each subsystem, we then recover the classical statement that the entropy of the universe cannot decrease. 
Most studies attempting to define a microscopic analog of the thermodynamic entropy follow somewhat similar lines. 

The above discussion meant to emphasize some of the basic principles involved in a general formulation of the 2nd law. 
Often, however, one does not have access to such ``luxuries''; that is, one does not have access to the full global dynamics, but only to an effective description, in terms of e.g. a master equation.
As a consequence,  Eq.~\eqref{Qgen_Sigma} or the top-down approaches may not be applicable. 
In situations such as this, several principles have been applied in the past to define entropy production. 

The most widely used, by far, is to \emph{postulate} that the entropy flux should be $\Phi = Q_E/T$, from which one then recovers $\Sigma = \Delta S_S + \Phi$ [Eq.~(\ref{intro_sigma_clausius})]. 
This approach is both simple and effective. 
It also has a neat interpretation at the trajectory level~\cite{Breuer2003}. 
But it has two shortcomings. First, it only holds for thermal baths and it is not at all obvious how to extend it to non-equilibrium reservoirs. 
And second, one may run into difficulties concerning what is in fact the heat $Q_E$, as discussed in Sec.~\ref{sec:classical}.

Fluctuation theorems greatly resolve these difficulties. 
In this case, entropy production is defined as the ratio between the path probabilities of a forward and time-reversed (backward) trajectory~\cite{Crooks1998}. 
These definitions are usually regarded as being  fundamental. 
However, they require knowledge of the full path probability, which is not always available, or can be hard to obtain~\cite{Spinney2012}. 
Moreover, as discussed in Sec.~\ref{sec:qgen_FT}, the backward trajectory is not uniquely defined, contrary to what was initially believed.

Finally, there is also the more pragmatic approach of simply manipulating $\Delta S$ and trying to identify a term which resembles an entropy production, such as the Schnackenberg approach discussed in Sec.~\ref{sec:pauli},  which is extremely popular in stochastic thermodynamics. 
This may seem rather \emph{ad hoc}, at first, but can lead to interesting results because, often, the ``correct'' formula really stands out. 
Moreover, it allows one to define entropy production for arbitrary open system dynamics, even those that are not generated by physical processes.

Many open questions still remain. 
However, as we have shown in this review, the last two decades have seen remarkable progress in our understanding of the basic ingredients and principles that should be involved in this endeavour. 
In particular, the community's appreciation of what the $2^{\rm nd}$ law \emph{should} represent, as well as the questions it should address, has evolved significantly. 
In light of the exciting advances on the experimental manipulation of coherent quantum systems, we believe these new foundations will play a significant role in our understanding of many potential future applications, as well as in the explanation of fundamental questions.


\acknowledgments

We acknowledge fruitful discussions and collaborations on the topics of this paper with colleagues in the following, certainly non-exhaustive, list: O. Abah, G. Adesso, M. Barbieri, G. Barontini, A. Bassi, A. Belenchia, F. Bernards, M. Brunelli, B. Cakmak, R. R. Camasca, S. Campbell, M. Campisi, M. Carlesso, L. Celeri, M. A. Ciampini, F. Ciccarello, M. A. Cipolla, S. Clark, N. E. Comar, L. Correa, G. De Chiara, M. Garc\'{i}a D\'{i}az, S. Donadi, T. Donner, A. Ferraro, L. Fusco, J. Garrahan, G. Gasbarri, M. G. Genoni, S. Gherardini, B. O. Goes, J. Goold, G. Guarnieri, S. Huelga, A. Imparato, N. Kiesel, I. Lesanovsky, S. Lorenzo, E. Lutz, L. Mancino,  M. Mitchison, K. Modi, O. A. D. Molitor, \"{O}. M\"{u}stecapl{\i}o\u{g}lu, G. M. Palma, J. Pekola, M. B. Pereira, M. B. Plenio, R. Puebla, Á. Rivas, M. Rossi, A. Sanpera, J. P. Santos, A. Schliesser, F. L. Semi\~ao, R. M. Serra, P. Sgroi, R. R. Soldati, H. Ulbricht, R. Uzdin, B. Vacchini, A. Varizi, V. Vedral, Q. Wu, G. Zicari, K. Zyczkowski. Some of them have provided very useful feedback on the manuscript, for which we are grateful. In particular, we thank  A. Varizi for his extremely careful reading.
This work was supported by H2020, through the Collaborative Project TEQ (Grant Agreement No.  766900), the S\~{a}o Paulo Research Foundation (FAPESP) (grant nr. 2018/12813-0 and 2017/50304-7), the DfE-SFI Investigator Programme (Grant No. 15/IA/2864), the Leverhulme Trust Research Project Grant UltraQute (grant nr.~RGP-2018-266), COST Action CA15220, the Royal Society Wolfson Research Fellowship scheme (RSWF\textbackslash R3\textbackslash183013) and International Mobility Programme, the UK EPSRC (grant nr.~EP/T028106/1), and the SPRINT programme supported by FAPESP and Queen's University Belfast.

\;
\bibliography{RMP}

\end{document}